\documentclass[preprint,11pt]{elsarticle}

\usepackage[T1]{fontenc}
\usepackage{charter}
\usepackage{wrapfig}
\usepackage{arydshln}

\usepackage[colorlinks,citecolor=blue,linktoc=all,linkcolor=cyan]{hyperref}

\usepackage{amssymb,amsbsy,amsmath,amsfonts}
\usepackage{mathrsfs} 
\numberwithin{equation}{section}
\numberwithin{table}{section}
\numberwithin{figure}{section}
\usepackage{nicefrac}
\usepackage{upgreek}
\usepackage{esint}
\usepackage{slashed}
\usepackage{multirow}
\usepackage{caption}

\biboptions{numbers,sort&compress}

\DeclareMathAlphabet{\mathantt}{OML}{antt}{l}{it}
\DeclareMathAlphabet{\mathpzc}{OT1}{pzc}{m}{n}

\def\beq{\begin{equation}}
\def\eeq{\end{equation}}
\def\bea{\begin{eqnarray}}
\def\eea{\end{eqnarray}}
\def\beqa{\begin{equation}\begin{array}{l}}
\def\eeqa{\end{array}\end{equation}}
\def\eqlab#1{\label{eq:#1}}
\def\figlab#1{\label{fig:#1}}

\def\seclab#1{\label{sec:#1}}
\def\eref#1{(\ref{eq:#1})}
\def\Eqref#1{Eq.~(\ref{eq:#1})}
\def\figref#1{Fig.~\ref{fig:#1}}

\def\Figref#1{Fig.~\ref{fig:#1}}

\def\Tabref#1{Table \ref{tab:#1}}
\def\Secref#1{Sect.~\ref{sec:#1}}

\def\boxfrac#1#2{\mbox{$\frac{#1}{#2}$}}
\def\half{\mbox{$\frac{1}{2}$}}
\def\thalf{\mbox{$\frac{3}{2}$}}
\def\quarter{\mbox{$\frac{1}{4}$}}
\def\third{\mbox{$\frac{1}{3}$}}
\def\sixth{\mbox{$\frac{1}{6}$}}
\def\nh{\nicefrac{1}{2}}
\def\nth{\nicefrac{3}{2}}

\def\barr{\left(\begin{array}{c}}
\def\earr{\end{array}\right)}
\def\bmat{\left(\begin{array}{cc}}
\def\emat{\end{array}\right)}
\def\al{\alpha}
\def\be{\beta}
\def\ga{\gamma} \def\Ga{{\it\Gamma}}
\def\de{\delta} \def\De{\Delta}
\def\veps{\varepsilon}  \def\eps{\epsilon}

\def\la{\lambda}

\def\si{\sigma} \def\Si{{\it\Sigma}}
\def\th{\theta}  
\def\w{\omega}  
\def\vfi{\varphi}

\def\pa{\partial}
\def\ra{\rightarrow}

\def\nn{\nonumber}
\def\dd{\mathrm{d}}
\def\cO{\mathcal{O}}

\def\aa{\mathantt{a}}
\def\rk{\mathantt{k}}
\def\xx{\mathscr{x}}

\def\scA{\mathscr{A}}
\def\scO{\mathscr{O}}

\def\cA{{\mathcal A}}
\def\cB{{\mathcal B}}
\def\cC{{\mathcal C}}

\def\zZ{\mathpzc{Z}}
\def\zT{\mathpzc{T}}
\def\zP{\mathpzc{P}}

\def\zp{\mathpzc{p}}

\DeclareMathOperator\im{Im}
\DeclareMathOperator\re{Re}

\def\ol#1{\overline{#1}}

\def\XXint#1#2#3{{\setbox0=\hbox{$#1{#2#3}{\int}$}     \vcenter{\hbox{$#2#3$}}\kern-.5\wd0}}

\DeclareMathOperator\arccosh{arccosh}

\def\br{\boldsymbol{r}}
\def\bn{\boldsymbol{n}}
\def\bp{\boldsymbol{p}}

\def\bq{\boldsymbol{q}}

\def\bQ{\boldsymbol{Q}}

\journal{Progress of Particle and Nuclear Physics}

\begin{document}

\begin{frontmatter}


\title{Nucleon Polarizabilities: 
from Compton Scattering to Hydrogen Atom}


\author[L1]{Franziska Hagelstein}
\author[L2]{Rory Miskimen}
\author[L1]{Vladimir Pascalutsa}
\address[L1]{ Institut f\"ur Kernphysik and PRISMA Excellence Cluster, Johannes
Gutenberg-Universit\"at Mainz, D-55128 Mainz, Germany}
\address[L2]{Department of Physics, University of Massachusetts, Amherst, 01003 MA, USA}

\begin{abstract}
We review the current state of knowledge of the nucleon polarizabilities
and of their role in nucleon Compton scattering and in hydrogen spectrum. 
We discuss the basic concepts, the recent lattice QCD calculations and 
advances in chiral effective-field theory.  
On the experimental side, 
we review the ongoing programs aimed
to measure the nucleon (scalar and spin) 
polarizabilities via the Compton scattering processes, with real
and virtual photons. 
A great part of the review is devoted to the
general constraints based on unitarity, causality,
discrete and continuous symmetries, which result in 
model-independent relations involving 
nucleon polarizabilities. We (re-)derive
a variety of such relations and discuss their empirical
value. 
The proton polarizability effects
are presently the major sources of uncertainty in the assessment
of the muonic hydrogen Lamb shift and hyperfine structure.
Recent calculations of these effects are  reviewed here
in the context of the ``proton-radius puzzle''.
 We conclude with summary plots
of the recent results and prospects for
the near-future work.
\end{abstract}


\end{frontmatter}
\thispagestyle{empty}
\tableofcontents

\section{Introduction}
The concept of {\em polarizabilities}, common in optics and  classical electrodynamics, was extended to the nucleon
in the 1950s \cite{Klein:1955qy,Baldin:1960}, together with the first observations of Compton scattering (CS) on the proton \cite{Oxley:1955zz,Oxley:1958zz,Pugh:1957zz,Hyman:1959zz,Bernardini:1960wya,Goldansky:1960}.
Since then, the CS process, with real (RCS) or virtual (VCS) photons, became the main experimental tool
in studying the nucleon polarizabilities, with dedicated experiments completed at: Lebedev
Institute (Moscow) \cite{Goldansky:1960,Baranov:1974ec}, MUSL (Illinois) \cite{Fed91}, SAL (Saskatoon)~\cite{Hallin:1993ft,MacGibbon95}, LEGS (Brookhaven) \cite{Leg01},  Bates (MIT) \cite{Bourgeois2011}, MaxLab (Lund) \cite{Myers14}, MAMI (Mainz) \cite{Zieger:1992jq,Olm01,Galler01,Wolf01,Cam02,Roche:2000,Janssens:2008qe}, and 
Jefferson Laboratory (Virginia) \cite{Laveissiere:2004}.

In recent years, the nucleon  polarizabilities
have advanced to the avantgarde of hadron physics.
They are a major source of uncertainty
in the muonic-hydrogen determination of the proton charge radius \cite{Pohl:2010zza} and Zemach radius \cite{Antognini:1900ns}, and 
hence are a prominent part of
the ``proton-radius puzzle'' \cite{Bernauer:2014cwa}. They play an important role in the
controversy of the electromagnetic (e.m.) contribution 
to the proton-neutron mass difference \cite{WalkerLoud:2012,Erben:2014hza,Gasser:2015dwa}.  
Several issues involving the nucleon polarizabilities have emerged from the ongoing `spin physics program' at the Jefferson Laboratory (JLab), which is mapping out the spin structure 
functions of the nucleon \cite{Slifer:2007bn,Chen:2008ng,Chen:2011zzp}. The various moments of these structure functions are related to 
the forward {\em spin  polarizabilities}, with one of 
them, $\de_{LT}$, being notoriously difficult to understand
within the chiral effective-field theory ($\chi$EFT)~\cite{Bernard:2012hb,Lensky:2014dda}.  
The currently operating photon beam facility MAMI 
has established a dedicated experimental program to disentangle the nucleon polarizabilities through the low-energy RCS with polarized beams \cite{Dow12,Pascalutsa13} 
and targets \cite{Hornidge12,Martel15}; a complementary
program, at even lower energy, is planned at HIGS (Duke)
\cite{Weller:2009zza,Ahm10}. A new experimental program
is being developed for the upcoming high-intensity
electron beam facility MESA (Mainz).
The recent theory advances include: (partially)
unquenched lattice QCD calculations \cite{Engelhardt:2007aa,Detmold:2009dx,Detmold2010,Primer:2013pva,Hall:2013dva,Lujan:2014kia,Chang:2015qxa}; novel $\chi$EFT calculations of CS~\cite{Lensky:2008re,Len10,Lensky:2012ag,McG13,Lensky:2014efa,Blin:2015era,Lensky:2015awa} 
and of the polarizability
effects in hydrogenic atoms \cite{Alarcon:2013cba,Peset:2014jxa}; 
development and evaluation of 
model-independent relations
involving the nucleon polarizabilities \cite{Sibirtsev:2013cga,Hall:2014lea,Pascalutsa:2014zna,Gryniuk:2015aa}.  
These are the topics of this review.\footnote{For other recent
reviews (more focused on a subset) of  these topics see:
\citet{Drechsel:2002ar}
(sum rules and fixed-$t$ dispersion relations for CS), \citet{Schumacher:2005an} (RCS experiments), \citet{Kuhn:2008sy} (spin structure functions and sum rules), \citet{Phillips:2009af} (few-nucleon $\chi$EFT,
neutron polarizabilities),
\citet{Griesshammer:2012we} ($\chi$EFT and RCS experiments), \citet{Guichon:1998xv} (VCS and generalized polarizabilities),
\citet{Holstein:2013kia} (pion, kaon, nucleon polarizabilities), \citet{Pohl:2013yb}, \citet{Carlson:2015jba}, \citet{Karshenboim:2015} (proton-radius puzzle).} 

The paper is organized as follows.  
\Secref{theory} outlines the basic concepts as well
as discusses the current efforts to calculate
the nucleon polarizabilities from first principles:
lattice QCD (\Secref{theory1}) and 
$\chi$EFT (\Secref{theory2}). \Secref{formal}
describes the way polarizabilities appear in the CS
processes, while~\Secref{experiment} discusses the
way they are extracted from the CS experiments. 
\Secref{SRs} is devoted to dispersive sum rules, i.e., a variety of model-independent relations derived
from general properties of
the forward doubly-virtual CS amplitude. 
They involve the wealth of inelastic electron-scattering data into the polarizability studies,
and their data-driven evaluations are discussed in~\Secref{theory3}. 
In \Secref{hydrogen} we present an overview of nucleon structure
contributions to the hydrogen Lamb shift and hyperfine structure.
The reviewed results for nucleon polarizabilities and
their effect in muonic hydrogen are collected in the summary plots 
in \Secref{conclusion}. The reader interested
in only a brief survey of the field may skip to that section.
Finally, the Appendices
contain the expressions for the Born contribution
to CS amplitudes (\ref{sec:appBorn}), a derivation of 
generic dispersion relations  (\ref{sec:disprel}),
and a collection of the  most important formulae (\ref{sec:appindex}).

The remainder of this section contains the notations
and conventions used throughout the paper.


\subsection{Notations and Conventions}
\seclab{definitions}
\begin{itemize}
\item We use the natural units, $\hbar=c=1$, and the following 
notation for the well-established parameters, along with their 
Particle Data Group (PDG) values \cite{Agashe:2014kda}:
\begin{description}
\item[$\al$] the fine-structure constant, $\al = 1/137.035999074(44)$.
\item[$\hbar c$] conversion constant, $\hbar c=197.326 9718(44) \, \mathrm{MeV}\,\mathrm{fm}$.
\item[$m$]   lepton mass, $(m_e, m_\mu) \simeq (0.5109990, 105.65837)\,\mathrm{MeV}$.
\item[$m_\pi$] pion mass, $(m_{\pi^0}, m_{\pi^\pm})\simeq (134.977, 139.570)\,\mathrm{MeV}$.
\item[$M$]   nucleon mass, $(M_p, M_n) \simeq (938.272, 939.565 )\,\mathrm{MeV}$.
\item[$\varkappa $]   nucleon anomalous magnetic moment, 
$(\varkappa_p, \varkappa_n) \simeq (1.7929,\, -1.9130 )$.
\item[$f_{\pi} $] pion decay constant, $f_{\pi}=92.21(14)\,\mathrm{MeV}$.
\item[$g_A $] nucleon axial charge, $g_A=1.2723(23)$.
\end{description}
\item Other frequently used notation:
\begin{description}
\item[$s$, $t$, $u$] Mandelstam variables.
\item[$\nu$, $\w_B$, $\w$] photon energy in the lab, Breit, 
and center-of-mass reference frames. 
\item[$\vartheta$, $\th_B$, $\th$] scattering angle in the lab, Breit, 
and center-of-mass reference frames. 
\item[$\dd\varOmega_L$, $\dd\varOmega_{cm}$] element of the solid angle in the lab
and center-of-mass reference frames. 
\item[$Q^2=-q^2$] momentum transfer, photon virtuality.
\item[$\tau=Q^2/4M^2$] dimensionless momentum-transfer variable.
\item[$x=Q^2/2M\nu$] Bjorken variable.
\item[$F_1(Q^2)$, $F_2(Q^2)$] Dirac and Pauli form factors.
\item[$G_E(Q^2)$, $G_M(Q^2)$] electric and magnetic Sachs form factors. $G_E=F_1-\tau F_2$, $G_M=F_1+F_2$.
\item[$f_{1,2} (x, Q^2)$] unpolarized structure functions.
\item[$g_{1,2} (x, Q^2)$] polarized (or spin) structure functions.
\item[$\la_\ga$, $\la_\ga'$] helicities of the incident and scattered photon.
\item[$\la_N$, $\la_N'$] helicities of the incident and scattered nucleon.
\item[$\si(\nu)$, $\dd\si/\dd\varOmega$] unpolarized total and differential cross sections.
\item[$\si_T$, $\si_L$] unpolarized absorption cross section 
of the transverse ($T$) or longitudinal ($L$) photon.
\item[$\si_{TT}$, $\si_{LT}$] doubly-polarized  photoabsorption cross sections [see below \Eqref{VVCSunitarity}].
\item[$\zZ e$, $\varkappa$] 
charge and anomalous magnetic moment of the nucleon 
($\zZ = 1$ for proton, $\zZ = 0$ for neutron).
\item[$Ze$, $\kappa$] 
charge and anomalous magnetic moment of the nucleus
(for hydrogen: $Z=1$, $\kappa=\varkappa_p$).
\item[$\mu$] magnetic moment of the 
nucleon or nucleus, $\mu=\zZ+\varkappa=Z(1+\kappa)$,
in units of nuclear magneton.
\item[$a$, $m_r$] Bohr radius and reduced mass, $a^{-1}=Z\al m_r$, $m_r=mM/(m+M)$.

\end{description}
\item Salient conventions: 
\begin{itemize}[$\square$]
\item Metric: $g_{\mu\nu} = {\rm diag}(+1,-1,-1,-1)$. Levi-Civita symbol: $\eps_{0123}=+1=-\eps^{0123}$.
\item Scalar products: $\bp\cdot \bq
=p_i \,q_i$, $p\cdot q = p_0\, q_0 - \bp\cdot \bq$,  $p\cdot T\cdot q = p_\mu T^{\mu\nu} q_\nu$,
$\slashed{p}=p\cdot \ga$.
\item  Pauli and Dirac matrices: 
\bea
&&  1_2 = \left( \begin{array}{cc} 1 & 0 \\ 
0 & 1 \end{array} \right),\, \si^1 = \left( \begin{array}{cc} 0 & 1 \\ 
1 & 0 \end{array} \right),\, \si^2 = \left( \begin{array}{cc} 0 & -i \\ 
i & 0 \end{array} \right),\, \si^3 = \left( \begin{array}{cc} 1 & 0 \\ 
0 & -1 \end{array} \right), \nn\\
&& \gamma^0 = \left( \begin{array}{cc} 1_2& 0 \\ 
0 & -1_2 \end{array} \right),\,\,\,
 \gamma^i = \left( \begin{array}{cc} 0 & \sigma^i \\ 
 -\sigma^i  & 0 \end{array} \right),\,\,\,
 \ga^5 = i\gamma^0\gamma^1\gamma^2\gamma^3=\left( \begin{array}{cc} 0 & 1_2 \\ 
1_2 & 0 \end{array} \right), \\
&& 
\ga_{\mu\nu}=\half\left[\ga_\mu,\ga_\nu\right]=-\nicefrac{i}{2}\, \epsilon_{\mu \nu \al \be}\gamma^\al \gamma^\be \gamma^5,\, \ga_{\mu\nu\al}=\half(\ga_\mu\ga_\nu\ga_\al - \ga_\al\ga_\nu\ga_\mu)=-i \epsilon _{\mu \nu \al \be} \gamma^\be \gamma ^5,\nn\\
&&\ga_{\mu\nu\al\be} = 
\half\left[\ga_{\mu\nu\al},\ga_\be\right]=i \eps_{\mu\nu\al\be}\ga^5,
\eea
satisfying: 
$ \half [\si_i,\si_j]=i\veps_{ijk}\si_k$, $\{\ga_\mu,\ga_\nu\}=2g_{\mu\nu}$, $\{\ga_\mu,\ga^5\}=0$.


\item[$\square$] Helicity spinors:
\beq 
\eqlab{hel}
  u_{\la}(\bp)  =  \left( \begin{array}{c} \sqrt{E_p + M} \\ 
                         2 \la \sqrt{E_p - M} \end{array} \right)
              \otimes  \chi_{\la} (\theta,\vfi) , 
\eeq 
with $E_p = \sqrt{M^2 + \bp^2}$;  $\la=\pm 1/2$ the helicity (i.e., the spin projection onto $\bp$); $\theta, \vfi$  the spherical coordinates of $\bp$; and the two-component Pauli 
spinors:
\beq
\eqlab{chi}
 \chi_{\nicefrac12}(\theta,\vfi) = \left( \begin{array}{c} 
\cos (\theta /2) \\
     e^{i\vfi}\sin (\theta /2) \end{array} \right) , \,\,\,
 \chi_{-\nicefrac12}(\theta,\vfi)  =  \left( \begin{array}{c} 
                                -e^{-i\vfi}\sin (\theta /2)  \\
                                 \cos (\theta /2) \end{array} \right).
\eeq
The helicity spinors satisfy the
following relations, for $p=(E_p, \bp)$:
\beq
\bar u_{\la'}( \bp )\, u_{\la} (\bp)
= 2M \delta_{\la' \,\la} ,\quad  \sum\nolimits_{\la} u_{\la} ( \bp ) \, 
\bar u_{\la}(\bp)  = \slashed{p} + M , \quad (\slashed{p}-M) u_{\la}(\bp) =0.
\eeq 
\item Photon polarization vector, $\veps_{\la_\ga} (q)$, for a photon with four-momentum $q$
 and helicity $\la_\ga=-1,0,1$.
\begin{enumerate}[a)]
\item for real photon moving along the $z$-axis,  $q=(\nu,0,0,\nu)$, 
there are only transverse polarizations,
\begin{subequations}
\bea
&&\text{i) circularly polarized photons:} \quad\veps^\mu_{\pm 1} = \mbox{$\frac{1}{\sqrt{2}}$} \big(0,\mp 1, -i,0\big),\\
&&\text{ii) linearly polarized photons:} \quad\veps^\mu (\phi) = \big(0,\cos \phi, \sin \phi,0\big),
\eea
\end{subequations}
\item for virtual photon moving along the $z$-axis,
$q=(\nu,0,0,\vert \bq \vert)$, there is, in addition, the longitudinal polarization:
\beq
\veps_{0}^\mu  = \frac{1}{\sqrt{q^2}} \big( \vert \bq \vert, 0 , 0,
\nu\big), \quad \mbox{with $\vert \bq \vert=\sqrt{\nu^2-q^2}$.}
\eeq
\end{enumerate}
The transversality, orthonormality and completeness conditions are: 
\beq
q\cdot \veps_{\la_\ga}(q)  = 0\,,\, \quad  \veps_{\la_\ga'}^\ast(q)
\cdot \veps_{\la_\ga}(q) = - \de_{\la_\ga'\,\la_\ga}\,,\, \quad 
\sum_{\la_\ga=\pm 1, 0} \veps_{\la_\ga}^{\ast\mu} \veps_{\la_\ga}^\nu= -g^{\mu\nu} + \frac{q^\mu q^\nu}{q^2} \,.
\eeq
Note that for a spacelike photon ($q^2=-Q^2<0$), the longitudinal polarization vector is antihermitian,
$\veps^\ast_0 = -\veps_0$, and as the result the
above orthonormality and completeness conditions are
not satisfied. This is why one often defines
the longitudinal polarization vector for a spacelike
photon as $\veps_0' \equiv i\veps_0= \nicefrac{1}{Q}( \vert \bq \vert, 0 , 0,
\nu)$. The two definitions are connected by a gauge transformation.
\end{itemize}
\end{itemize}

\section{Basic Concepts and Ab Initio Calculations}
\seclab{theory}

\subsection{Na\"{i}ve Picture}
\begin{wrapfigure}[32]{r}{0.5\textwidth} 
  \centering 
  \vspace{-3mm}
       \includegraphics[width=0.24\textwidth]{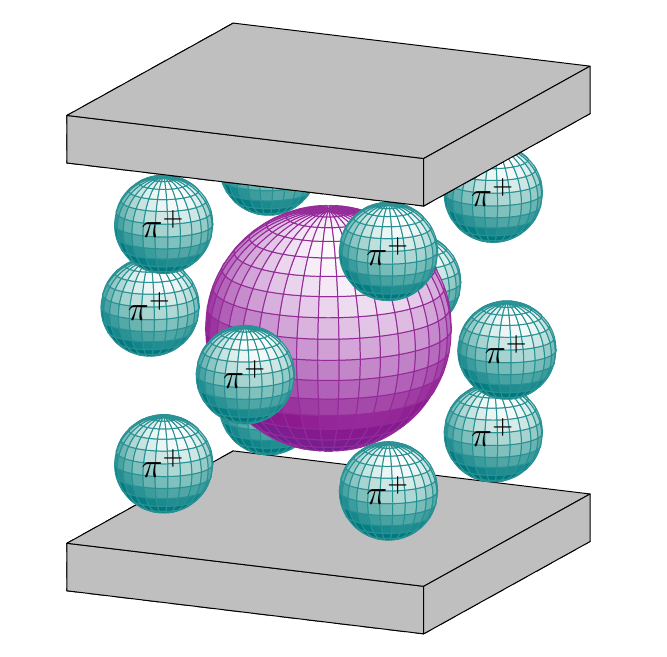}
       \includegraphics[width=0.24\textwidth]{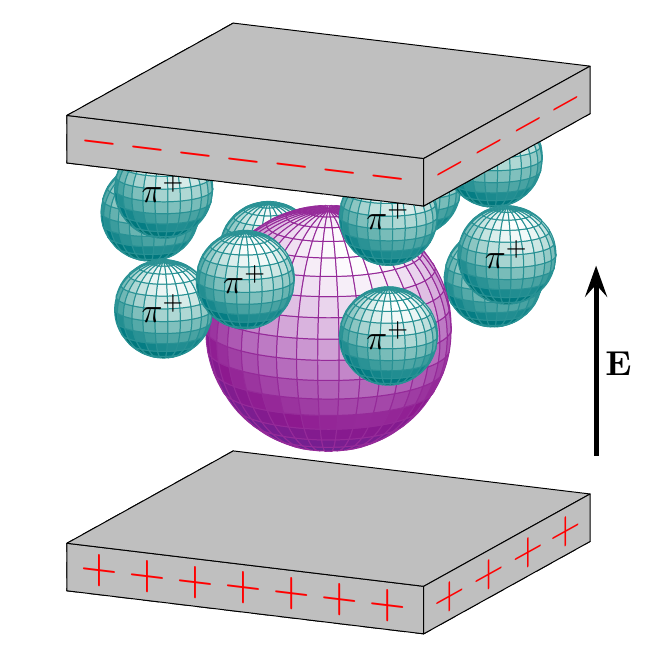}
 \caption{Naive view of the proton, consisting of a pion cloud and
 a quark core, placed between the plates of a parallel plate capacitor. The left (right) figure shows the capacitor discharged (charged). Plot courtesy of Phil Martel.
 \label{fig:capacitor}}
 \vspace{4mm}
        \includegraphics[width=0.24\textwidth]{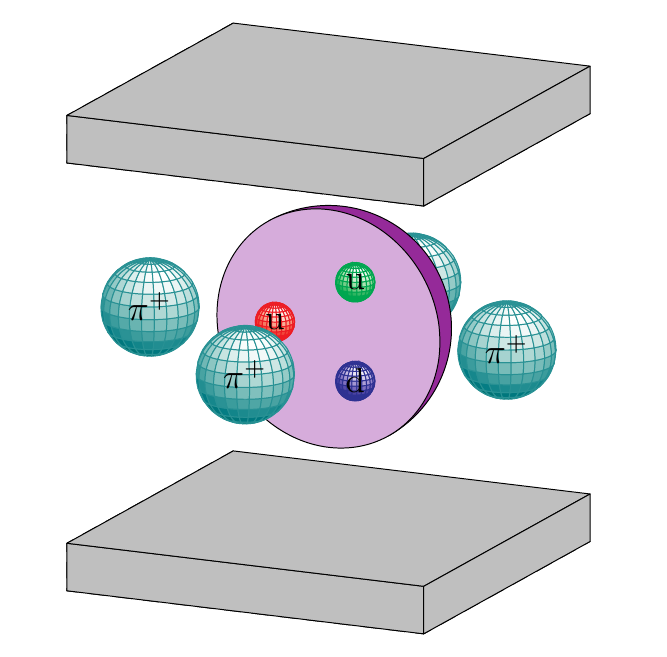}
       \includegraphics[width=0.24\textwidth]{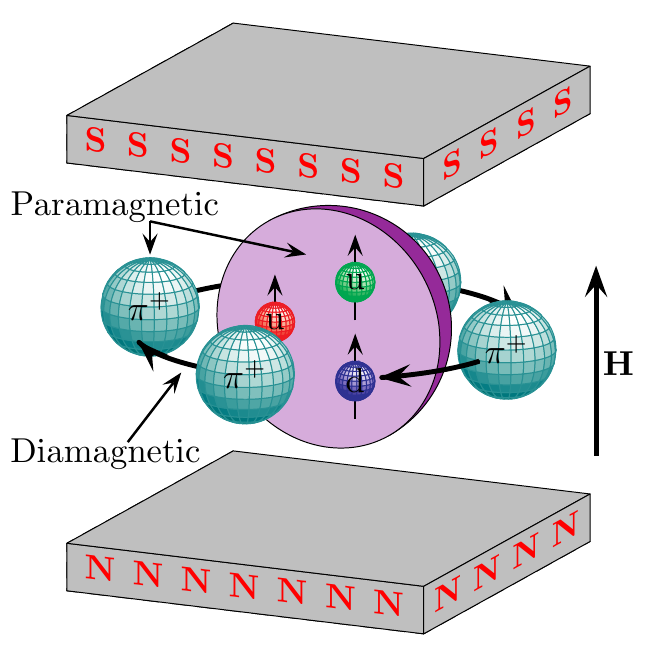}
 \caption{Naive view of the proton, consisting of a pion cloud and
 a quark core,  placed between the poles of a magnet. The left (right) figure shows the external magnetic field turned off (on).  Plot courtesy of Phil Martel.
 \label{fig:magnet}}
\end{wrapfigure}
A polarizability, by definition, quantifies the response of a system to an external electromagnetic (e.m.) field, or more precisely,
the e.m.\ moments induced in response to a moderate e.m.\ field. An evident picture is
provided by an atom immersed in a homogeneous electric field. The atomic nucleus and the electron cloud displace in opposite directions, thus creating an electric dipole moment proportional
to the field strength, with the proportionality coefficient $\al_{E1}$, the electric dipole polarizability. The polarizability mechanism in the nucleon is less obvious,  but, very roughly, one can replace  the electron cloud by the 
``pion cloud'' and the nucleus by a ``quark core'' to have
a similar picture, see \figref{capacitor}. An analogous
representation of the magnetic polarizability, $\beta_{M1}$,  is displayed
in \figref{magnet}. 

 This na\"{i}ve interpretation is realized, in a way, in $\chi$EFT where the (renormalized) pion loops can be thought of as the effect of the pion cloud, 
while the $\Delta(1232)$-resonance excitation and the low-energy  constants (LECs) are the effect of the quark core. 
In the case of the magnetic dipole
polarizability $\be_{M1}$,  the {\it diamagnetic} contribution of the pion cloud is competing against the {\it paramagnetic} contribution of the quark-core excitation,
see  \figref{magnet}. The two contributions
are largely canceling each other, leaving the nucleon with a relatively small magnetic polarizability, cf.~\Secref{theory2} for details. 

Other intuitive pictures of the nucleon polarizabilities emerge in quark models \cite{Capstick:1992tx,Capstick:1992qi, Hecking:1981,Schafer:1984hw,Weiner:1985ih}, the Skyrme model \cite{Scoccola:1990yh,Chemtob:1987ut,Golli:1993np,Schwesinger:1992qw,Broniowski:1992vj,Broniowski:1991tg}, and the Nambu-Jona-Lasinio model \cite{Nikolov:1993ty}.
All of them point out the large paramagnetic contribution due to the
nucleon-to-$\Delta$(1232) $M1$ transition. 

While for the atoms the polarizabilities are of order of the atomic volume, the nucleon being 
much tighter bound (nearly $99\%$ of its mass coming from the binding force) has polarizabilities which 
are about three orders of magnitude smaller than its volume.  It is 
customary to use the units of $10^{-4}\, \mathrm{fm}^3$ 
for the dipole polarizabilities of the nucleon. 

The critical electric field strength needed to induce any appreciable polarizability of the nucleon can be estimated as the ratio of the average  energy level spacing in the nucleon to the size of the nucleon, i.e., E$_\mathrm{crit.} \approx 100\, \mathrm{MeV}/(e\, \mathrm{fm}) = 10^{23}\, \mathrm{Volt}/\mathrm{m}$. Static electric field strengths of this intensity are not available in a laboratory, and will never be available. However, a classical estimate of the electric field strength of a $100\,\mathrm{MeV}$ photon Compton scattering from the nucleon is approximately $10^{23}\, \mathrm{Volt}/\mathrm{m}$. Given the absence of static e.m.\ fields of the required immensity, the CS process is currently
the best available tool for accessing the nucleon polarizabilities experimentally, cf.~\Secref{experiment}.

In the rest of this section we introduce the nucleon polarizabilities
and discuss their calculation from first principles. We shall
focus on describing the efforts to compute the nucleon polarizabilities
in lattice QCD and chiral EFT. In the latter case, calculations of the
CS observables will be discussed too.

It is worthwhile noting that is a number of sophisticated theoretical
approaches, other than lattice QCD and chiral EFT, applied to the nucleon polarizabilities and  low-energy CS.  They include: the
fixed-$t$ dispersion relations \cite{Hearn:1962zz,Lvov:1996xd,Drechsel:1999rf,Pas07},
effective-Lagrangian models with \cite{Kondratyuk:2000kq,Kondratyuk:2001qu,Gasparyan:2010xz,Gas11}
and without \cite{Pascalutsa:1995vx,Feuster:1998cj,Zhang:2013uaa}
causality constraints, 
the Dyson--Schwinger equation approach to QCD \cite{Eichmann:2012mp}.
The first one in this list is very popular in the extractions
of polarizabilities from CS data, and will be  mentioned frequently in other
chapters of this review.
\subsection{Defining Hamiltonian}
The response in the energy of the system due to 
polarizability effects is described
by an effective Hamiltonian, which usually 
is ordered according to the number of
spacetime derivatives of the e.m.\ 
field $A_\mu(x)$~\cite{Bab98,Holstein:1999uu},
\begin{subequations}
\bea 
\mathcal{H}_{\mathrm{eff}}^{(2)} &=& -4\pi\left(\half\,\alpha_{E1} \,\boldsymbol{E}^2 + 
\half \,\beta_{M1}\, \boldsymbol{H}^2 \right), 
\eqlab{H2eff}\\
\mathcal{H}_{\mathrm{eff}}^{(3)} & = & -4\pi \left(
\half\, \gamma_{E1E1}\,\boldsymbol{\sigma}\cdot(\boldsymbol{E}\times\dot{\boldsymbol{E}})+\half\, \gamma_{M1M1}\,\boldsymbol{\sigma}\cdot(\boldsymbol{H}\times\dot{\boldsymbol{H}}) 
-\gamma_{M1E2}\,E_{ij}\sigma_{i}H_{j}+\gamma_{E1M2}\,H_{ij}\sigma_{i}E_{j}\right) ,
\eqlab{H3eff}\quad\\
\mathcal{H}_{\rm eff}^{(4)} &=& - 4\pi
    \left(\half \,\alpha_{E1\nu}\, \dot{\boldsymbol{E}}^2 + \half\, \beta_{M1\nu}\, \dot{\boldsymbol{H}}^2\right)
    - 4\pi \left(\mbox{$\frac{1}{12}$} \,\alpha_{E2}\, E_{ij}^2 + \mbox{$\frac{1}{12}$}\,  \beta_{M2}\, H_{ij}^2\right),\eqlab{H4eff}
\eea 
\eqlab{Heff}
\end{subequations}
where the electric ($\boldsymbol{E}$) and magnetic
($\boldsymbol{H}$) fields are expressed
in terms of the e.m.\ field tensor, 
$F_{\mu\nu} =\pa_\mu A_\nu - \pa_\nu A_\mu$, 
as:  $E_i = F_{0i}$, $H_i = \half \eps_{ijk} F_{jk}$. Furthermore, the following shorthand notation is used:
\beq 
E_{ij} = \half (\nabla_i E_j + \nabla_j E_i), \qquad
H_{ij} = \half (\nabla_i H_j + \nabla_j H_i).
\eeq 

The 3\textsuperscript{rd}-order term depends on the
nucleon spin via the Pauli matrices $\boldsymbol{\sigma}$, and the
corresponding polarizabilities are called
the {\em spin polarizabilities} \cite{Ragusa:1993rm}.
They
have no analog in classical electrodynamics, 
but evidently they describe the coupling of the
induced e.m.\ moments with the nucleon spin.  
Unlike the scalar polarizabilities,
they are not invoked by static e.m.\ fields. 

The above Hamiltonian is quadratic
in the e.m.\ field. This means that the polarizabilities
can directly be probed in the CS process. The expansion
in derivatives of the e.m.\ field translates then into
the low-energy expansion.  The polarizabilities
thus appear as coefficients in the low-energy
expansion of the CS amplitudes, cf.~\Secref{formal}.

As noted above, the scalar dipole polarizabilities
are measured in units of $10^{-4}\, \mathrm{fm}^{3}$. In general,
the nucleon polarizabilities are measured in
units $10^{-4}\, \mathrm{fm}^{n+1}$, where
$n$ is the order at which they appear.

\subsection{Lattice QCD}
\seclab{theory1}

Presently all of the lattice QCD calculations of
nucleon polarizabilities use 
the background-field method~\cite{Bernard:1982yu,Fiebig:1988en}, which amounts to measuring the shift in the mass 
spectrum upon applying a classical background field.
On a given configuration, one multiplies the SU$(3)$ gauge fields by a U$(1)$ gauge field. The U$(1)$ links are given by
\beq 
	U_\mu(x) = \exp\left[ie_q\, \aa A_\mu(x)\right],
\eeq 
where $e_q$ is the quark charge and $\aa$ is the lattice
spacing.

The case of a constant magnetic field is the simplest
to illustrate. For the field with a magnitude $H$ pointing in the $+z$-direction, the usual choice is
$A_\mu(x,y,z,t)= \aa H x\,\delta_{\mu y}$. 
The problem with this choice is that due to the condition that the gauge links $U_\mu$ must be periodic, the field is continuous only if $e_q \,\aa^2 H=2\pi n/L$, with integer $n$. 
The minimal value of $H$ is thus severely limited by the size of the lattice, although
an improvement to $H \sim 1/L^2$ behavior is easily achieved (see, e.g., Ref.~\cite{Aubin:2008qp}).

One can calculate a baryon two-point function which behaves for large time in the usual manner
\beq\label{eq:2pt}
	C(t) \sim  e^{-M(H)\, t} + \ldots\ ,
\eeq 
but with the exponential damping governed by a field-dependent mass \cite{Alexandru:2008sj}
\beq \label{eq:mB}
	M(H) = M_0 - \mu_z H - \half  \beta_{M1} H^2 + O(H^3)\ ,
\eeq 
where $M_0$ is the mass with no field and $\mu_z$ is
the projection of the magnetic moment.
One may cancel the odd terms by considering 
$M(H) + M(-H)$ and fit the remaining
$H$-dependence by adjusting the value of the magnetic polarizability.

Implementation of the electric field is 
somewhat more tricky and has led to an overall
sign mistake in the value of the electric polarizability (which affects, e.g., Ref.~\cite{Engelhardt:2007aa} as well as many of the earlier calculations).
In the proton case, one in addition 
needs to take care of the Landau levels, 
which thus far has only been done by the
NPLQCD collaboration \cite{Chang:2015qxa}.
Implementation of the varying fields needed
to compute the spin polarizabilities is considered
in Ref.~\cite{Detmold:2006vu}.

In the background field
method one obviously assumes that the
Taylor expansion in the field strength is quickly convergent. 
The non-analyticity due to the pion production  
induced by the background field may however become
a problem. This problem is similar to the one
encountered in experiment, where to see the signal
in the CS observables one needs energies 
approaching the pion-production threshold. 

Another difficulty of this method is the inclusion
of the background-field effect on the ``sea''. 
Most of the calculations to date assume the sea
quarks to be neutral. Studies of the
charged sea-quark contributions have been done
in, e.g., Refs.~\cite{Engelhardt:2007aa,Freeman:2014kka}.

\begin{figure}[hbt] 
  \centering 
\includegraphics[width=0.6\textwidth]{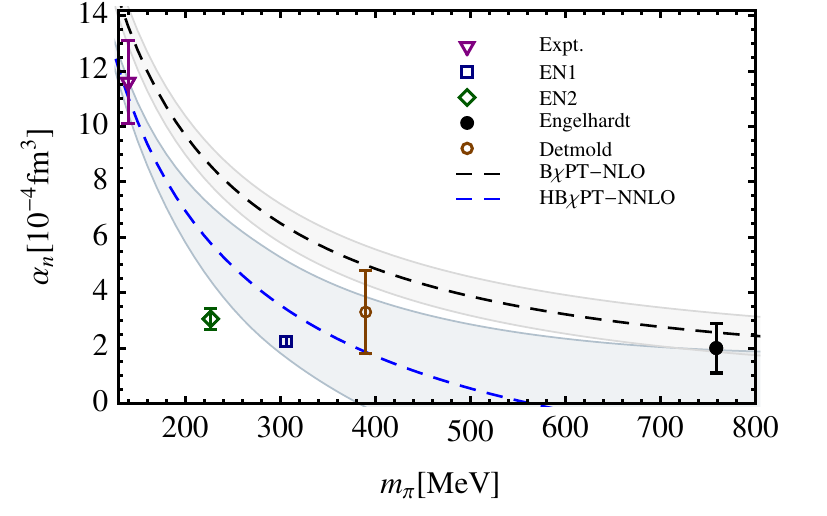}
\caption{Lattice QCD results for the electric
polarizability of the neutron, at unphysical values
of pion mass: \citet{Engelhardt:2007aa}, \citet{Detmold:2009dx,Detmold2010}, and
\citet{Lujan:2014kia} on two different
ensembles ('EN1' and 'EN2').
The curves with error band are the predictions
of the baryon \cite{Len10} and heavy-baryon \cite{Griesshammer:2015ahu} $\chi$PT.
The HB$\chi$PT result is a fit at the physical
pion mass. Plot courtesy of Andrei Alexandru.
 \label{fig:GWUalpha}}
\end{figure}

\begin{wrapfigure}{r}{0.5\textwidth}
  \centering 
       \includegraphics[width=0.33\textwidth,angle=90]{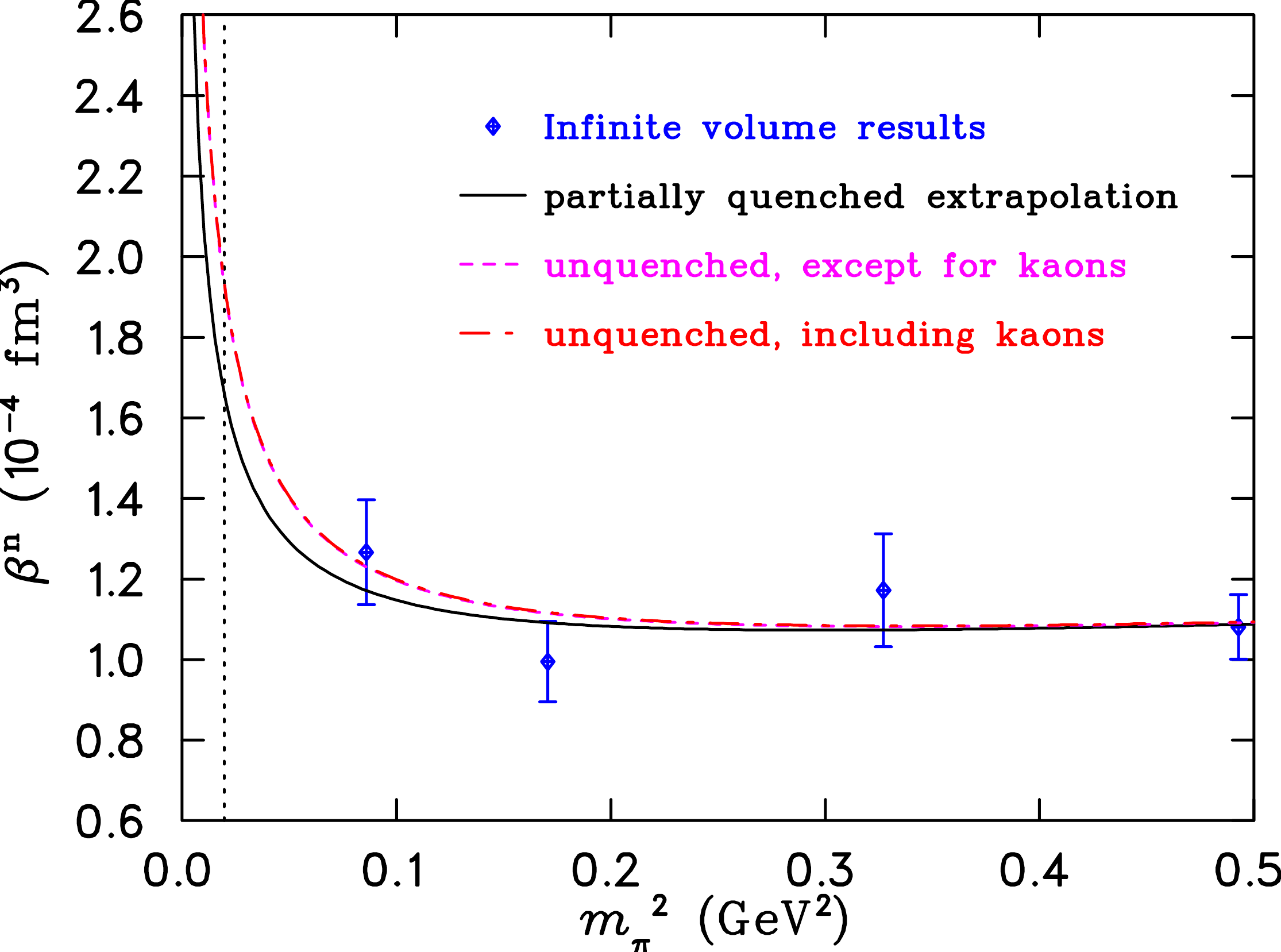}
 \caption{Results of \citet{Hall:2013dva} for the
 magnetic polarizability of the neutron. Plot courtesy of Jonathan Hall.
 \label{fig:ADLbeta}}
\end{wrapfigure}
Despite these concerns, the recent lattice
results are very encouraging. Most of the results
are obtained for the neutron electric polarizability,
see \Figref{GWUalpha}. The lightest-pion results, 
indicated as EN1 and EN2 therein,  are from  
the GWU group \cite{Lujan:2014kia}. They have recently
received substantial finite-volume corrections, moving them upwards, right onto the HB$\chi$PT curve~\cite{Lujan:2014qga}.

For the magnetic polarizability
we refer to the recent work of the Adelaide CSSM group
\cite{Primer:2013pva,Hall:2013dva} which used the PACS-CS $2+1$ flavor gauge field configurations \cite{Aoki:2008sm} and performed an extrapolation to the physical pion mass 
and infinite volume,
see \Figref{ADLbeta}. Their extrapolated result is \cite{Hall:2013dva}: $\beta_{M1}^{(n)} = 1.93 (11)_{\mathrm{stat}} (11)_{\mathrm{sys}} \times 10^{-4}\,\mathrm{fm}^3,$ and can be directly compared with experiment and other theoretical results in \Figref{alphabeta_n}. Very recently, the NPLQCD Collaboration obtained results for the
electric and magnetic polarizabilities of both proton and neutron (and a number of light nuclei), albeit at a relatively large pion mass of 800 MeV \cite{Chang:2015qxa}.

As an alternative to the background-field method, one may consider
a four-point function calculation of the Compton tensor for spacelike photons. At present this has
only been done for the light-light system (light-by-light scattering) \cite{Green:2015sra}. This method would avoid the problem 
with the non-analytic behavior of the mass spectrum in the 
background field, 
but would require an extrapolation to the real-photon point. On the
other hand, it would be a direct calculation of the polarizabilities at a finite photon virtuality $Q^2$,
which experimentally is accessed through
the dispersive sum rules (cf.~\Secref{SRs}), and is required
in the atomic and two-photon-exchange calculations 
(cf.~\Secref{hydrogen}). In practice, such 
a direct calculation of doubly-virtual CS on the nucleon
would be very challenging due to the usual problems of 
the noise-to-signal ratio and excited-state contamination.
The CS on the pion would certainly be a better place to start.

\subsection{Chiral EFT}
\seclab{theory2}

The chiral effective-field theory
($\chi$EFT), also referred to as the chiral perturbation theory ($\chi$PT) is
a low-energy EFT of QCD, see Refs.~\cite{Weinberg:1978kz,Gasser:1983yg,Gasser:1987rb} 
for the seminal papers. It is a quantum field theory with 
the Lagrangian written in terms of hadronic fields, in contrast to QCD which is
written in terms of quark and
gluon fields.  The $\chi$EFT, however, holds the promise
to match the QCD description of the low-energy phenomena, where by low-energy one
assumes the relative energy in the hadronic system 
to be well below 1 GeV. In our case of CS processes, this means that the energy $\nu$ and the momentum transfer $Q$  with which the photon probes 
the nucleon are much smaller than 1 GeV. At these energies the pion
interactions are suppressed. More precisely, the
interaction goes with pion 4-momenta, which
are relatively small for small pion energy. A perturbative expansion of scattering amplitudes
in pion momenta is called the chiral expansion.

In contrast to QCD, $\chi$EFT is non-renormalizable in the usual sense.
However, to any finite order in the EFT expansion, all the divergencies
are absorbed by renormalizations of a finite number of parameters, 
called low-energy constants (LECs). The renormalized LECs are to be matched
to QCD: in practice, either extracted from the lattice QCD results or fit to experimental data. 

The number of unknown LECs grows quickly with the 
order (or precision) to which one wants to compute. For this reason, 
one might be tempted to dismiss such theory as having practically no predictive power.
This indeed would be the case if the LECs are treated as entirely free parameters, 
i.e., allowed to take arbitrary values. 
They are not --- their effect must be of {\it natural size}~\cite{Georgi:1994qn}, which simply speaking means that the LECs may only have an effect consistent with the estimate based on {\it power counting} (i.e., in most cases
the  {\it naive dimensional analysis}). When a certain LEC effect is unnaturally large, hence exceeds the expectation
and/or requires the ``promotion to a lower order'', the EFT should  be revised 
to include the missing low-energy physics. 
On the other hand, when {\it naturalness} is implemented, the EFT is predictive and the uncertainty due to 
neglect of the higher-order corrections can be estimated.

The calculations with no divergencies, and/or no new constants to be fit,
are genuine predictions of $\chi$EFT. Such examples are quite rare, however the
calculation of nucleon polarizabilities presents one of them. 
The leading-order [$O(p^3)$] contribution to
nucleon polarizabilities is predictive, as
there are no LECs renormalizing the
polarizabilities [until $O(p^4)$]. This case
therefore presents a great testing ground of the $\chi$EFT framework. 

To begin with, one can clearly see 
here that rather different predictions are obtained depending on whether 
the so-called {\it heavy-baryon (HB) expansion} \cite{Jenkins:1990jv} is 
employed \cite{Butler:1992ci,Bernard:1995dp} or not \cite{Bernard:1991rq,Bernard:1991ru}. For example,
for the proton dipole polarizabilities $\{\al_{E1}, \,
\be_{M1}\}$ one obtains 
$\{12.2,\, 1.2\}$ in 
HB$\chi$PT, versus $\{6.8,\, -1.8\}$ in $\chi$PT.
The uncertainty on such a leading order prediction 
can be quite large and hence this discrepancy 
might not look as bad at first. The discrepancy
deepens at the next order, i.e.,
with the inclusion of the $\De(1232)$ as an
explicit degree of freedom. The $\De$ contributions
to the nucleon polarizabilities come out to be large
in HB$\chi$PT~\cite{Hemmert:1996rw}
and ought to be canceled eventually by the LECs which are 
``promoted'' from higher orders, cf.~\citet{Hildebrandt:2003fm,Griesshammer:2012we}.
This problem is
discussed at length in Refs.~\cite{Len10,Hall:2012iw}.
For a more general overview of the $\chi$PT in the single-baryon
sector (B$\chi$PT) and the current status of the theory,
see \citet{Geng:2013xn}.

The $\De(1232)$-resonance plays
a prominent role in the modern
formulation of $\chi$PT in the baryon sector.
Its excitation energy,
\beq 
\varDelta=M_\De - M_N \simeq 293 \; \mbox{MeV},
\eeq
is relatively low, and the $\De$ must be included
explicitly in the $\chi$PT Lagrangian. The construction of HB$\chi$PT (semi-relativistic) Lagrangians 
with $\De$'s, and decuplet fields in general,
was considered in Refs.~\cite{Jenkins:1990jv,Butler:1992ci}.
The manifestly Lorentz-invariant B$\chi$EFT Lagrangians with the spin-3/2 $\De$ fields
were reviewed in \cite[Sect.\ 4]{Pascalutsa:2006up}.

Concerning the power counting of the $\De$ contributions, \citet{Hemmert:1996xg} 
coined the term ``$\epsilon$-expansion'',
whereas a different
counting (``$\delta$-expansion'') was
subsequently proposed by~\citet{Pascalutsa:2003aa}. 
The two power-counting schemes differ in how much the excitation energy $\varDelta$ weighs in,
compared with the pion mass $m_\pi$. 
In the $\epsilon$-expansion they are the same
($\varDelta \sim m_\pi$), while in
the $\delta$-expansion $m_\pi \ll \varDelta$.
The main advantage of the latter is that it provides
a systematic counting of the $\De$-pole contributions, which go as $1/(p-\varDelta)$ where
$p$ is the typical energy or momentum. Indeed,
as $p$ is of the same order as $m_\pi$ and $\varDelta$, the $\eps$ counting implies that these contributions
are always overwhelmingly important. In practice, however, the $\eps$-expansion counts these propagators as $1/p$. This works
for the energies well below the resonance, but
in the $\De$-resonance region these contributions
are dominating and the power counting should reflect that. The $\de$ counting does just that transition. When
$p\sim m_\pi$, the propagator $1/(p-\varDelta)$ counts as $1/\varDelta$. When $p\sim \varDelta$
(the resonance region), the $\De$-pole contributions
are summed yielding the dressed propagator $1/(p-\varDelta - \Si)$, with
$\Si$ the self-energy of the $\De$, and the 
resulting dressed propagator counts as 
$1/\Si \sim 1/p^3$, since usually $\Si\sim p^3$.

Thus, the $\de$-expansion is an EFT with 
a hierarchy of two low-energy scales and as such
has two regimes, $p\sim m_\pi$ (low-energy) and
$p\sim \varDelta$ (resonance), where the $\De$
contributions count differently.
For definiteness, the following powers are assigned
to the low-energy scales in the two regimes:
\begin{enumerate}
\item low-energy: $m_\pi \sim p$, $\varDelta \sim 
p^{1/2}$.
\item resonance: $m_\pi \sim p^2$, $\varDelta \sim 
p$.
\end{enumerate}
Hence, e.g., the propagator 
$1/(p-\varDelta - \Si)$ is of $O(p^{-1/2}) $
in the first region and of $O(p^{-3}) $
in the second.

The present state-of-the-art calculations of
CS observables, within HB$\chi$PT \cite{McG13,Griesshammer:2012we} and B$\chi$PT \cite{Lensky:2008re,Len10,Lensky:2015awa}, employ the 
$\de$-expansion. Other applications of the 
$\de$-expansion include: the forward
VVCS \cite{Lensky:2014dda,Alarcon2015}, 
pion-nucleon scattering~\cite{Alarcon:2011zs},
pion photo- \cite{Blin:2014rpa} and 
electro-production \cite{Pascalutsa:2005ts,Pascalutsa:2006aa}, radiative pion photoproduction \cite{Pascalutsa:2004je,Pascalutsa:2007wb}. 
For the VVCS case, there is a significant discrepancy between
two B$\chi$PT calculations, based on the
$\eps$ \cite{Bernard:2012hb} and the
$\de$ \cite{Lensky:2014dda,Alarcon2015} counting schemes,
for some of the forward spin polarizabilities,
see \Figref{gamma0} and \Figref{deltaLT}. 
This discrepancy is not yet completely 
understood.

\begin{wrapfigure}[16]{l}{0.45\textwidth}
  \centering 
   \includegraphics[width=0.4\textwidth]{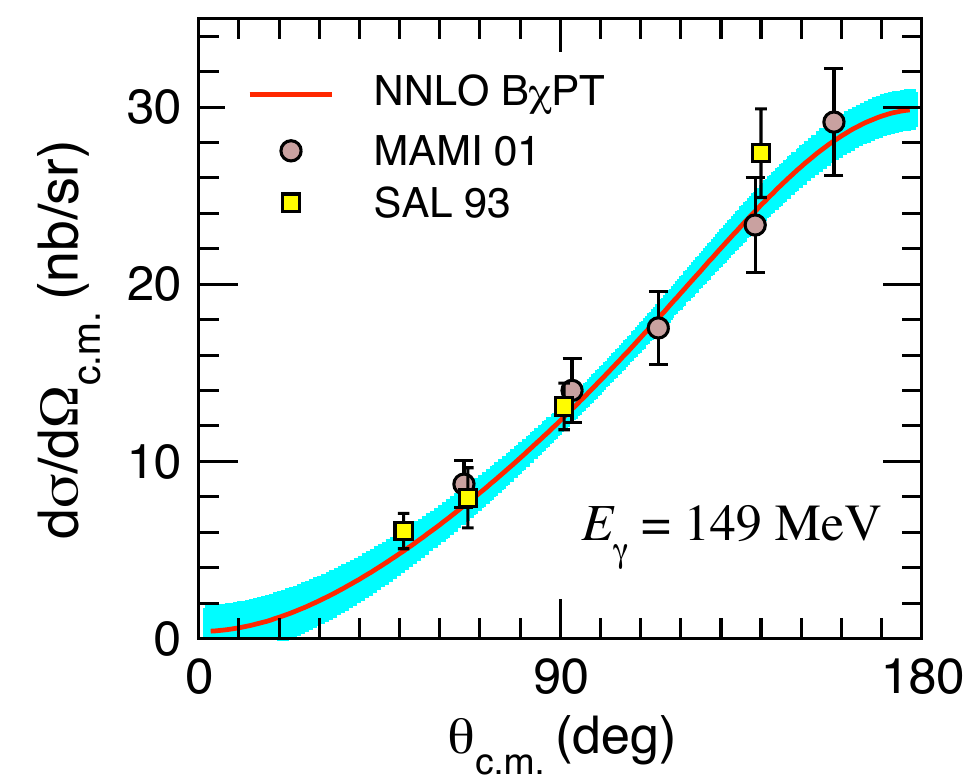}
 \caption{ Results of the B$\chi$PT
 calculation \cite{Len10} for the differential
 cross section of proton Compton scattering 
 (red curve with the band), compared with the
 experimental data from SAL \cite{Hallin:1993ft} and MAMI \cite{Olm01}.
 \label{fig:RCScs}}
\end{wrapfigure}

Coming back to RCS, the B$\chi$PT calculation of \citet{Len10}
are done to next-to-next-to-leading order (NNLO),
i.e.~to $O(p^{7/2})$, in the low-energy region. To this order, 
these are genuine predictions of B$\chi$PT
in the sense that all the parameters are
determined from elsewhere; the LECs intrinsic to polarizabilities
do not enter until $O(p^4)$. A very good
description of the CS experimental data is nevertheless
observed ---
a typical description of the unpolarized
angular distribution is shown in \Figref{RCScs}.

The numerical composition of the various contributions
to the dipole polarizabilities of the proton
is given in B$\chi$PT by (in units of $10^{-4}\,$fm$^3$):
\begin{subequations}
\begin{eqnarray}
\eqlab{alChPT}
 \alpha_{E1}^{(p)}&=& \underbrace{6.8}_{O(p^3)} + \,(\underbrace{-0.1 + 4.5}_{O(p^{7/2})}) = 11.2\,,\\
\beta_{M1}^{(p)}&=&\underbrace{-1.8}_{O(p^3)} +\, (\underbrace{ 7.1-1.4}_{O(p^{7/2})}) =3.9  \,,
\end{eqnarray}
\end{subequations}
where the first number in $O(p^{7/2})$ comes from
the $\De$-resonance excitation while the second comes from
the $\pi\De$ loops. One sees that the $\De$-excitation
is mainly affecting the magnetic polarizability and is
of paramagnetic nature (i.e., positive contribution
to $\beta_{M1}$). This is expected from the
first nucleon excitation which predominantly is
of $M1$ type. The pion loops, playing here the 
role of the ``pion cloud'', induce the diamagnetic
effects (i.e., negative $\beta_{M1}$).

These B$\chi$PT results can be contrasted with the
corresponding HB$\chi$PT calculation \cite{Hemmert:1996rw}:
\begin{subequations}
\eqlab{abHBChPT}
\bea 
  \alpha_{E1}^{(p)}(\mbox{HB}) &=& 12.2 \,\mbox{($\pi$N loop)} + 0 \,\mbox{($\Delta$ pole)} + 8.6 \, \mbox{($\pi\Delta$ loop)}  = 20.8\,,\\
 \beta_{M1}^{(p)}(\mbox{HB})&=& 1.2 \,\mbox{($\pi$N loop)}  
 + 12 \,\mbox{($\Delta$ pole)} +1.5 \,  \mbox{($\pi\Delta$ loop)} = 14.7  \,.
\eea 
\end{subequations}
Here the chiral loops give a much larger (than in B$\chi$PT) effect in the 
electric polarizability, while in the magnetic they even have an opposite sign. 
As the result, both polarizabilities come out to be way above  their empirical
values. As noted above, this discrepancy is usually corrected
by promoting the higher-order [$O(p^4)$] LECs, at the expense of 
violating the naturalness requirement. 

Recently, the B$\chi$PT framework of Ref.~\cite{Len10} has been extended
to the $\De$-resonance region \cite{Lensky:2015awa}, with
an update on the predictions for
$\al_{E1}$ and $\be_{M1}$ (included in the above numbers for the proton). Predictions
for the spin and higher-order polarizabilities
have also been obtained. 
The results for the
scalar polarizabilities of the proton 
are presented in \Tabref{polsscalar}, where they are
compared with the $\chi$PT results obtained by fitting the experimental RCS cross sections. The fitting in Refs.~\cite{Griesshammer:2012we,Lensky:2014efa} is done using LECs from the orders beyond NNLO. The fact that the B$\chi$PT fit \cite{Lensky:2014efa} and prediction \cite{Lensky:2015awa} agree,
within the uncertainties, indicates that the LEC effect (which is
the only difference between the two calculations)
is of natural size.

\begin{table}[tb]
\footnotesize
\centering 
\caption{Predictions for the {\it proton} static dipole, quadrupole, and dispersive polarizabilities, in units of $10^{-4}$~fm$^{3}$ (dipole) and $10^{-4}$~fm$^{5}$ (quadrupole and dispersive),
compared with the $\chi$PT-based fits dipole polarizabilities to RCS database. The
latest PDG values for dipole polarizabilities are shown too.
}
\begin{tabular}{|c|c|c|c|c|c|c|}
\hline
  Source      &$\alpha_{E1}$    & $\beta_{M1}$     &$\alpha_{E2}$  &$\beta_{M2}$ &$\alpha_{E1\nu}$ &  $\beta_{M1\nu}$
  \\
  \hline
HB$\chi$PT fit~\cite{Griesshammer:2012we}
              & $10.65\pm 0.50$ & $3.15\pm 0.50$   & $\cdots $         & $\cdots $           & $\cdots $ & $\cdots $   \\
B$\chi$PT fit~\cite{Lensky:2014efa}
              & $10.6\pm 0.5 $  & $3.2\pm 0.5$     & $\cdots $         & $\cdots $         & $\cdots $  & $\cdots $   \\
B$\chi$PT NNLO \cite{Lensky:2015awa}        & $ 11.2\pm0.7 $  &$3.9\pm0.7$& $17.3\pm3.9$       & $-15.5\pm3.5$    & $-1.3\pm 1.0$ & $7.1\pm 2.5$   \\
PDG~\cite{Agashe:2014kda}
             & $11.2\pm 0.4$   & $2.5\pm 0.4$     &     $\cdots $    &  $\cdots $     & $\cdots $   & $\cdots $  \\
\hline
\end{tabular}
\label{tab:polsscalar}
\end{table}

A number of predictions for polarized CS observables,
emphasizing the role of the chiral loops, are given in Ref.~\cite{Lensky:2015awa} as well, see e.g., \Figref{sigma3_sensitivity}
below. 
The corresponding results for the proton spin polarizabilities  are shown in
Table~\ref{Tab:SPs_theory}. 

A brief summary of the $\chi$PT results for the nucleon polarizabilities is given in \Secref{conclusion}. The HB$\chi$PT
calculations therein have recently been reviewed by \citet{Griesshammer:2012we}.

\section{Polarizabilities in Compton Scattering}
\seclab{formal}
\subsection{Compton Processes} 
The CS processes, represented by \Figref{CSgen}, can be classified according to
the photon virtualities, $q^2$ and $q^{\prime\,2}$, while the target particle  (hereby the nucleon) is on the mass shell: $p^2 = p^{\prime\,2} = M^2$. 
The Mandelstam variables for this two-body scattering process are:
\begin{subequations}
\bea 
s&=& (p+q)^2 = M^2 + 2 p\cdot q +q^2 = (p'+q')^2,\\
u&=& (p'-q)^2 = M^2 - 2 p'\cdot q +q^2 = (p-q')^2,\\
t&=& (p-p')^2 = 2M^2 - 2 p\cdot p' = (q-q')^2 . 
\eea
\end{subequations}
\begin{wrapfigure}[11]{l}{7.3cm}
\centering
  \includegraphics[width=6.7cm]{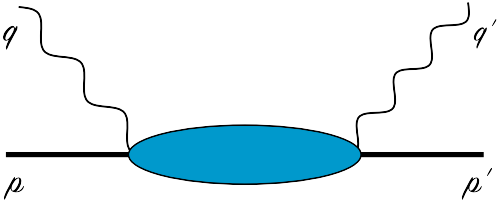}\\
  \caption{The Compton scattering off the nucleon: $\gamma(q)+ N(p)\rightarrow \gamma(q')+ N(p')$. \figlab{CSgen}}
\end{wrapfigure}
Their sum is as usual given by the sum of invariant masses squared: $s+t+u = 2M^2 +q^2 +q^{\prime\,2}$. Throughout the paper we use the following
kinematical invariants,
\bea 
&& \nu = p\cdot q/M , \quad \nu' = p\cdot q'/M, 
\eea 
which in the lab frame become the energy  of, respectively, the incoming and outgoing  photon.

In the most general case, the initial and final photons are virtual, with  
different virtualities, $q^2\neq q^{\prime 2}$. In reality, this 
situation may occur in the dilepton electro-production, $e^-N\to e^-N\, e^+e^-$, the $N\bar N$
production in $e^+e^-$ collisions, or in the two-photon-exchange contribution to lepton-nucleon scattering, discussed in \Secref{hydrogen} in the context of atomic calculations.

Denoting the photon helicity by $\la_\ga=\pm 1, 0$  and the nucleon helicity by $\la_N = \pm 1/2$, there are obviously $3\times3\times 2\times 2= 36$ helicity amplitudes, $T_{\la_\ga'\la_N'\la_\ga \la_N}(s,t) $, describing this process. Discrete symmetries, such as parity and time reversal, reduce the number of independent helicity amplitudes by more than a half, as will be discussed in more detail 
below.  

The Feynman amplitude $T^{\mu\nu}$ describing this process is a
rank-2 tensor-spinor
which depends on the four-momenta $q$, $q'$, $p$, $p'$.
Due to momentum conservation, three of them are independent, e.g.: $q$, $q'$, and $P=\half (p+p')$. The helicity amplitudes are expressed in terms of the Feynman amplitude as:
\begin{equation}
\eqlab{helamp}
T_{\la_\ga'\la_N'\la_\ga \la_N} 
= \bar u_{\la_N'} (\bp') \, \veps^\ast_{\la_\ga'}(q')\cdot T(q',q,P)\cdot \veps_{\la_\ga}(q)
\, u_{\la_N}  (\bp)\,  ,
\end{equation}
with the nucleon spinors and photon polarization vectors defined 
in \Secref{definitions}. A consequence of the e.m.\ gauge invariance is 
\begin{equation}
q^\prime_\mu T^{\mu\nu}(q',q,P) = 0 =
q_\nu T^{\mu\nu}(q',q,P),
\end{equation}
valid for on-shell nucleons and arbitrary photon virtualities.
The Lorentz decomposition of the  Feynman amplitude in terms
of the invariant amplitudes $\scA_i$, 
\begin{equation}
T^{\mu\nu}(q',q,P) =  e^2 \sum_i \scO_i^{\mu\nu} \scA_i (\nu', q^{\prime\,2}, \nu, q^2),
\end{equation}
contains $18$ terms, after the constraints due to parity, time reversal and gauge invariance  are taken into account \cite{Tarrach:1975tu}. 
For off-forward VVCS with $q^{\prime\,2} = q^2$, this number reduces to $12$;
for forward VVCS, to 4. 
For the rest of this section we restrict 
ourselves to the RCS, i.e., the case
where  both photons are real ($q^{\prime\,2} = q^2=0$).
The case where one of the photons is virtual (VCS and timelike CS)
is briefly discussed in \Secref{experiment}. The forward
doubly-virtual CS appears prominently in \Secref{SRs} and
\ref{sec:hydrogen}. 

\subsection{Helicity Amplitudes}
\seclab{RCS}

Consider the classic CS: the elastic scattering of a real photon, or real CS (RCS). 
The RCS on a spin-$1/2$ target
is described by the helicity amplitudes, $T_{\la_\ga'\la_N'\la_\ga \la_N}(s,t) $, subject to the
following parity ($\zP$) and time-reversal ($\zT$) constraints:
\begin{subequations}
\bea
\zP : && T_{-\la_\ga'\,-\la_N'\,-\la_\ga\, -\la_N}(s,t) = (-1)^{\la_\ga'-\la_N'-\la_\ga+\la_N} \,T_{\la_\ga'\la_N'\la_\ga \la_N}(s,t),
\\
\zT : && T_{\la_\ga \la_N \la_\ga' \la_N'}(s,t) = (-1)^{\la_\ga'-\la_N'-\la_\ga+\la_N} \,T_{\la_\ga'\la_N'\la_\ga \la_N}(s,t).
\eea
\end{subequations}
These constraints reduce the number of independent amplitudes from $2\times 2\times 2\times 2=16$ to $6$. The amplitudes in this case depend only on the combined (or total) helicities, $H = \la_\ga - \la_N$, $H' = \la_\ga' - \la_N'$, which 
run through $\pm \nh$, $\pm \nth$, and the above constrains read:
\beq 
\eqlab{symrels}
T_{H'H}(s, t) \stackrel{\zP}{=} (-1)^{H'-H} T_{-H'\, -H}(s, t) \stackrel{\zT}{=} (-1)^{H'-H} T_{HH'}(s, t).
\eeq

The six independent amplitudes are usually chosen as follows \cite{Hearn:1962zz}:
\beq
\begin{aligned}
( 8\pi s^{1/2})\,  \Phi_1 &\equiv    T_{-\nh\, -\nh} =  T_{+\nh\, +\nh},\\
( 8\pi s^{1/2})\,  \Phi_2 &\equiv    T_{-\nh\, +\nh}  =  - T_{+\nh\,-\nh}, \\
( 8\pi s^{1/2})\,  \Phi_3 &\equiv   T_{-\nh\,+\nth} = T_{+\nh\,-\nth} = T_{+\nth\, -\nh}=T_{-\nth\, +\nh},\\
( 8\pi s^{1/2})\,  \Phi_4 &\equiv   T_{-\nh\,-\nth} = -T_{+\nh\,+\nth} = T_{+\nth\, +\nh}=-T_{-\nth\, -\nh},\\
( 8\pi s^{1/2})\,  \Phi_5 &\equiv   T_{+\nth\, +\nth} = T_{-\nth\,-\nth},\\
( 8\pi s^{1/2})\,  \Phi_6 &\equiv   T_{-\nth\, +\nth} = - T_{+\nth\,-\nth}.
\end{aligned}
\eeq
Their normalization is chosen such that, given the unpolarized cross section 
element, 
\beq
\eqlab{csdef}
\dd\si^\text{unpol.} = \frac{\dd t}{16\pi (s-M^2)^2} \,\frac14  \sum_{H H'} \big| T_{H'H}\big|^2 \,,
\eeq
the unpolarized angular distribution in the center-of-mass frame is:
\begin{equation}
\eqlab{csexpr}
\frac{\dd\si^\text{unpol.}}{\dd\varOmega_{cm}} = \frac{1}{64\pi^2 s} \, \frac14 \sum_{H H'} \big|T_{H'H}\big|^2
= \half |\Phi_1|^2 + \half |\Phi_2|^2 + |\Phi_3|^2 +  |\Phi_4|^2 +
\half |\Phi_5|^2 + \half |\Phi_6|^2 .
\end{equation}
Incidentally, the conversion to the lab frame goes as:
\begin{equation}
\frac{\dd\si}{\dd\varOmega_L } = \frac{\dd\varOmega_{cm}}{\dd\varOmega_L}
\frac{\dd\si}{\dd\varOmega_{cm}}  = \Big(\frac{\nu'}{\nu}\Big)^2\!\frac{s}{M^2}
\frac{\dd\si}{\dd\varOmega_{cm}}.
\end{equation}
As a simple illustration consider the RCS in tree-level QED, 
described by the following Feynman amplitude:
\begin{wrapfigure}[6]{l}{6.5cm}
  \centering
  \includegraphics[scale=0.17]{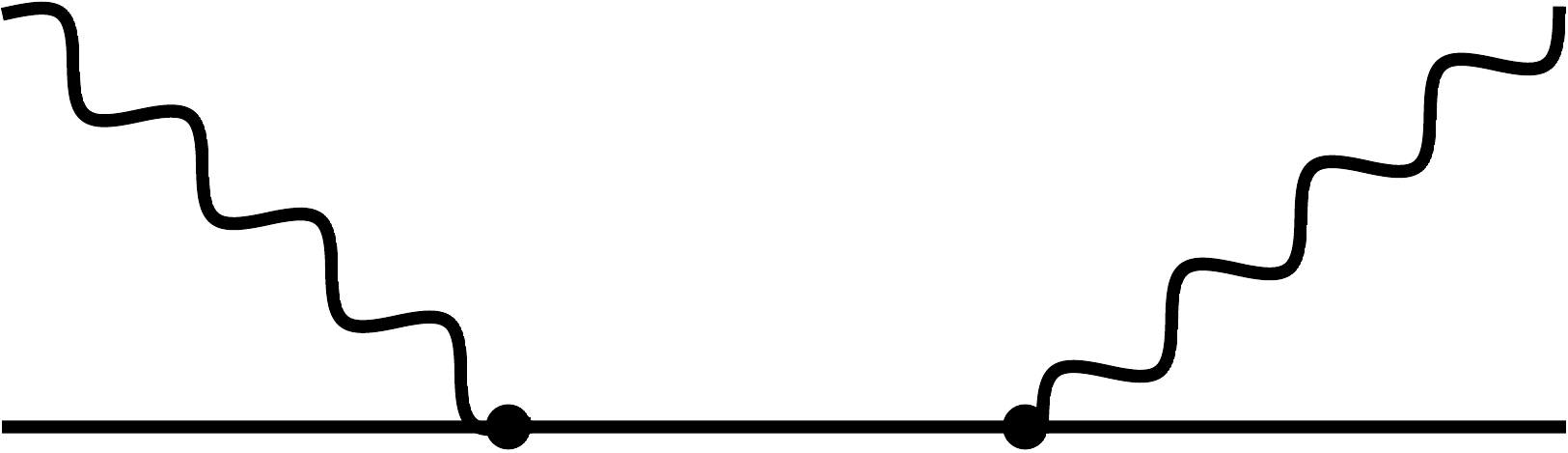}
   \hfill  \includegraphics[scale=0.17]{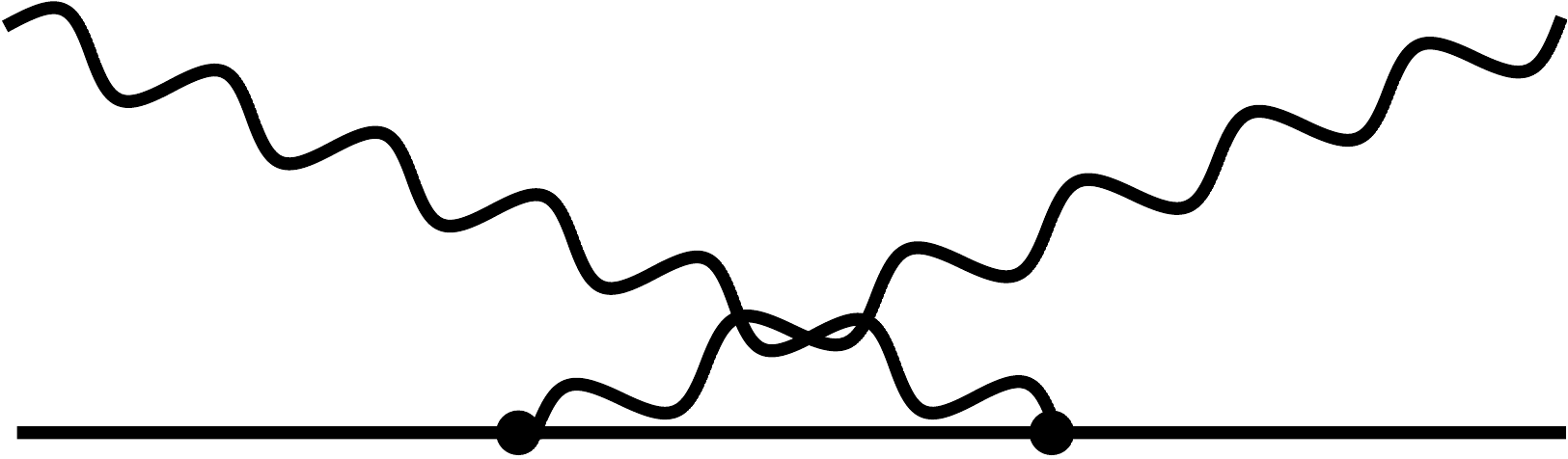}
  	\caption{Tree-level CS graphs.}  
	\figlab{TreeLevelCS}
\end{wrapfigure} 
\beq
T^{(1)\mu\nu} = -e^2\left(\ga^\mu\frac{\slashed{p}+\slashed{q}+M}{s-M^2} \ga^\nu
+ \ga^\nu\frac{\slashed{p}\,'-\slashed{q}+M}{u-M^2} \ga^\mu \right).
\eeq
Using \Eqref{helamp} and definitions from \Secref{definitions}, 
one easily obtains the following 
expressions for the helicity amplitudes:\footnote{This, 
up to a phase convention which flips the sign of $\Phi_{2,4,6}$, agrees with 
\citet[Eq.~(5)]{Tsai:1972vn}.}
\beq
\begin{aligned}
8\pi s^{1/2}\,  \Phi_1^{(1)} &\equiv T_{+1\,+\nh\;+1\,+\nh}^{(1)}=-\pi\al \frac{(M\eta -2t \nu) \eta^{1/2}}{
M\nu^2\nu'}  ,\\
8\pi s^{1/2}\,  \Phi_2^{(1)}&\equiv T_{-1\,-\nh\;+1\,+\nh}^{(1)}=\pi\al \frac{(-t)^{3/2}}{\nu^2\nu'}   , \\
8\pi s^{1/2}\,  \Phi_3^{(1)}&\equiv T_{-1\,+\nh\;+1\,+\nh}^{(1)}= \pi\al \frac{t\eta^{1/2}}{\nu^2\nu'} , \\
8\pi s^{1/2}\,  \Phi_4^{(1)}&\equiv T_{+1\,-\nh\;+1\,+\nh}^{(1)}=\pi\al\frac{ (-t)^{1/2}\eta}{\nu^2\nu'}, \\
8\pi s^{1/2}\,  \Phi_5^{(1)}&\equiv T^{(1)}_{-1\,+\nh\;-1\,+\nh}=-\pi\al \frac{\eta^{3/2}}{\nu^2\nu'}   , \\
 8\pi s^{1/2}\,  \Phi_6^{(1)} &\equiv  T_{-1\, +\nh\; +1\, -\nh}^{(1)}= -\pi\al \frac{(-t)^{3/2} s}{M^2\nu^2\nu'} ,
\end{aligned}
\eeq
where the superscript on the amplitudes indicates that they are of first order in $\al$, and the kinematical invariants are:
\begin{equation}
\eqlab{invdefs1}
\eta=\frac{M^4-su}{M^2}, \quad \nu =\frac{s-M^2}{2M}, \quad \nu'=\frac{M^2-u}{2M}.
\end{equation}
Substituting these expressions in \Eqref{csexpr}, one
obtains the Klein-Nishina cross section~\cite{Klein:1929}:
\begin{equation}
\frac{\dd\si^{\mathrm{KN}}}{\dd\varOmega_L }=
\frac{\al^2}{2M^2} \Big(\frac{\nu'}{\nu}\Big)^2
\left(\frac{\nu'}{\nu}+\frac{\nu}{\nu'}-\sin^2\vartheta \right),
\end{equation}
where  $\sin^2\vartheta
= 1-(1+t/2\nu\nu')^2 = -  t \eta/(2\nu\nu')^2$.

The same steps can be done for polarized observables. For instance, for the linearly polarized
photon beam we have:
\beq
\eqlab{beamas}
\frac{\dd\si}{\dd\varOmega_{cm} } \Si_3  \equiv  
\frac12 \left( \frac{\dd\si_{||}}{\dd\varOmega_{cm}} - \frac{\dd\si_{\perp}}{\dd\varOmega_{cm}}\right)
= -\re \big[ (\Phi_1+\Phi_5)\Phi_3^\ast +(\Phi_2-\Phi_6)\Phi_4^\ast\big],
\eeq
where $\Si_3$ is the beam asymmetry. The tree-level QED result is given by:
\beq
\frac{\dd\si_{||}}{\dd\varOmega_L} - \frac{\dd\si_{\perp}}{\dd\varOmega_L}   = -\frac{\al^2}{M^2} \left(\frac{\nu'}{\nu}\right)^2 \sin^2\vartheta.
\eeq
 This result holds  (in tree-level QED) for RCS on a particle with spin  $0, \,1/2,\, 1$, and it might hold for higher spins as well.
Other polarized observables are considered in \Secref{polO}.
 
\subsection{Multipole Expansion}
A better use of the rotational and discrete symmetries is made by the partial-wave
expansion in the center-of-mass system: 
\begin{subequations}
\bea
T_{H'H}(\w, \th) &= & \sum_{J=\nh}^\infty (2J+1) \, T^J_{H'H}(\w) \, d_{HH'}^J (\th) ,\\
T^J_{H'H}(\w) &=& \frac12 \int_{-1}^1 \dd(\cos \th)\, T_{H'H}(\w, \th)\, d_{HH'}^J (\th),
\eea
\end{subequations}
where $J$ is total angular momentum and
$d^J(\th) $ are the Wigner $d$-functions.\footnote{For the $d$-functions we use the conventions of Edmonds
(also used by Davydov,
or Varshalovich). In the other popular convention (Rose, Wigner or Landau and Lifshitz) the sign of $\th $ is the opposite and hence, due to the property $d_{HH'}^J (-\th) = d_{H'H}^J (\th)$, the helicities appearing on the $d$-functions would be interchanged.}  The partial wave-amplitudes
$T^J$ and the $d$-functions satisfy the symmetry relations \eref{symrels} separately. 

Assuming the parity to be a good quantum number, it is convenient to form the partial-wave amplitudes with definite $J^\zP$:
\begin{subequations}
\bea 
\cA^{J}_\zp &=& \Phi_1^J + \zp\, \Phi_2^J = \frac{1}{8\pi s^{1/2}} \Big( 
T^J_{+\nh\, +\nh} - \zp\, T^J_{+\nh\, -\nh} \Big),\\
\cB^{J}_\zp &=& - \zp\, \Phi_3^J -  \Phi_4^J = \frac{1}{8\pi s^{1/2}} \Big( 
T^J_{+\nh\, +\nth} - \zp\, T^J_{+\nh\, -\nth} \Big) ,\\
\cC^{J}_\zp &=& \Phi_5^J + \zp\, \Phi_6^J =\frac{1}{8\pi s^{1/2}} \Big( 
T^J_{+\nth\, +\nth} - \zp\, T^J_{+\nth\, -\nth} \Big),
\eea 
\end{subequations}
where $\zp = \pm$ is the parity eigenvalue. Note that in the above partial-wave expansion neither $H$ nor $H'$ exceed
$J$, hence the amplitudes $\cB$ and $\cC$ are only defined for $J\geq \nth$. 

The conventional multipole amplitudes ({\em multipoles} for short) are denoted as \cite{Ritus:1957,Pfeil:1974ib}
\beq
f^{\ell \mp }_{\rho'\rho}(\w), \,\,\mbox{with $\ell = J \pm \nh  $,  and $ \rho', \rho = E,\,  M$}.
\eeq 
The combination of $\rho$ and $\ell$ reflects the photon multipolarity 
(e.g., $E1$ the electric dipole).
The multipoles are expressed in terms of the partial-wave amplitudes as follows.
\begin{subequations}
\begin{description}
\item{For $J=\nh$:}
\begin{equation}
f_{\substack{ EE \\ MM }}^{1-} = \frac12 \cA^{J=\nh}_\pm, \quad
f_{\rho'\rho}^{0+} =0. 
\end{equation}
\item{For $J\geq \nth$:}
\bea 
f_{\substack{ EE \\ MM }}^{(J+\nh)-} &=& \frac{2}{(2J+1)^2}\left( \cA^{J}_\pm
- 2\mbox{$\sqrt{\frac{J-\nh}{J+\nth}}$} \, \cB^{J}_\pm + 
\mbox{$\frac{J-\nh}{J+\nth}$}\,  \cC^{J}_\pm \right), \\
f_{\substack{ EE \\ MM }}^{(J-\nh)+} &=& \frac{2}{(2J+1)^2}\left( \cA^{J}_\mp
+ 2\mbox{$\sqrt{\frac{J+\nth}{J-\nh}}$} \, \cB^{J}_\mp + 
\mbox{$\frac{J+\nth}{J-\nh}$}\,  \cC^{J}_\mp \right),\\
f_{\substack{ EM \\ ME }}^{(J-\nh)+} &=& \frac{2}{(2J+1)^2}\left( -\cA^{J}_\mp
- \mbox{$\frac{2}{\sqrt{(J+\nth)(J-\nh)}}$} \, \cB^{J}_\mp +  \cC^{J}_\mp \right).
\eea 
\end{description}
\end{subequations}
The inverse relation can be written as (in shorthand notation, $f_{EE\pm MM} = f_{EE}\pm f_{MM}$):
\begin{subequations}
\bea 
\Phi_{\substack{ 1\\ 2}}^{J} &=&\quarter  \big\{ (J+\nth)^2
f_{EE\pm MM}^{(J+\nh)-} \pm (J-\nh)^2 f_{EE\pm MM}^{(J-\nh)+}
\mp 2 (J+\nth)(J-\nh) f_{EM\pm ME}^{(J-\nh)+}\big\},\\
\Phi_{\substack{ 3\\ 4}}^{J} &=&\quarter  \sqrt{(J+\nth)(J-\nh)} \big\{ (J+\nth)
f_{EE\mp MM}^{(J+\nh)-} \pm (J-\nh) f_{EE\mp MM}^{(J-\nh)+}
\mp 2 f_{EM\mp ME}^{(J-\nh)+}\big\},\\
\Phi_{\substack{ 5\\ 6}}^{J} &=&\quarter (J+\nth)(J-\nh) \big\{
f_{EE\pm MM}^{(J+\nh)-} \pm f_{EE\pm MM}^{(J-\nh)+}
\pm 2 f_{EM\pm ME}^{(J-\nh)+}\big\}.
\eea 
\end{subequations}

As an illustration we once again consider the tree-level QED, and for 
greater simplicity take the zero-energy limit, $\w=0$. According 
to the low-energy theorem (LET) \cite{Thirring:1950fj, Low:1954kd,Gell_Mann:1954kc}, the tree-level result in this limit
is exact and we may omit the label indicating the order of $\al$.
The zero-energy helicity amplitudes are thus given by:
\beq
\Phi_{\substack{ 1\\ 2}} (0,\th) = \mp\mbox{$\frac{\al}{M} \left(\frac{1}{2}\pm \frac{1}{2} \cos\th\right)^{\nth}$}\!, \,
\Phi_{\substack{ 3\\ 4}} (0,\th) 
=\mbox{$\frac{\al}{M} \left(\frac{1}{2}\mp \frac{1}{2} \cos\th\right) \left(\frac{1}{2}\pm \frac{1}{2} \cos\th\right)^{\nh}$}\!, 
\, \Phi_{\substack{ 5\\ 6}} (0,\th)  
=\pm \Phi_{\substack{ 1\\ 2}}  (0,\th),
\eeq
while the non-vanishing partial-wave amplitudes are:
\begin{subequations}
\bea
\Phi_{\substack{ 1\\ 2}}^{J=\nh}  (0) &=& 
\mp\frac{\al}{2M} \int_{-1}^1 \dd(\cos \th)\, \left(\frac{1\pm\cos\th}{2}\right)^{3/2}\! d_{+\nh\, \pm\nh}^{\nh} (\th) = -\frac{\al}{3M}, \quad \Phi_{\substack{ 1\\ 2}}^{J=\nth}  (0)=\mp\frac{\al}{12M}, \\
\Phi_{\substack{ 3\\ 4}}^{J=\nth}  (0) &=& \mp
\frac{\al}{2M} \int_{-1}^1 \dd(\cos \th)\,\frac{1\mp\cos\th}{2} \left(\frac{1\pm\cos\th}{2}\right)^{1/2}\! d_{+\nh\, \mp\nth}^{\nth} (\th) = \mp \frac{\al}{4\sqrt{3}M},\\
\Phi_{\substack{ 5\\ 6}}^{J=\nth}  (0) &=& -
\frac{\al}{2M} \int_{-1}^1 \dd(\cos \th)\, \left(\frac{1\pm\cos\th}{2}\right)^{3/2}\! d_{+\nth\, \pm\nth}^{\nth} (\th) =\mp \frac{\al}{4M}.
\eea
\end{subequations}
The parity-conserving amplitudes assume the following values:
\beq
\cA^{J=\nh}_\pm = \frac{\al}{M} \left( -\boxfrac{2}{3} \atop 0 \right), \, 
\cA^{J=\nth}_\pm = \frac{\al}{M} \left(0\atop  -\sixth \right), \, 
\cB^{J=\nth}_\pm = \frac{\al}{M} \left(0\atop  -\boxfrac{1}{2\sqrt{3}} \right),\,
\cC^{J=\nth}_\pm = \frac{\al}{M} \left(0\atop  -\half \right),
\eeq
and as the result, there are only two non-vanishing multipoles which happen to be equal (at $\w=0$):
\beq
f_{EE}^{1-}(0) = f_{EE}^{1+}(0) = -\, \frac{\al}{3M}.
\eeq

\subsection{Tensor Decompositions}
For various microscopic calculations, as well as for the general low-energy expansion (LEX), it is convenient to isolate the Lorentz
structure of the amplitude by decomposing it  into a set of tensors. 
There are several 
neat decompositions described in the literature, we only consider two of them here. 
The first --- perhaps the earliest one --- is the following decomposition into
a non-covariant set of $6$ (minimal number) tensors:
\begin{subequations}
\eqlab{helAmp}
\beq
\bar u^{\,\prime} (\veps' \cdot T \cdot \veps)  u =  2 M e^2 \,  \hat A^T (s, t)\, 
 \chi^{\,\prime} \veps_i' \, \hat O_{ij}\,  \veps_j\,  \chi, 
\eeq 
with $\hat A$ and $\hat O$ being respectively the
arrays of the scalar complex amplitudes and tensors:
\bea
\hspace{-1.5cm}&& \hat A(s,t) = \big\{A_1, \, \cdots, \, A_6 \big\} (s,t), \\
\hspace{-1.5cm}&& \hat O_{ij} = \big\{ \de_{ij}, \, n_i n_j',\, i \eps_{ijk} \si_k,\, \de_{ij}
 i \eps_{klm} \si_k n_l' n_m , \,i \eps_{k lm}\si_k ( \de_{il} n_m n_j' -  
 \de_{jl} n_i n_m'), \,i \eps_{k lm}\si_k ( \de_{il} n_m' n_j' -  
 \de_{jl} n_i n_m) \big\},\qquad
\eea
where $\bn$ and $\bn'$ are the directions of the incoming and outgoing photons.
\end{subequations}

The second decomposition considered here is a covariant, overcomplete  set of 8  tensors~\cite{Pascalutsa:2003aa}: 
\begin{subequations}
\eqlab{CovarAmp}
 \beq
 \bar u^{\,\prime} (\veps' \cdot T \cdot \veps)  u  = e^2 \hat \scA^T (s,t)\, \bar u^{\,\prime} \hat \scO^{\mu \nu}  u \,  \mathcal{E}_{\mu}^{\prime }  \mathcal{E}_{\nu},
 \eeq
 with\footnote{Here we correct the typos of Refs.~\cite{Pascalutsa:2003aa,McG13} made in the expressions
for $\scO_6$ and $\scO_8$, respectively.} 
 \bea
\hspace{-1.5cm}&& \hat \scA(s,t) = \big\{\scA_1, \, \cdots, \, \scA_8 \big\} (s,t), \\
\hspace{-1.5cm}&& \hat \scO^{\mu\nu} = \big\{ -g^{\mu \nu}, \; q^{\mu} q^{\prime\,\nu},\; 
-\gamma^{\mu \nu},\; g^{\mu \nu} (q' \cdot \gamma \cdot q),\; q^{\mu} q'_{\alpha} \gamma^{\alpha \nu}-\gamma^{\alpha \mu} q_{\alpha} q'^{\nu},\;q^{\mu} q_{\alpha} \gamma^{\alpha \nu}-\gamma^{\alpha \mu} q'_{\alpha} q'^{\nu},\nn \\  
\hspace{-1.5cm}&& \qquad\qquad 
q^{\mu} q^{\prime\,\nu}(q' \cdot \gamma \cdot q) ,\; 
- i \gamma_5 \epsilon^{\mu \nu \alpha \beta} q'_{\alpha} q_{\beta}\big\}, \\
\hspace{-1.5cm}&& \mathcal{E}_{\mu}  = \veps_{\mu} - \frac{P \cdot \veps}{P \cdot q} \, q_{\mu},
\; \mathcal{E}_{\mu}'  = \veps_{\mu}' - \frac{P \cdot \veps'}{P \cdot q} \, q_{\mu}',
\; P_\mu = \half (p + p')_\mu, \; P\cdot q=P\cdot q' = M \xi.
 \eqlab{epsilon_b}
\eea 
\end{subequations}
This decomposition is manifestly gauge-invariant, because the vectors $\mathcal{E}$ are.
It can be reduced \cite{Lensky:2015awa} to any of the covariant sets with the minimum number of tensors, such as that of \citet{Hearn:1962zz} or L'vov \cite{Bab98}. Nevertheless, the overcomplete set is better suited for practical calculations of Feynman diagrams using computer algebra, since 
simple Gordon-like identities are sufficient for the decomposition. Another advantage is that it readily applies to the forward VVCS case, see \Secref{VVCS}.

The correct relation between the amplitudes $A_{1,\ldots,6}$ and $\scA_{1,\ldots,8}$ was given by \citet{McG13}: 
\beq
\begin{aligned}
\eqlab{eq:relations_Ai_covarAi}
A_1& = \frac{\eps_B}{M} \scA_1 + \frac{\omega_B t }{2 M} \scA_4, \\
A_2 &= \frac{\eps_B \w_B^2 }{M} \scA_2 + \frac{\omega_B^3}{M} \left(
\scA_5 + \scA_6  -  \half t \scA_7 \right),  \\
A_3 &=  \frac{\eps_B }{M} \scA_3 - \frac{M^2 \eta\, t}{4M^2-t} 
\left(\frac{\scA_5 + \scA_6 }{2M(\eps_B +M)} - \scA_7\right) - 
\frac{\omega_B t }{2 M} \scA_8, \\
A_4 &= \w_B^2 \scA_4, \\
A_5 &= \w_B^2 \scA_5 +  \frac{\w_B^2}{2M (\eps_B + M)} \Big[\half \scA_3 
+ \frac{M^2\eta}{4M^2-t}  \left(\scA_5 + \scA_6 \right) \Big] 
-\w_B^2 (\omega_B^2 + \half t) \scA_7 + \frac{\omega_B^3}{2 M} \scA_8,  \\
A_6 &= \w_B^2 \scA_6 -  \frac{\w_B^2}{2M (\eps_B + M)} \Big[\half \scA_3 
+ \frac{M^2\eta}{4M^2-t}  \left(\scA_5 + \scA_6 \right) \Big] + \omega_B^4 \scA_7 - \frac{\omega_B^3}{2 M} \scA_8, 
\end{aligned}
\eeq
where $\eps_B$ and $\w_B$ are the nucleon and photon energies in the 
Breit frame (defined by $\bp'=-\bp$). These kinematical variables, along with $\eta$,
can be expressed in terms of Mandelstam invariants, cf.~\ref{sec:appindex}. Thus, although obtained in the Breit frame, this relation is 
Lorentz invariant.

Both sets of amplitudes have a definite parity under the photon crossing
(i.e., $\veps\leftrightarrow \veps'$, $q\leftrightarrow q'$, hence $s\leftrightarrow u$,
etc.). Writing the amplitudes as functions of $\xi$ and $t$, the crossing symmetry implies:
\begin{subequations}
\bea 
 &&A_{1,2} (-\xi, t) = A_{1,2} (\xi, t), \quad\; A_{3,\ldots,6} 
(-\xi, t) = -A_{3,\ldots,6}  (\xi, t),\\
 &&\scA_{1,2,8} (-\xi, t) = \scA_{1,2,8} (\xi, t), \quad \scA_{3,\ldots,7} 
(-\xi, t) = -\scA_{3,\ldots,7} (\xi, t).
\eea
\end{subequations}

\subsection{Unitarity Relations}
\subsubsection{Optical Theorem}
Derived from unitarity, the optical theorem establishes the relation
between the imaginary part of the forward CS amplitude and the total
photoabsorption cross section, and in our case of the nucleon target reads:
\beq 
\im T_{H'H}(\nu, 0) = \nu \, \si_H (\nu) \, \de_{HH'},
\eeq
where $\de_{HH'}$ is the Kronecker symbol. The cross section
$\si_H$ corresponds with 
the absorption of a circularly polarized photon on
a longitudinally polarized target, with their combined helicity given by $H$.
In terms of the invariant amplitudes we have:
\begin{subequations}
\bea 
&&\im A_1(\nu,0) = \im \scA_1(\nu,0) = \frac{\nu}{4\pi\al}  \si_T(\nu),  \\
&&\im A_3(\nu,0) = \im \scA_3(\nu,0) = \frac{\nu}{4\pi\al}
\si_{TT}(\nu),  
\eea
\end{subequations}
where $\si_T = (\si_{1/2} + \si_{3/2})/2$ is the unpolarized cross section and $\si_{TT} = (\si_{1/2} - \si_{3/2})/2$.
The remaining amplitudes do not contribute in the forward scattering, and as such
are not constrained by the optical theorem.

\subsubsection{Watson's Theorem Extended}

 At low energies there are further unitarity constraints for the nucleon CS. They are less strict, since they hold in a limited energy range and to leading order in $\al$. At the same time, they are more stringent, since they 
apply to all the multipole 
 and partial-wave amplitudes, and hence are not limited to the forward kinematics. 

Below the two-pion threshold, one is limited to the channel space
spanned by  the $\pi N$ and $\ga N$ states, and hence the following four processes:
\beq
\begin{array}{ll}
\pi N\ra \pi N ,\,\,  & \pi N \ra\ga N, \,\,\\
\ga N \ra \pi N, \,\,& \ga N \ra \ga N .
\end{array}  
\eeq
To have exact unitarity, in this channel space, we set up a 
linear coupled-channel integral equation:
\beq
\eqlab{channel}
\bmat T_{\pi\pi} & T_{\pi \ga}\\
T_{\ga \pi} & T_{\ga\ga} \emat
=\bmat V_{\pi\pi} & V_{\pi \ga}\\
V_{\ga \pi} & V_{\ga\ga} \emat +
 \bmat V_{\pi\pi} & V_{\pi \ga}\\
V_{\ga \pi} & V_{\ga\ga} \emat
\bmat G_{\pi} & 0\\
0& G_{\ga} \emat
\bmat T_{\pi\pi} & T_{\pi \ga}\\
T_{\ga \pi} & T_{\ga\ga} \emat,
\eeq 
where $T$ and $V$ are suitably normalized amplitudes and 
potentials of pion-nucleon  scattering ($\pi\pi$), 
pion photoproduction ($\pi\ga$),
absorption ($\ga\pi$), and nucleon CS ($\ga\ga$). 
The propagators $G_\pi$ and $G_\ga$ are, respectively, 
the pion-nucleon and photon-nucleon two-particle propagators.
With the assumption of hermiticity of the potential and time-reversal
symmetry, which relates the $\ga\pi$ and $\pi\ga$
amplitudes, the above equation leads to the unitary
$S$-matrix, $S_{fi}=\de_{fi}+2i T_{fi}$.

Neglecting the iterations of the potential involving photons 
(which amount to small radiative corrections), the coupled-channel
equation reduces to:
\begin{subequations}
\bea
\eqlab{trunc}
\eqlab{eq1}
T_{\pi \pi} &=& V_{\pi\pi} + V_{\pi\pi} G_{\pi} T_{\pi\pi},\\
\eqlab{eq2}
T_{\pi\ga} &=& V_{\pi\ga} + T_{\pi\pi} G_{\pi} V_{\pi\ga},\\
T_{\ga\pi} &=& V_{\ga\pi} + V_{\ga\pi} G_{\pi} T_{\pi\pi},\\
T_{\ga\ga} &=& V_{\ga\ga} + V_{\ga\pi} G_{\pi} T_{\pi\ga}. 
\eea
\end{subequations}
Only the first of these is an integral equation, the rest are obtained
by a one-loop calculation. 

After the partial-wave expansion, the solution for the
$\pi N$ amplitude can be written as:
\beq 
T_{\pi \pi }^{IJ\zp} = \frac{ K^{IJ\zp} }{1 - i \rk  K ^{IJ\zp}} = 
\frac{1}{\rk} e^{i \de^I_{\ell\zp}}
\sin \de^I_{\ell\zp}, 
\eeq 
where $K^{IJ\zp}$ is the `$K$-matrix' with definite isospin $I$, total angular momentum $J$, and parity $\zp$; the corresponding $\pi N$ phase-shift is $\de^I_{\ell\zp} = \arctan (\rk K^{IJ\zp}) $, which is a function of the 
$\pi N$ relative momentum $\rk$. The latter is given by
\beq 
\rk= \frac{1}{2s^{1/2} } \big[ \big(s-(M+m_\pi)^2\big)\, \big((s-(M-m_\pi)^2\big) \big]^{1/2}. 
\eeq
Note that we have neither specified $V_{\pi\pi}$,  
nor solved \Eqref{eq1};  we have merely written it
in the  manifestly unitary form. 

Continuing to the pion photoproduction channel, we obtain the statement of
the celebrated Watson's theorem \cite{Watson:1954uc}:
\beq
T^{IJ\zp}_{\ga\pi} = \big|T^{IJ\zp}_{\ga\pi}\big|\, e^{i\de^I_{\ell\zp}} ,
\eeq
with $\ell = J-\nicefrac{\zp}{2}$.
The phase of the photoproduction amplitudes is thus identical to the $\pi N$ phase-shift, for each set of good quantum numbers.

Extending these arguments to the Compton channel, we obtain
\beq 
\eqlab{WatsonRCS}
T_{\ga\ga}^{J\zp} = V_{\ga\ga}^{J\zp} + \sum_{I=1/2}^{3/2} \big|T^{IJ\zp}_{\ga\pi}\big|^2\,\big(
-\tan \de^I_{\ell\zp} + i \big).
 \eeq 
There are two interesting results here. The first is that the imaginary part
of the partial-wave RCS amplitude is given by the isospin sum of
the photoproduction amplitudes squared. In this case the sum over the isospin
is equivalent to the sum over the charged states, hence, e.g., for the proton
\beq 
\im T_{\ga p\to \ga p}^{J\zp} = \sum_{I=1/2}^{3/2} \big|T^{IJ\zp}_{\ga\pi}\big|^2  = 
\big|T^{J\zp}_{\ga p \to \pi^0 p }\big|^2 + 
\big|T^{J\zp}_{\ga p \to \pi^+ n }\big|^2\,.
\eeq

The second result concerns the $\De(1232)$ resonance, which is the only
resonance occurring between the one- and two-pion production thresholds.
Recall that, in the $\pi N$
scattering, this resonance occurs in the P$33$ partial wave (i.e., $I=3/2=J$, $\ell = 1$, $\zp=+$). 
The position, $M_\De$,  of such an elastic
resonance is identified with the phase-shift crossing $90^{\circ}$.
This means the tangent terms in \Eqref{WatsonRCS} blow up and can only
be canceled by a singularity in $V_{\ga\ga}$. 
Near the resonance position the $K$-matrix takes the form
\beq 
K^{\mathrm P33} \approx \frac{M_\De \varGamma_\De}{\rk (s - M_\De^2) },
\eeq
where $\varGamma_\De$ is the resonance width, and hence the
cancellation is achieved when
\beq 
\lim_{s\to M_\De^2} \big[ (s-M_\De) V_{\ga\ga}^{3/2+}\big] = 
M_\De \varGamma_\De \lim_{s\to M_\De^2} \big|T^{\mathrm P33}_{\ga\pi}\big|^2 .
\eeq
Thus, unitarity provides a stringent relation among the $\De(1232)$-resonance 
parameters occurring in the different processes. As a consequence, the 
$\De(1232)$-resonance contribution to polarizabilities is constrained too.

These results apply as well to the multipole amplitudes.
In particular, the imaginary parts of the Compton multipoles, between
the one- and two-pion production thresholds, are given by the 
pion photoproduction multipoles:
\begin{subequations}
\bea
&& \im f_{EE}^{\ell\pm} = \rk
\sum\nolimits_c\big|E^{(c)}_{(\ell\pm 1)\mp}\big|^2,  \quad \im f_{MM}^{\ell\pm}=\rk\sum\nolimits_c\big|M^{(c)}_{\ell\pm}\big|^2\,,
\eqlab{mult-f1}\\
& & \im f_{EM}^{(\ell\pm 1) \mp}=  \im f_{ME}^{\ell\pm}=\mp\, \rk\sum\nolimits_c\re\big(E^{(c)}_{\ell\pm}M^{(c)*}_{\ell\pm}\big)\,,
\eqlab{eq:mult-f2}
\eea
\end{subequations}
where the sum is over the charged $\pi N$ states, i.e: $c=\pi^0 p,\,\pi^+n$
and $c=\pi^0 n,\,\pi^- p$ for the proton and neutron RCS, respectively.
As mentioned above, an equivalent result is obtained by summing over the isospin
states.

\subsection{Expansion in Static vs.\
Dynamic Polarizabilities }

The celebrated LET
for RCS~\cite{Thirring:1950fj, Low:1954kd,Gell_Mann:1954kc}
can be extended to include higher-order terms, parametrized in 
terms of polarizabilities~\cite{Klein:1955qy}. It is customary to separate out the 
Born contribution by writing
\beq
T = T^{\mathrm{Born}} + \ol T,
\eeq
such that  $T^{\mathrm{Born}}$ is the Born contribution specified in~\ref{sec:appBorn}. In the low-energy limit, it yields the classic LET. 
The rest (non-Born), $\ol T$, is expanded in powers of energy
with coefficients given by {\it static} polarizabilities. For example,
the LEX of the non-Born part of the 6 invariant amplitudes of the
decomposition \eref{helAmp}
goes as
(in the Breit frame): 
\beq
\label{eq:anb}
\begin{aligned}
\al \bar A_1(\w_B ,t) &=   \omega_B^2 \big[ \alpha_{E1}+ \beta_{M1} + 
\omega_B^2\, (\alpha_{E1\nu}+ \beta_{M1\nu})\, \big] + \half t 
\big( \beta_{M1} + \w_B^2 \beta_{M1\nu} \, \big) \\
&+\;  \w_B^4 \boxfrac{1}{12} ( \alpha_{E2}+\beta_{M2}) 
+\half t (4\w_B^2+t)\boxfrac{1}{12} \beta_{M2}+ O(\omega_B ^6), \\ 
\al \bar A_2 (\w_B ,t)&=   - \w_B^2 \big( \beta_{M1} + \w_B^2 \beta_{M1\nu}
\big) +
\w_B^4 \boxfrac{1}{12} ( \alpha_{E2}-\beta_{M2}) -
t \w_B^2\boxfrac{1}{12} \beta_{M2}+ O(\omega_B^6),\\
\al \bar A_3(\w_B ,t) &= -  \omega_B^3 \, \big[\gamma_{E1E1}+\gamma_{E1M2} + z\, (\gamma_{M1E2}+\gamma_{M1M1})  \big] + O(\omega_B^5),  \\ 
\al \bar A_4(\w_B ,t) &=  \omega_B^3\, (\gamma_{M1E2}-\gamma_{M1M1})   + O(\omega_B^5),   \\
\al \bar A_5(\w_B ,t) &=   \omega_B^3 \, \gamma_{M1M1}+ O(\omega_B^5),\\
\al \bar A_6 (\w_B ,t) &= \omega_B^3 \,  \gamma_{E1M2}+ O(\omega_B^5).
\end{aligned}
\eeq
Certainly, the convergence
radius of such a Taylor expansion is limited by the first singularity, which in the nucleon case is set by the pion-production branch cut (neglecting the small effects
from radiative corrections). 
An expansion which  extends beyond the pion-production threshold is the multipole expansion.  The relation between the two expansions (i.e., polarizability vs.\
multipole) is as follows.

One can divide out the Born contribution in the multipole amplitudes,
$f = f^{\mathrm{Born}} + \bar f$. The non-Born part of the multipoles is then used
to define the {\em dynamic} polarizabilities as~\cite{Guiasu:1978dz}:
\begin{subequations}
\label{eq:pol-f1}
\bea
\left( \begin{array}{c}
\alpha_{E\ell}(\omega) \\ 
\beta_{M\ell}(\omega) \end{array}\right) &=&
\frac{[\ell (2 \ell - 1)!!]^2}{\omega^{2\ell}} \left[ (\ell+1)\bar f^{\ell+}_{\substack{ EE \\ MM }}(\w)+\ell \bar f^{\ell-}_{\substack{ EE \\ MM }}(\w) \right] \eqlab{dynalphabeta},\\
\gamma_{\substack{E\ell E\ell\\ M\ell M\ell} }(\omega)&=&\frac{2 \ell - 1}{\omega^{2\ell+1}}\left[ \bar f^{\ell+}_{\substack{ EE \\ MM}}(\w) - \bar f^{\ell-}_{\substack{ EE \\ MM}}(\w)\right], \eqlab{dyngammaEE}\\
\gamma_{\substack{ E\ell M(\ell+1)\\ M\ell E(\ell+1)}}(\omega)&=&2^{2 - \ell} \frac{(2 \ell + 1)!!}{\omega^{2\ell+1}} \bar f^{\ell+}_{\substack{EM\\ME}}(\w).
\eqlab{dyngammaEM}
\eea
\end{subequations}
\begin{figure}[bt]
\centering 
 \includegraphics[width=0.85\columnwidth]{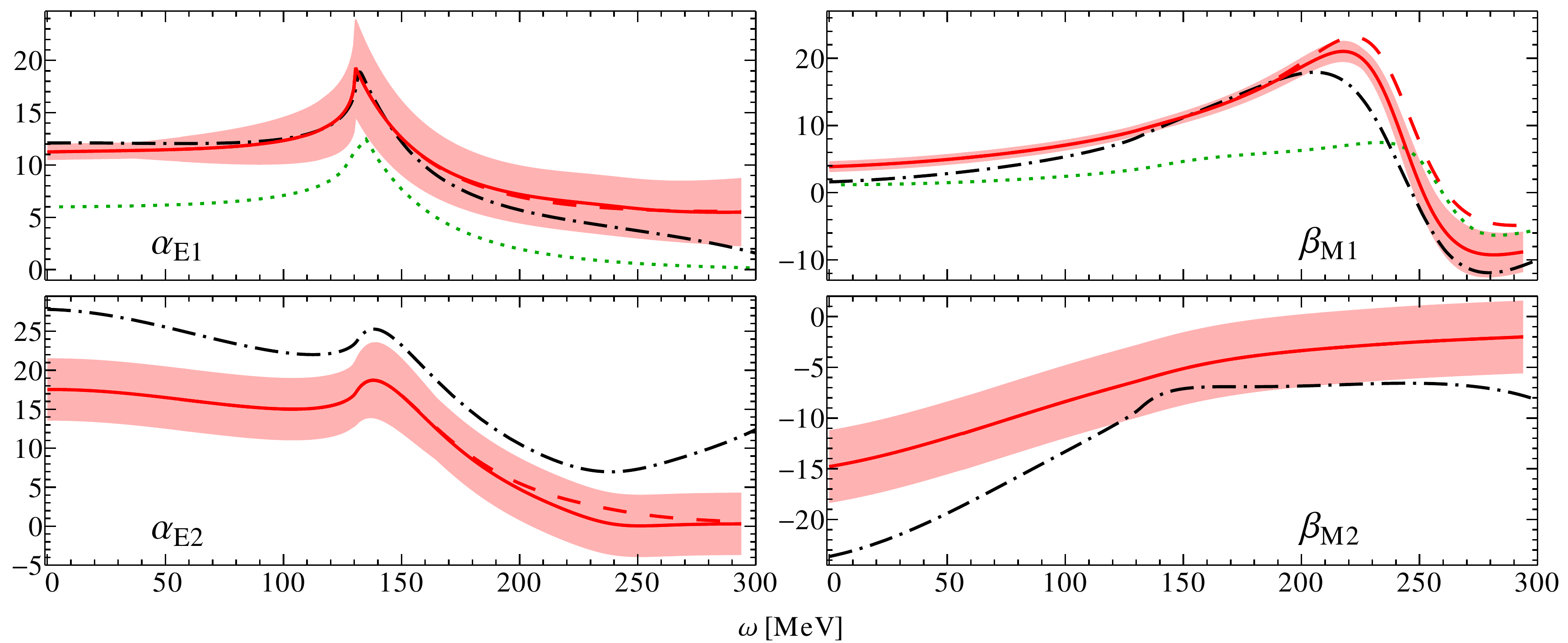}
\caption{  The scalar dipole and quadrupole dynamic polarizabilities of the {\it proton}, in units of $10^{-4}\mathrm{fm}^3$ and
$10^{-4}\mathrm{fm}^5$, respectively. 
The curves are the results of the B$\chi$PT calculation of \citet{Lensky:2015awa}
 (red bands),
compared with the results of the DR calculation of~\citet{Hildebrandt:2003fm} (black dot-dashed)
and with the results of~\citet{Aleksejevs:2010zw,Aleksejevs:2013cda} (green dotted, not shown for the quadrupole polarizabilities).}
\label{fig:pols_scalar}
\end{figure}
 Given that the low-energy behavior of the non-Born part of multipoles goes as
\beq
\bar f^{\ell\pm}_{\substack{ EE \\ MM }}\sim  \omega^{2\ell}, \qquad \bar f^{\ell+}_{\substack{ EM \\ ME }}\sim  \omega^{2\ell+1},
\eeq 
the low-energy limit of \eref{dynalphabeta} and \eref{dyngammaEM} is straightforward and corresponds with the static polarizabilities. 
The limit of \eref{dyngammaEE} needs more care, but the matching 
to the static polarizabilities is possible as well~\cite{Lensky:2015awa}. 
As an illustration,
\Figref{pols_scalar} shows the plots of the scalar
dynamic polarizabilities of the proton from Ref.~\cite{Lensky:2015awa}.

\subsection{Polarized Observables}
\seclab{polO}

Besides the unpolarized differential cross section, given by \Eqref{csdef}, and the linearly-polarized photon beam asymmetry $\Si_3$, \Eqref{beamas}, 
there is a number of observables that depend on polarization of the (nucleon) target. Here we only consider the case when the polarizations of the final particles (scattered photon and 
recoiled nucleon) are not observed.

We consider the photon, traveling along the $z$-axis, with a linear polarization  and 
$P^\gamma_{T}$ at the angle $\phi$ with respect to the
scattering plane $xz$, and the right-handed circular polarization $P^\gamma_{R}$.
The degree of the target polarization along the $x$-, $y$-, $z$-direction 
is denoted as $P_x$, $P_y$, $P_z$ respectively. 
In this case the polarized cross-section element is given by
\bea  
\dd \si & = & \dd \si^{\mathrm{unpol.}} \big[ 1 + P^\gamma_{T}\Si_3 \cos2\phi 
+ P_x \big( P^\gamma_{R} \Si_{2x} + P^\gamma_{T}\Si_{1x} \sin2\phi \big) \nn\\
&&  + \, P_y \big( \Si_{y} + P^\gamma_{T}\Si_{3y} \cos 2\phi \big)
+ P_z \big( P^\gamma_{R} \Si_{2z} + P^\gamma_{T}\Si_{1z} \sin2\phi \big) \big],
\eea
where $\dd \si^{\mathrm{unpol.}}$ stands for the unpolarized cross section and $\Si$'s denote the various asymmetries. This notation for asymmetries is motivated by \citet{Bab98}.  
The conversion to the standard notation adopted in meson photoproduction 
(see, e.g., Appendix A of \citet{Worden1972253}) is as follows:
\bea
&& \Si_3 = -\Sigma,  \quad \Si_{2x} = F, \quad \Si_{1x} = -H,\quad \Si_y= T, \quad \Si_{3y} = -P,  \quad \Si_{2z} = -E, \quad \Si_{1z} = G.
\eea

Similar to \Eqref{beamas} for $\Si_3$, we may express
these asymmetries in terms of the specific polarized cross sections and 
in terms of the helicity amplitudes:\footnote{Here it is important that, 
in the center-of-mass frame, the nucleon travels in the $-z$ direction.
Hence, its polarization along $z$-axis corresponds with helicity $-\nicefrac12$, and so on.}
\begin{subequations}
\bea 
\frac{\dd\si}{\dd\varOmega_{cm} } \Si_{2x}  &\equiv  &\frac12 \bigg( \frac{\dd\si^{R}_{x}}{\dd\varOmega_{cm}} - 
\frac{\dd\si^{L}_{x}}{\dd\varOmega_{cm}}\bigg)=\re [\Phi_4 (\Phi_1-\Phi_5)^{*} - (\Phi_2+\Phi_6) \Phi_3^{*}] ,\\
\frac{\dd\si}{\dd\varOmega_{cm} } \Si_{2z}  &\equiv  &\frac12 \bigg( \frac{\dd\si^{R}_{z}}{\dd\varOmega_{cm}} - 
\frac{\dd\si^{L}_{z}}{\dd\varOmega_{cm}}\bigg)= - \half \big(|\Phi_1|^2 + |\Phi_2|^2 -|\Phi_5|^2- |\Phi_6|^2\big),\\
\frac{\dd\si}{\dd\varOmega_{cm} } \Si_{1x}  &\equiv  &
\frac12 \bigg( \frac{\dd\si^{\nicefrac{\pi}{4}}_{x}}{\dd\varOmega_{cm}} - 
\frac{\dd\si^{-\nicefrac{\pi}{4} }_{x}}{\dd\varOmega_{cm}}\bigg)
= \im [\Phi_1^\ast \Phi_2+ \Phi_5^\ast \Phi_6 ],\\
\frac{\dd\si}{\dd\varOmega_{cm} } \Si_{1z}  &\equiv  &
\frac12 \bigg( \frac{\dd\si^{\nicefrac{\pi}{4}}_{z}}{\dd\varOmega_{cm}} - 
\frac{\dd\si^{-\nicefrac{\pi}{4} }_{z}}{\dd\varOmega_{cm}}\bigg)
= - \im [\Phi_3 (\Phi_1-\Phi_5)^{*} - (\Phi_2+\Phi_6) \Phi_4^{*}], \\
\frac{\dd\si}{\dd\varOmega_{cm} } \Si_y  &\equiv  &
\frac12 \bigg( \frac{\dd\si_{y}}{\dd\varOmega_{cm}} - \frac{\dd\si_{-y}}{\dd\varOmega_{cm}}\bigg)
= - \im [\Phi_4 (\Phi_1+\Phi_5)^{*} + (\Phi_2-\Phi_6) \Phi_3^{*}] ,\\
\frac{\dd\si}{\dd\varOmega_{cm} } \Si_{3y}  &\equiv  &
\frac12 \bigg( \frac{\dd\si_{y}^0}{\dd\varOmega_{cm}} - 
\frac{\dd\si_{y}^{\nicefrac{\pi}{2}}}{\dd\varOmega_{cm}}\bigg)
= \frac{\dd\si}{\dd\varOmega_{cm} } \Si_{3} + 
 \im [\Phi_1^\ast \Phi_2 + 2\Phi_3^\ast \Phi_4 -  \Phi_5^\ast \Phi_6 ] .
\eea 
\end{subequations}
The superscript on $\si$ indicates here the photon polarization: 
right ($R$), left ($L$) for the circular polarization, or
the value of $\phi$ for the linear polarization. 
The subscript indicates the  nucleon spin polarization. 
The expressions in terms of the helicity amplitudes assume parity conservation.
Terms proportional to the imaginary part are negligible below the pion-production
threshold (because the imaginary part of the 
amplitudes is suppressed by an $\al$).

Some of these polarized observables
have been measured for the proton RCS, which brings
us to the following section.

\section{Compton Scattering Experiments}

\seclab{experiment}

\subsection{Low-Energy Expansion of Observables}

The CS processes remain to be the only method of accessing the nucleon polarizabilities experimentally.
The nucleon probed by long-wave photons reveals its
structure in the manner of a multipole expansion. 
At first, one distinguishes the electric charge and magnetic dipole moment, contributions which comprise the
low-energy theorem for CS \cite{Low:1954kd,Gell_Mann:1954kc}.  Further terms (in multipole or energy expansion) can be
described in terms of polarizabilities, of which $\al_{E1}$ and $\be_{M1}$
play the leading role. 
For instance,  the low-energy expansion (LEX) of the unpolarized CS cross section, truncated at $O(\nu^2)$, is given by:  
\begin{flalign}\label{Eq:LEX}
\frac{\dd\sigma}{\dd\varOmega_L} = \frac{\dd\sigma^\mathrm{Born}}{ \dd\varOmega_L} -
\nu \nu^{\prime} \left(\frac{\nu^{\prime} }{ \nu}\right)^2 \frac{2\pi\al}{M} \left[ \left( \alpha_{E1} + \beta_{M1} \right)\left(1+z \right)^2 + \left(\alpha_{E1} - \beta_{M1} \right)\left(1-z \right)^2 \right], 
\end{flalign} 
where $z=\cos\vartheta$, and $\nu$ ($\nu^{\prime}$) is the energy of the incoming (scattered) photon.  

The simplicity of the formalism is appealing. However, the region of its applicability is unclear a priori. It is only clear that the convergence radius of such a LEX is limited
by the nearest singularity, which in this case is at pion production threshold (neglecting the small e.m.\ corrections). This is illustrated by the
dynamical polarizabilities in \Figref{pols_scalar}. Their LEX begins with 
the static value, which is the value at zero energy ($\w=0$). Clearly, 
approximating the dynamic polarizabilities by a constant (the static value)
only works well at energies
far below the pion-production threshold ($\w \ll m_\pi $).

\begin{wrapfigure}[19]{r}{0.55\textwidth}
 \vspace{-2mm}
\centering
\includegraphics[width=0.5\textwidth]{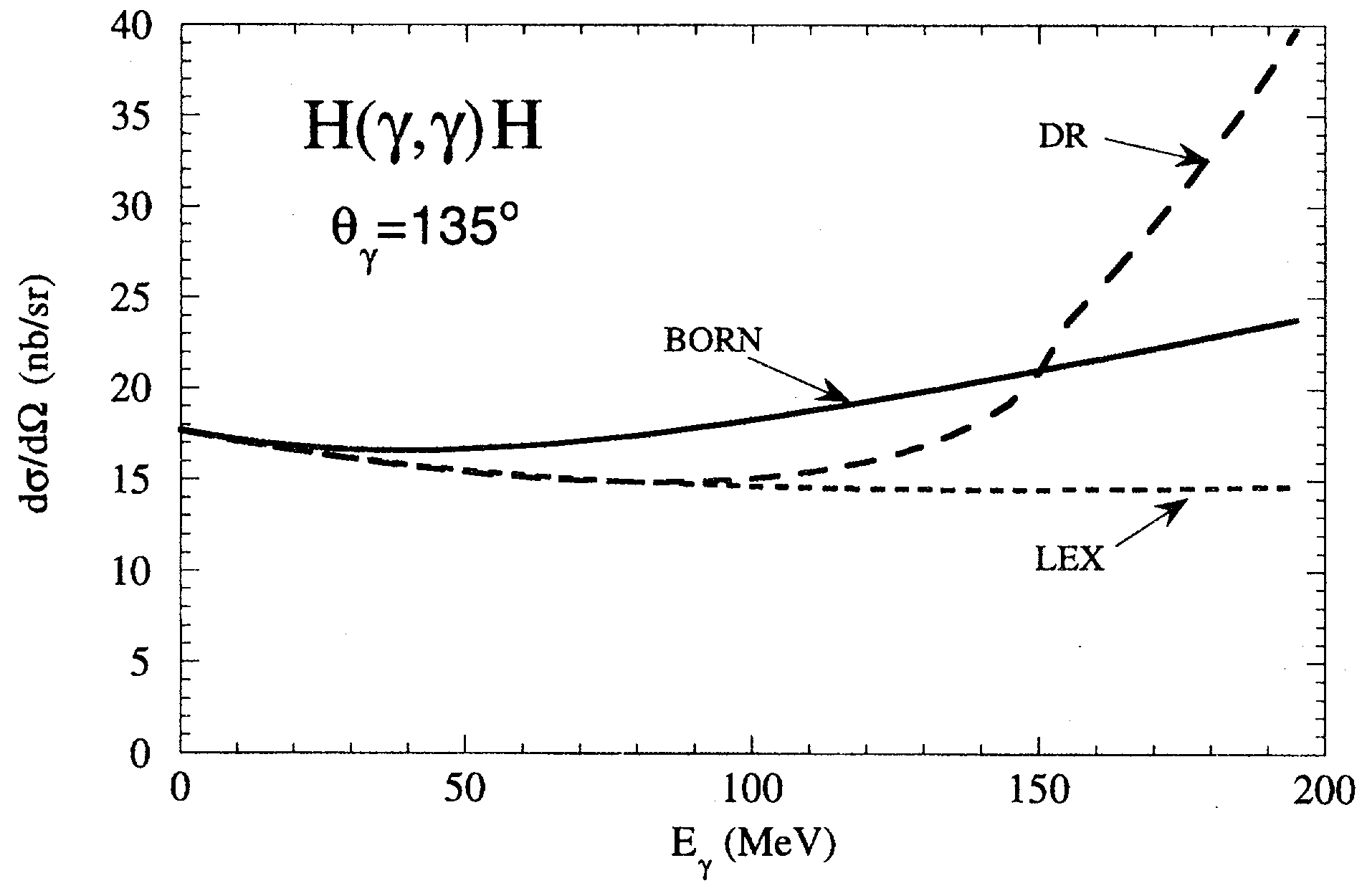}
\caption{CS cross sections  calculated in the Born approximation (solid), the 
leading-order LEX (dotted) and a dispersion model calculation (dashed). Plot reproduced from \citet{MacGibbon95}.
  \figlab{alphabetacurves}}
\end{wrapfigure}
Further insight can be obtained by comparing the LEX of the cross section with experimental data and  the results of calculations that extend beyond the
pion threshold.
Figure~\ref{fig:alphabetacurves} shows a calculation of the Born and LEX cross sections for CS on the proton at fixed angle and function of beam energy \cite{MacGibbon95}.  There is increased sensitivity to the polarizabilities at higher incident energies.  However, at higher energies there is also increased sensitivity in terms of order  $O(\nu^3)$ and higher.  The importance of the higher order terms is indicated by the dashed curve, which is a dispersion model calculation valid to all orders in $\nu$, albeit with unclear model-dependencies.  One sees that above $100\,\mathrm{MeV}$, the lowest-order LEX is not adequate. The higher-order
terms may extend its applicability, but will depend
on the spin polarizabilities and other,  
higher-order  polarizabilities. 

Nonetheless, the LEX is instructive for understanding how the polarizabilities affect the observables.
One can see, for example, that at forward angles the polarizability 
effect enters as $\alpha_{E1} + \beta_{M1}$, while at the backward angles as $\alpha_{E1} - \beta_{M1}$. Hence, one can in principle use the angular distribution
to disentangle $\alpha_{E1}$ and $\be_{M1}$.

The leading LEX expressions for CS with linearly polarized photons are obtained
by \citet{Max89} who finds [rearranging his Eqs. (19) and (20)]:
\begin{subequations}
\label{Eq:Max1}
\bea
\left(\frac{\dd\sigma_{\perp}}{ \dd \varOmega_L} - \frac{\dd\sigma^\mathrm{Born}_{\perp} }{ \dd\varOmega_L}\right) 
-\left(\frac{\dd\sigma_{||} }{\dd \varOmega_L} - \frac{\dd\sigma^\mathrm{Born}_{||} }{ \dd\varOmega_L}\right) 
&=&- \frac{8\pi \al  }{ M} \left( \frac{\nu^{\prime} }{ \nu} \right)^2 \nu \nu^{\prime} \, \alpha_{E1} \sin^2\vartheta 
\eqlab{Max1a}\\
\cos^2\vartheta \left(\frac{\dd\sigma_{\perp} }{\dd \varOmega_L} - \frac{\dd\sigma^\mathrm{Born}_{\perp} }{ \dd\varOmega_L}\right) 
-\left(\frac{\dd\sigma_{||}}{ \dd \varOmega_L} - \frac{\dd\sigma^\mathrm{Born}_{||} }{ \dd\varOmega_L}\right) 
&=&\frac{8\pi \al  }{M} \left( \frac{\nu^{\prime} }{ \nu} \right)^2 \nu \nu^{\prime} \,\beta_{M1} \cos\vartheta \sin^2\vartheta \eqlab{Max1b}
\eea
\end{subequations}
 where $\dd\sigma_{\perp}/\dd\varOmega_L$ and $\dd\sigma_{||}/\dd\varOmega_L$ are the differential cross sections for CS perpendicular and parallel to the plane of incident photon polarization. By taking weighted differences of $\sigma_{\perp}$ and $\sigma_{||}$, it is possible to measure $\beta_{M1}$ separately from  $\alpha_{E1}$. Hence, in separating the electric and magnetic polarizabilities
the beam polarization can serve as an alternative
to the angular distribution.

The issue here is that the cross section differences in parentheses on the left sides of the above equations are relatively small, at the $10\%$ level compared to the cross sections. To measure polarizabilities with a precision of $\approx 10\%$ requires measurement of the absolute cross sections $\sigma_{\perp}$ and $\sigma_{||}$ with uncertainties of $\approx 1\%$.   This is a severe challenge for CS experiments, both in statistics and systematic errors. 

It is much easier in this respect to measure the linear polarization asymmetry, 
\begin{flalign}\label{Eq:sigma_3}
\Si_3 = \frac{ \dd\sigma_{||} - \dd\sigma_{\perp} }{ 
\dd\sigma_{||} + \dd\sigma_{\perp} }.
\end{flalign}
The asymmetry measurements do not critically depend on knowing the incident photon flux, target thickness, and photon and recoil proton detection efficiencies.  To $O(\nu^2)$ this asymmetry depends only on $\beta_{M1}$ and is independent of $\alpha_{E1}$
\cite{Pascalutsa13}:
\begin{flalign}\label{Eq:krupina}
\Si_3 = \Si_3^\mathrm{Born} - 
\frac{4 M \omega_B^2 \cos\theta_B \sin^2\theta_B }{
 (1+\cos^2\theta_B)^2 }\, \alpha^{-1}\beta_{M1}
\end{flalign}
where $\w_B$ and $\theta_B $ are the photon energy and scattering angle in the Breit (brick-wall) frame, see Eqs.~\eref{kin1} and \eref{kin3} for their relations to invariants.  Calculations at NNLO in B$\chi$PT indicate that the range of applicability of Eq.~(\ref{Eq:krupina}) is as high as $100\,\mathrm{MeV}$. 
However, the sensitivity to $\beta_{M1}$ in Eq.~(\ref{Eq:krupina}) is weak, at the order of $\Delta \Si_3 / \Delta \beta_{M1}   \approx 0.02 $. To measure $\beta_{M1}$ at the level of $\pm 0.5 $ (in the usual units) would require uncertainty in $\Si_3$ to reach $\pm 0.01$. To attain this statistical precision, photon intensities at least an order of magnitude greater than those available with standard photon tagging techniques would be required. 

Nevertheless, the LEX formula (\ref{Eq:krupina}) paves the way
for a more sophisticated analysis, awaiting the first data for
$\Si_3$ below the pion production threshold. 
The LEX formula for the doubly-polarized asymmetries, $\Si_{2x}$ and $\Si_{2z}$ can be found in \citet{Krupina:2014nfa}.


To summarize, a model-independent extraction of polarizabilities from
CS observables based on LEX is nearly impossible at the existing facilities.
On one hand, the sensitivity must be substantial enough for a good signal at the given level of experimental accuracy, which drives
the experiment to higher energy. On the other
hand, at higher energies the LEX applicability is compromised. 
Therefore, although the LEX gives a valuable insight on the sensitivity 
of observables, it has been impractical in quantitative extractions. 

A more practical and common approach
is to extract polarizabilities by fitting the CS data using a systematic theoretical
framework, such as $\chi$PT~\cite{Beane:2002wn,Beane:2004ra,Lensky:2008re,Len10,Lensky:2012ag,Griesshammer:2012we,McG13,Lensky:2014efa} or fixed-$t$ 
dispersion relations (DRs)~\cite{Lvov:1993fp,Lvov:1996xd,Drechsel:1999rf,Pasquini:2001yy,Pas07}.
Table \ref{Tab:calculations} presents a listing of several calculations that have been or could be used for fitting cross section and asymmetry data from pion threshold up to the $\Delta(1232)$ region. 
\begin{table}[bt]
\footnotesize
  \caption{Theoretical frameworks used in the analysis of CS cross sections and asymmetries on proton and nuclear targets.}
 \label{Tab:calculations}
\centering
  \begin{tabular}{|c|c|c|c|}
    \hline
    Target & Energy range & Theoretical model & Reference \\ \hline
   Proton & 2$\pi$ threshold & Fixed-$t$ dispersion calculation &  Drechsel et al.~\cite{Drechsel:1999rf} \\ 
    Proton & $\Delta(1232)$ region & Chiral Lagrangian   &   Gasparyan et al.~\cite{Gas11}  \\ 
    Proton & $\Delta(1232)$ region &  HB$\chi$PT with $\Delta(1232)$  &  McGovern et al.~\cite{McG13} \\ 
    Proton & $\Delta(1232)$ region & B$\chi$PT with $\Delta(1232)$ &   
    \citet{Len10} \\
     Deuteron & $\approx 130\,\mathrm{MeV}$ & $\chi$EFT &   Grie{\ss}hammer et al.~\cite{Griesshammer:2012we} \\  
     $^3$He & 
     $\approx 130\,\mathrm{MeV}$ &  $\chi$EFT &  Shukla et al.~\cite{Shukla09} \\ \hline
        \end{tabular}
\end{table}
By fitting data with several theoretical models, it is possible to obtain an estimate for the model dependence of the result. This was, for example, the approach taken by Martel et al.~\cite{Martel15} for their analysis of double-polarized CS in the $\Delta(1232)$ region.\footnote{A limitation in fitting large numbers of data points with a fitting program such as {\sc MINUIT} is that many recursive calls are made to the subroutine calculating the observable. If the calculation of the observable is based upon numerical integrations, such as the fixed-$t$ DR code of \citet{Drechsel:1999rf}, then execution times can stretch into days. 
\citet{Martel15} handled this problem assuming a linear dependence of the observable on the polarizabilities: 
\begin{flalign}\label{Eq:linear1}
\cO_i(\{P\})=\cO_i(\{P_0\})+ \sum_{j=1}^6 \frac{\partial \cO_i(\{P_0\})  }{\partial P_j }( P_j - P_{0j}),
\end{flalign}
where $\cO_i$ is the observable, $\{P\}$ is the set of six polarizabilities (two scalar and four spin polarizabilities), and $\{P_0\}$ is the set of starting "guesses" for the polarizabilities.   The standard expression for $\chi^2$ is given by:
\begin{flalign}\label{Eq:linear2}
\chi^2= \sum_{i=1}^{N_\mathrm{data}} \left( \frac{\cO_i^\mathrm{data} - 
\cO_i(\{P\})  }{ \sigma_i } \right)^2 + \sum_{j=1}^6 \left( \frac{P_j^\mathrm{data} -P_j  }{ \sigma_{P_j} } \right)^2,
\end{flalign}
where the second term in $\chi^2$ allows for the possibility of introducing constraints on the polarizabilities (e.g., the sum-rule constraints on $\alpha_{E1}+\beta_{M1}$ and $\gamma_0$). 
Substituting Eqs.~(\ref{Eq:linear1}) into (\ref{Eq:linear2}), and setting $\partial \chi^2/\partial P_j = 0$ leads to a linear equation, 
$C_i = D_{ij}P_j$,
with 
\bea\label{Eq:linear4}
C_i&=&{P_i^\mathrm{data} \over \sigma_{P_i}^2} +\sum_{j=1}^{N_\mathrm{data}}{1 \over \sigma^2_j} \Big[\cO_j^\mathrm{data}-\cO_j(\{P_0\})+\sum_{k=1}^6 {\partial \cO_j(\{P_0\})\over \partial P_k}P_{0k} \Big]
{\partial \cO_j(\{P_0\}) \over \partial P_i}, \quad i=1,\dots,6\;,
\nonumber \\
D_{ij}&=& {\delta_{ij} \over \sigma^2_{P_i}} + \sum_{k=1}^{N_\mathrm{data}} {1\over \sigma^2_k} {\partial \cO_k(\{P_0\}) \over \partial P_i} {\partial \cO_k(\{P_0\}) \over \partial P_j}, \quad i, j=1,\dots,6\;.\nn
\eea
One can solve it for $\{P\}$, as $P=D^{-1}C$.
After the first iteration the substitution  $\{P\} \rightarrow \{P_0\}$ is made,  $\cO_i(\{P_0\})$,  $\partial \cO_i(\{P_0\}) / \partial P_j$ and $\{P\}$ are reevaluated, and the fit repeated until the set $\{P\}$ differs from $\{P_0\}$ by less than one standard deviation. 
It was found that this methodology is very efficient in fitting large data sets with computationally intensive codes. }


\subsection{Determination of Scalar Polarizabilities}

\subsubsection{Proton}
The first CS measurement of the proton  polarizabilities using a tagged photon beam was by Federspiel et al.~\cite{Fed91} at the University of Illinois MUSL-2 microtron.    Experiments prior to this  used bremsstrahlung beams, and were limited by systematic errors due to uncertainties in the incident photon energy. The most recent published results for CS on the proton are by the LEGS group \cite{Leg01}, and the TAPS at MAMI setup \cite{Olm01}, both in 2001.  Their results for $\alpha_{E1}$ and $\beta_{M1}$ are in agreement. 

After publication of the LEGS and TAPS results, activity in this area slowed. Then in 2010-2012  new $\chi$PT calculations of CS \cite{Len10,Griesshammer:2012we} showed that $\beta_{M1}$ is larger than the PDG average of that time, $\beta_{M1}^{(p)} = (1.9 \pm 0.5)\times 10^{-4}\,\mathrm{fm}^3$ \cite{Beringer:1900zz}, by $+1$ to $+3$ of the standard deviations. 
In 2014 the CS global analysis of~\citet{McG13} was included in the PDG average, and the current (2014) PDG values are \cite{Agashe:2014kda}:
\bea
\al_{E1}^{(p)}=(11.2\pm 0.4) \times 10^{-4}\,\mathrm{fm}^3, \quad \beta_{M1}^{(p)}=(2.5\mp 0.4) \times 10^{-4}\,\mathrm{fm}^3.
\eea
Nevertheless, the uncertainty quoted in the PDG average 
should still be taken with a grain of salt because the data sets are not treated consistently: e.g., results from the analysis of specific experiments are averaged with results from global analyses.  
The summary plots in \Secref{conclusion} show the real state of affairs for $\alpha_{E1}$ and $\beta_{M1}$ of the proton, cf.~Figs.~\ref{fig:alphabeta_p} and \ref{fig:alphaVSbeta} (left panel). It not as certain as 
the PDG average portrays it. There is certainly a room for improvement. 

The thrust of new proton polarizability measurements is to use linearly polarization as an analyzer to  measure $\alpha_{E1}$ and $\beta_{M1}$
separately, and independently of the Baldin sum rule value for $\alpha_{E1} + \beta_{M1}$.  Programs to measure the proton polarizabilities with linearly polarized photons are currently underway at the Mainz Microtron (MAMI) and at the High Intensity Gamma-Ray Source (HIGS). 

\begin{figure}
\centering
\includegraphics[scale=0.6]{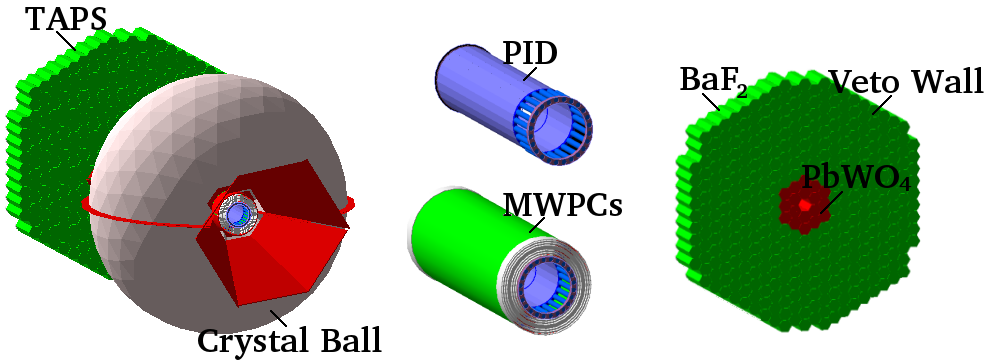}
\caption{The Crystal Ball and TAPS detectors at the Mainz microtron MAMI. \label{fig:CB_TAPS}}
\end{figure}

\begin{figure}[tbh]
\centering
\begin{minipage}{0.49\textwidth}
\centering
\includegraphics[scale=0.47]{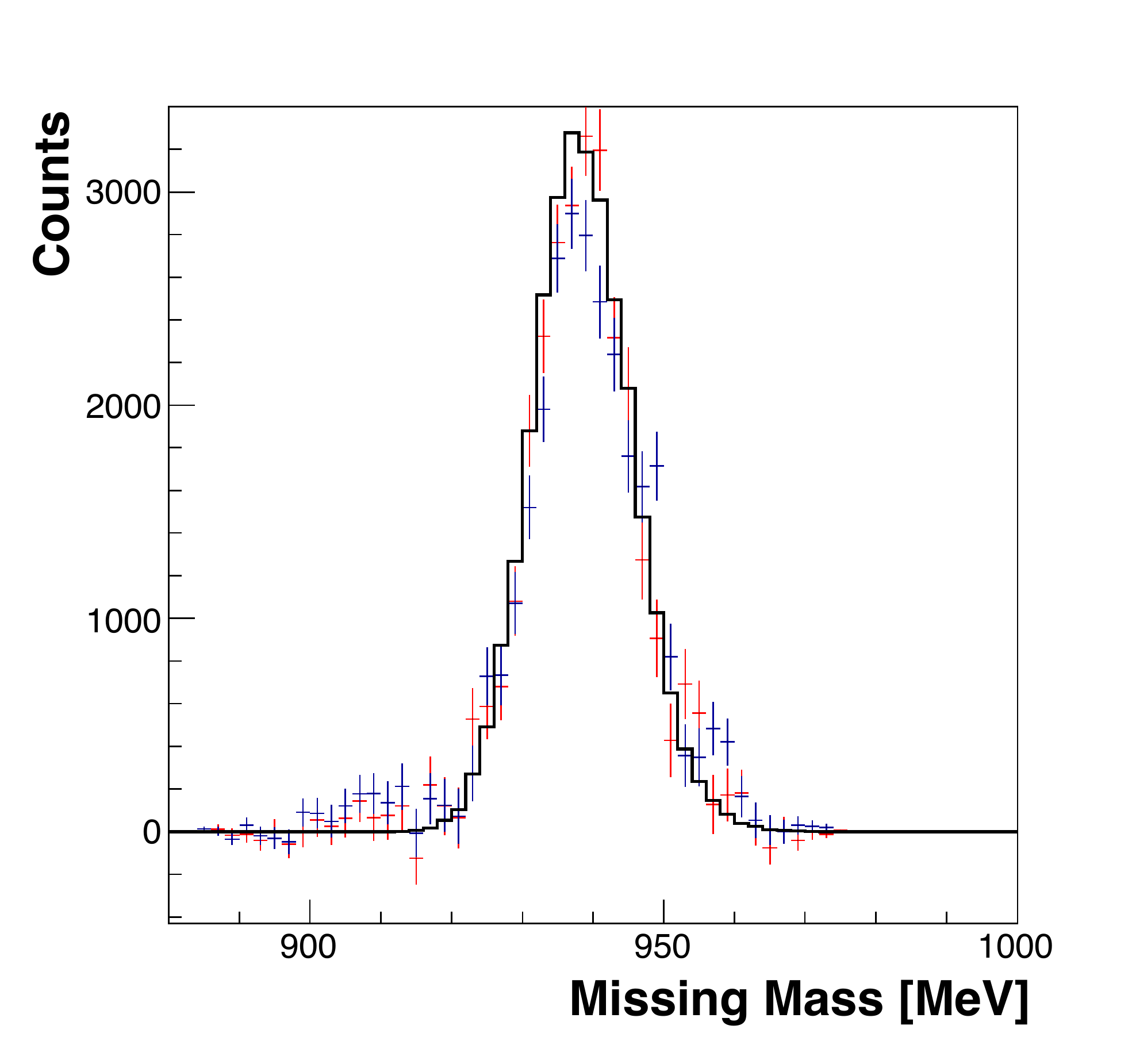}
\caption{Missing mass distribution for the Mainz proton scalar polarizability measurement for incident photon energies from $120$ to $140\,\mathrm{MeV}$, and angular range  $60^\circ < \vartheta < 150^\circ$ \cite{Dow12}. Data with parallel photon polarization is blue, and perpendicular polarization is red. The black curve represents the simulated distribution. Plot courtesy of Vahe Sokhoyan. 
 \label{fig:alpha_beta_mm}}
\end{minipage}\hfill
\begin{minipage}{0.49\textwidth}
\centering
\includegraphics[scale=0.65]{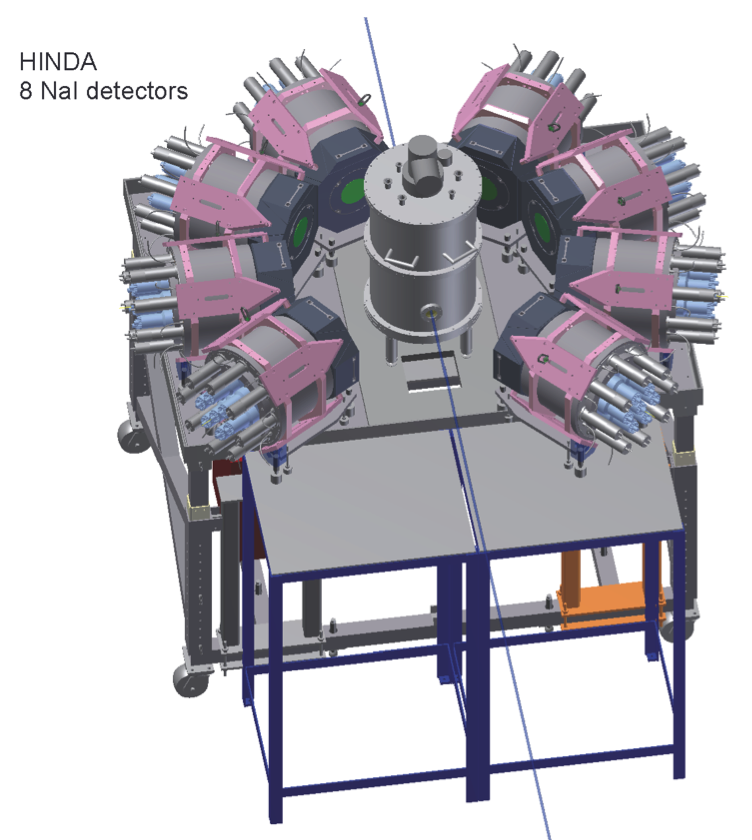}
\caption{Diagram of the HINDA NaI array at HIGS.\figlab{Hinda}} 
\end{minipage}
\end{figure}

Preliminary data \cite{Dow12} with linearly polarized incident photons have been taken by the A2 collaboration at the tagged photon facility \cite{McG08} at MAMI \cite{Kai08}. In this experiment a diamond radiator is used to produce linearly polarized coherent bremsstrahlung \cite{Loh94} with a peak polarization of approximately  $75\%$. The target is a $10\, \mathrm{cm}$ long liquid hydrogen target, and Compton scattered photons are detected in the Crystal Ball \cite{Sta01} and TAPS detectors \cite{Nov91},
both of which are outfitted with charged particle identification systems \cite{Tar08}.
The Crystal Ball, TAPS, and the charged particle system internal to the Crystal Ball (the PID scintillator array and MWPC), are shown in \figref{CB_TAPS}.  The solid angle coverage for Compton-scattered photons is approximately $97\%$ of $4\pi$. The incident photon energies for the measurement range from $80$ to $140\,\mathrm{MeV}$. Recoil proton detection is not required because the backgrounds are sufficiently low. The missing mass distribution, shown in \figref{alpha_beta_mm}, reveals there is relatively little background in the measurement. The goals for this experiment are to obtain precision measurements of the cross sections $\dd\sigma_{\perp}$ and $\dd\sigma_{||}$, the asymmetry $\Si_3$, and the unpolarized cross section by running on an amorphous photon radiator. 

At HIGS \cite{Weller:2009zza} there are plans \cite{Ahm10} to take CS data on the proton with linearly polarized photons in 2016. The incident photon beam at HIGS is exceptionally well suited for a LEX based CS analysis for $\alpha_{E1}$ and $\beta_{M1}$; energies up to $100\,\mathrm{MeV}$ are available, the beam energy is monochromatic with a spread of $\approx 4.5 \%$, photon intensities on the target are $\approx 10^7 \,\nicefrac{\gamma\text{'s}}{s}$, and the linear polarization is approximately $100\%$ \cite{myers12}. Figure \ref{fig:Hinda} shows the HINDA NaI array that has been developed for CS experiments.  Each module has a $25.4\,\mathrm{cm}$ in diameter and $30.5\,\mathrm{cm}$ long NaI core, surrounded by active NaI shields that are $7.5\,\mathrm{cm}$ thick and $30.5\,\mathrm{cm}$ long. 

\subsubsection{Neutron}
\seclab{NeutronExp}
As a ``free'' neutron target does not exist, there are no truly model-independent means to determine the neutron's polarizabilities.  
Obtaining neutron polarizabilities from the analysis of deuteron CS data requires accurate effective-field theory calculations. The favored approach is to use elastic scattering data, not quasi-free d$\left(\gamma, \gamma^\prime\text{n}\right)$p data, because (i) the elastic process is theoretically less complicated than the quasi-free scattering, and treatable through effective-field theory calculations, and (ii) elastic scattering has a larger cross section and greater sensitivity to the polarizabilities than the quasi-free process. The latter point can be understood by noting that for CS on a free neutron there is no Thomson term in the scattering amplitude, because the neutron is neutral, and the polarizability effect goes as $O(\nu^4)$. For elastic scattering there is a Thomson term, because the deuteron is charged, and the polarizability effect goes as $O(\nu^2)$. The disadvantage of elastic scattering is that it places a premium on the utilization of large NaI detectors with sufficient energy resolution to resolve elastic and inelastic scattering. Furthermore, the elastic scattering is sensitive to the isoscalar polarizabilities, and the proton contribution must be subtracted. 

Until recently there has been a paucity of CS data on the deuteron. For example, the analysis of
Grie{\ss}hammer et al.~\cite{Griesshammer:2012we} was based on three data sets with a total of 37 data points.
The focus of new studies has been to stage experiments that can obtain relatively high statistics with wide kinematic coverage, utilize large NaI detectors for optimal energy resolution, and to use targets with $A>2$.  In this section recent progress in this area is outlined. 

New data from Lund has recently been published for elastic CS on the deuteron \cite{Myers14}, nearly doubling the effective number of world data points and extending the energy range by $20\,\mathrm{MeV}$ to higher energies. 
The large-volume, segmented NaI(Tl) detectors,
BUNI, CATS, and DIANA were
used to detect Compton-scattered photons in the experiment. These detectors are composed of a
NaI(Tl) core surrounded by optically isolated, annular
NaI(Tl) segments. The cores of the BUNI and CATS
detectors each measures $26.7\,\mathrm{cm}$ in diameter, while the
core of the DIANA detector measures $48.0\,\mathrm{cm}$. The
depth of all three detectors is greater than $20$ radiation
lengths. The annular segments are $11\,\mathrm{cm}$ thick on the
BUNI and CATS detectors and $4\,\mathrm{cm}$ thick on the DIANA
detector.  Figures \ref{fig:alphabeta_n} and \ref{fig:alphaVSbeta} (right panel) show the present state of affairs for $\alpha_{E1}$ and $\beta_{M1}$ of the neutron.

\begin{figure}
\centering 
     \includegraphics[width=9cm]{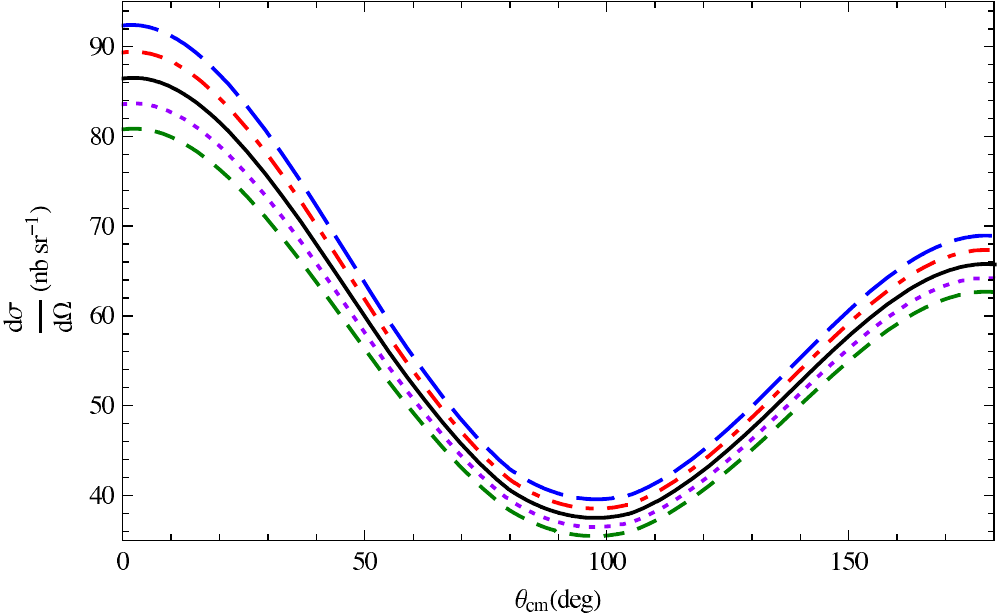}
  \hfill  \includegraphics[width=9cm]{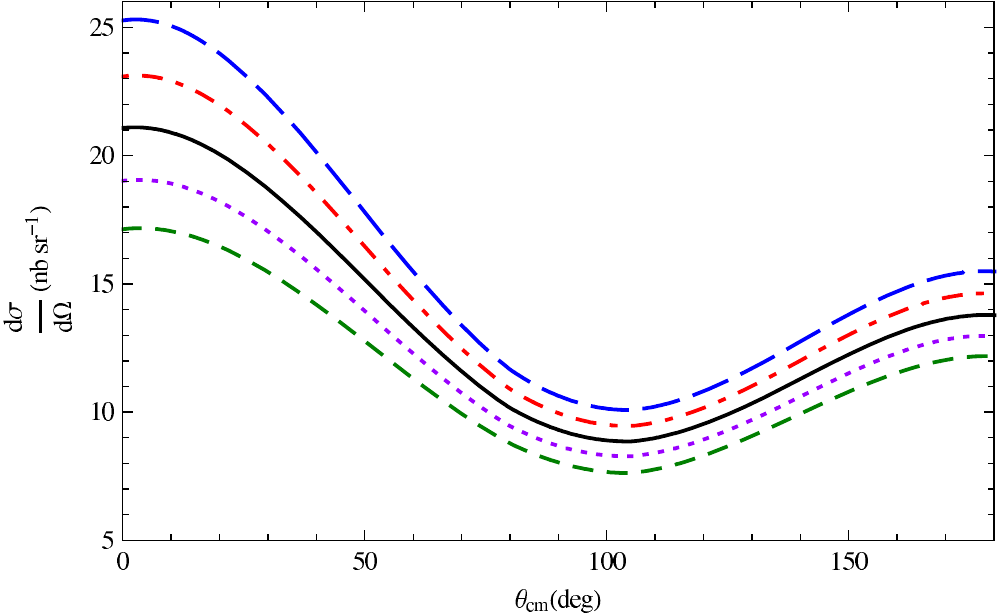}
 \caption{Sensitivity of the differential cross sections for CS on $^3$He in the c.m.\ frame at NLO
in $\chi$EFT without explicit $\Delta(1232)$ \cite{Shukla09} at $60\,\mathrm{MeV}$ (left panel) and $120\,\mathrm{MeV}$ (right panel). Solid (black) curve: central value $\alpha^n_{E1}= 12.2$; long-dashed (blue): $\alpha^n_{E1}-4$;
dot-dashed (red): $\alpha^n_{E1}-2$; dotted (magenta): $\alpha^n_{E1}+2$; dashed (green): $\alpha^n_{E1}+4$; in units of $10^{-4}\,\mathrm{fm}^3$. Plots reproduced from \citet{Shukla09}.
 \label{fig:3He}}
\end{figure}

There are compelling reasons to consider CS on $Z>1$ nuclei as an attractive route to the neutron polarizabilities; the Thomson cross section goes at $Z^2$, and there is a better ratio of elastic to incoherent scattering for $A>2$. As a test of the effective interaction theories used to analyze the data, it is also important to demonstrate that polarizabilities obtained from  deuterium are in agreement with results from other nuclei. Figure \ref{fig:3He} shows the results of a NLO $\chi$EFT calculation without $\Delta(1232)$ degrees of freedom at $60\,\mathrm{MeV}$ (left panel) and $120\,\mathrm{MeV}$ (right panel) for CS on $^3$He.   The curves show appreciable sensitivity to the neutron polarizabilities, especially at the higher energy.  

There are plans \cite{Annand13}  to measure elastic CS on $^3$He and $^4$He with an active, gaseous helium target using the Crystal Ball and TAPS detectors at the Mainz microtron MAMI. The operating principle of the target is that ionizing particles produce UV scintillation light in the helium, and by the addition of small amounts of N$_2$ as a wavelength shifter, and a photo-detector coupled to the target cell, the target can operate as a detector.   Elastic scattering is separated from incoherent processes by detecting the recoil helium nucleus in coincidence with the Compton-scattered photon. Development of this target is in progress, and test runs are anticipated in 2016.

\subsection{Determination of Proton Spin Polarizabilities}
Compared to the situation for the proton scalar polarizabilities, relatively little is known experimentally about the spin polarizabilities. 
Prior to the advent of single-polarized and double-polarized CS asymmetry measurements, only two linear combinations of the polarizabilities were known.  One combination is the forward spin polarizability: 
\begin{flalign}\label{Eq:gamma_0}
\gamma_0 =  -\gamma_{E1E1} -\gamma_{E1M2}-\gamma_{M1M1} - \gamma_{M1E2},
\end{flalign} 
fixed by the GTT sum rule \eref{FSP}. The results
of the GTT sum rule evaluation are summarized in 
Table \ref{Tab:GDH}.
The other combination is the backward spin polarizability $\gamma_\pi$: 
\begin{flalign}\label{Eq:gamma_pi}
\gamma_{\pi} = -\gamma_{E1E1} - \gamma_{E1M2} + \gamma_{M1M1} + \gamma_{M1E2}.
\end{flalign}

The forward (backward) spin polarizability, according to its name, appears 
in the
spin-dependent CS amplitude at forward (backward)
kinematics. More specifically, in both kinematics the CS amplitude
splits into a spin-independent and spin-dependent part, i.e.:
\begin{subequations}
\bea 
\label{Eq:forward_Compton}
{1 \over 8 \pi M} T(\nu, \vartheta=0) 
&=&  f(\nu) \, \boldsymbol{\varepsilon}^{\prime *} \cdot \boldsymbol{\varepsilon} +  g(\nu)\,  i \boldsymbol{\sigma} \cdot \boldsymbol{\varepsilon}^{\prime *} \times \boldsymbol{\varepsilon} ,\\
\label{Eq:backward_Compton}
{1 \over 8 \pi M} T(\nu, \vartheta=\pi) 
&=& \frac{1}{\sqrt{1+2\nu/M}} \Big[ \tilde f(\nu) \, \boldsymbol{\varepsilon}^{\prime *} \cdot \boldsymbol{\varepsilon} +  \tilde g(\nu)\,  i \boldsymbol{\sigma} \cdot \boldsymbol{\varepsilon}^{\prime *} \times \boldsymbol{\varepsilon} \Big],
\eea 
\end{subequations}
and the low-energy expansion for the
scalar amplitudes goes as follows \cite{Bab98}:
\begin{subequations}
\bea 
f(\nu)  &=&  -\frac{ \zZ^2\al}{M}
+  (\al_{E1} +\be_{M1} ) \nu^2 
+O(\nu^4), \\
 g(\nu) &=&  -\frac{ \al \varkappa^2 }{2M^2}\nu + \gamma_0 \nu^3
+O(\nu^5), \\
 \tilde f(\nu)  &=&  \Big(1+\frac{\nu}{M}\Big)
\Big[ -\frac{ \zZ^2\al }{M}
+  (\al_{E1} -\be_{M1} ) \frac{\nu^2}{1+2\nu/M}
+\ldots\Big], \\
 \tilde g(\nu)&=&  \big[-\half \varkappa^2 +
(\zZ+\varkappa)^2\big]
\frac{\al \nu}{M^2}+ \gamma_\pi \frac{\nu^3}{1+2\nu/M}
+\ldots 
\eea 
\end{subequations}
Hence, $\ga_0$ and $\ga_\pi$ appear at $O(\nu^3)$ in the LEX of the
spin-flip amplitude at, respectively, the forward and backward scattering angle.

Figure \ref{fig:backward} shows the sensitivity of backward angle CS cross sections in the $\Delta(1232)$ region to $\gamma_\pi$. The most widely accepted value for $\gamma_\pi$  is actually an average of three measurements at MAMI performed with different detector configurations: TAPS \cite{Olm01}, LARA \cite{Galler01,Wolf01}, and SENECA \cite{Cam02}. All three of the measurements agree within their statistical and systematic errors, and the average value is \cite{Cam02}: 
\begin{equation}\label{Eq:gamma_pi_value} 
\gamma_{\pi} = (8.0 \pm 1.8) \times 10^{-4}\, \mathrm{fm}^4, 
\end{equation}
where the error includes statistical and estimated model uncertainties. Here we use the standard convention of excluding the t-channel $\pi^0$-pole contribution\footnote{In the literature the result with the $\pi^0$-pole excluded is sometimes referred to as $\bar \gamma_\pi$ (see, e.g., Ref.~\cite{Leg01}).}. The latter is evaluated as (e.g., Ref.~\cite{Pascalutsa:2003zk}):
\beq
\gamma^{\pi^0\mathrm{pole}}_\pi =-\frac{2\al g_A}{(2\pi f_{\pi})^2m_{\pi^0}^2}=\left(-46.04\pm0.16\right)\times 10^{-4}\, \mathrm{fm}^4.
\eeq
The $\pi^0$-pole contribution is also excluded from each of the spin polarizabilities: $\gamma_{E1E1}$, $\gamma_{M1M1}$, $\gamma_{E1M2}$ and $\gamma_{M1E2}$ \cite{Bab98}. However, this contribution cannot affect $\gamma_{0}$ and other forward polarizabilities, because the $\pi^0$-pole diagram vanishes in the forward direction.


The LEGS result \cite{Leg01}, 
\beq
\gamma_\pi= \left(18.81 \pm 2.27_{\,\mathrm{stat+syst}}+ \left[^{+2.24}_{-2.10}\right]_{\,\mathrm{model}}\right)\times 10^{-4}\, \mathrm{fm}^4,
\eeq
is in disagreement with the Mainz result, despite the good agreement between LEGS and~\citet{Olm01} for $\alpha_{E1}$ and $\beta_{M1}$. As seen in \figref{backward}, the main cause for this disagreement  is the discrepancy between measured cross sections at backward angles at energies above pion threshold. Figure \ref{fig:gammaPi} summarizes 
the results for the backward spin polarizability of the proton.

\begin{figure}[tbh]
\centering
\begin{minipage}{0.54\textwidth}
\centering
\includegraphics[scale=0.775]{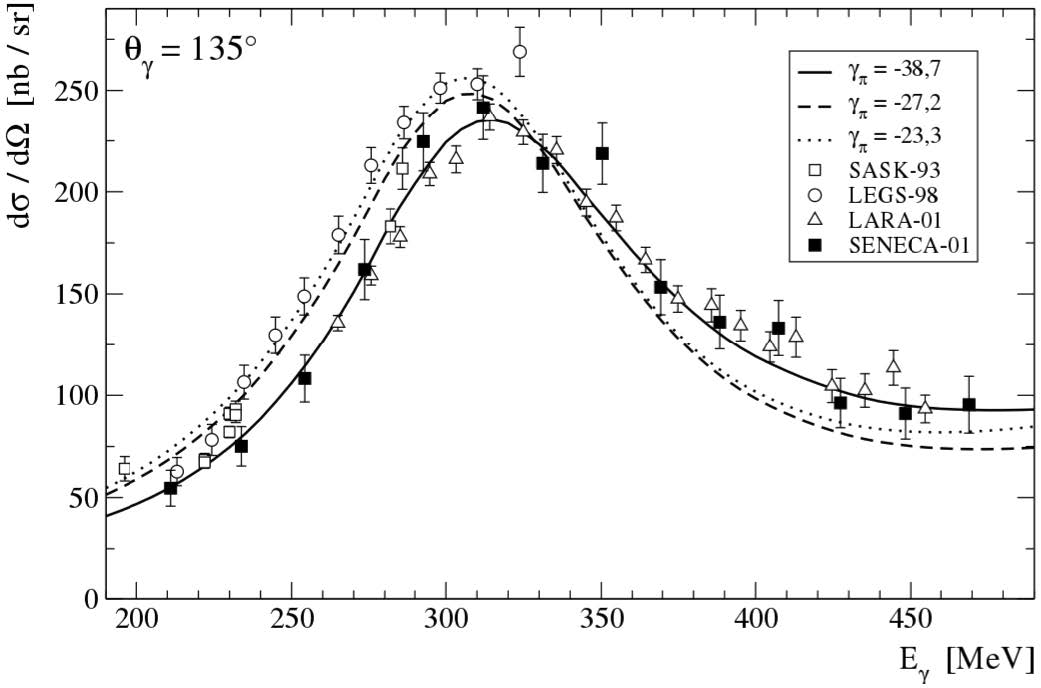}
\caption{Differential cross section of the proton RCS for the center-of-mass angle of $135^\circ$ in the $\Delta(1232)$ region \cite{Cam02}. The curves correspond to different values of $\gamma_\pi$, which here include the t-channel $\pi^0$-pole contribution.
Plot reproduced from \citet{Cam02}.  \figlab{backward}}
\end{minipage}\hfill
\begin{minipage}{0.44\textwidth}
\centering
\includegraphics[scale=0.54]{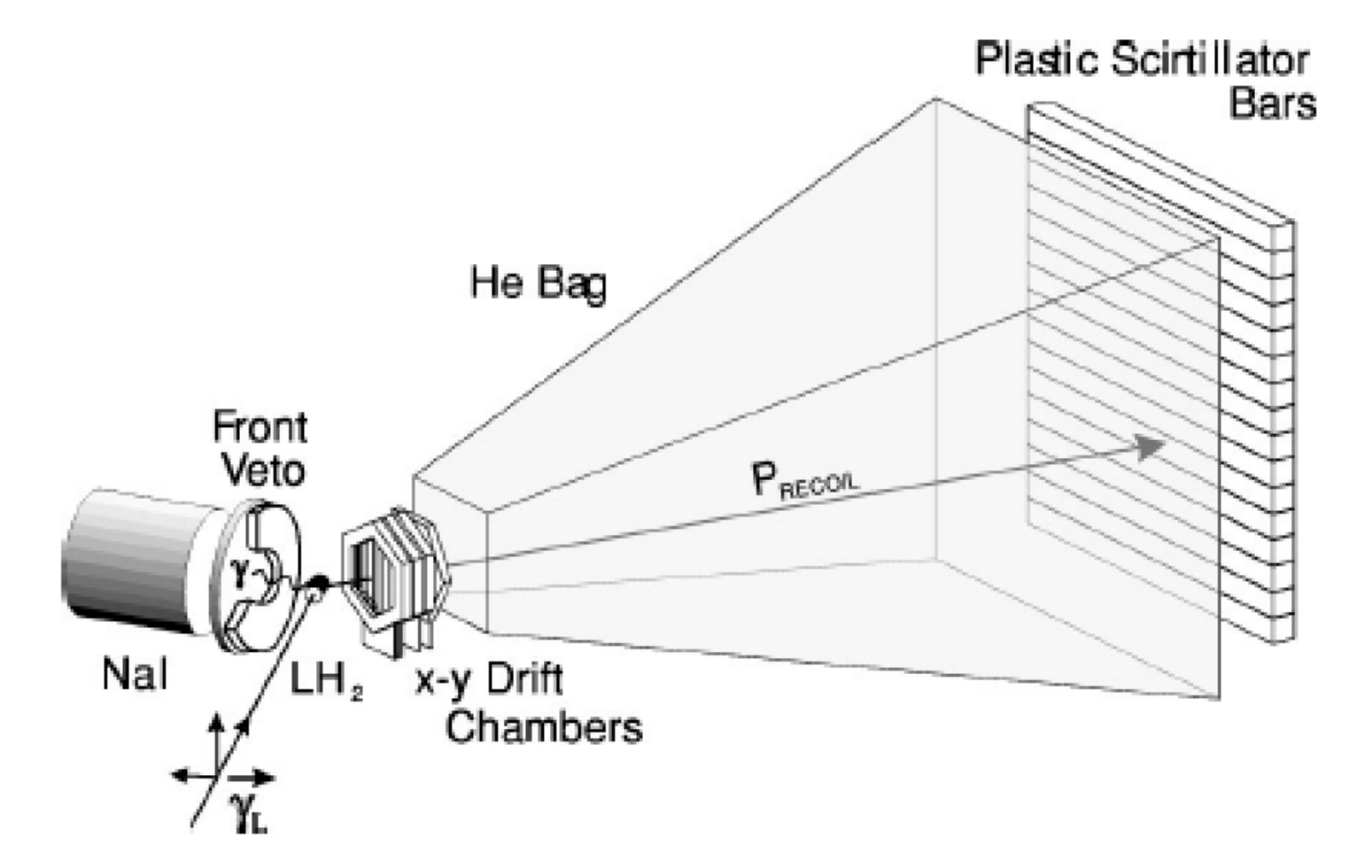}
\caption{CS configuration for the LEGS detectors.  \label{fig:LEGS}}
\end{minipage}
\end{figure}

Single and double-polarized CS asymmetries have sensitivity to the spin polarizabilities \cite{Pas07,Krupina:2014nfa}. Measurements of this type provide essentially the only means by which the four lowest order spin polarizabilities $\gamma_{E1E1}$, $\gamma_{M1M1}$, $\gamma_{E1M2}$
and $\gamma_{M1E2}$ can be individually separated. The linear polarization asymmetry $\Si_3$ in the $\Delta(1232)$ region was measured at LEGS \cite{Leg01}, and a new program of single and double polarized CS measurements is currently underway at  MAMI \cite{Hornidge12}.

\subsubsection{Linearly Polarized Photons}

\begin{figure}[bht]
 \includegraphics[width=\columnwidth]{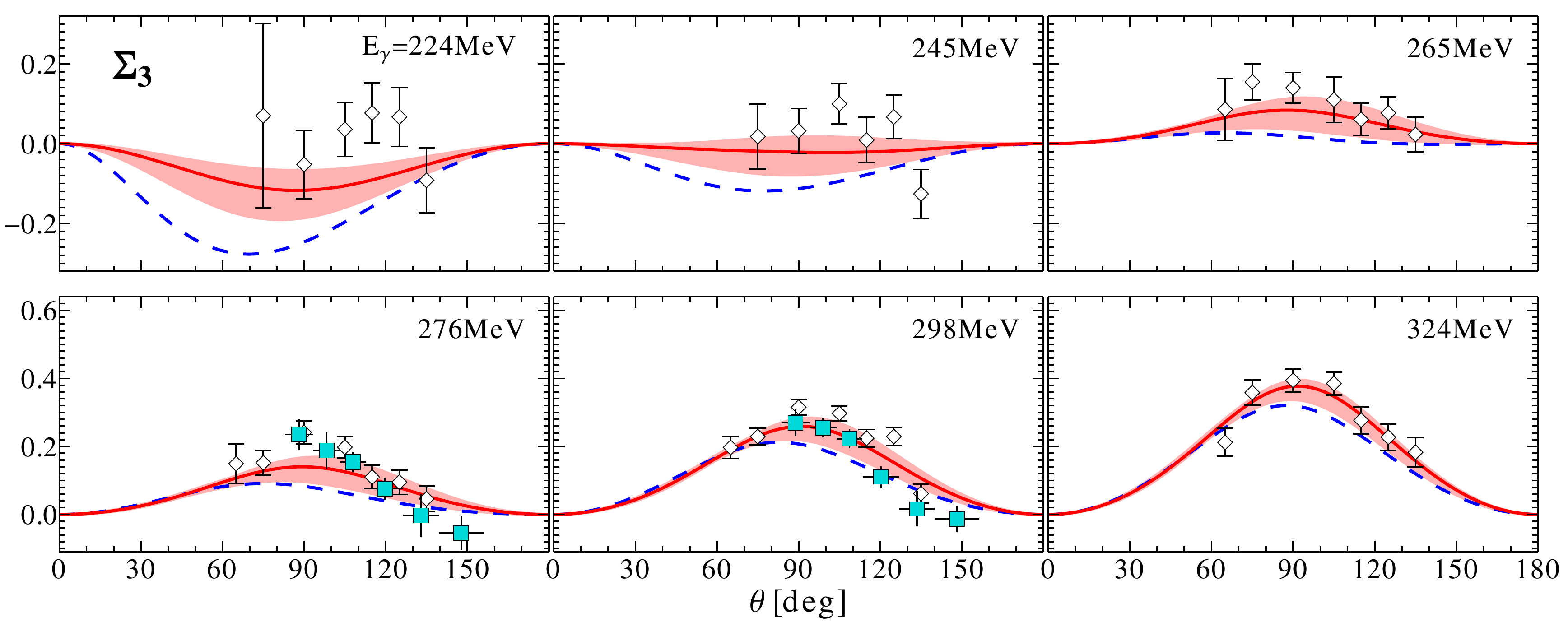}
\caption{The CS beam asymmetry $\Sigma_3$ as function of the c.m.\ angle at different values of the beam energy $E_\gamma$. The experimental data are
from LEGS~\cite{Leg01} (open diamonds) and MAMI~\cite{Collicott15} (cyan squares).
The bands represent the NNLO B$\chi$PT result of~\citet{Lensky:2015awa}. 
The blue dashed lines represent their calculation with only the 
nucleon-, pion-, and Delta-pole graphs included (chiral loops are switched off).
}
\label{fig:sigma3_sensitivity}
\end{figure}

The linear polarization asymmetry, $\Si_3$, is defined in Eq.~(\ref{Eq:sigma_3}),
see also \Eqref{beamas}. The first measurements of this asymmetry  were by the LEGS collaboration \cite{Leg01} in the $\Delta(1232)$ region, see Fig.~\ref{fig:sigma3_sensitivity}. Their experimental setup, shown in \figref{LEGS}, is fairly typical of CS experiments in the $\Delta(1232)$ region, where the Compton photon is detected in a NaI detector, and the recoil proton is detected in a spectrometer arm specifically designed for recoil detection.   Detecting the recoil proton is necessary to suppress background from $\pi^0\rightarrow \gamma \gamma$;   the ratio of $\pi^0$ photoproduction to CS in the $\Delta(1232)$ region is approximately $100$:$1$. In the LEGS measurement precision wire chambers are used to define the trajectory of the proton, and time-of-flight over $4\,\mathrm{m}$ and energy loss in scintillator paddles are used to establish particle type and momentum.

In the LEGS analysis two spin polarizability combinations are fit to their data:
\begin{subequations}
\label{Eq:gamma_13_14}
\bea
\gamma_{13} &\equiv& -\gamma_{E1E1} + \gamma_{E1M2} = \left(3.94 \pm 0.53_{\,\mathrm{stat+syst}} +\left[^{+0.20}_{-0.18}\right]_{\,\mathrm{model}}\right)\times 10^{-4}\, \mathrm{fm}^4,\\
\gamma_{14} &\equiv& -\gamma_{E1E1} - \gamma_{E1M2} - 2\gamma_{M1M1} =-\left(
2.20 \pm 0.27_{\,\mathrm{stat+syst}} +\left[^{+0.05}_{-0.09}\right]_{\,\mathrm{model}}\right)\times 10^{-4}\, \mathrm{fm}^4.\qquad
\eea
\end{subequations}

\begin{figure}
\centering
\includegraphics[scale=0.55]{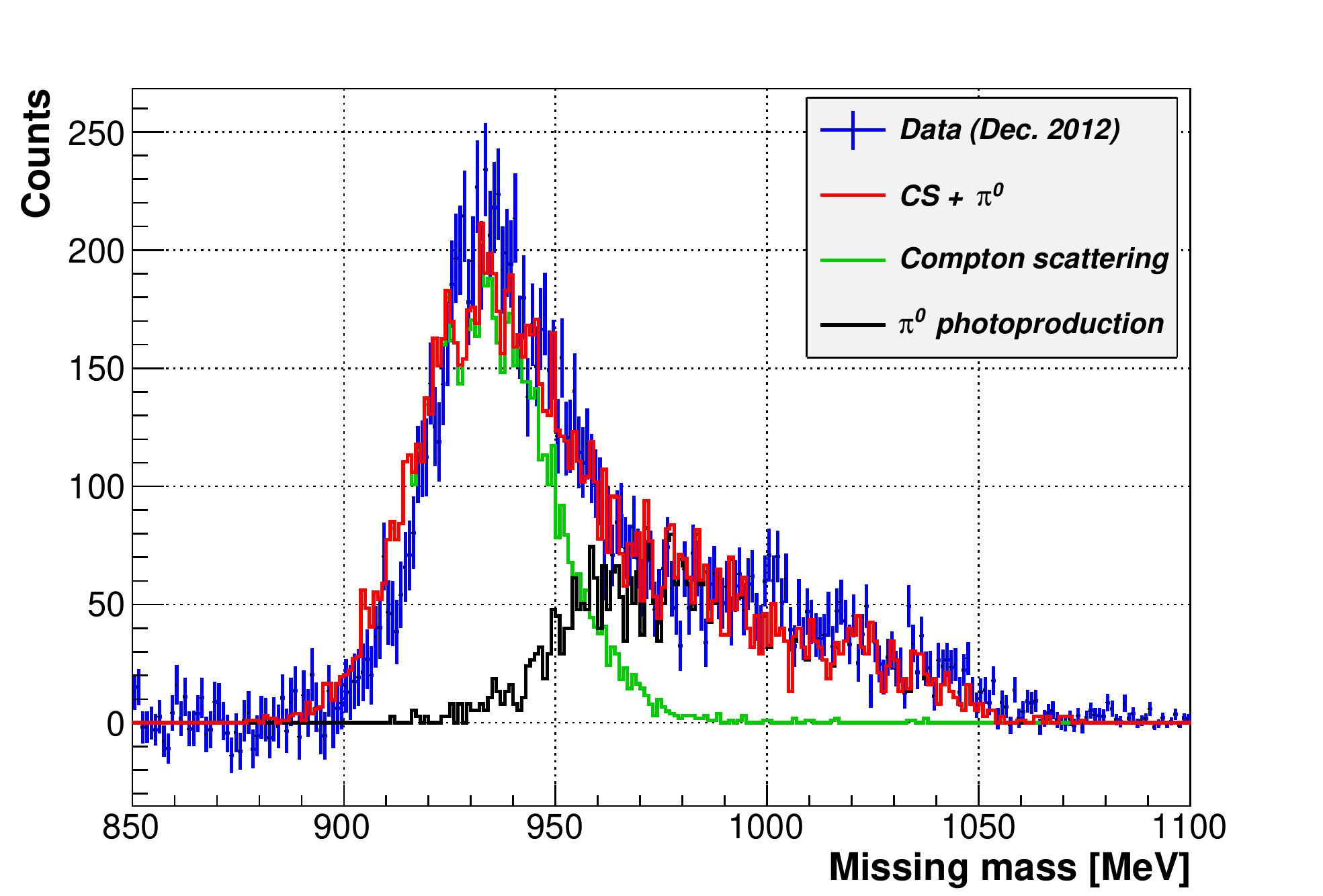}
\caption{ Missing mass distribution for the Mainz $\Si_3$ experiment for incident photon energy $277 \pm 10\,\mathrm{MeV}$, and $90^\circ - 100^\circ$. The blue histogram is data, the green curve shows the simulated response for CS, the black curve is a simulation of $\gamma p \rightarrow \pi^0 p$ events that satisfy the exclusivity requirements for CS, and the red curve is the sum of simulated CS and $\pi^0$ background. Plot courtesy of Cristina Collicott. \label{fig:sigma3_mm}}
\end{figure}

New data for the $\Si_3$ asymmetry in the $\Delta(1232)$ region have recently been taken by the Mainz A2 collaboration at the microtron MAMI using the Crystal Ball and TAPS detectors \cite{Collicott15}.    In this experiment $\pi^0$ events are suppressed by making use of the hermeticity of the detector,  requiring that only one neutral and one charged track are present in the event. Additional background suppression is provided by imposing a co-planarity and opening angle cut of $15^{\circ}$ on the direction of the recoil proton relative to  the momentum transfer direction $\boldsymbol q$ defined by the  incident and final photons. A missing mass distribution with cuts applied is shown in \figref{sigma3_mm}  for an incident photon energy of $277\,\mathrm{MeV}$. In contrast to the missing mass distribution shown in \figref{alpha_beta_mm}, CS experiments in the $\Delta(1232)$ region typically have prominent background due to $\gamma p \rightarrow \pi^0 p$. 
For the Mainz analysis a conservative cut is placed on the missing mass distribution to limit $\pi^0$ background to the few percent level.

\subsubsection{Circularly Polarized Photons and Transversely Polarized Target}
\begin{figure}[tbh] 
  \centering 
\begin{minipage}[t]{0.49\textwidth}
    \centering 
       \includegraphics[width=\textwidth]{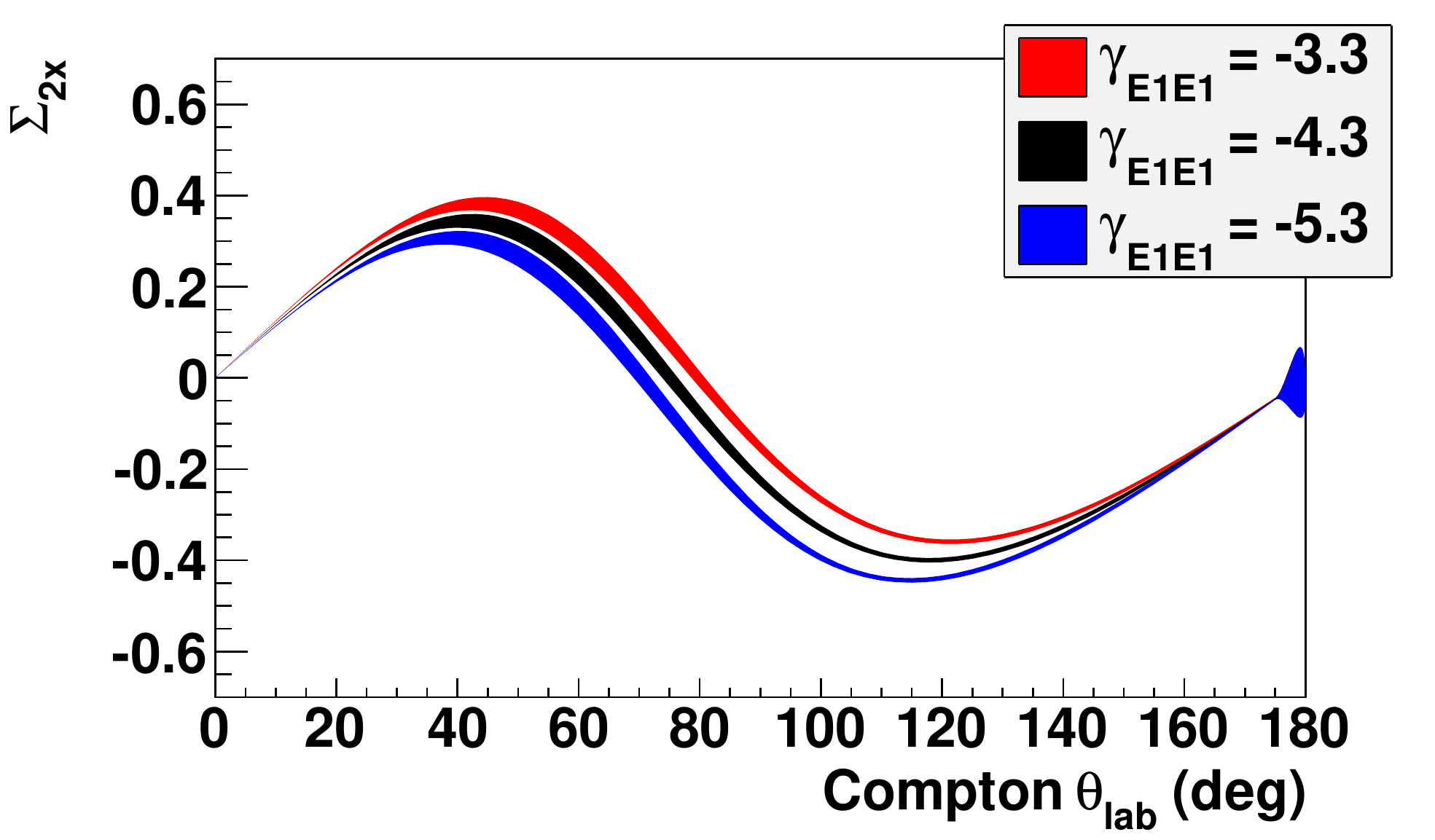}
\end{minipage}
\begin{minipage}[t]{0.49\textwidth}
    \centering 
       \includegraphics[width=\textwidth]{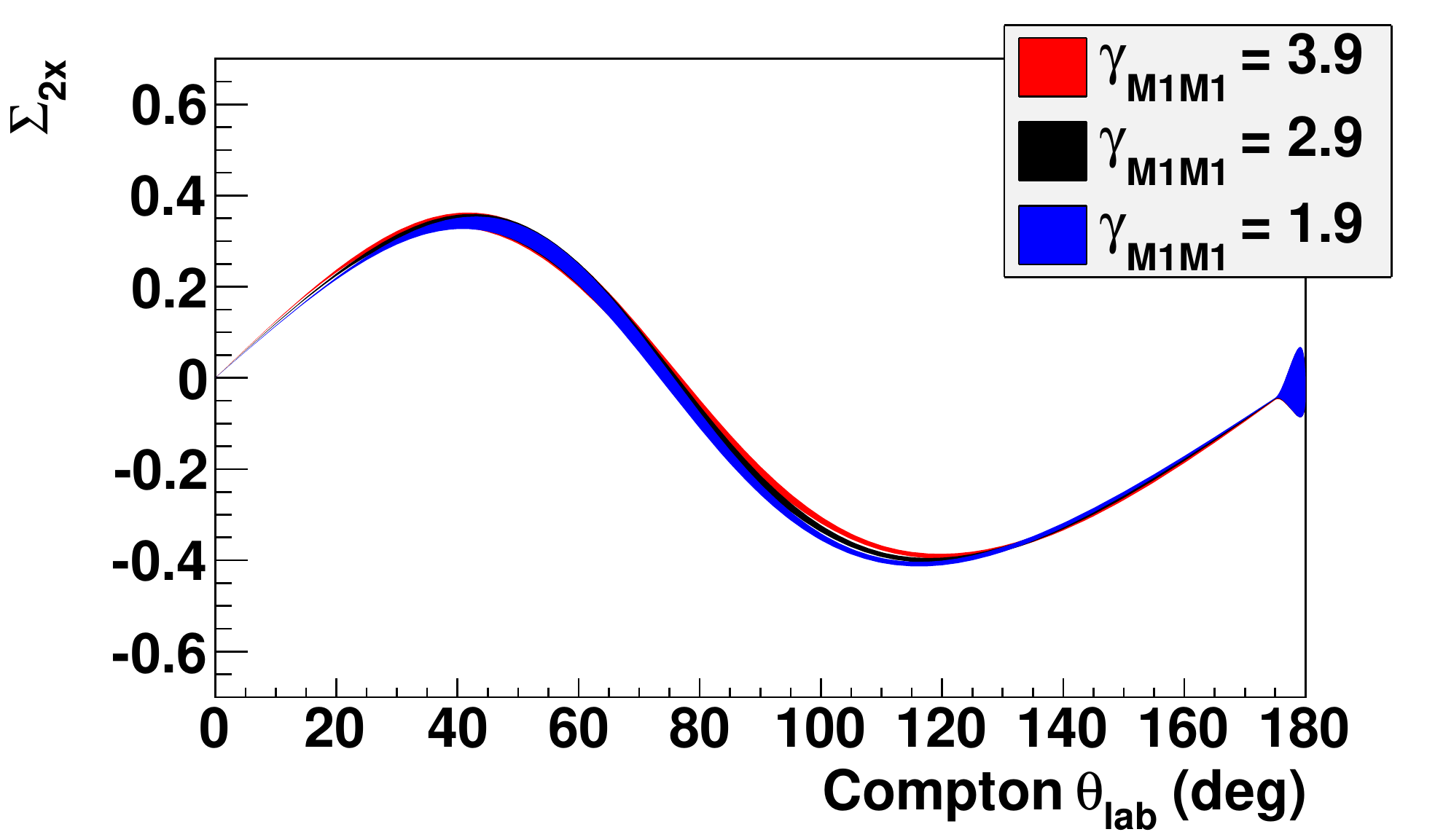}
       \end{minipage}
 \caption{The asymmetry $\Si_{2x}$ for $E_{\gamma} = 290\,\mathrm{MeV}$. The curves are from a dispersion theory calculation \cite{Pas07} with $\alpha_{E1}$, $\beta_{M1}$, $\gamma_{0}$, and $\gamma_{\pi}$ held fixed at their experimental values. The left plot has $\gamma_{M1M1}$ fixed at $2.9$, and the red, black, and blue bands are for $\gamma_{E1E1}$ equal to $-3.3$, $-4.3$, and $-5.3$, respectively. The width of each band represents the propagated errors from $\alpha_{E1}$, $\beta_{M1}$, $\gamma_{0}$, and $\gamma_{\pi}$ combined in quadrature. The right plot has $\gamma_{E1E1}$ fixed at $-4.3$, and the red, black, and blue bands are for $\gamma_{M1M1}$ equal to $3.9$, $2.9$, and $1.9$, respectively (in units of $10^{-4} \mathrm{fm}^4$). Plot courtesy of Phil~Martel.
 \label{fig:sigma2x_sensitivity}}
\end{figure}

The relevant double-polarized CS asymmetry is defined as
\beq \label{Eq:Sigma_2x}
\Si_{2x} = \frac{  \dd\sigma^R_{x}  -   \dd\sigma^L_{x} }{\dd\sigma^R_{x}  +  \dd\sigma^L_{x}  },
\eeq 
where $\dd\sigma^{R(L)}_{x}$ is the differential cross section for  transverse target polarization in the $x$-direction, and for right (left) circularly polarized photons. 
Data for the $\Si_{2x}$ asymmetry in the $\Delta(1232)$ region have recently been published  by the Mainz A2 collaboration \cite{Martel15}. Figure~\ref{fig:sigma2x_sensitivity} shows the expected sensitivity of $\Si_{2x}$ to the spin polarizabilities.    The left figure shows significant sensitivity to $\gamma_{E1E1}$, and the right figure shows little sensitivity to $\gamma_{M1M1}$. Based on measurements of $\Si_{2x}$ at angles $\vartheta \approx 90^{\circ}$, it is possible to uniquely identify $\gamma_{E1E1}$  \cite{Miskimen09}. 

The target for this experiment was a frozen spin butanol target \cite{Tho11}, approximately $2\,\mathrm{cm}$ long, where the protons in the butanol are polarized by dynamic nuclear polarization \cite{Cra97}.
Proton polarizations were typically $90\%$ with relaxation times on the order of $1000$ hours. 
To remove systematic effects, the direction of polarization was reversed several times, typically once per week of experiment running time. 
To remove backgrounds from interactions of the photon beam with the material of the cryostat and non-hydrogen nucleons in the butanol target and He bath, separate data were taken using a carbon foam target, POCOFoam  \cite{Pocofoam}, with density $0.55\, \mathrm{g}/\mathrm{cm}^3$  inserted into the cryostat. 
The density of the carbon foam was such that a cylinder of identical geometric size to the butanol target provided a close approximation to the number of non-hydrogen nucleons in the butanol target, allowing for a simple $1$:$1$ subtraction accounting only for differences in luminosity. 

Event selection is similar to that described for the Mainz $\Si_3$ analysis. 
Even with the exclusivity selection, accidental subtraction, and opening angle requirement, backgrounds persist into the missing-mass spectrum, similar if not worse than those shown in \figref{sigma3_mm}.
Typically these backgrounds
 are from $\pi^{0}$ events, where a low-energy decay photon escaped detection by passing up or down the beam-line, or through the gap between the CB and TAPS. An estimate of the $\pi^{0}$ background was made by measuring the rate of good $\pi^{0}$ events where both decay photons are detected, but one of the photons is detected in a detector region  adjacent to a region with reduced or zero acceptance. 
The subtraction of backgrounds is done separately for each helicity state, as the $\pi^{0}$ backgrounds themselves result in non-zero asymmetries.
After removing the background contributions, the final missing-mass distribution is shown in \figref{MMaSubt}.
A simulation of the CS lineshape shows good agreement between data and calculation for the Compton peak. 
A relatively conservative integration limit of $940 \,\mathrm{MeV}$ was used in the analysis.
\begin{table}[hbt]
\footnotesize
\centering
\caption{Predictions for proton spin polarizabilities compared with experimental extractions (in units of $10^{-4}\,\mathrm{fm}^{4}$).\label{Tab:SPs_theory}}
\begin{tabular}
{|c|c|c|c|c|c|c|c|c|c|c|}
\hline
&$O(p^{4})_{b}$ & $O(\epsilon^3)$  & $O(p^{4})_{a}$ & K-matrix  & HDPV & DR& $L_{\chi}$ & HB$\chi$PT & B$\chi$PT &  Experiment  \\
& \cite{Gel01}&\cite{Hem98} & \cite{Kum00} & \cite{Kondratyuk:2001qu} & \cite{Holstein:1999uu}& \cite{Bab98}&  \cite{Gas11}&\cite{McG13}  & \cite{Lensky:2015awa}&  \\\hline

$\gamma_{E1E1}$ & $-1.9$ & $-5.4$ & $-1.3$ & $-4.8$ & $-4.3$ & $-5.6$ & $-3.7$ & $-1.1 \pm 1.8_{\,\mathrm{th}}$ & $-3.3\pm0.8$ &$-3.5 \pm 1.2$  \cite{Martel15} \\ 

$\gamma_{M1M1}$ & $0.4$ & $1.4$ & $3.3$ & $3.5$ & $2.9$ & $3.8$ & $2.5$ & $2.2 \pm 0.5\pm 0.7_{\,\mathrm{th}}$ & $2.9\pm1.5$& $3.16 \pm 0.85$  \cite{Martel15} \\ 

$\gamma_{E1M2}$ & $0.7$ & $1.0$ & $0.2$ & $-1.8$ & $-0.02$ & $-0.7$ & $1.2$ & $-0.4 \pm 0.4_{\,\mathrm{th}}$ & $0.2\pm0.2$& $-0.7 \pm 1.2$  \cite{Martel15} \\ 

$\gamma_{M1E2}$ & $1.9$ & $1.0$ & $1.8$ & $1.1$ & $2.2$ & $2.9$ & $1.2$ & $1.9 \pm 0.4_{\,\mathrm{th}}$ & $1.1\pm0.3$& $1.99 \pm 0.29$  \cite{Martel15} \\ 

$\gamma_{0}$ & $-1.1$ & $1.9$ & $-3.9$ & $2.0$ & $-0.8$ & $-0.4$ & $-1.2$ & $-2.6$ & $-0.9\pm1.4$ & $-.90 \pm .08 \pm .11$  \cite{Pasquini:2010zr}\\ 

$\gamma_{\pi}$ & $3.5$ & $6.8$ & $6.1$ & $11.2$ & $9.4$ & $13.0$ & $6.1$ & $5.6$ & $7.2\pm1.7$& $8.0 \pm 1.8$  \cite{Cam02} \\ \hline
\end{tabular}
\end{table}

The measured asymmetries are plotted in \figref{AsyE1}.
The curves are from the fixed-$t$ DR model of \citet{Pas07} for values of $\gamma_{E1E1}$ ranging from\footnote{Whenever the units are omitted, it
 is understood that the scalar (spin) polarizabilities 
 are measured in units of $10^{-4}\,\mathrm{fm}^3$ ($\mathrm{fm}^4$).} $-6.3$ to $-2.3$, but with $\gamma_{M1M1}$ fixed at  $2.9$ \cite{Holstein:1999uu}.
The width of each band represents the propagated errors using $\alpha_{E1} = 12.16 \pm 0.58$ and $\beta_{M1} = 1.66 \pm 0.69$, as well as $\gamma_{0}$ and $\gamma_{\pi}$ from Table \ref{Tab:SPs_theory}, combined in quadrature.
The curves graphically demonstrate the sensitivity of the asymmetries to $\gamma_{E1E1}$, showing a preferred solution of $\gamma_{E1E1} \approx -4.3 \pm 1.5$. 

\begin{figure}[tbh]
\centering
\begin{minipage}{0.46\textwidth}
\centering
\includegraphics[scale=0.43]{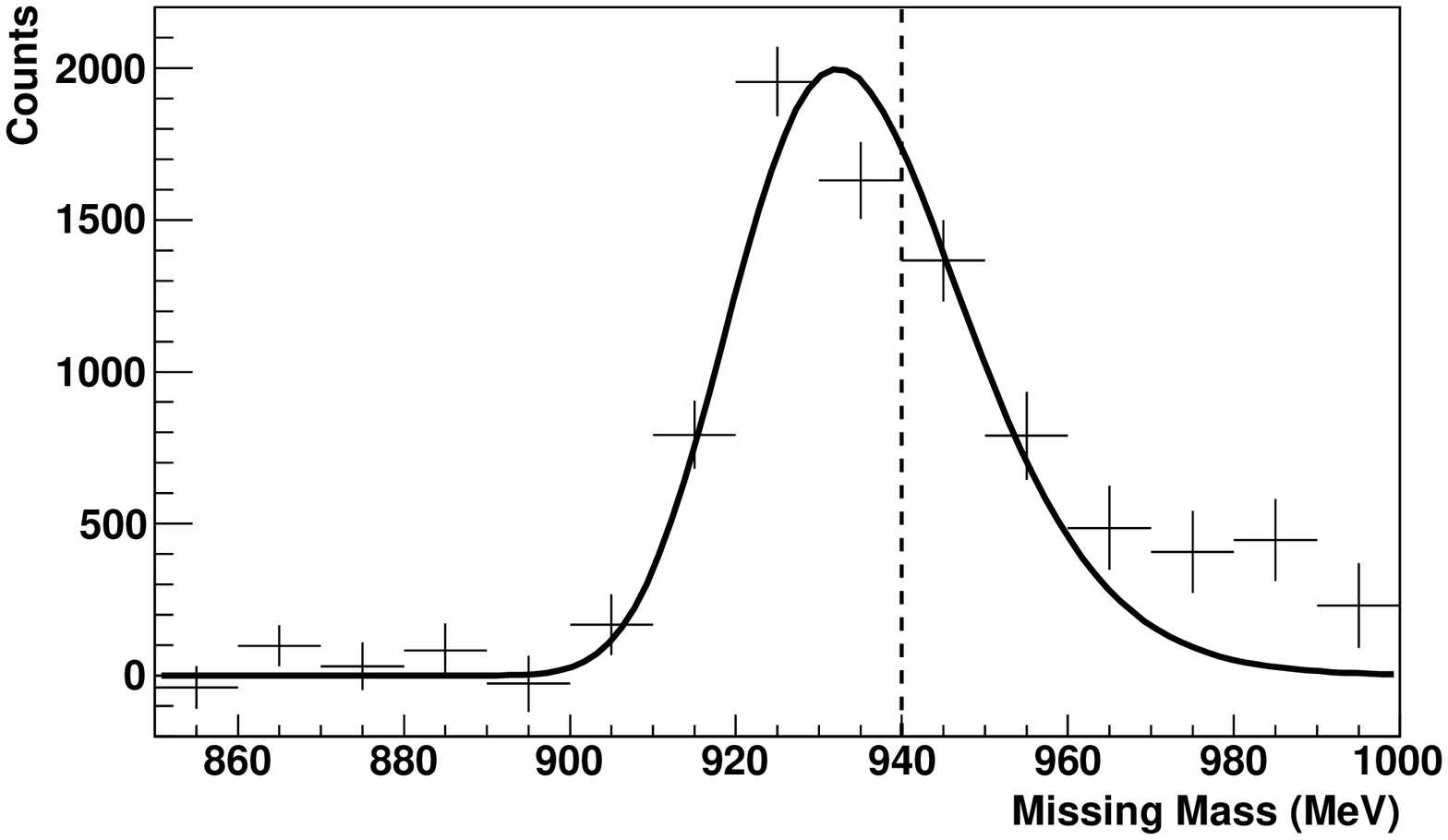}
\caption{Missing-mass spectrum after removal of backgrounds for $E_{\gamma} = 273-303\,\mathrm{MeV}$, and $\vartheta = 100-120^{\circ}$. The solid line is the CS lineshape determined from simulation. The dashed line indicates the upper integration limit used in the analysis. Plot courtesy of Phil~Martel.  \figlab{MMaSubt}}
\end{minipage}\hfill
\begin{minipage}{0.52\textwidth}
\centering
\includegraphics[scale=0.45]{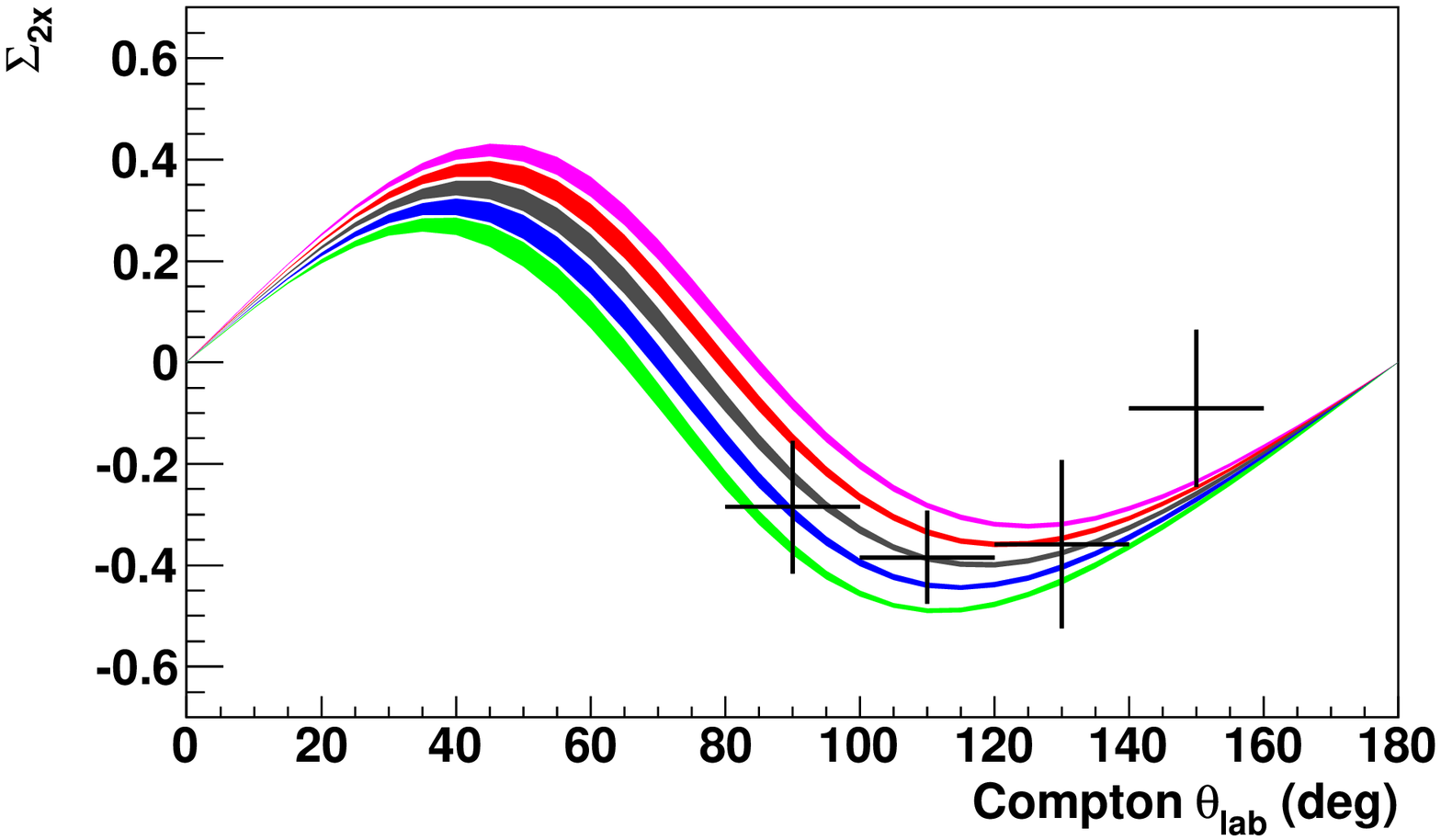}
\caption{$\Si_{2x}$ for $E_{\gamma} = 273-303\,\mathrm{MeV}$. The curves are from a dispersion theory calculation \cite{Pas07} with $\alpha_{E1}$, $\beta_{M1}$, $\gamma_{0}$, and $\gamma_{\pi}$ held fixed at their experimental values, and $\gamma_{M1M1}$ fixed at $2.9$ \cite{Holstein:1999uu}. From bottom to top, the green, blue, brown, red and magenta bands are for $\gamma_{E1E1}$ equal to $-6.3$, $-5.3$, $-4.3$, $-3.3$, and $-2.3$, respectively. The width of each band represents the propagated errors from $\alpha_{E1}$, $\beta_{M1}$, $\gamma_{0}$, and $\gamma_{\pi}$ combined in quadrature. Plot courtesy of Phil~Martel. \figlab{AsyE1}}
\end{minipage}
\end{figure}

Martel et al.~\cite{Martel15} performed a global analysis of single and double-polarized CS data in the $\Delta(1232)$ region \cite{Leg01,Martel15} 
using the DR model of \citet{Pas07}, and the NNLO B$\chi$PT calculation of \citet{Len10} amended with some higher-order LECs which are then fitted to the
data.
Only $\Si_3$ asymmetry points below double-pion photoproduction threshold were used in the analysis. 
In the fitting procedure, $\alpha_{E1} - \beta_{M1}$, $\alpha_{E1}+\beta_{M1}$,  $\gamma_{E1E1}$, $\gamma_{M1M1}$, $\gamma_{0}$, and $\gamma_{\pi}$ were fitted to the asymmetry data sets, and to known constraints on $\alpha_{E1}+\beta_{M1}$, $\alpha_{E1}-\beta_{M1}$, $\gamma_{0}$, and $\gamma_{\pi}$.
The constraint $\alpha_{E1} - \beta_{M1} = (7.6 \pm 0.9)\times 10^{-4}\,\mathrm{fm}^3$ is taken from the analysis of Ref.~\cite{Griesshammer:2012we}.

\begin{table}[!ht]
\footnotesize
  \caption{Results from fitting $\Si_{2x}$ and $\Si_{3}$ asymmetries using either a dispersion model calculation (Disp) \cite{Pas07} or a B$\chi$PT calculation \cite{Len10}. Spin polarizabilities are given in units of $10^{-4}\, \mathrm{fm}^4$.} 
  \label{Tab:Fit}
\centering
  \begin{tabular}{|c|c|c|c|c|}
    \hline
    $\Si_{2x}$ \cite{Martel15} & $\Si_3$ \cite{Leg01} & Model & $\gamma_{E1E1}$ & $\gamma_{M1M1}$ \\ \hline
   \checkmark  & & Disp & $-4.6 \pm 1.6$ & $-7 \pm 11$ \\ 
     & \checkmark  & Disp & $-1.4 \pm 1.7$ & $3.20 \pm 0.85$ \\ 
   \checkmark & \checkmark & Disp  & $-3.5 \pm 1.2$ & $3.16 \pm 0.85$ \\ 
    \checkmark & \checkmark & B$\chi$PT & $-2.6 \pm 0.8$ & $2.7 \pm 0.5$ \\ \hline
  \end{tabular}
\end{table}

Table \ref{Tab:Fit} shows results from data fitting.
The first two columns give the data sets used for fitting, the third column shows the model used, and the fourth and fifth columns show the results for $\gamma_{E1E1}$ and $\gamma_{M1M1}$.   
The first data row confirms the graphical analysis of Figures~\ref{fig:sigma2x_sensitivity} and \ref{fig:AsyE1}, that the $\Si_{2x}$ data prefer a solution $\gamma_{E1E1} \approx -4.6\times 10^{-4}\,\mathrm{fm}^4$, and the data by themselves have little predictive power for $\gamma_{M1M1}$. The second data row confirms the graphical analysis of \figref{sigma3_sensitivity}, that the $\Si_3$ data have reasonable sensitivity to $\gamma_{M1M1}$, and markedly less sensitivity to  $\gamma_{E1E1}$. 
The third row shows the results from the combined fit of $\Si_{2x}$ and $\Si_{3}$ data using the dispersion model \cite{Pas07}, and the fourth row shows the combined fit using the B$\chi$PT calculation \cite{Len10}.
Within the uncertainties, the results for $\gamma_{E1E1}$ and $\gamma_{M1M1}$ from the two model fits are in agreement,
indicating that the model dependence of the polarizability fitting is comparable to, or smaller than, the statistical errors.

The last column of Table~\ref{Tab:SPs_theory} displays the 
results of \citet{Martel15} for all four spin polarizabilities,  obtained from the combined analysis of $\Si_{2x}$, $\Si_{3}$ using the DR model of \citet{Pas07}. The empirical  results for $\gamma_{0}$ and $\gamma_{\pi}$ shown therein have also been used  in the analysis of \citet{Martel15}. There is a generally good agreement between the extracted spin polarizabilities and the $\chi$PT calculations \cite{McG13,Lensky:2014efa} shown in the table. The other
calculations lack an uncertainty estimate, which makes them harder
to judge. Nevertheless, it is rather clear from the table that 
the fixed-$t$ dispersive framework \cite{Holstein:1999uu,Pas07} and causal K-matrix modeling \cite{Kondratyuk:2001qu,Gas11} agree rather well with the experiment as well.

\subsubsection{Circularly Polarized Photons and Longitudinally Polarized Target}
We finally consider the following double-polarized CS asymmetry: 
\beq\label{Eq:Sigma_2z}
\Si_{2z} = {  \dd\sigma^R_{z}  -   \dd\sigma^L_{z}   \over 
  \dd\sigma^R_{z}  +  \dd\sigma^L_{z}  },
\eeq 
where $\dd\sigma^{R(L)}_{z}$ is the differential cross section for right (left) circularly polarized photons to scatter from a nucleon target polarized in the incident beam direction. Note that the value of $\Si_{2z}$ at the 
zero scattering angle is well-known from the sum rules for the forward CS amplitudes, see \Secref{theory3}.

\begin{figure}[htb] 
  \centering 
\begin{minipage}[t]{0.45\textwidth}
    \centering 
       \includegraphics[width=\textwidth]{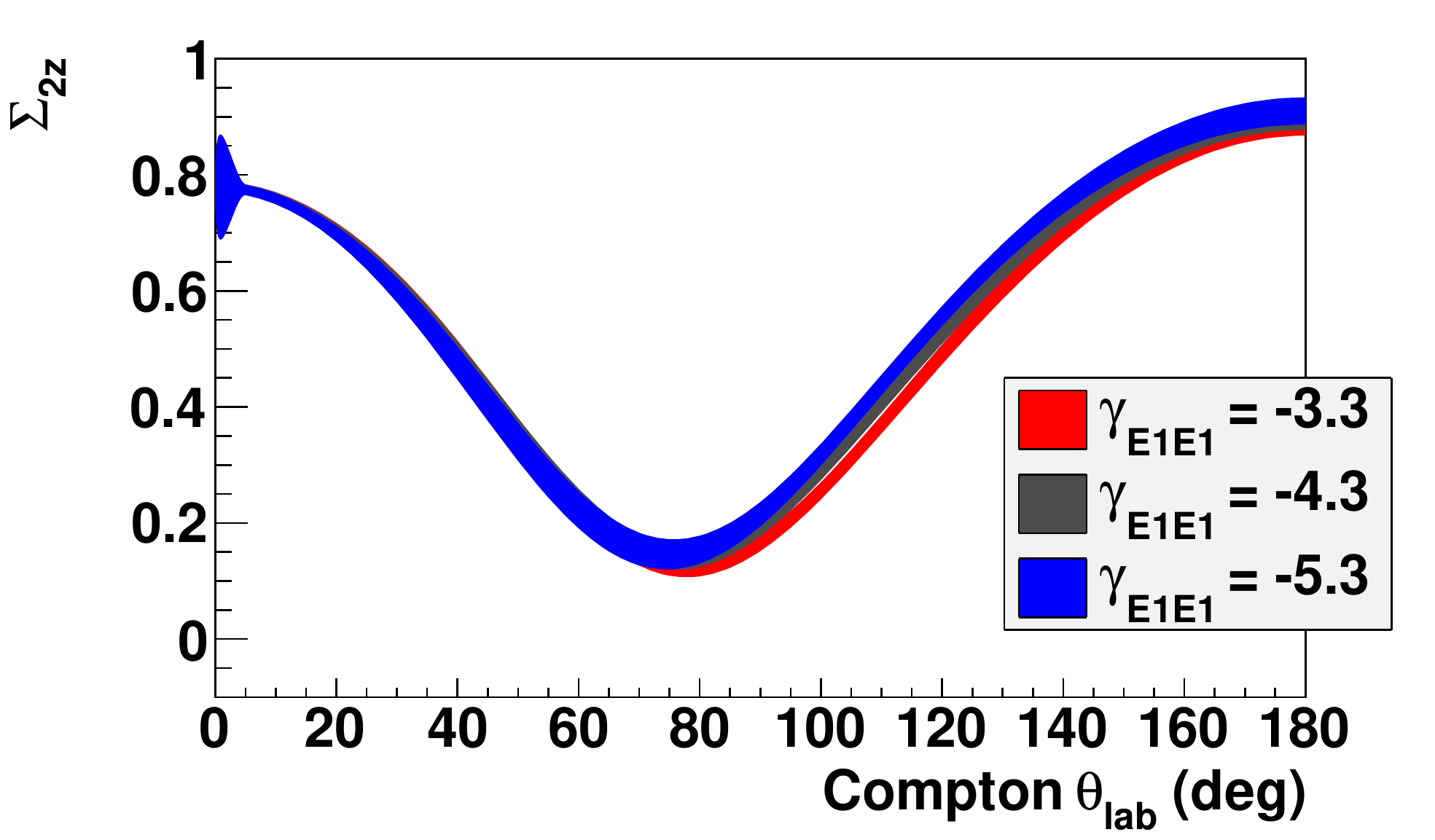}
\end{minipage}
\begin{minipage}[t]{0.45\textwidth}
    \centering 
       \includegraphics[width=\textwidth]{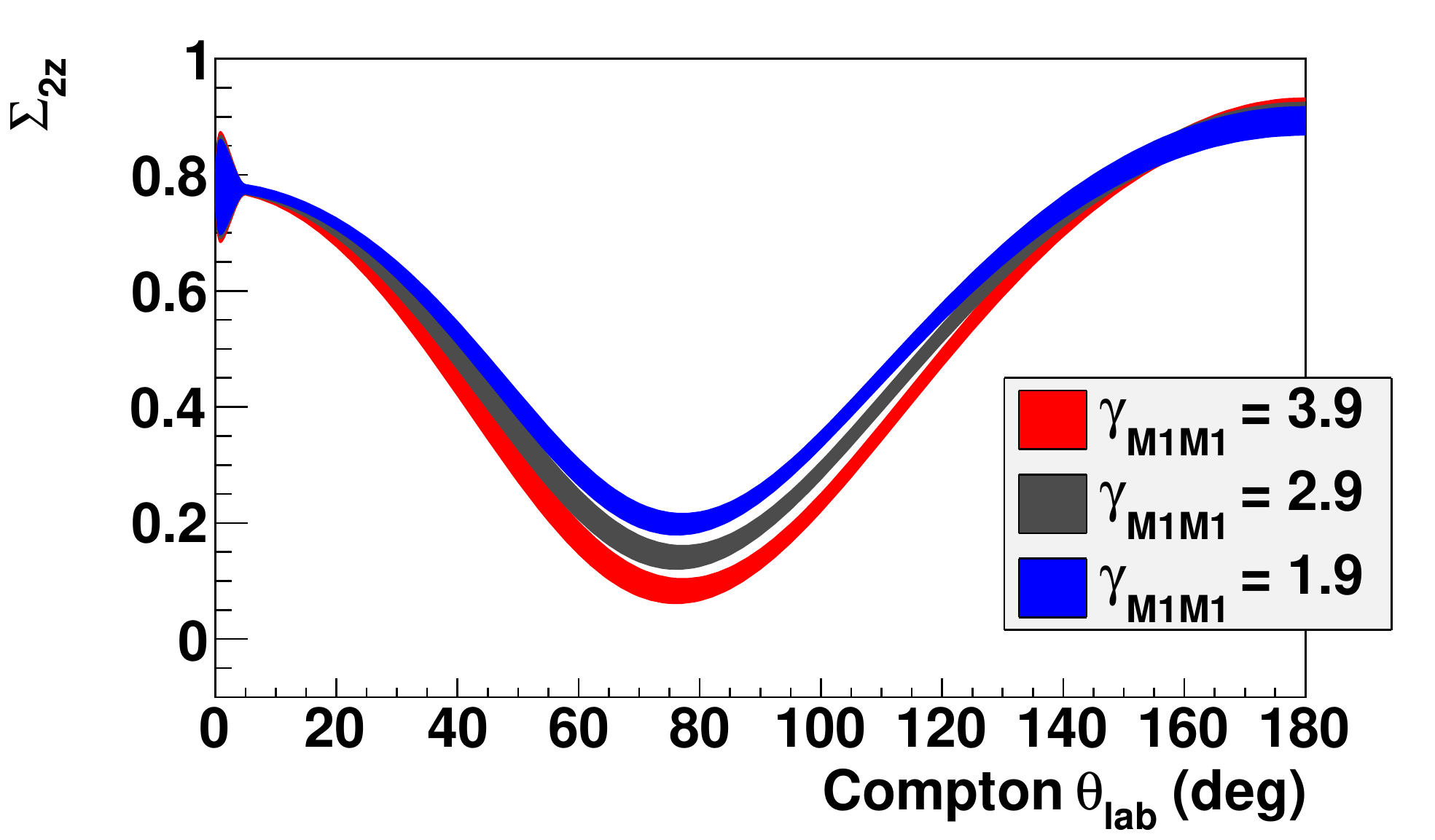}
       \end{minipage}
 \caption{The asymmetry $\Si_{2z}$ for $E_{\gamma} = 290\,\mathrm{MeV}$. The curves are explained in \figref{sigma2x_sensitivity}. Plot courtesy of Phil~Martel.
\label{fig:sigma2z_sensitivity}}
\end{figure}

Figure~\ref{fig:sigma2z_sensitivity} shows the sensitivity of $\Si_{2z}$ to the spin polarizabilities. The definition of the curves is identical to that of \figref{sigma2x_sensitivity}.   The left panel shows little sensitivity to $\gamma_{E1E1}$, while the right panel shows significant sensitivity to $\gamma_{M1M1}$. A measurement of $\Si_{2z}$ will thus compliment the information obtained from the $\Si_{2x}$ asymmetry.  
Data taking on the $\Si_{2z}$ asymmetry started at MAMI in 2014, and  
continued in 2015. 

\begin{figure}[htb]
\centering
\includegraphics[width=10cm]{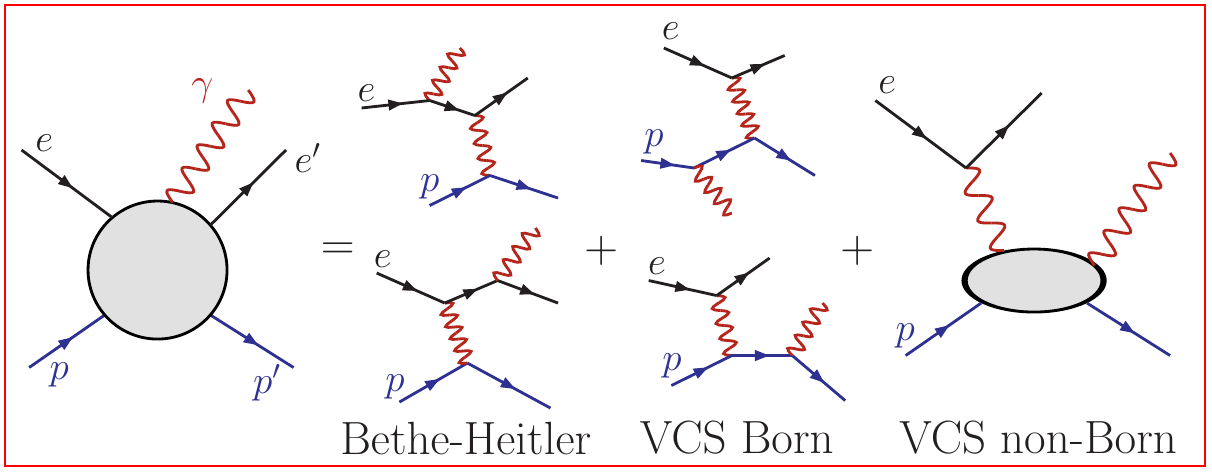}
\caption{The photon electroproduction process
giving access to the VCS amplitude. \figlab{VCS_interactions}}
\end{figure}

\subsection{Virtual Compton Scattering}
Although the electric, magnetic and even the spin polarizabilities of the proton are now known with reasonable accuracy from CS experiments, relatively little is known about the distribution of polarizability density inside the nucleon. To measure a polarizability density it is necessary to use the VCS reaction \cite{Guichon:1995}, where the incident photon is virtual. The VCS reaction is sensitive to the generalized electric and magnetic polarizabilities $\alpha_{E1}(Q^2)$ and $\beta_{M1}(Q^2)$.\footnote{Note that the connection between the scalar GPs (i.e., the VCS polarizabilities discussed in this section), and the VVCS polarizabilities (e.g., the generalized Baldin sum rule of \eref{genBaldinSR})
is not known  at finite $Q^2$. A low-$Q$ relation for some of the spin GPs
exists, cf.~\Eqref{S12sr3}. Alternative expansions of VVCS and VCS in generalized polarizabilities were proposed in Refs.~\cite{Gorchtein:2008aa}
and \cite{Gorchtein:2009wz}.} 

\subsubsection{Response Functions and Generalized Polarizabilities}
\seclab{VCSPol}

The relationship between VCS cross sections and the polarizabilities is most easily seen in the LEX of the unpolarized VCS cross section \cite{Guichon:1995}:
\beq\label{Eq:VCS_cross}
\dd^5\sigma^{VCS} = \dd^5\sigma^\mathrm{BH+Born}+ \vert\bq^{\prime}\vert\, \varPhi\, \Psi_0(\vert\bq\vert,\epsilon,\theta,\phi) + O(\vert\bq'\vert^2), 
\eeq
where  $\vert\bq\vert (\vert\bq^{\prime}\vert)$ is the absolute value of the incident (final) photon three-momentum in the photon-nucleon c.m.\ frame,  $\epsilon$ is the photon polarization-transfer parameter, $\theta (\phi)$  is the c.m.\ polar (azimuthal) angle for the outgoing photon, and  $\varPhi$ is a phase space factor. Note that $\dd^5\sigma^\mathrm{BH+Born}$ is the cross section for the Bethe-Heitler + Born amplitudes only, i.e., no polarizability information is contained, and it is exactly calculable from QED and the nucleon form factors (FF). The Bethe-Heitler and Born diagrams for the VCS reaction are shown in \figref{VCS_interactions}. The polarizabilities enter the cross section expansion at order $O(\vert\bq^{\prime}\vert)$ through the term  $\Psi_0$, given by \cite{Guichon:1998xv}: 
\beq\label{Eq:VCS_GPs}
\Psi_0(\vert \bq\vert,\epsilon,\theta,\phi)=V_1 \bigg[ P_{LL}(\vert\bq \vert)- \frac{P_{TT}(\vert\bq \vert)}{\epsilon} \bigg] + V_2 \sqrt{\eps(1+\eps)} P_{LT}(\vert\bq \vert),
\eeq
where $P$'s are the response functions of unpolarized VCS, and $V$'s are
functions of kinematical variables, the $\eps$-dependence is written out explicitly. In the limit of $\vert\bq\vert \rightarrow 0$, $P_{LL} \propto \alpha_{E1}$, $P_{TT} \propto \gamma_{E1M2}$, and $P_{LT} \propto \beta_{M1}$.
Therefore, response function $P_{LL}(\vert\bq \vert)$ is proportional to $\alpha_{E1}(Q^2)$, 
$P_{LT}(\vert\bq \vert)$ is proportional to $\beta_{M1}(Q^2)$  plus a spin polarizability term, and  $P_{TT}(\vert\bq \vert)$  is proportional to spin polarizabilities. 

There have been two analysis techniques utilized to obtain the response functions $P_{LL}(\vert\bq \vert)- P_{TT}(\vert\bq \vert)/\epsilon$ and $P_{LT}(\vert\bq \vert)$ from VCS cross sections. The first technique is the LEX, cf.\ Eqs.~(\ref{Eq:VCS_cross}) and (\ref{Eq:VCS_GPs}).  In the LEX analysis, the response functions are fitted to the $\theta$, $\phi$ and $\vert \bq'\vert$ dependence of VCS cross sections at fixed $\vert\bq\vert$ and $\epsilon$.  To find the generalized polarizabilities $\alpha_{E1}(Q^2)$ and $\beta_{M1}(Q^2)$, a theoretical calculation of $P_{TT}$ and the spin polarizability contributions to $P_{LT}$ must be utilized, and the predictions subtracted from the experimental results for $P_{LL}- P_{TT}/\epsilon$ and $P_{LT}$.  VCS experiments have generally operated in kinematic regions where the spin polarizability contributions are small, but not negligible.   For example, in the kinematics of the MIT-Bates VCS experiment \cite{Bourgeois2011}, it is  estimated that the spin polarizability contribution to $P_{LL}-  P_{TT}/\epsilon$ is $8\%$, and the contribution to $P_{LT}$ is $31\%$ \cite{Hemmert:1997at,Hemmert:1999pz}.

The second technique uses the VCS dispersion model \cite{Drechsel:2002ar}.  
In this analysis, the
VCS amplitudes obtained from the MAID $\gamma^* p \rightarrow \pi N$   multipoles \cite{Drechsel:1998hk} are held fixed, and the two unconstrained asymptotic contributions to the VCS amplitudes are fit to experimental data at fixed $Q^2$. For data fitting, a dipole ansatz has traditionally been used \cite{Drechsel:2002ar} to parametrize the asymptotic contributions: 
\begin{subequations}
\label{Eq:GP_Q2}
\bea
\alpha_{E1}(Q^2) -\alpha_{E1}^{\pi N}(Q^2) &=& { \alpha_{E1} - \alpha_{E1}^{\pi N} \over (1+Q^2/\Lambda^2_{\alpha})^2},
 \\
\beta_{M1}(Q^2) -\beta_{M1}^{\pi N}(Q^2) &=& { \beta_{M1} - \beta_{M1}^{\pi N} \over (1+Q^2/\Lambda^2_{\beta})^2}.
\eea
\end{subequations}
In the first equation, $\alpha_{E1}$  is the experimental electric polarizability from RCS, $\alpha_{E1}^{\pi N}$ is the calculated $\pi N$  contribution to the electric polarizability at $Q^2=0$ , and $\alpha_{E1}^{\pi N}(Q^2)$  is the calculated $\pi N$  contribution to the electric polarizability at finite $Q^2$.  The definitions are the same for the magnetic polarizability. The only free parameters in Eq.~(\ref{Eq:GP_Q2}) are  $\Lambda_{\alpha}$ and $\Lambda_{\beta}$.  
The advantage of parameterizing the generalized polarizabilities this way is that once the parameters $\Lambda_{\alpha}$  and $\Lambda_{\beta}$ are fixed from fitting VCS cross sections,  the formalism has  predictive power for the response functions and generalized polarizabilities at other $Q^2$ values, provided of course the dipole assumption in Eq.~(\ref{Eq:GP_Q2}) is valid. 
Once the parameters $\Lambda_{\alpha}$  and $\Lambda_{\beta}$  are determined, the generalized polarizabilities are calculated from Eq.~(\ref{Eq:GP_Q2}), and the response functions $P_{LL}-P_{TT}/\epsilon$ and  $P_{LT}$ are found by summing the asymptotic terms with calculated  spin polarizability contributions.

\begin{figure}
\begin{minipage}[t]{0.49\textwidth}
   \centering
\includegraphics[scale=0.44]{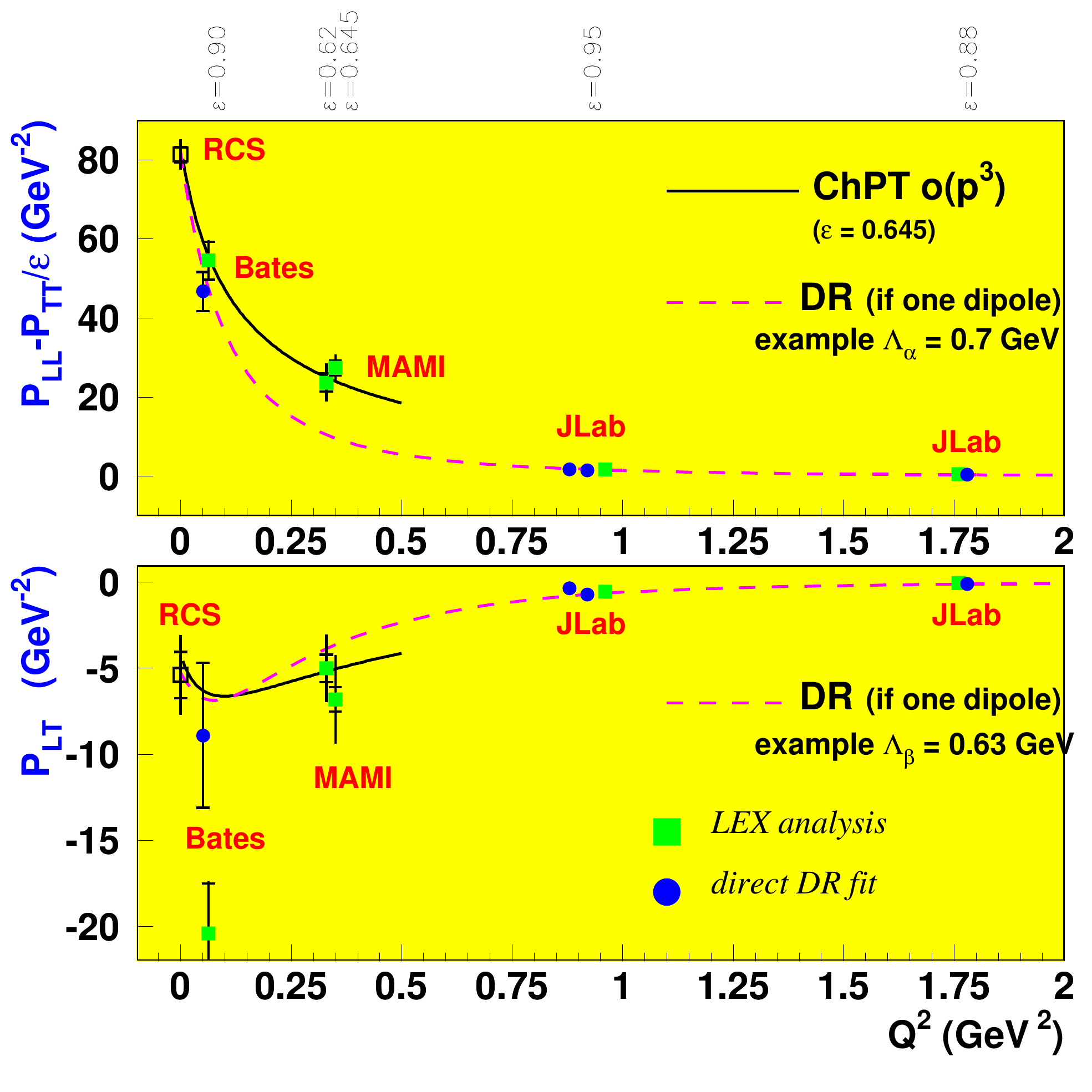}
\caption{VCS response functions from: \cite{Bourgeois2011}, Mainz 2000 \cite{Roche:2000}, Mainz 2008 \cite{Janssens:2008qe}, and JLab  \cite{Laveissiere:2004}. RCS points correspond with the older
values for polarizabilities~\cite{Schumacher:2005an}. The solid curves are $O(p^3)$  HB$\chi$PT calculation \cite{Hemmert:1999pz} with $\epsilon = 0.9$.  The dashed curve is a dispersion-model fit \cite{Drechsel:2002ar} to the RCS and MIT-Bates data points. Plot courtesy of Helene~Fonvieille. \figlab{structure_functions}}
 \end{minipage}
 \hfill
\begin{minipage}[t]{0.49\textwidth}
\centering
\includegraphics[scale=0.46]{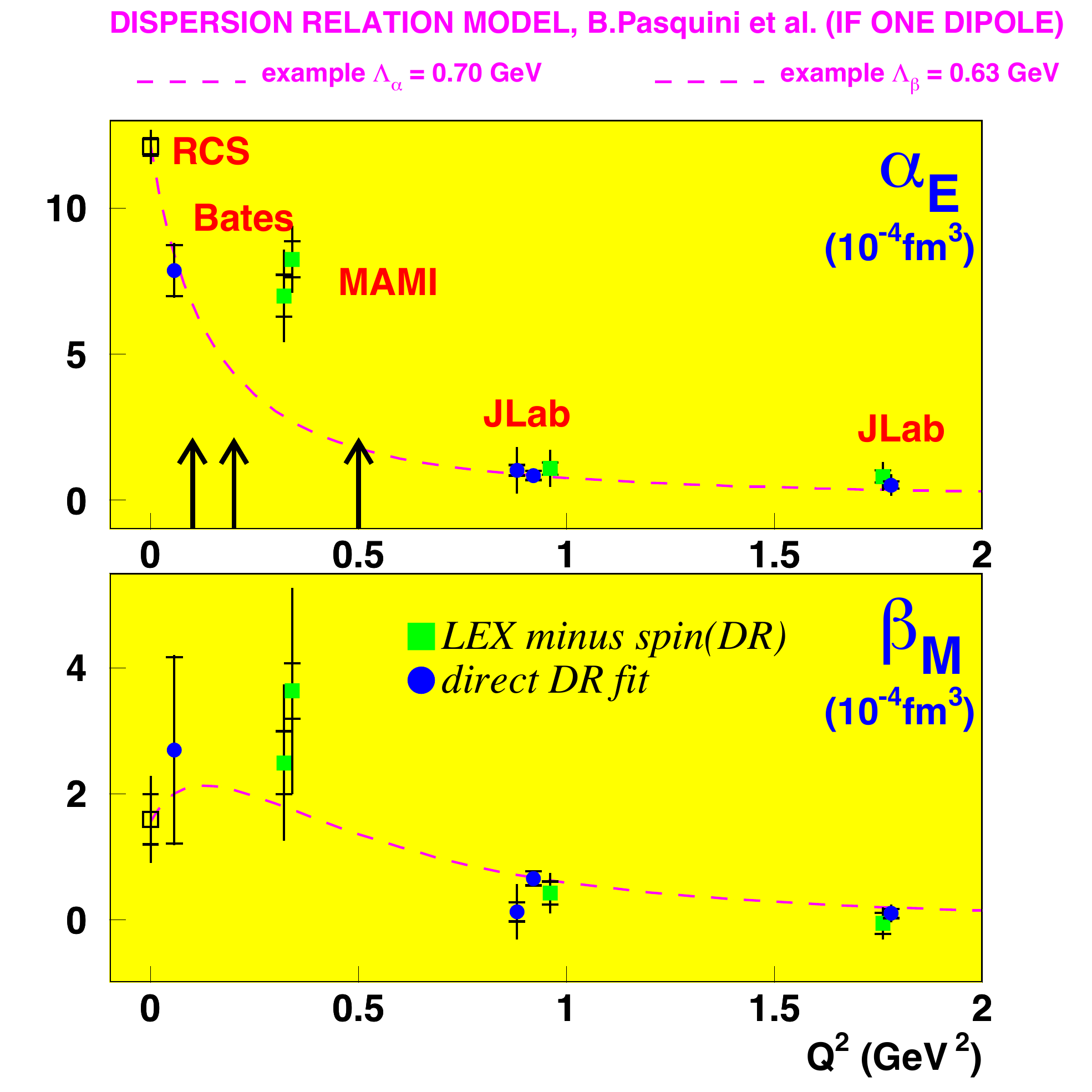}
        \caption{The generalized polarizabilities $\alpha_{E1}(Q^2)$ (upper panel) and $\beta_{M1}(Q^2)$ (lower panel). The curves and data references are the same as in  \figref{structure_functions}. The arrows indicate
        $Q^2$ of forthcoming MAMI data.
        Plot courtesy of Helene~Fonvieille.
        \figlab{VCS_GP_alpha}}
 \end{minipage}
 \end{figure}
 

The VCS response functions $P_{LL}(\vert \bq \vert) - P_{TT}(\vert \bq \vert)/\epsilon$ and $P_{LT}(\vert \bq \vert)$ are plotted in \Figref{structure_functions}  for the three lowest $Q^2$ VCS experiments: MIT-Bates \cite{Bourgeois2011}, Mainz \cite{Roche:2000} \cite{Janssens:2008qe}, and JLab \cite{Fonvieille:2012cd}. The dashed curves in \figref{structure_functions} are the dispersion model calculations   assuming the dipole choice of Eq.~(\ref{Eq:GP_Q2}), and the fitted values for $\Lambda_{\alpha}$  and $\Lambda_\beta$ that by construction make the dispersion calculations go directly through the RCS and JLab data points.  As shown in  \figref{structure_functions}, the dipole ansatz of Eq.~(\ref{Eq:GP_Q2}) allows for a unified description of all low-$Q^2$ VCS response function measurements, with the exception of the Mainz  $P_{LL} - P_{TT}/\epsilon$,
and to a lesser extent $P_{LT}$, measurements at $Q^2= 0.33\, \mathrm{GeV}^{2}$. 

The generalized polarizabilities $\alpha_{E1}(Q^2)$ and $\beta_{M1}(Q^2)$ are shown in \figref{VCS_GP_alpha}. 
The dashed curves in the figures are from the dispersion model calculation using the same $\Lambda_\alpha$  and $\Lambda_\beta$   as  shown in \figref{structure_functions}.
Similar to the situation shown in \figref{structure_functions}, the dipole assumption of Eq.~(\ref{Eq:GP_Q2}) also allows for a unified description of all low-$Q^2$ polarizability measurements, with the exception of the Mainz 
measurements at $Q^2= 0.33\, \mathrm{GeV}^{2}$. 

The $\pi N$  contribution to the electric polarizability is positive at low $Q$, but quickly decreases and crosses 0 at $Q^2 \approx 0.1\, \mathrm{GeV}^{2}$.   Having a negative contribution to electric polarizability is not unphysical. 
A negative electric polarizability occurs in a class of materials known as ferroelectrics, where the internal electric field of the material is stronger than the applied external field. 
The $\pi N$  contribution to the magnetic polarizability is paramagnetic (positive) in HB$\chi$PT, as discussed in \Secref{theory2}. The asymptotic contribution is diamagnetic (negative) and mimics the effect of the LECs. 

\subsubsection{Radial Distribution of the Electric Dipole Polarizability}
\begin{wrapfigure}[38]{l}{0.45\textwidth}
   \centering 
  \includegraphics[scale=1.1]{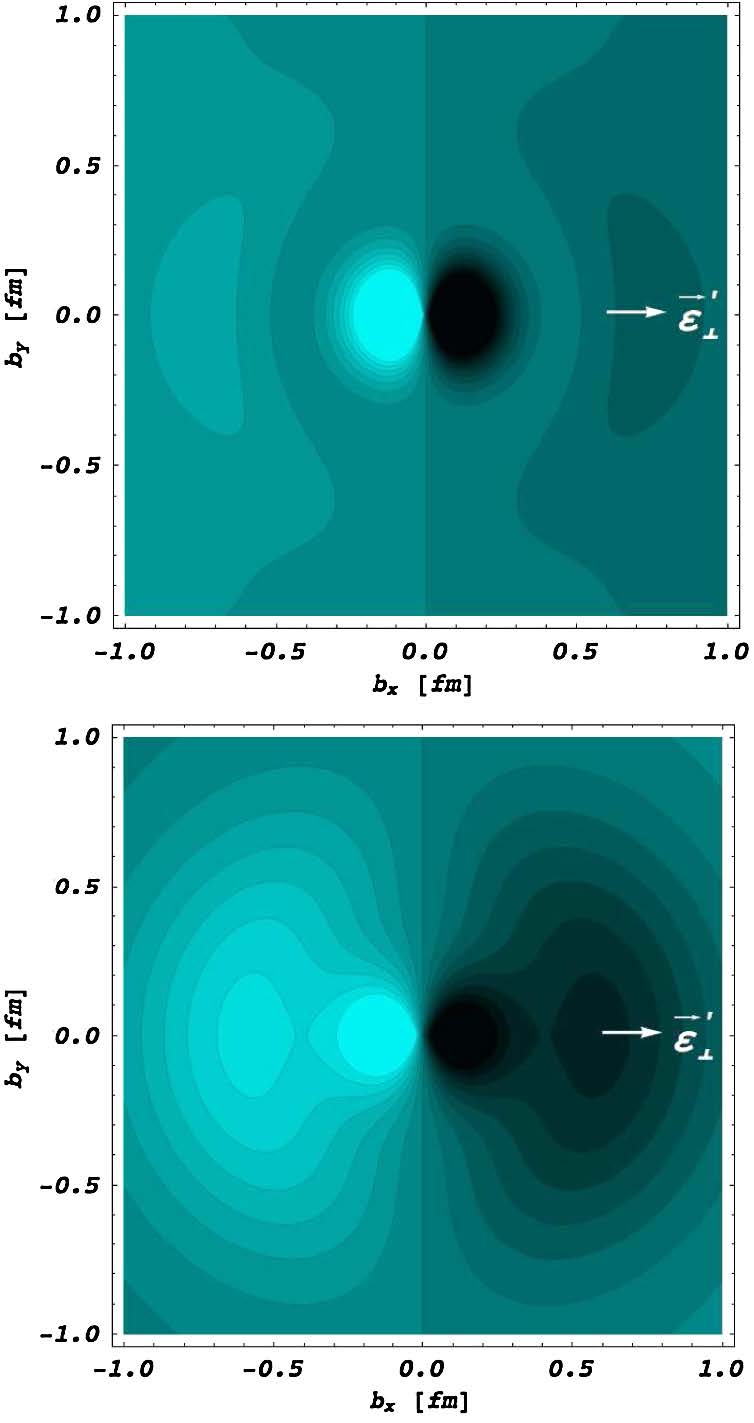}
\caption{The induced polarization of the proton for a transverse electric field.   White denotes positive induced polarization, and black denotes negative induced polarization.   The upper panel is for the dispersion fit for $\alpha_{E1}(Q^2)$ shown in \figref{VCS_GP_alpha}. The lower panel is for $\alpha_{E1}(Q^2)$ adjusted to fit the Mainz data. Plot courtesy of Marc~Vanderhaeghen. \figlab{induced_polarization}}
\end{wrapfigure} 

The mean-square electric polarizability radius, defined as
\begin{flalign}\label{Eq:rms}
\langle r^2 \rangle_{\al_{E1}} = - {6 \over \alpha_{E1} } {\dd \alpha_{E1}(Q^2) \over \dd Q^2 } \bigg|_{Q^2=0}\;,
\end{flalign}
was extracted for the proton by \citet{Bourgeois2011} using the DR
fit to the MIT-Bates data. Their result, 
\begin{flalign}\label{Eq:rms_value}
\langle r^2 \rangle_{\al_{E1}} = \left(2.02+ \left[^{+0.39}_{-0.59}\right]\right)\, \mathrm{fm}^2,
\end{flalign}
where the error is statistical only, is in good agreement with the HB$\chi$PT prediction  \cite{Hemmert:1996gr} of $1.7\,\mathrm{fm}^2$.  

The mean-square polarizability radius is thus significantly larger than the 
proton mean-square charge radius (which is about $0.77\,\mathrm{fm}^2$) demonstrating the dominance of mesonic effects in the electric polarizability. The additional e.m.\ vertex in the polarizability diagram relative to the FF diagram increases the range of the interaction by approximately a factor of two as compared to the charge FF.  Also of interest is the uncertainty principle estimate for the mean square radius of the pion cloud, $\langle r^2 \rangle \approx (1 /m_\pi)^2 = 2\, \mathrm{fm}^2$, which is in better agreement with $\langle r^2 \rangle_{\al_{E1}}$. 

As seen in \figref{VCS_GP_alpha},  there is poor agreement at $Q^2 = 0.33\, \mathrm{GeV}^{2}$ between the measured values of $\alpha_{E1}(Q^2)$ and the dispersion model fit to the RCS and MIT-Bates data points.  This discrepancy has been analyzed by Gorchtein et al.~\cite{Gorchtein:2009qq} using a light-front interpretation of the generalized polarizabilities. This formalism provides a way to calculate  deformations of the quark charge
densities when an external e.m.\ field is applied. They found that adding a Gaussian term to Eq.~(\ref{Eq:GP_Q2}) to improve agreement with  data at $Q^2=0.33\,\mathrm{GeV}^{2}$, also gives the proton 
a pronounced structure in its induced polarization at large
transverse distances,  $0.5$ to $1\,\mathrm{fm}$. This is vividly shown in \figref{induced_polarization}, where the bottom panel shows the effect of adjusting $\alpha_{E1}(Q^2)$ to fit the Mainz data points. 
Clearly, additional VCS data in the low to intermediate $Q^2$ region $0.06$ to $0.5\,\mathrm{GeV}^{2}$ are needed to confirm this prediction. New VCS data have recently been taken by the Mainz A1 collaboration at $Q^2 \approx 0.1$, $0.2$ and $0.5 \,\mathrm{GeV}^{2}$, and this data are currently under analysis \cite{Fonvieille2015}.

\subsection{Timelike Compton Scattering}

The process of {\it dilepton photoproduction} from protons by cosmic microwave
background radiation, $\gamma p \to  p \ell^+ \ell^-$, is one of the main
mechanisms for depletion of cosmic-ray energy in the universe.  It is
dominated by the Bethe-Heitler (BH) process, \Figref{BetheHeitler},  which can be accurately
calculated from QED alone.  In certain kinematics, however, the nuclear
component---represented by timelike VCS---dominates.  Until now there has been
only one lab-based experiment dedicated to dilepton photoproduction off the proton~\cite{Alvensleben:1973mi}.
Conducted in the
70s, it aimed to measure the forward CS
amplitude $f(\nu)$ at $\nu\approx 2.2\,\mathrm{GeV}$, and to verify the Kramers-Kronig relation for the proton, cf. \Secref{theory3}. Recently, \citet{Pauk:2015oaa} made a proposal to study the ratio between photoproduction of $e^-e^+$ and $\mu^-\mu^+$ pairs on a proton target in the limit of very small momentum transfer. This measurement would serve as a test of lepton universality and is of interest to the proton-radius puzzle.

Without distinguishing electrons from positrons, the cross-section element (for the $e^+e^-$ photoproduction) can be
written as an incoherent sum of BH and VCS cross-sections, the
interference terms drop out because of the charge-conjugation symmetry:
\beq
\dd\si = \dd \si^{\mathrm{BH}}+ \dd\si^{\mathrm{VCS}}\,.
\eeq

\begin{figure}
\begin{minipage}[t]{0.49\textwidth}
   \centering
\includegraphics[scale=1.0]{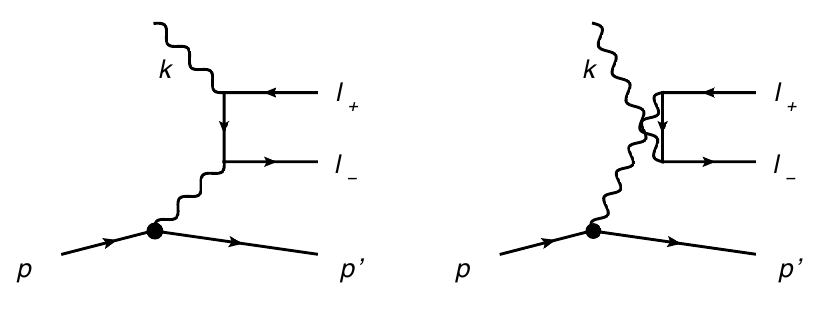}
\caption{Bethe-Heitler process in dilepton photoproduction, $\gamma p \rightarrow l^- l^+ p$.  \figlab{BetheHeitler}}
 \end{minipage}
 \hfill
\begin{minipage}[t]{0.49\textwidth}
\centerline{\includegraphics[scale=0.5]{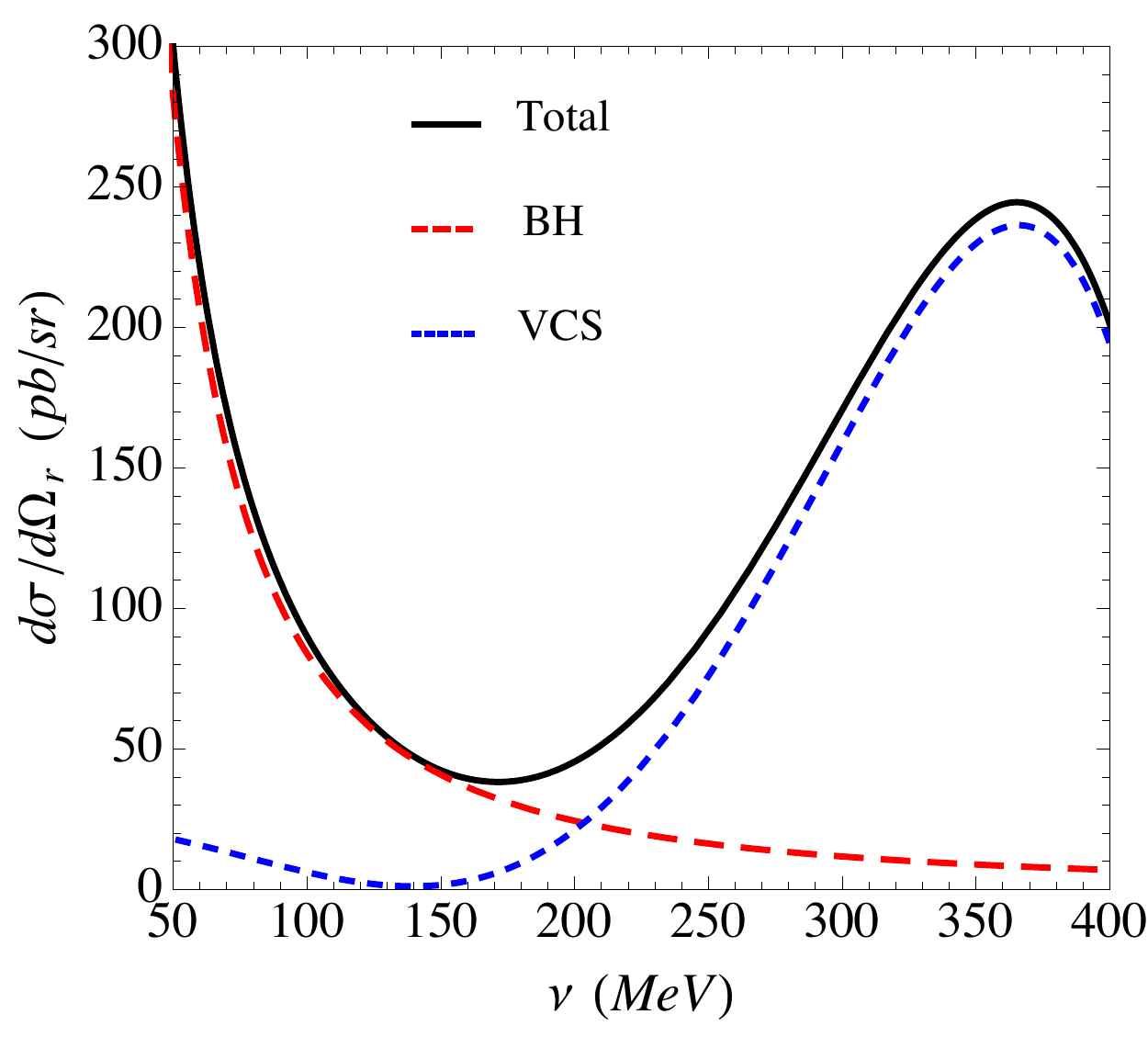}}
\caption{The differential cross-section of $e^+e^-$ photoproduction off the proton
as a function of the photon beam energy for forward-recoil kinematics.
\figlab{recoildist}}
 \end{minipage}
 \end{figure}

In this case VCS can be accessed only in the regions where the BH mechanism is
suppressed.  In the distribution over the solid angle of the recoil proton,
$\varOmega_r$, the leading contribution to BH is given by the two diagrams shown in \Figref{BetheHeitler}, which
are known to cancel in the forward kinematics \cite{Motz:1969ti}. That is
to say that $\dd\si^{\mathrm{BH}}/\dd\varOmega_r$ is smallest when $\vartheta_r=0$, where
for pointlike proton it takes a particularly simple form:
\beq
\frac{ \dd\si^{\mathrm{BH}}}{ \dd\varOmega_r }(\nu, \, \vartheta_r=0) = 
\frac{\al^3}{8\pi^4 \nu^4} \left\{ \left[4m_e^2+\nu^2\right] \, K\left( \mbox{$\sqrt{1-\frac{4m_e^2}{\nu^2}}$}\right)
- 2\nu^2 \, E\left( \mbox{$\sqrt{1-\frac{4m_e^2}{\nu^2}}$}\right)\right\},
\eeq 
with $E$ and $K$ being the elliptic integrals, and $m_e$ the electron mass.
On the other hand, if we neglect for a moment the momentum-transfer
dependence of the VCS process, the VCS cross section factorizes into the RCS
cross section and a factor responsible for the pair production:
\beq
\frac{\dd\si}{ \dd\varOmega_r \, \dd M_{ee}}
=\frac{ \dd\si^{\mathrm{RCS}}}{ \dd\varOmega_r} \, \frac{\al}{3\pi M_{ee}^2}
\sqrt{ 1-\frac{4m_e^2}{M_{ee}^2} } \left(1+\frac{2m_e^2}{M_{ee}^2} \right)
\left[\left(1+\frac{M_{ee}^2}{2M\nu} \right)^2 -\frac{  M_{ee}^2 (M^2+2M\nu) }{M^2\nu^2} \right]^{3/2},
\eeq
where $M_{ee}$ is the invariant mass of the lepton pair.
 
The resulting angular distributions for BH and VCS, as well as their sum, are
shown as functions of photon energy in 
\Figref{recoildist}. The figure
clearly shows that for beam energies above $200\,\mathrm{MeV}$ the CS off
proton is the dominant mechanism. Any substantial deviation from these
predictions can be interpreted as the timelike momentum-transfer dependence
of the Compton process, and hence attributed to the aforementioned effects of
the timelike e.m.\ structure of the nucleon.

\section{Sum Rules}
\seclab{SRs}
\seclab{sumrules3}

\begin{figure}[tbh]
\centering
       \includegraphics[width=6.25cm]{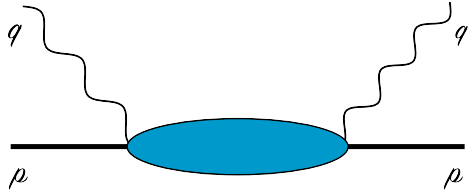}

\caption{Forward Compton scattering: $N(p)+\gamma(q)\rightarrow N(p)+\gamma(q)$, with either real or virtual photons.  \label{fig:CSgeneric}}
\end{figure}

The fundamental relation between light absorption and scattering, encompassed for example
in the celebrated Kramers--Kronig relation,
is manifested 
in a variety of model-independent relations.
They allow us to express certain linear combinations of
polarizabilities in terms of weighted energy integrals of total photoabsorption cross sections, or equivalently,
in terms of the moments of structure functions \cite{GellMann:1954db,Baldin:1960,Drechsel:2002ar,Kuhn:2008sy}.  They all are
derived from the analyticity, unitarity and 
symmetry properties of the forward CS amplitude,
depicted in \Figref{CSgeneric}. In general,
the photons are virtual, with spacelike virtuality
$q^2<0$. The corresponding amplitude is then
referred to as the forward doubly-virtual Compton scattering (VVCS) amplitude. In what follows
we consider its properties, sketch the derivation
of the sum rules, and discuss their empirical 
consequences.

\subsection{Forward Doubly-Virtual Compton Scattering Amplitude}
\seclab{VVCS}
In the forward kinematics ($t=0$), the Lorentz
structure of the VVCS amplitude is decomposed
in four independent tensor forms:
\bea
T^{\mu \nu}(q,p) & = &  
\left( -g^{\mu\nu}+\frac{q^{\mu}q^{\nu}}{q^2}\right)
T_1(\nu, Q^2)  +\frac{1}{M^2} \left(p^{\mu}-\frac{p\cdot
q}{q^2}\,q^{\mu}\right) \left(p^{\nu}-\frac{p\cdot
q}{q^2}\, q^{\nu} \right) T_2(\nu, Q^2)\nn \\
&& -   \frac{1}{M}\gamma^{\mu \nu \al} q_\al \,S_1(\nu, Q^2) -  
\frac{1}{M^2} \big( \gamma^{\mu\nu} q^2 + q^\mu \gamma^{\nu\al} q_\al  -  q^\nu \gamma^{\mu\al}
q_\al \big) S_2(\nu, Q^2),
\eqlab{fVVCS}
\eea
with $T_{1,2}$ the spin-independent and $S_{1,2}$ the spin-dependent invariant amplitudes,
functions of $\nu = (s-u)/4M$ and $Q^2=-q^2$.
This decomposition is explicitly gauge invariant and splits into symmetric
and antisymmetric parts, 
\begin{equation}
T^{\mu \nu} = T^{\mu \nu}_S+T^{\mu \nu}_A,
\end{equation}
which, respectively, {\it do not} and {\it do} depend on the nucleon spin. 
Given that the vector indices are to be contracted with either the polarization vector,
satisfying $q\cdot \veps=0$, or with another gauge-invariant tensor, the terms containing
$q^\mu$ or $q^\nu$ can be omitted, hence,\footnote{It is customary to write the spin-dependent amplitude with the help of 
the nucleon spin four-vector $s^\al$, satisfying $s^2=-1$ and $s\cdot p=0$:
$$
T^{\mu \nu}_A (q,p) = 
\frac{i}{M}\,\epsilon^{\mu\nu\alpha\beta}\,q_{\alpha}
s_{\beta}\, S_1(\nu, Q^2) + \frac{i}{M^3}\,\epsilon^{\mu\nu\alpha\beta}\,q_{\alpha}
(p\cdot q\ s_{\beta}-s\cdot q\ p_{\beta})\, S_2 (\nu, Q^2).
$$
}
\begin{subequations}
\bea
T^{\mu \nu}_S(q,p) & = & -g^{\mu\nu}\,
T_1(\nu, Q^2)  +\frac{p^{\mu} p^{\nu} }{M^2} \, T_2(\nu, Q^2), \eqlab{VVCS_TS}\\
T^{\mu \nu}_A (q,p) & = &-\frac{1}{M}\gamma^{\mu \nu \al} q_\al \,S_1(\nu, Q^2) +
\frac{Q^2}{M^2}  \gamma^{\mu\nu} S_2(\nu, Q^2).\eqlab{VVCS_TA}
\eea
\end{subequations}
One immediate observation is that the symmetry under photon crossing translates into the following
conditions, for real $\nu$:
\begin{subequations}
\eqlab{VVCScrossing}
\bea 
&& T_1 ( -\nu, Q^2) =  T_1 ( \nu, Q^2) ,\quad T_2 ( -\nu, Q^2) =  T_2 ( \nu, Q^2), \\
&& S_1 ( -\nu, Q^2) =  S_1 ( \nu, Q^2), \quad S_2 ( -\nu, Q^2) = -\, S_2 ( \nu, Q^2).
\eea
\end{subequations}
Hence, $S_2$ is odd with respect to the sign reflection of $\nu$, the other amplitudes are even. We will often
consider the combination  $\nu S_2$, such that it has
the same crossing properties as the other
amplitudes.

The Born contribution to these amplitudes is well known (cf.\ \ref{sec:appBorn}) and given by~\cite{Drechsel:2002ar}:
\begin{subequations}
\eqlab{T12Born}
\bea
T_1^{\mathrm{Born}}(\nu, Q^2) &=&\frac{4\pi \alpha}{M}\bigg\{\frac{Q^4\big[F_1(Q^2)+F_2(Q^2)\big]^2}{Q^4-4M^2\nu^2}-F_1^2(Q^2)\bigg\}\,,
\eqlab{T1Born}\\
T_2^{\mathrm{Born}}(\nu, Q^2)&=&\frac{16\pi \alpha M Q^2}{Q^4-4M^2\nu^2}\bigg\{F_1^2 (Q^2)+\frac{Q^2}{4M^2} F_2^2(Q^2)\bigg\}\,,
\eqlab{T2Born}
\\
S_1^{\mathrm{Born}}(\nu, Q^2) &=& \frac{2\pi \alpha}{M}
\bigg\{\frac{4M^2 Q^2\big[F_1(Q^2)+F_2(Q^2)\big]F_1(Q^2)}{Q^4-4M^2\nu^2}-F_2^2(Q^2)\bigg\}\,, \eqlab{S1Born}\\
S_2^{\mathrm{Born}}(\nu, Q^2)&=&
-\, \frac{8 \pi \alpha M^2 \nu}{Q^4-4M^2\nu^2}\big[ F_1 (Q^2)+F_2(Q^2)\big] F_2(Q^2) \,,
\eqlab{S2Born}
\eea
\end{subequations}
where $F_1$ and $F_2$ are the elastic Dirac and Pauli FFs of the nucleon, which are normalized to $F_1(0)=\zZ$ and $F_2(0)=\varkappa$. Introducing the Bjorken variable, 
$x= Q^2/2M\nu$, we recall that the physical region in electron scattering corresponds with $x\in [0,1]$. Obviously the Born graphs exhibit the
pole at $\nu = Q^2/2M\equiv \nu_\mathrm{el}$, or equivalently $x=1$. This nucleon-pole part of the
Born contribution is isolated below, see \Eqref{VVCSpole}. 
The non-Born part of the full amplitudes will be denoted 
as $\ol T_i$,  $\ol S_i$.

\subsection{Unitarity and Relation to Structure Functions}
The optical theorem relates the absorptive parts of the forward VVCS amplitudes to 
the nucleon structure functions\footnote{The unpolarized structure functions $f_1$ and $f_2$ are the standard $F_1$ and $F_2$. However, the latter notation is
reserved here for the Dirac and Pauli FFs respectively.}, or equivalently, to the cross sections of virtual-photon absorption $\ga^\ast N \to X$:\footnote{The flux factor for virtual photons which goes into these cross sections is rather arbitrary, cf.\ \cite{Drechsel:2002ar} for common choices. Our expressions correspond to the flux factor choice $K=\nu$.
Expressions in terms of the structure functions are
not affected by the choice of the flux factor. }
\begin{subequations}
\eqlab{VVCSunitarity}
\bea
\im T_1(\nu,Q^2)&=&\frac{4\pi^2\al}{M}f_1(x,Q^2)=\nu\,\sigma_T(\nu,Q^2), \eqlab{ImT1} \\
\im T_2(\nu,Q^2)&=&\frac{4\pi^2\al}{\nu}f_2(x,Q^2)=\frac{Q^2  \nu }{\nu^2+Q^2}\left[\sigma_T+\sigma_L\right](\nu,Q^2), \eqlab{ImT2}\\
\im S_1(\nu,Q^2) &=& \frac{4\pi^2 \alpha}{\nu} \, g_1(x,Q^2) = 
\frac{M  \nu^2 }{\nu^2+Q^2}\left[\frac{Q}{\nu}\sigma_{LT}  + \sigma_{TT}\right](\nu,Q^2), \eqlab{ImS1}\\
\im S_2(\nu,Q^2) & =&  \frac{4\pi^2 \alpha M}{\nu^2} \, g_2(x, Q^2)  
= \frac{M^2  \nu }{\nu^2+Q^2}\left[\frac{\nu}{Q}\sigma_{LT}  - \sigma_{TT}\right](\nu,Q^2), \eqlab{ImS2}
\eea
\end{subequations}
where the cross sections are defined as: $\sigma_T=\nicefrac12\, (\sigma_{1/2}+\sigma_{3/2})$ and $\sigma_{TT}=\nicefrac12\, (\sigma_{1/2}-\sigma_{3/2})$ for transversely polarized photons, and $\sigma_L=\nicefrac12\, (\sigma_{1/2}+\sigma_{-1/2})$ for longitudinal photons, where the subscript on the right-hand side (\textit{rhs}) indicates the total helicity of the $\gamma^\ast N$ state. The cross section $\sigma_{LT}$ 
corresponds with a simultaneous helicity
change of the photon (from longitudinal to transverse) and the nucleon (spin-flip) such that the
total helicity is conserved. 
These unitarity relations hold in the physical region, where the Bjorken variable is confined to the unit interval.

\begin{figure}[tbh]
\centering
\begin{minipage}{0.49\textwidth}
\centering
       \includegraphics[width=4.5cm]{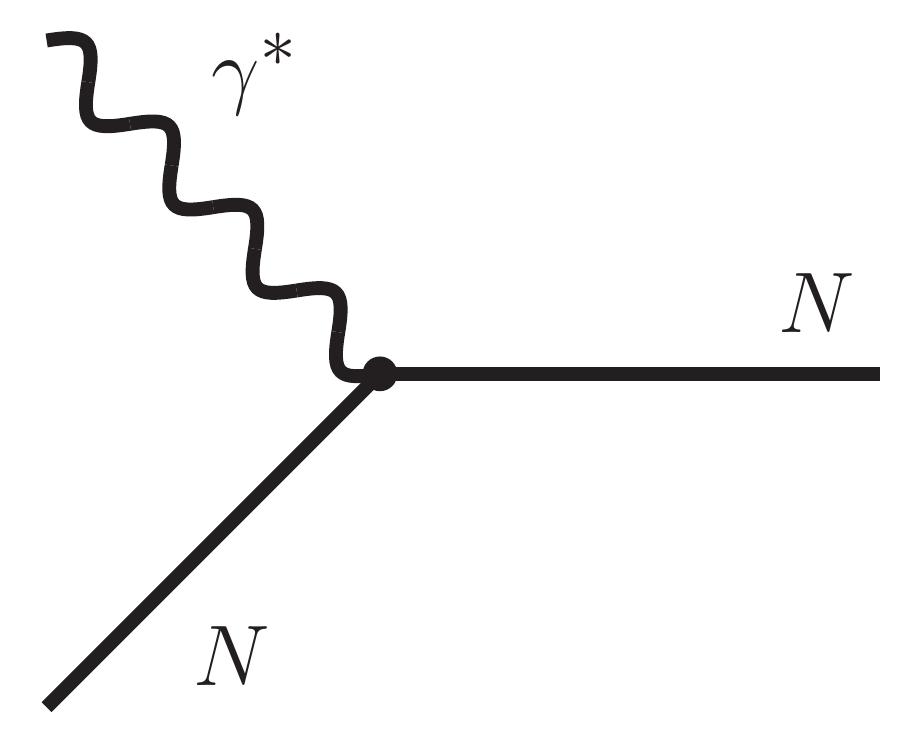}

      (a)
    \end{minipage}\hfill
\begin{minipage}{0.49\textwidth}
\centering
  \includegraphics[width=4.55cm]{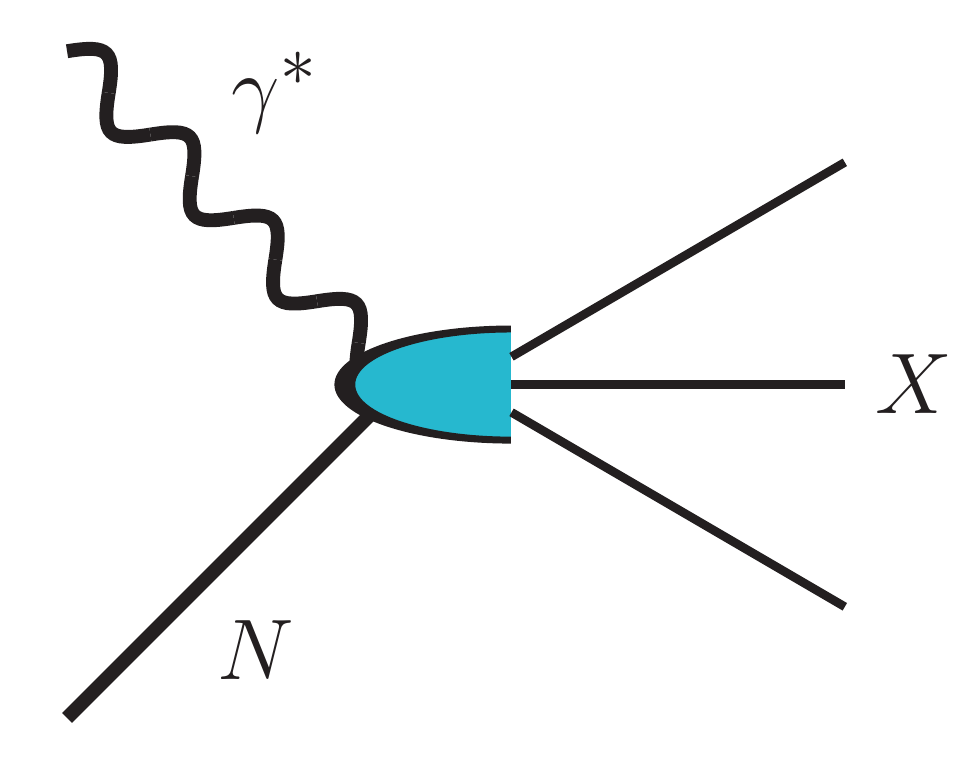}
  
  (b)
\end{minipage}
\caption{(a) `Elastic' and (b) `inelastic' part of the photoabsorption cross section. (a) is related to the `pole' contribution, whereas (b) is related to the `non-pole' contribution. \label{fig:elastic}}
\end{figure}

Figure \ref{fig:elastic} illustrates the photoabsorption process. The structure functions describing the purely elastic scattering, shown in the left panel, are given in terms of the elastic FFs:
\begin{subequations}  
\eqlab{elstructure}
\bea 
f_1^{\mathrm{el}}(x,Q^2) & =  & \frac{1}{2}\,  G^2_M(Q^2)\,  \delta( 1 - x ),\eqlab{f1elastic} \\
f_2^{\mathrm{el}}(x,Q^2) & = & \frac{1}{1 + \tau} \, \big[G^2_E(Q^2) + \tau G^2_M(Q^2) \big]\, \delta( 1 - x), \eqlab{f2elastic}\\
g_1^{\mathrm{el}}(x,Q^2) & = &  \frac{1}{2}\,  F_1(Q^2) \,G_M(Q^2) \, \delta(1 - x), \eqlab{g1elastic}\\
g_2^{\mathrm{el}}(x,Q^2) & = & -  \frac{\tau}{2}\, F_2(Q^2)\, G_M(Q^2)\,  \delta(1 - x)\eqlab{g2elastic},
\eea
\end{subequations}
where  $\tau = Q^2/4M^2$, and 
the electric and magnetic Sachs FFs are
linearly related to the Dirac and Pauli FFs as: 
\bea 
&& G_E(Q^2)  = F_1(Q^2) - \tau F_2(Q^2), 
\quad G_M(Q^2) = F_1(Q^2) + F_2(Q^2).
\eea
Furthermore, $\de(x)$ is the Dirac delta-function, such that 
\beq
\de(1-x) = \nu_{\mathrm{el}}\,  \de(\nu-\nu_{\mathrm{el}}), \quad \mbox{with}\;\,  \nu_{\mathrm{el}} \equiv  2M\tau.
\eeq 

In the limit, $Q^2\to \infty$, fixed $x$, the structure functions are related to the parton distribution functions. We are, however, interested in the limit where $Q$ and $\nu$ are small.
In this case the VVCS amplitudes can on one hand  be expanded in terms of polarizabilities and e.m.\ radii, and on the other in terms of moments of structure functions. This expansion and
the resulting relations between the static e.m.\ properties of the nucleon and
the moments of structure functions will be discussed further 
below. Before that, we need to establish 
the dispersion relations (DRs) for 
the forward VVCS amplitudes.

\subsection{Analyticity and Dispersion Relations}
\seclab{DR}

We next consider the analytic structure of the VVCS amplitudes $T_i$ and $S_i$
in the complex plane of $\nu$. 
We have already seen that the
Born contribution contains the nucleon pole at 
the kinematics of elastic scattering, $\nu=\nu_{\mathrm{el}} $.
The inelastic processes are manifested in the
branch cuts, starting at the first particle-production threshold $\nu_0$ and extending
to infinity. Due to crossing symmetry, the analytic
structure for negative real $\nu$ is similar.
In any case, the physical singularities are confined to the real axis. Elsewhere in the complex plane  
the amplitudes are {\it analytic} (or, holomorphic) functions. The latter requirement is inferred
by micro-causality, a fundamental postulate
of special relativity which states that all the signals propagate strictly within the light-cone. 

For the subsequent discussion it is important that 
the VVCS amplitudes
do not have a pole in the limit $Q^2 \to 0$, then $\nu\to 0$. 
Such a pole can only come from the nucleon propagator entering the
Born contribution. From Eq.~\eref{T12Born} we see that the pole is absent from all the amplitudes except $S_2$. We therefore will write down the DR
for $\nu S_2$, which is pole free for real photons.

These analytic properties, together with the
crossing symmetry conditions from Eq.~\eref{VVCScrossing}, are well-known to result in the following DRs (cf.~\ref{sec:disprel} for the derivation):
\begin{subequations}
\eqlab{genDRs1}
\bea 
T_i ( \nu, Q^2) &=& \frac{2}{\pi} \int_{\nu_{\mathrm{el}}}^\infty \dd \nu' \, 
\frac{\nu' \im T_i ( \nu', Q^2)}{\nu^{\prime\,2} -\nu^2 - i 0^+} ,\\
S_1 ( \nu, Q^2) &=&  \frac{2}{\pi} \int_{\nu_{\mathrm{el}}}^\infty \dd \nu' 
\, \frac{\nu' \im S_1 ( \nu', Q^2)}{\nu^{\prime\,2} -\nu^2 - i0^+ },\\
\nu S_2 ( \nu, Q^2) &=&  \frac{2}{\pi} \int_{\nu_{\mathrm{el}}}^\infty \dd \nu' 
\, \frac{\nu^{\prime\, 2} \im S_2 ( \nu', Q^2)}{\nu^{\prime\,2} -\nu^2 - i0^+ }\eqlab{S2DR},
\eea 
\end{subequations}
where $0^+$ is an infinitesimally small positive number. 
As emphasized in the derivation of these relations,
they are only valid provided the ``good''
behavior of these amplitudes for large $\nu$. 
It turns out that for $T_1$ the 
above {\it unsubtracted} DR is not warranted
and at least one subtraction is required.
We postpone a detailed discussion of this issue 
till \Secref{SRforRCS} and \Secref{sumrules} while continuing to deal here
with the unsubtracted DR.  

Substituting the unitarity relations,
\Eqref{VVCSunitarity}, into \Eqref{genDRs1} we have:\footnote{Using that, with $x=\nicefrac{\nu_\mathrm{el}}{\nu}$, $\nu_\mathrm{el}=\nicefrac{Q^2}{2M}$, the change of the integration variable from $\nu$ to $x$ goes as: \beq
\int_{\nu_\mathrm{el}}^\infty \frac{\dd\nu}{\nu^n} f(\nu,Q^2)  = \left(\frac{2M}{Q^2}\right)^{n-1} \! \int_0^1 \dd x\, 
x^{n-2} f(x,Q^2). \eqlab{xnuConv}
\eeq} 
\begin{subequations}
\eqlab{genDRs}
\bea 
T_1 ( \nu, Q^2) &=& \frac{8\pi\al}{M} \int_{0}^1 
\frac{\dd x}{x} \, 
\frac{f_1 (x, Q^2)}{1 - x^2 (\nu/\nu_{\mathrm{el}})^2 - i 0^+} = \frac{2}{\pi} \int_{\nu_{\mathrm{el}}}^\infty\! \dd \nu' \, 
\frac{\nu^{\prime\,2} \si_T ( \nu', Q^2)}{\nu^{\prime\,2} -\nu^2 - i 0^+},\\
T_2 ( \nu, Q^2) &=& \frac{16\pi\al M}{Q^2} \int_{0}^1 
\!\dd x\, 
\frac{f_2 (x, Q^2)}{1 - x^2 (\nu/\nu_{\mathrm{el}})^2  - i 0^+} =\frac{2Q^2}{\pi} \int_{\nu_{\mathrm{el}}}^\infty\! \dd \nu' \, 
\frac{\nu^{\prime\,2}[ \si_T+\si_L] ( \nu', Q^2)}{(\nu^{\prime\,2}+Q^2)(\nu^{\prime\,2} -\nu^2- i 0^+)} \eqlab{T2dr},\\
S_1 ( \nu, Q^2) &=&  \frac{16\pi\al M}{Q^2} \int_{0}^1 
\!\dd x\, 
\frac{g_1 (x, Q^2)}{1 - x^2 (\nu/\nu_{\mathrm{el}})^2  - i 0^+}=\frac{2M}{\pi} \int_{\nu_{\mathrm{el}}}^\infty \!\dd \nu' \, 
\frac{\nu^{\prime\,3}\big[ \frac{Q}{\nu'}\si_{LT}+\si_{TT}\big] ( \nu', Q^2)}{(\nu^{\prime\,2}+Q^2)(\nu^{\prime\,2} -\nu^2- i 0^+)}\eqlab{S1DR},\\
\nu S_2 ( \nu, Q^2) &=& \frac{16\pi\al M^2}{Q^2} \int_{0}^1 
\!\dd x\, 
\frac{g_2 (x, Q^2)}{1 - x^2 (\nu/\nu_{\mathrm{el}})^2  - i 0^+} = \frac{2M^2}{\pi} \int_{\nu_{\mathrm{el}}}^\infty \!\dd \nu' \, 
\frac{\nu^{\prime\,3}\big[ \frac{\nu'}{Q}\si_{LT}-\si_{TT}\big] ( \nu', Q^2)}{(\nu^{\prime\,2}+Q^2)(\nu^{\prime\,2} -\nu^2- i 0^+)}.\qquad
\eqlab{nuS2}
\eea 
\eqlab{SRgen}
\end{subequations}
Substituting here the elastic
structure functions,
\Eqref{elstructure}, we obtain the
nucleon-pole contribution:
\begin{subequations}
\eqlab{VVCSpole}
\bea
T_1^{\mathrm{pole}}(\nu, Q^2) &=&\frac{4\pi \alpha}{M}\frac{\nu_{\mathrm{el}}^2\,  G_M^2(Q^2)}{\nu_{\mathrm{el}}^2-\nu^2
-i 0^+ }\,,
\eqlab{T1pole}\\
T_2^{\mathrm{pole}}(\nu, Q^2)&=&\frac{8\pi \alpha\,  \nu_{\mathrm{el}} }{\nu_{\mathrm{el}}^2-\nu^2
-i 0^+ }\frac{G^2_E(Q^2) + \tau G^2_M(Q^2)}{1 + \tau} \,,
\eqlab{T2pole}
\\
S_1^{\mathrm{pole}}(\nu, Q^2) &=& \frac{4 \pi \alpha \, \nu_{\mathrm{el}}}{\nu_{\mathrm{el}}^2-\nu^2
-i 0^+ } F_1(Q^2)\, G_M(Q^2)\,, \eqlab{S1pole}\\
\left[\nu S_2\right]^{\mathrm{pole}}(\nu, Q^2)&=&-\frac{2 \pi \alpha\,  \nu_{\mathrm{el}}^2}{\nu_{\mathrm{el}}^2-\nu^2
-i 0^+ } F_2(Q^2)\, G_M(Q^2)\,. \eqlab{S2pole}
\eea 
\end{subequations}
These pole terms vanish in the limit $Q^2 \to 0$,
then $\nu\to 0$, as required. 

We are now in a position to derive the various sum rules
arising from low-energy and/or low-momentum expansion
of the CS amplitudes. The above DRs clearly show that the expansion in energy $\nu$ is an expansion
in the moments of structure functions.
For example, the Burkhardt-Cottingham (BC) sum rule \cite{Burkhardt:1970ti}
arises from taking 
the low-energy limit, $\nu\to 0$, of the relation \eref{nuS2} for $
\nu S_2$: 
\beq
0 = \int_{0}^1\dd x\,  g_{2}(x,\,Q^2),
\eqlab{BCsumrule}
\eeq
valid for any $Q^2 >0$. 
Note that, although the unitarity
relations are valid in the physical region only, the DRs
can be valid outside of the physical region. The photon
virtuality must nevertheless be spacelike, $Q^2>0$,
in order to exclude the particle production off the 
external photons.

Subtracting the DR \eref{nuS2} at $\nu=0$, and using the BC sum rule,  we obtain:
\beq 
\eqlab{DRS2woBC}
S_2 ( \nu, Q^2) = \frac{64\pi\al M^4\nu}{Q^6}  \int_{0}^1 
\!\dd x\,
\frac{x^2 g_2 (x, Q^2)}{1 - x^2 (\nu/\nu_{\mathrm{el}})^2  - i 0^+} = \frac{2M^2\nu}{\pi} \int_{\nu_{\mathrm{el}}}^\infty \!\dd \nu' \, 
\frac{\nu^{\prime}\big[ \frac{\nu'}{Q}\si_{LT}-\si_{TT}\big] ( \nu', Q^2)}{(\nu^{\prime\,2}+Q^2)(\nu^{\prime\,2} -\nu^2- i 0^+)}.
\eeq
This expression could be obtained immediately by writing 
the DR for $S_2$, rather than $\nu S_2$, but then we would not
have established the BC sum rule. Substituting in here the elastic $g_2$, we find that the pole and Born part of $S_2$ coincide, see \Eqref{S2subpole}.

\subsection{Sum Rules for Real Photons}
\seclab{SRforRCS}
We start with considering the model-independent
of \eref{genDRs} for the case of $Q^2=0$.
The amplitudes $T_2$ and $S_2$ drop out, and so do
the cross sections containing longitudinal photons.
We thus have:
\begin{subequations}
\eqlab{realDRs}
\bea 
T_1 ( \nu, 0) &=&  \frac{2}{\pi} \int_0^\infty\! \dd \nu' \, 
\frac{\nu^{\prime\,2} \si_T ( \nu')}{\nu^{\prime\,2} -\nu^2 - i 0^+},\\
S_1 ( \nu, 0) &=&  \frac{2M}{\pi} \int_0^\infty \!\dd \nu' \, 
\frac{\nu^{\prime\,} \si_{TT}( \nu')}{\nu^{\prime\,2} -\nu^2- i 0^+}.
\eqlab{S1dr}
\eea 
\end{subequations}
The cross sections $\si_T$ and $\si_{TT}$ are,
respectively, the unpolarized and helicity-difference
photoabsorption cross sections: $\nicefrac12\, (\si_{1/2}\pm
\si_{3/2})$. The amplitudes $T_1(\nu, 0) $ and $S_1(\nu,0)$ are (up to overall factors) identical to the RCS amplitudes $\scA_1(\nu,0)$ and 
$\scA_3(\nu,0)$ introduced in \Secref{formal}, 
and hence the above DRs apply to the latter amplitudes too.

The low-energy expansion of the amplitudes goes as:
\begin{subequations}
\bea 
\frac{1}{4\pi} T_1(\nu, 0) & = & -\frac{ \zZ^2\al}{M}
+  (\al_{E1} +\be_{M1} ) \nu^2 
+ \big[\alpha_{E1\nu} + \beta_{M1\nu} + \nicefrac{1}{12}\,(\alpha_{E2} + \beta_{M2}) \big]\,\nu^4
+O(\nu^6), \\
\frac{1}{4\pi} S_1(\nu, 0) & = & -\frac{ \al \varkappa^2}{2M} + M\gamma_0 \nu^2 + M\bar \gamma_0 \nu^4
+O(\nu^6), 
\eea 
\end{subequations}
where the $O(\nu^0)$ terms represent the low-energy theorem (LET) \cite{Low:1954kd,Gell_Mann:1954kc}; the scalar polarizabilities $\al_E$ and $\be_M$
are introduced in \Secref{theory}; the forward spin polarizabilities  
$\ga_0$, $\bar \ga_0$ are linear combinations
of spin polarizabilities, e.g.: 
\beq 
\ga_0 = - (\ga_{E1E1}+ \ga_{M1M1}+\ga_{E1M2}
+\ga_{M1E2}).
\eeq 

The {\em rhs}  of \Eqref{realDRs}
can also be Taylor expanded in $\nu^2$ and each term
matched to the low-energy expansion of the amplitude on 
the left-hand side ({\em lhs}). 
We however  run immediately into the following difficulty. At $\nu=0$ (the 0\textsuperscript{th} order in $\nu$),
the relation for $T_1$ yields an apparently wrong result:
\beq -\zZ^2 \al/M = (2/\pi) \int_0^\infty  \dd \nu \,\sigma_T(\nu).\eeq 
The {\em lhs} is negative definite whereas
the {\em rhs} is positive definite. The empirical
knowledge of the photoabsorption cross section
for the nucleon
shows in addition that the integral on the {\em rhs}
diverges. This invalidates the unsubtracted DR
for $T_1$. A common choice is to make
a subtraction at $\nu=0$, and use the LET to
obtain:
\beq 
\eqlab{subT1dr}
T_1 ( \nu, 0) =  -\frac{4\pi \zZ^2\al}{M} + 
\frac{2\nu^2}{\pi} \int_0^\infty\! \dd \nu' \, 
\frac{ \si_T ( \nu')}{\nu^{\prime\,2} -\nu^2 - i 0^+}.
\eeq 
The integral now converges and its evaluation for
the proton will be discussed in \Secref{theory3}.

Matching the low-energy expansion of $T_1$
at $O(\nu^2)$, one obtains the Baldin sum rule
\cite{Baldin:1960}:
\beq
\eqlab{BaldinSR}
\alpha_{E1} + \beta_{M1} = \frac{1}{2\pi^2} \int_{0}^\infty \!\dd\nu\, \frac{\si_T(\nu)}{\nu^2}.
\eeq
At $O(\nu^4)$, we obtain a sum rule for a linear combination of the energy slope of the dipole polarizabilities ($\alpha_{E1\nu}$, $\beta_{M1\nu}$) and the quadrupole  
polarizabilities ($\alpha_{E2}$, $\beta_{M2}$) \cite{Gryniuk:2015aa}:
\beq
\label{4thSR}
\alpha_{E1\nu} + \beta_{M1\nu} + \nicefrac{1}{12}\,(\alpha_{E2} + \beta_{M2}) = \frac{1}{2\pi^2} \int_{0}^\infty \!\dd\nu \,\frac{\si_T(\nu)}{\nu^4},
\eeq
referred to as the
4\textsuperscript{th}-order Baldin sum rule.

Considering the low-energy expansion 
of $S_1$, at the 0\textsuperscript{th} order 
one obtains the celebrated
Gerasimov-Drell-Hearn (GDH) sum rule~\cite{Gerasimov:1965et,Drell:1966jv,Hosoda01081966}:
\begin{equation}
\frac{ \al }{M^2}\varkappa^2= -\frac{1}{\pi^2}\int _{0}^\infty \!\dd\nu\,\frac{\sigma_{TT}(\nu)}{\nu },
\eqlab{GDH}
\end{equation}
which expresses the anomalous magnetic moment $\varkappa$
in terms of an energy-weighted integral of
the helicity-difference photoabsorption cross section.
This is probably the best studied sum rule. 
It directly
demonstrates the idea of expressing a purely quantum
effect, which is the anomalous magnetic moment,
in terms of a classical quantity, which is the cross section.
The perturbative verifications of the GDH sum rule
in QED and other quantum field theories provide
further insight into quantum dynamics (see, e.g., Refs.~\cite{Weinberg1970,Altarelli:1972nc,Dicus:2000cd,Pascalutsa:2004ga,Holstein:2005db}).

At $O(\nu^2)$ one arrives at the  
forward spin polarizability (FSP) sum rule, also referred to as the Gell-Mann, Goldberger and Thirring (GGT) sum rule \cite{GellMann:1954db}:
\begin{equation}
\eqlab{FSP} 
\gamma_0=\frac{1}{2 \pi^2}\int _{0}^\infty\!\dd\nu\,\frac{\sigma_{TT}(\nu)}{\nu ^3},
\end{equation}
while at $O(\nu^4)$ one obtains the
higher-order FSP sum rule \cite{Pasquini:2010zr}:
\begin{equation}
\eqlab{HOFSP} 
\bar\gamma_0=\frac{1}{2 \pi^2}\int _{0}^\infty\!\dd\nu\,\frac{\sigma_{TT}(\nu)}{\nu ^5}.
\end{equation}

The numerical 
evaluation of these sum rules based on empirical  photoabsorption cross sections is discussed in \Secref{theory3}.

\subsection{Relations at Finite $Q$}
\seclab{sumrules}
The main idea in the derivation of sum rules is 
to use the unitarity relations in combination with the DRs, Eq.~(\ref{eq:genDRs}), and then expand
the left- and right-hand sides in the photon energy $\nu$ and the virtuality
$Q^2$. In so doing, one expresses the static e.m.\ properties
of the nucleon (e.g., magnetic moment, charge radius, polarizabilities), 
which appear as coefficients
in the low-momentum expansion of the VVCS amplitudes, in terms of the moments
of its structure functions. In what follows, we derive a number of such sum rules and relations for the spin-independent and spin-dependent properties of the nucleon.

\subsubsection{Spin-Independent Relations}
\seclab{sumrulesSpinIndep}
As we have established, for $Q^2=0$ 
the convergence properties 
of the $T_1$ amplitude are such that 
its DR requires one subtraction. It is customary to
choose $\nu=0$ as the subtraction point, leading to:
\begin{eqnarray}
T_1(\nu,Q^2)=T_1(0,0)+\frac{2}{\pi}\left\{\int^\infty_{\nu_{\mathrm{el}}} \dd\nu'\,  \frac{\nu' \im T_1(\nu',Q^2)}{\nu^{\prime\,2}-\nu^2- i 0^+}-\int^\infty_0 \dd\nu'\, \frac{\im T_1(\nu',0)}{\nu'- i 0^+}\right\}.
\end{eqnarray}
The subtraction term, in accordance with the classic LET of \citet{Low:1954kd}, \citet{Gell_Mann:1954kc}, is given by the Thomson term:
\beq
T_1(0,0)=-4\pi \zZ^2\al/M, \eqlab{LETT1}
\eeq
while the rest of the amplitude $T_1(\nu,Q^2)$ 
could completely be determined by an integral of
$\im T_1 = (4\pi^2 \al /M) f_1$. 
That would be quite remarkable, because we
could calculate $T_1(0,Q^2)$ and then for instance
take its non-Born piece which goes as:
\beq
\eqlab{dreamT1}
\ol T_1 (0,Q^2)= 4\pi \beta_{M1}\, Q^2+O(Q^4),
\eeq
and extract the magnetic polarizability. 
In other words, we could have a sum rule
for $\beta_{M1}$ and for $\alpha_{E1}$ separately,
rather than together as in the Baldin sum rule.

Unfortunately, this appears to be not possible, as
the DR requires a subtraction at each $Q^2$,
cf.~\cite{Sucher:1972qh,Bernabeu:1974mb,Lvov:1998vg}.
In this case $T_1 (0,Q^2)$ is an unknown 
subtraction function,
 and the corresponding
DR reads:
\bea  
\eqlab{T1DR}
T_1 ( \nu, Q^2) & = & T_1(0,Q^2) +\frac{2\nu^2}{\pi} \int_{\nu_{\mathrm{el}}}^\infty \dd \nu'
\frac{\im T_1 ( \nu', Q^2)}{\nu'( \nu^{\prime\,2} -\nu^2 - i 0^+)} \nn,\\
&=& T_1(0,Q^2) + \frac{32\pi\al M\nu^2}{Q^4} \int_{0}^1 
\dd x 
\frac{x\, f_1 (x, Q^2)}{1 - x^2 (\nu/\nu_{\mathrm{el}})^2 - i 0^+},\eqlab{T1subDRx}\\
&=& T_1(0,Q^2) + \frac{2\nu^2}{\pi } \int_{\nu_{\mathrm{el}}}^\infty\! \dd \nu' \, 
\frac{\si_T ( \nu', Q^2)}{\nu^{\prime\,2} -\nu^2 - i 0^+}\,.\nn
\eea 
For the other spin-independent amplitude
we shall continue to use the unsubtracted DR of \Eqref{T2dr}.

In the real-photon limit, an immediate observation
is that
\beq 
\eqlab{T1T2relation}
\frac{\pa}{\pa Q^2} T_2 ( \nu, Q^2)\Big|_{Q^2=0}
=\frac{T_1(\nu,0)-T_1(0,0)}{\nu^2} =
\frac{2}{\pi} \int_0^\infty\! \dd \nu' \, 
\frac{\si_T ( \nu')}{\nu^{\prime\,2} -\nu^2 - i 0^+}\,.
\eeq 

For finite $Q$, the pole part of the amplitudes, $T_i^\mathrm{pole}$
in \Eqref{VVCSpole}, satisfies the DR with the elastic part of the structure functions. We therefore consider just the non-pole parts and their determination from the inelastic
structure functions.

The low-energy, low-momentum expansion 
of the spin-independent amplitudes is given by: 
\begin{subequations}
\bea
\frac{1}{4\pi}\Big[T_1-T_1^\mathrm{pole}\Big](\nu,Q^2)&=&-
\, \frac{ \zZ^2 \al}{M}+\left(\frac{\zZ\al}{3M}\langle r^2\rangle_1 +\beta_{M1}\right)Q^2 + \left(\alpha_{E1}+\beta_{M1}\right)\nu^2\nn\\
&&+ \big[\alpha_{E1\nu} + \beta_{M1\nu} + \nicefrac{1}{12}\,(\alpha_{E2} + \beta_{M2}) \big]\,\nu^4 +\ldots ,\eqlab{T1LEX}\\
\frac{1}{4\pi}\Big[T_2-T_2^\mathrm{pole}\Big](\nu,Q^2)&=& \left(\alpha_{E1}+\beta_{M1}\right)Q^2+
 \big[\alpha_{E1\nu} + \beta_{M1\nu} + \nicefrac{1}{12}\,(\alpha_{E2} + \beta_{M2}) \big]\,Q^2 \nu^2+ \ldots .\eqlab{T2LEX}
\eea
\end{subequations}
where,
in case of $T_1$, the Thomson term and the Dirac radius $\langle r^2\rangle_1=-6\,\nicefrac{\dd}{\dd Q^2}\,F_1(Q^2)\vert_{Q^2=0}$, come from the
non-pole part of the Born contribution.

The fact that the same combination of
polarizabilities enters in both amplitudes
follows from \Eqref{T1T2relation}.
We thus see for instance that the Baldin
sum rule can equivalently be written as:
\beq
\alpha_{E1}+\beta_{M1}= \lim_{Q^2\to 0} 
\frac{8 \al M}{Q^4}\int_0^{x_0} \! \dd x \, x\, f_1(x,Q^2) 
= \lim_{Q^2\to 0} 
\frac{4 \al M}{Q^4}\int_0^{x_0} \!\dd x \,  f_2(x,Q^2) ,\eqlab{altBaldinSR}
\eeq
where $x_0$ is the inelastic threshold. 
At the next order in $\nu^2$,
we have the 4\textsuperscript{th}-order
sum rule:
\beq
\alpha_{E1\nu} + \beta_{M1\nu} + \nicefrac{1}{12}\,(\alpha_{E2} + \beta_{M2}) = \lim_{Q^2\to 0} 
\frac{32 \al M^3 }{Q^6}\int_0^{x_0} \! \dd x \, x^3\, f_1(x,Q^2) 
= \lim_{Q^2\to 0} 
\frac{16 \al M^3}{Q^6}\int_0^{x_0} \!\dd x \, x^2\, f_2(x,Q^2) .\eqlab{alt4orderSR}
\eeq
These sum rules can be generalized to
finite $Q$, and the usual choice 
is to do that using $f_1$. For instance, the 
generalization of the Baldin sum rule
reads
 \cite{Drechsel:2002ar}:
\beq
\alpha_{E1}(Q^2)+\beta_{M1}(Q^2)=\frac{8 \al M}{Q^4}\int_0^{x_0} \dd x \, x\, f_1(x,Q^2).\eqlab{genBaldinSR}
\eeq
It was evaluated in Ref.~\cite{Liang:2004tk}, and more recently in 
Refs.~\cite{Sibirtsev:2013cga,Hall:2014lea} 
using an
improved empirical parametrization
of the structure function $f_1$. 

In general, the relation
in \Eqref{T1T2relation} implies that the longitudinal
structure function: 
\beq 
f_L(x,Q^2) = -2x f_1(x,Q^2) + f_2(x,Q^2),
\eeq 
which is known to vanish for asymptotically large $Q^2$
(Callan--Gross relation), also 
vanishes for low $Q^2$, and its moments go as:
$
\lim_{Q^2\to 0} 
Q^{-4-2n}\int \! \dd x \, x^{2n}\, f_L(x,Q^2)
=0$. 

It is natural to consider the combination, 
$\tilde f_L\equiv f_L + (2M x/Q)^2 f_2 =Q^2 \si_L(\nu,Q^2)/4\pi^2\al$, and 
define a longitudinal polarizability as:\footnote{This definition
differs from the original one \cite{Drechsel:2002ar} by a factor $1/Q^2$, and
as the result, $\al_L(0)$ is not vanishing here.}
\beq 
\alpha_L(Q^2) = \frac{4\al M}{Q^6} \int_0^{x_0} \dd x \,  \tilde f_L(x,Q^2)
= \frac{1}{2\pi^2} \int_{\nu_0}^\infty \!\dd\nu\, \frac{\si_L(\nu,Q^2)}{ Q^2\nu^2}.
\eeq 
At low $Q^2$ this quantity is easily described in B$\chi$PT, but not in 
HB$\chi$PT, cf.~\cite[Fig.~3]{Lensky:2014dda}.
As all the quantities involving the longitudinal polarization, it is fairly insensitive to the $\De(1232)$-resonance excitation.

\subsubsection{Spin-Dependent Relations}
\seclab{sumrulesSpinDep}
The sum rule derivation for the spin-dependent amplitudes proceeds in the same steps.
As noted earlier, the DR for $\nu S_2$ in the limit $\nu\to 0$ leads to the 
BC sum rule \cite{Burkhardt:1970ti}, see \Eqref{BCsumrule}. This sum rule
implies the following relation between the elastic and inelastic part of $S_2$:
\beq
I_2(Q^2)\equiv \frac{2M^2}{Q^2}\int_0^{x_0}\dd x\, g_2(x,Q^2)=\frac{1}{4}F_2(Q^2)G_M(Q^2).
\eeq

We next consider the simultaneous expansion of the
non-pole parts in $\nu$ and $Q^2$ \cite{Drechsel:1998zm, Pascalutsa:2014zna}:
\begin{subequations}
\eqlab{S12lex}
\bea
\eqlab{S1lex}
\frac{1}{4\pi}\Big[S_1-S_1^\mathrm{pole}\Big](\nu,Q^2)&=&
 \frac{\alpha}{2M } \,  \varkappa^2 \left[-1+\frac{1}{3}Q^2\langle r^2 \rangle_2 \right]
+   M  \gamma_0 \,  \nu^2  \,+\,  M\, Q^2 \Big\{\gamma_{E1 M2}\nn\\ 
&&- 3 M \alpha \big[ P^{\prime(M1, M1)1} (0)+  P^{\prime (L1, L1)1}(0) \big]\Big\}+O(\nu^4,\nu^2Q^2,Q^4),\\
\frac{\nu}{4\pi} \Big[S_2-S_2^\mathrm{pole}\Big](\nu, 0) &=&   - M^2 \nu^2  \Big\{ \ga_0 +\ga_{E1E1}\nn\\
&&- 3 M \alpha \big[ P^{\prime(M1, M1)1} (0)-  P^{\prime (L1, L1)1}(0) \big]\Big\}+O(\nu^4), \eqlab{S2lex}
\eea
\end{subequations}
where $\varkappa$ is the nucleon anomalous magnetic moment; 
$\langle r^2\rangle_2=-6/\varkappa\;\nicefrac{\dd}{\dd Q^2}\,F_2(Q^2)\vert_{Q^2=0}$ is the mean-square Pauli radius;  
$\gamma_{E1 M2}$ and $\ga_{E1E1}$ are the spin polarizabilities;
$\gamma_0$ is the forward spin polarizability; and $P$s are the generalized polarizabilities (GPs) coming from 
the VCS, see \Eqref{GPslope} below. 
In case of $S_1$, the first term originates from
the difference between the Born and pole amplitudes, whereas
polarizabilities affect the non-Born part of the amplitudes only.

On the {\it rhs} of the DRs for $S_1$ and $S_2$ we have an expansion
in terms of moments of the spin structure functions $g_1$ and $g_2$. 
The 0\textsuperscript{th} moment of $g_1$ is related to the 
generalized GDH integrals:
\begin{subequations}
\eqlab{genInt}
\bea
\label{I1sr}
I_1(Q^2)  &=&  \frac{2M^2}{Q^2}\!\int_0^{x_0}\! \dd x\, g_1(x,Q^2),\\
I_A(Q^2)&=&\frac{2M^2}{Q^2}\int_0^{x_0}\dd x\,  g_{TT}(x,Q^2)
=\frac{M^2}{4\pi^2\al} \int_{\nu_0}^\infty \! \frac{\dd\nu}{ \nu}
\si_{TT}(\nu,Q^2) , \label{IAsr}
\eea
\end{subequations}
with $g_{TT}=g_1- (4M^2x^2/Q^2)g_2$. In the limit, $Q\to 0$, $\nu\to 0$,
they yield the GDH sum rule of \eref{GDH}: 
\beq 
-\quarter \varkappa^2 =  I_1(0) =  I_A(0).
\eeq 
The 2\textsuperscript{nd} moments appear in the following generalization
of the forward spin polarizabilities
\cite{Drechsel:2002ar}:
\bea
\eqlab{genFSP}
\gamma_0(Q^2)&=&\frac{16\al M^2}{Q^6}\int_0^{x_0} \dd x \, x^2 \,
g_{TT}(x,Q^2) = \frac{1}{2 \pi^2}\int _{0}^\infty\!\frac{\dd\nu}{\nu ^3}
\sigma_{TT}(\nu,Q^2) , \\
\delta_{LT}\,(Q^2)&=&\frac{16\al M^2}{Q^6}\!\int_{0}^{x_0}\! \dd x \, x^2 
[ g_1+g_2](x,Q^2)= \frac{1}{2 \pi^2}\int _{0}^\infty\!\frac{\dd\nu}{\nu^2 Q}
\sigma_{LT}(\nu,Q^2) ,
\label{deltaLT}
\eea
which evidently satisfy the following relations at $Q^2=0$:
\bea 
\gamma_0&=&\lim_{Q^2\to 0} \frac{16\al M^2}{Q^6}\int_0^{x_0} \dd x \, x^2 \,
g_1(x,Q^2),\\
\de_{LT}&=&\ga_0+\lim_{Q^2\to 0}\frac{16\al M^2}{Q^6} \int_{0}^{x_0}\dd x\, x^2\,  g_{2}\,(x,\,Q^2).
\eqlab{dLTg0}
\eea 
The first of these is simply the GTT sum rule given in \Eqref{FSP}.
At large $Q^2$, where the Wandzura--Wilczek relation 
\cite{Wandzura:1977qf} [quoted in \Eqref{WWrelation} below] is applicable and the elastic
contributions can be neglected, one can
show that~\cite{Drechsel:2002ar}:  $\delta_{LT}(Q^2) = \third \gamma_0(Q^2)$.

From the $Q^2$ term in the expansion of $S_1$, and the $\nu^2$ term
in the expansion of $S_2$,  one obtains the following 
relations involving the GPs \cite{Pascalutsa:2014zna}:
\begin{subequations}
\eqlab{S12sr3}
\bea
\al I_1^\prime(0) &=&  \boxfrac{1}{12} \al \varkappa^2  \langle r^2 \rangle_2 
+  \boxfrac{1}{2} M^2 \gamma_{E1 M2} 
- \boxfrac{3 }{2}  \al M^3\, \big[ P^{\prime(M1, M1)1} (0)+  P^{\prime (L1, L1)1}(0) \big]
\,, 
\label{S1sr3}
\\
\delta_{LT} &=& 
- \gamma_{E1E1} 
+ 3 \al M \, \big[ P^{\prime(M1, M1)1} (0)-  P^{\prime (L1, L1)1}(0) \big] . 
\label{S2sr3}
\eea
\label{qsqrgdhsr}
\end{subequations}
The momentum derivatives of the GPs are given by:
\bea
\eqlab{GPslope}
&&  P^{\prime\,  (M1, M1)1}(0) \pm  P^{\prime\,  (L1, L1)1}(0) 
 \equiv \frac{\dd}{\dd \bq^2 } \Big[
 P^{ (M1, M1)1}(\bq^2) \pm  P^{ (L1, L1)1}(\bq^2 )  \Big]_{\bq^2 =0},
 \eea
with $\bq^2$ being the initial photon c.m.\ three-momentum squared. The superscript indicates the multipolarities, $L1(M1)$ denoting electric (magnetic) dipole transitions of the initial and final photons, and `$1$' implies that these transitions involve the spin-flip of the nucleon, cf.\ \cite{Guichon:1995, Guichon:1998xv}. An empirical implication of these relations,
in the context of the so-called ``$\de_{LT}$-puzzle", 
is briefly considered  in \Secref{conclusion}. 

Another combination of the 2\textsuperscript{nd} moments of spin structure functions,
i.e.:
\beq
\bar d_2(Q^2)=\int_0^{x_0}\dd x\,x^2\left[3 g_2(x,Q^2)+2g_1(x,Q^2)\right],\\
\eeq
is of interest in connection to the concept of
{\it color polarizability} \cite{Filippone:2001ux}. 
In terms of the above-introduced quantities it reads:
\bea
\bar d_2(Q^2)=\frac{Q^4}{8M^4}\left\{\frac{M^2 Q^2}{\al}\delta_{LT}(Q^2)+\left[I_1(Q^2)-I_A(Q^2)\right]\right\},
\eea
and goes as $Q^6$ for low $Q$.

\subsection{Empirical Evaluations of Sum Rules}
\seclab{theory3}

\begin{figure}[b!]
\vspace{1mm}
\centering
	\includegraphics[width=16cm]{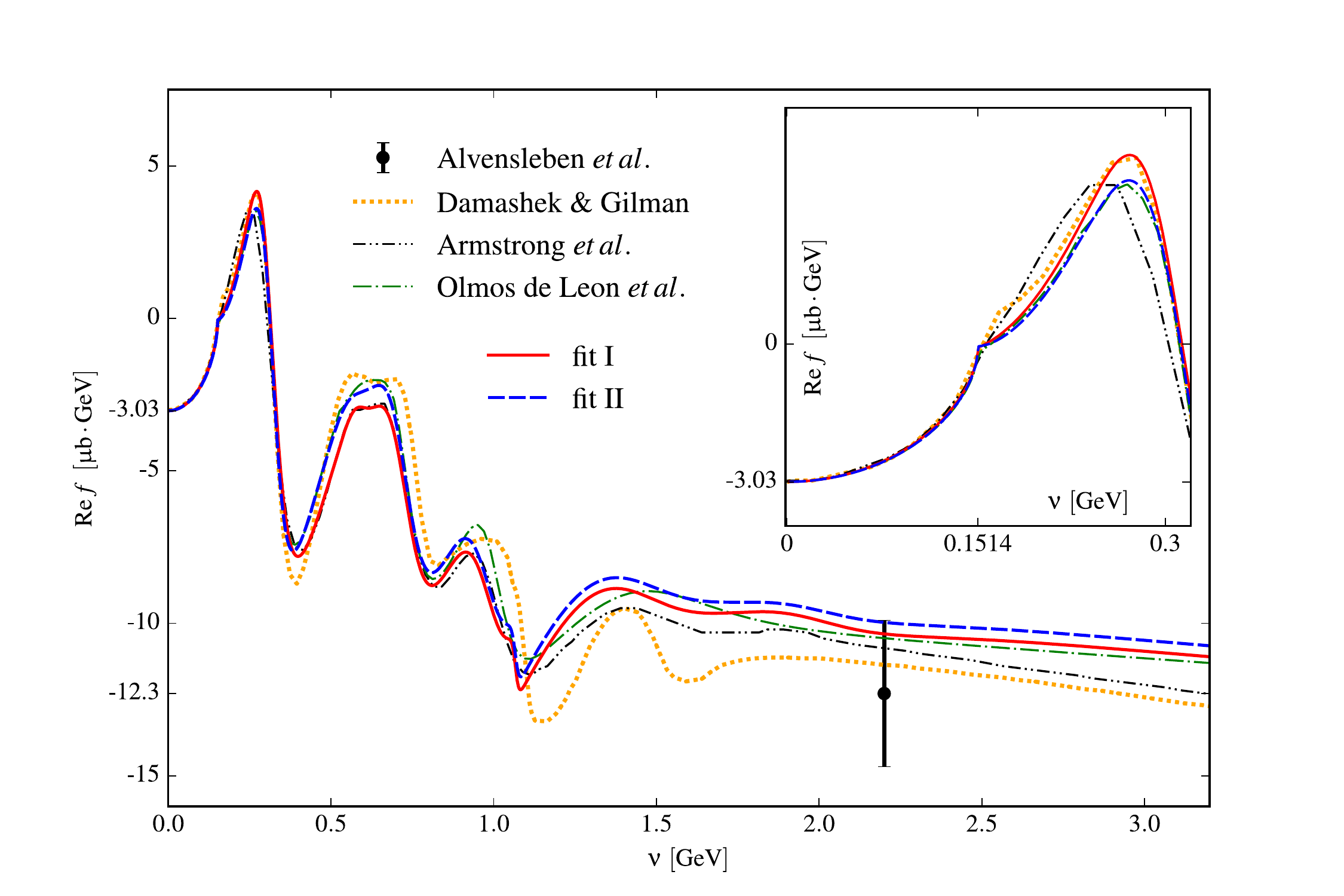}
	\caption{Amplitude $ f (\nu)$ for the proton obtained from \Eqref{KK2} 
    using different fits of the total photoabsorption cross section
    \cite{Damashek:1969xj,Armstrong,Olm01,Gryniuk:2015aa} (fit I \& II refer to the results of Ref.~\cite{Gryniuk:2015aa}). The experimental point is
    from   DESY \cite{Alvensleben:1973mi}.}
	\label{fig:Re_f}
\end{figure}

\begin{figure}[thb]
\centering
	\includegraphics[width=11cm]{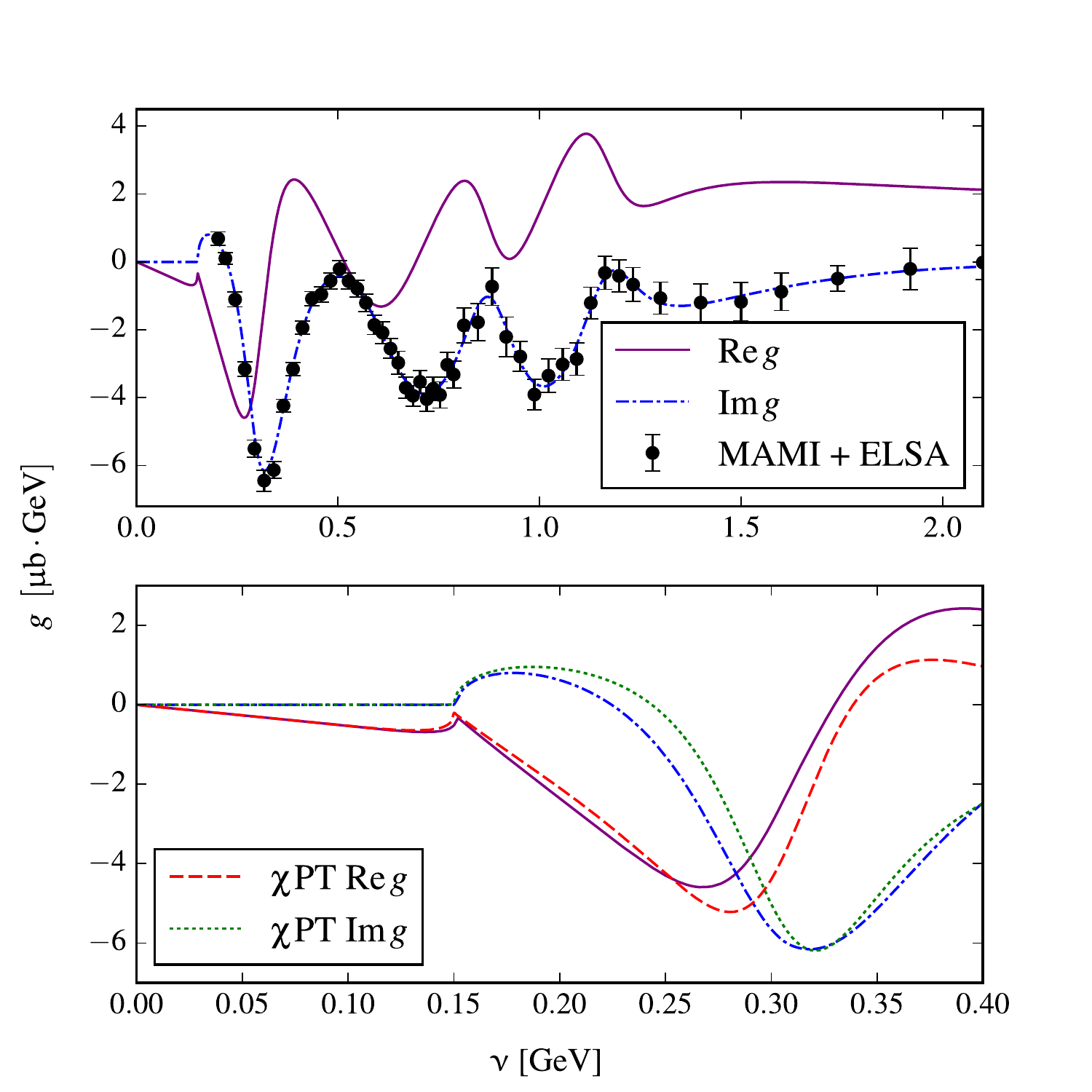}
	\caption{ Spin-dependent amplitude $g(\nu)$ obtained from 
    Eq.~(\ref{gOnceSubtr}).
The lower panel shows also the 
B$\chi$PT predictions for this amplitude \cite{Len10,Lensky:2015awa}.}
	\label{fig:g}
\end{figure}

Recall that the forward RCS is described by two
scalar amplitudes, denoted here [and in Eq.~(\ref{Eq:forward_Compton})] as:
\beq 
f(\nu) \equiv \frac{T_1(\nu,0)}{4\pi} = \frac{\sqrt{s}}{2M} \big(
\Phi_1+\Phi_5 \big)\big|_{\th=0}, \qquad g(\nu) \equiv \frac{\nu S_1(\nu,0)}{4\pi M}
= \frac{\sqrt{s}}{2M} \big(
\Phi_1-\Phi_5 \big)\big|_{\th=0}, 
\eeq
where the helicity amplitudes $\Phi_i$ are introduced in \Secref{RCS}.
The corresponding DRs, Eqs.~\eref{subT1dr} and \eref{S1dr}, 
read then as follows:
\begin{subequations}
\bea
f(\nu) &=& -\frac{ \zZ^2\alpha}{M} + \frac{\nu^2}{2\pi^2 }\int_0^\infty\! \dd\nu'\,\frac{\si_T(\nu')}{\nu^{\prime \, 2}-\nu^2 - i0^+},
\eqlab{KK2} \\
g(\nu)& = & \frac{\nu}{2\pi^2}\int_0^\infty \!
\dd\nu'\,\frac{\nu' \si_{TT}(\nu')}{\nu^{\prime\,2}-\nu^2- i0^+}.
\label{gOnceSubtr}
\eea
\end{subequations}
Therefore, given the total unpolarized cross section $\si_T$ and the
helicity-difference cross-section $\si_{TT}$, the forward 
CS can be completely determined. The cross sections for the proton
are fairly well known. Their most recent fits and 
the evaluation of the integrals are performed by \citet{Gryniuk:2015aa}.
The corresponding results for the amplitudes are displayed 
in \Figref{Re_f} and \Figref{g}. The first figure shows also the results
of previous evaluations and an experimental point from 
the DESY 1973 experiment \cite{Alvensleben:1973mi}. In the second figure the upper panel shows the fit to $\im g$ together with the corresponding  result for the real part. The lower panel shows a comparison of these results with a B$\chi$PT calculation at lower energy.
Given these amplitudes, one can determine the two non-vanishing (in the forward limit) observables:
\beq 
\frac{\dd \si}{\dd\varOmega_L} \stackrel{\th=0}{=} |f|^2 + |g|^2, 
\quad \Si_{2z}\stackrel{\th=0}{=} -\frac{f g^\ast + f^\ast g}{|f|^2 + |g|^2 }.
\eeq 
The obtained $\Si_{2z}$ \cite{Gryniuk2015}, compared with the
B$\chi$PT predictions, demonstrates the importance of chiral dynamics
in this observable, cf.~\cite[Fig.\ 16]{Lensky:2015awa}.

One can also evaluate the various sum rules presented
in \Secref{SRforRCS}.
Evaluations of the sum rules deriving from $f(\nu)$ 
(i.e., Baldin sum rule, etc.) are gathered in Table \ref{Tab:sumruletest_proton}
for the proton and neutron, respectively. These results are summarized 
and compared to the state-of-art $\chi$PT results in Figures \ref{fig:alphabeta_p} and \ref{fig:alphabeta_n}.


\citet{Damashek:1969xj} initiated a  study of the high-energy behavior of 
the amplitude $f(\nu)$ for the proton. In addition to the Regge prediction, 
they found a constant contribution comparable in sign and magnitude to the Thomson term: $-\al/M\simeq -3.03\,\upmu \mathrm{b}\, \mathrm{GeV}$. This extra constant is assumed to correspond to a fixed $J=0$ Regge pole ($\al_i(t)=0$) \cite{Creutz:1968ds, Brodsky:1971zh}, originating from local photon interactions with the constituent quarks. Based on newer photoabsorption data, \citet{Gorchtein:2011xx} have
obtained a significantly smaller value $f(\infty)=(-0.72\pm0.35)\,\upmu \mathrm{b}\, \mathrm{GeV}$.

Recently, \citet{Gasser:2015dwa} have made a sum rule determination of the proton-neutron difference (isovector combination) of  the electric dipole polarizability:
\beq
\al_{E1}^{(p-n)}=-1.7(4)\times10^{-4}\,\mathrm{fm}^3.
\eeq
Their calculation is based on a Reggeon dominance assumption, which means there is no fixed pole in the isovector CS amplitude. It could be that the
fixed pole,  which is likely to be present for the proton, is canceled exactly
by the one of the neutron. However, this is yet to be verified. 
The fact that the above value is in agreement with the
empirical information on the isovector polarizability 
is certainly encouraging.

\begin{table}[hbt]
\footnotesize
\caption{
	Empirical evaluation of spin-independent sum rules for the proton and neutron.
	}
\label{Tab:sumruletest_proton}
\centering
	\begin{tabular}{|c|c|c|c|c|c|c|}
	\hline
	&  \multicolumn{2}{|c|}{Baldin SR} & \multicolumn{2}{|c|}{$4^\mathrm{th}$-order SR} &  $6^\mathrm{th}$-order SR & $\re f(\nu=2.2\,\mathrm{GeV}$) \\
	&  \multicolumn{2}{|c|}{[$10^{-4}\,\mathrm{fm}^3$]}  & \multicolumn{2}{|c|}{[$10^{-4}\,\mathrm{fm}^5$]} &  [$10^{-4}\,\mathrm{fm}^7$] & [$\upmu \mathrm{b}\,\mathrm{GeV}$] \\
\cline{2-7}
&Proton&Neutron&Proton&Neutron&Proton&Proton\\
	\hline
	 Gryniuk et al.~\cite{Gryniuk:2015aa}& $14.00\pm 0.20$& & $6.04 \pm 0.03 $ && $ 4.39 \pm 0.03 $ & $-10.18$ \\
	\citet{Armstrong} & &&&  & & $-10.8$\\
	\citet{Damashek:1969xj}      & $14.2\pm 0.3$ && &  &  & \\
	Schr{\"o}der \cite{Schroder:1977sn}            & $14.7\pm0.7$&$13.3\pm0.7$& $6.4$&$5.6$&  & \\
	Babusci et al.~\cite{Babusci}     & $13.69\pm0.14$ &$14.40\pm0.66$&  & & & \\
    \citet{Levchuk:1999zy}&$14.0(3\ldots 5) $&$15.2\pm0.4$&&&&\\
\citet{Olm01} & $13.8\pm0.4$   &  & && & $-10.5$\\
	MAID ($\pi$ channel)~\cite{MAID} & $11.63$  &$13.28$&   &  &  & \\
	SAID ($\pi$ channel)~\cite{SAID}         & $11.5$    &$12.9$&       &  &  & \\
	Alvensleben et al.~\cite{Alvensleben:1973mi} &          &  & && & $-12.3\pm2.4$\\
	\hline
	\end{tabular}
\end{table}

\begin{table}[hbt]
\footnotesize
\caption{
	Empirical evaluation of the GDH and GTT sum rules for the proton and neutron.}
\label{Tab:GDH}
	\centering 
	\begin{tabular}{|c|c|c|c|c|}
	\hline
\multirow{2}{*}{}&\multicolumn{2}{|c|}{GDH SR $[\upmu \mathrm{b}]$}& \multicolumn{2}{|c|}{$\gamma_0 \, [10^{-4} \,\mathrm{fm}^4]$}\\
\cline{2-5}
&  Proton & Neutron&  Proton & Neutron \\
    \hline
Sum rule value (\textit{lhs})& $205$     &$233$&& \\
GDH-Coll.\ \cite{Ahrens:2001qt,Tiator:2002zj, Dutz:2003mm,Helbing:2006zp}& $212\pm17$     & $225$&$-1.01\pm0.13$&\\
Gryniuk et al.~\cite{Gryniuk2015}&$204.5\pm9.4$&&$-0.93\pm 0.06$&\\
    Pasquini et al.~\cite{Pasquini:2010zr}&$210\pm 15$&&$-0.90\pm0.14$&\\
    Babusci et al.~\cite{Bab98} &&&$-1.5$&$-0.4$\\
    \citet{Schumacher:2011gs} &&&$-0.58\pm0.20$&$0.38\pm0.22$\\
MAID ($\pi$ channel)~\cite{MAID} & $-165.65$     &$-132.32$ &$-0.730$     &$-0.005$ \\
	SAID ($\pi$ channel)~\cite{SAID}         & $-187$           & $-137$& $-0.85$           & $-0.08$\\
	\hline
	\end{tabular}
\end{table}

The evaluations of the GDH and GTT sum rules, deriving from $g(\nu)$, 
are presented in Table \ref{Tab:GDH} and  \Figref{gamma0}.
These results became largely  possible due to the GDH-Collaboration
data for the helicity-difference photoabsorption cross section, 
in the region from $0.2$ to $2.9\,\mathrm{GeV}$ \cite{Ahrens:2000bc, Ahrens:2001qt, Dutz:2003mm,Dutz:2004zz}.

For the proton, the running GDH integral,
\begin{equation}
I^\mathrm{GDH}_{\mathrm{run}}(\nu_{\mathrm{max}})=\int _{\nu_0}^{\nu_{\mathrm{max}}} \,\frac{\dd\nu}{\nu }\left[\sigma_{\nicefrac{3}{2}}(\nu)-\sigma_{\nicefrac{1}{2}}(\nu)\right],
\end{equation}
with $\nu_0$ being the lowest particle-production threshold, effectively set by the pion-production threshold $\nu_\pi=m_\pi+(m_\pi^2+Q^2)/(2M)$, evaluates to \cite{Ahrens:2001qt}: 
\begin{equation}
I^\mathrm{GDH}_{\mathrm{run}}(2.9 \,\mathrm{GeV})=226\pm5_{\,\mathrm{stat}}\pm 12_{\,\mathrm{syst}} \,\upmu \mathrm{b},
\end{equation}
where the unmeasured low-energy region is covered by MAID \cite{MAID} and 
SAID \cite{SAID} analyses. The extrapolated result,
\begin{equation}
I^\mathrm{GDH}_{\mathrm{run}}(\infty)=212\pm6_{\,\mathrm{stat}}\pm 16_{\,\mathrm{syst}} \,\upmu \mathrm{b},
\end{equation}
is in agreement with the GDH sum rule value (obtained by substituting
the proton anomalous magnetic moment): $205\, \upmu \mathrm{b}$. The negative contribution to the integrand at higher energies is supported by a Regge parametrization of the polarized data, as well as by fits of deep-inelastic scattering (DIS) data \cite{Bianchi:1999qs, Simula:2001iy}.

The neutron cross section is extracted from the difference of deuteron and proton cross sections, see \citet{Arenhovel:2000jk} for critical discussion. 
Presently, the GDH integral for the neutron is estimated to be $225 \, \upmu\mathrm{b}$, which compares well to the sum rule value of $233\, \upmu \mathrm{b}$.
Table \ref{Tab:GDH} summarizes the GDH sum rule results for
the proton and neutron.

In future, one would like to measure the neutron cross sections 
based on $^3\text{He}$ targets. Since the proton spins are paired in the ground state, a polarized $^3\text{He}$ target is a good alternative to the non-existent free neutron target. The neutron spin structure is quite similar to the one of $^3\text{He}$. Therefore, in contrast to deuterium, the magnetic moment of $^3\text{He}$ is comparable to that of the neutron, with  the GDH integral equal to: $-496\, \upmu\mathrm{b}$. 
Below the pion-production threshold, the dominant channels are the two-
and three-body breakup reactions:  $^3\vec{\text{He}}\left(\vec{\gamma}, \text{n}\right)\text{d}$ and 
$^3\vec{\text{He}}\left(\vec{\gamma}, \text{n}\right)\text{pp}$. The latter
has been experimentally accessed at HIGS \cite{Laskaris:2013ehq}. At MAMI, the helicity-dependent total inclusive $^3\text{He}$ cross section is measured with circularly polarized photons in the energy range: $200<\nu<500\,\mathrm{MeV}$ \cite{Costanza:2012zza}. An estimate of the GDH sum rule for the neutron based on $^3\text{He}$ experiments has not yet been done.


\section{Proton Structure in (Muonic) Hydrogen}
\seclab{hydrogen}

An exciting development in the field of nucleon structure has come recently from atomic physics.
The CREMA collaboration discovery of the $2P-2S$
transitions in muonic hydrogen ($\mu$H) has led
to a precision measurement of the proton charge  
radius~\cite{Pohl:2010zza,Antognini:1900ns}. 
The resulting value is an
order of magnitude more precise than that from hydrogen spectroscopy (H) or electron-proton ($ep$) scattering. 
It also turned out to be substantially ($7\si$)
different from the CODATA value~\cite{Mohr:2012aa}, which had been the standard value based on H and $ep$ scattering. The latter discrepancy
is known as the {\it proton-radius puzzle} (see \Secref{puzzle}
for more details and references).

\begin{figure}[tbh]
\centering
\begin{minipage}{0.49\textwidth}
\centering
\includegraphics[width= 1.03\textwidth]{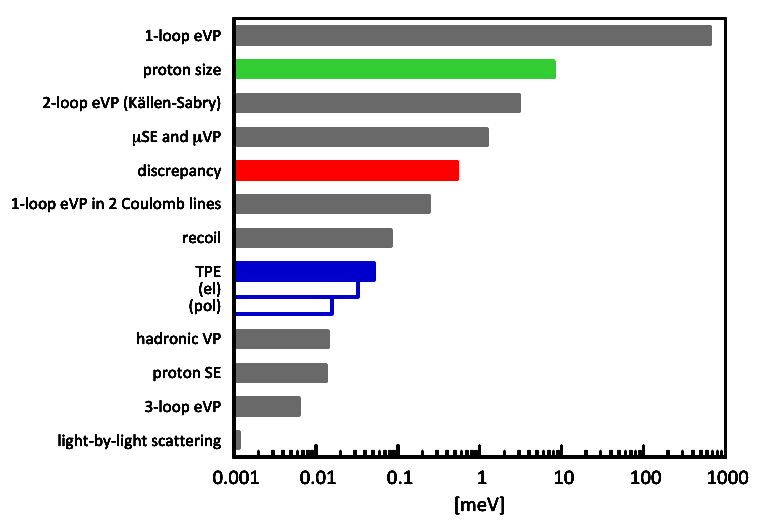}
                     \caption{The budget of the $\mu$H Lamb shift \cite{Antognini:2012ofa}. The TPE is displayed in blue; we give estimates for the elastic and polarizability contributions (unfilled bars), as well as for the total TPE contribution (solid bar). The proton radius discrepancy (shown in red) amounts to $0.31\, \mathrm{meV}$. The theoretical uncertainty is estimated as $0.0025 \,\mathrm{meV}$, cf.\ \Eqref{LS theory}. \label{fig:Pohl}}
\end{minipage}\hfill
\begin{minipage}{0.49\textwidth}
\centering
\includegraphics[width=\textwidth]{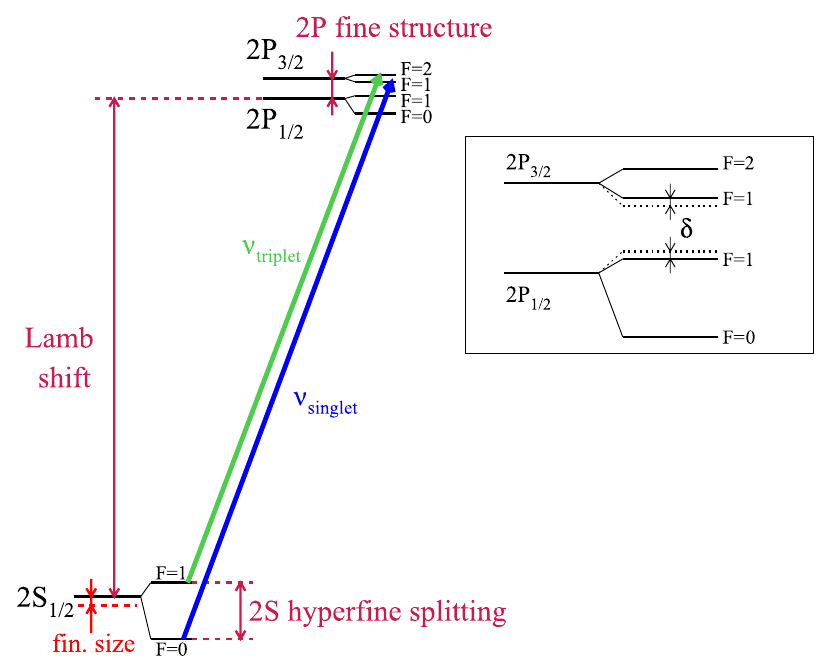}
\caption{$2S$ and $2P$ energy levels in $\mu$H. The two measured transition frequencies, $\nu_t$ \cite{Pohl:2010zza} and $\nu_s$ \cite{Antognini:1900ns}, are indicated together with the Lamb shift, fine and hyperfine structure, and finite-size effects. The main figure is drawn to scale. The inset zooms in on the $2P$ states. Here, the mixing of the $2P^{F= 1}$-levels shifts them by $\pm \delta$. Plot courtesy of Aldo~Antognini.\label{fig:Antognini}}
\end{minipage}
\end{figure}

In this section we examine the  proton structure effects
in hydrogen-like atoms.
While all the following formulae are applicable
to both H and $\mu$H, the numerics will
only be worked out for $\mu$H. 
As far as proton structure is concerned,
all the effects are much more pronounced in $\mu$H.\footnote{
In layman's terms, because the Bohr radius of
$\mu$H is about 200 times smaller than that of H,
the muon comes much closer to the proton, thus having
a ``better view'' (or, more precisely, spending considerably more time ``inside
the proton'', thus ``feeling'' less Coulomb attraction).} The proton charge radius, for example, is the second largest contribution to the $\mu$H Lamb shift (after the vacuum polarization due to
the electron loop in QED), cf.\ \Figref{Pohl}.

The proton structure effects are naturally divided into two categories:
\begin{enumerate}[(i)]
\item Finite-size  (or `elastic') effects, i.e., the effect of the elastic FFs, $G_E$ and $G_M$.
\item Polarizability\footnote{In the literature {\it polarizability} is sometimes called {\it polarization}. We prefer to reserve the 
latter for the proton spin polarization.} (or `inelastic')  effects, which basically is everything else.\footnote{In exceptional cases, `inelastic' may refer to only a part of the polarizability effect, as explained in Sect.~\ref{section:polcontrLS}.}
\end{enumerate}

Assuming the proton e.m.\  structure is confined within a femtometer radius, the finite-size effects can be expanded in the moments of charge
and magnetization distributions, $\rho_E(r)$ and
 $\rho_M(r)$, which are the Fourier transforms of the
 elastic FFs $G_E$ and $G_M$, respectively.
To $O(\al^5)$, the finite-size effects in the hydrogen Lamb shift and hyperfine splitting (HFS) are found as (omitting recoil) \cite{Eides:2000xc}:
\begin{subequations}
\eqlab{FSEs}
\bea
E_\mathrm{LS}&\equiv &E(2P_{1/2})-E(2S_{1/2})=-\frac{Z\al}{12a^3} \left[ R_E^2 - (2a)^{-1} R^3_{\mathrm{F} }\right] + O(\al^6),\eqlab{LambShift}\\
E_{\mathrm{HFS}}(nS)&\equiv&E(nS^{F=1}_{1/2})-E(nS^{F=0}_{1/2}) =E_F(nS)\left[1-2a^{-1} R_{\mathrm{Z}}\right]+ O(\al^6), \eqlab{HFS}
\eea
\end{subequations}
where $a=1/(Z\al m_r)$ is the Bohr radius, $Z$ is the nuclear charge ($Z=1$ for the proton), $E_F$ is the Fermi energy of the $nS$-level:
\beq
E_F(nS)=\frac{8Z \al}{3a^3}\frac{1+\kappa}{mM} \frac{1}{n^3}\,,
\eqlab{EFermi}
\eeq
and
the radii are defined as follows (for other notations, see \Secref{definitions}):
\begin{itemize}[$\square$]
\item Charge radius (shorthand for the root-mean-square (rms) radius of the charge distribution): 
\begin{subequations}
\beq 
\eqlab{moments}
R_E =  \sqrt{\langle r^2\rangle_E}, \quad \langle r^2\rangle_E
\equiv \int \! \dd \br\,  r^2  \rho_{E}(\br) = -6 \frac{\dd}{\dd Q^2} G_E(Q^2)\Big|_{Q^2= 0}\;;
\eeq
\item Friar radius (or, the 3\textsuperscript{rd} Zemach moment):
\beq
R_{\mathrm{F} } =\sqrt[3]{\langle r^3\rangle_{E(2)} }, \quad \langle r^3\rangle_{E(2)} \equiv  \frac{48}{\pi} \int_0^\infty \!\frac{\dd Q}{Q^4}\,
\Big[ G_E^2(Q^2) -1 +\third R^2_E\, Q^2\Big]; \eqlab{3rdZmoment}
\eeq
\item Zemach radius:
\beq 
R_{\mathrm{Z}}\equiv -\frac{4}{\pi}\int_0^\infty \frac{\dd Q}{Q^2}\left[\frac{G_E(Q^2)G_M(Q^2)}{1+\kappa}-1\right]. \eqlab{RZ}
\eeq
\end{subequations}
\end{itemize}
A derivation of these formulae will be given in \Secref{6.2}.

The proton polarizability effects
begin to contribute at order $(Z\al)^5 m_r^4$. 
The usual way of calculating these effects is 
through the two-photon exchange (TPE) diagram,  see \Secref{poleffect}. The elastic effects beyond the charge
radius (i.e., the contributions of Friar and Zemach radii),
together with some recoil corrections, 
are sometimes referred to as the 'elastic TPE'.
Therefore the TPE effect is split into the `elastic'
and `polarizability' contribution (see, e.g., \Figref{Pohl}).

\subsection{Charge and Zemach Radii from Muonic Hydrogen}

Figure~\ref{fig:Antognini} shows the $n=2$ energy-level scheme of $\mu$H and the measured transitions, i.e.:
\begin{subequations}
\bea
h \nu_t&=&E(2P_{3/2}^{F=2})-E(2S_{1/2}^{F=1}),\\
h \nu_s&=&E(2P_{3/2}^{F=1})-E(2S_{1/2}^{F=0}).
\eea 
\end{subequations}
The obtained experimental values for the Lamb shift and the HFS \cite{Pohl:2010zza,Antognini:1900ns,Antognini:2012ofa},
\begin{subequations}
\bea
E_{\mathrm{LS}}^{\mathrm{exp}}&=& \nicefrac{1}{4}\,h\nu_s+\nicefrac{3}{4}\,h\nu_t-E_{\mathrm{FS}}(2P)-\nicefrac{1}{8}\,E_{\mathrm{HFS}}(2P_{3/2})
-\nicefrac{1}{4}\,\delta=202.3706(23)\, \mathrm{meV},\\
E_{\mathrm{HFS}}^{\mathrm{exp}}(2S)&=&h\nu_s-h\nu_t+E_{\mathrm{HFS}}(2P_{3/2})-\delta=22.8089(51)\, \mathrm{meV}, 
\eea
\end{subequations}
thus rely on the theoretical calculation of the
fine and hyperfine splittings of the $2P$-levels \cite{Martynenko:2006gz}:
\begin{itemize}[$\square$]
\item $2P$ fine structure splitting: $ E_{\mathrm{FS}}(2P)=8.352082\,\mathrm{meV}$,
\item $2P_{3/2}$ hyperfine structure splitting: $ E_{\mathrm{HFS}}(2P_{3/2})=3.392588 \,\mathrm{meV}$,
\item $2P^{F = 1}$ level mixing: $\delta=0.14456 \,\mathrm{meV}$.
\end{itemize}

Furthermore, the  extraction of charge and Zemach radii from $\mu$H relies on the following 
theoretical description of the ($2P-2S$) Lamb shift and the $2S$ HFS~\cite{Antognini:2012ofa} (in units of $\mathrm{meV}$):
\begin{subequations}
\eqlab{muHtheory}
\bea
E_{\mathrm{LS}}^{\mathrm{th}}&=&206.0336(15)-5.2275(10)\,(R_E/\mathrm{fm})^2+E^{\mathrm{TPE}}_{\mathrm{LS}}\, , \quad \text{with} \,\,  E^{\mathrm{TPE}}_{\mathrm{LS}}=0.0332(20),\quad\eqlab{LS theory}\\
 E_{\mathrm{HFS}}^{\mathrm{th}}(2S)&=&22.9763(15)-0.1621(10)\,(R_\mathrm{Z}/\mathrm{fm}) +E^{\mathrm{pol}}_{\mathrm{HFS}}(2S)\, , \quad \text{with} \,\, E^{\mathrm{pol}}_{\mathrm{HFS}}(2S)=0.0080(26),\qquad
\eea
\end{subequations}
where $E^{\mathrm{TPE}}_{\mathrm{LS}}$
contains the Friar radius, recoil finite-size effects, and 
the polarizability effects; $E^{\mathrm{pol}}_{\mathrm{HFS}}(2S)$ is the HFS polarizability effect only. 
The precise numerical values of these TPE effects will 
be considered in \Secref{empir}. For review of
the QED effects we refer to Refs.~\cite{Pachucki:1996zza,Eides:2000xc,Borie:2012zz}.

Fitting the theory to experiment thus allows one to extract both the proton charge radius:  $R_E=0.84087(39) \,\mathrm{fm}$, and the Zemach radius: $R_\mathrm{Z}=1.082(37)\,\mathrm{fm}$.\footnote{The first $\mu$H measurement~\cite{Pohl:2010zza} had only determined $\nu_t$, and hence needed theory input
for the $2S$ HFS too: $E_{\mathrm{HFS}}(2S)=22.8148(78) \,\mathrm{meV}$ \cite{Martynenko:2004bt} (using $R_\mathrm{Z}=1.022 \,\mathrm{fm}$ \cite{Faustov:2001pn}).}
One caveat here,
as pointed out by~\citet{Karshenboim:2014maa, Karshenboim:2014vea}, is that this extraction 
relies on the Friar radius obtained from empirical FFs, which in turn have a different $R_E$ than extracted from the $\mu$H Lamb shift. This issue will be 
discussed in \Secref{consistency}.

\subsection{Finite-Size Effects by Dispersive Technique} \seclab{6.2}
\begin{wrapfigure}[13]{l}{0.2\textwidth}
\centering
\includegraphics[scale=0.7]{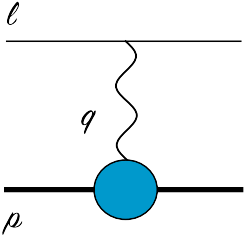}
    \caption{One-photon exchange graph with nucleon FFs, giving rise to FSE.  \label{fig:1gamma}}  
\end{wrapfigure}  
The usual derivation of the finite-size effects (FSE)
is done in terms of the charge and magnetization 
densities (see, e.g., \citet{Friar:1978wv}), which makes it difficult to derive
the relativistic corrections. We choose a different 
path \cite{Hagelstein:2015aa} and derive the Breit
potential from the manifestly Lorentz-invariant expression
for the Feynman diagram of \Figref{1gamma}, with the e.m.\ vertex of the nucleus given by:\footnote{Here we
introduce the e.m.\ FFs of a spin-1/2 nucleus. For hydrogen they are of course identical to the proton FFs. 
The Compton scattering formalism  
of the previous sections is 
applicable to spin-1/2 nuclei with $Z>1$, provided we replace the FFs as: $F_i \to Z F_i$,  and set $F_1(0)=1$, $F_2(0) =\kappa$, rather than $\zZ$, $\varkappa$ of the nucleon case. Likewise, we replace the structure functions: $f_i \to Z^2 f_i$, $g_i \to Z^2 g_i$.}
\beq
\Ga^\mu =Ze \ga^\mu F_1(Q^2) -\frac{Ze}{2M} \ga^{\mu\nu} q_\nu
F_2(Q^2). \eqlab{protonphotonvertex}
\eeq
The Dirac and Pauli FFs are then assumed to
fulfill the once-subtracted DRs:
\bea
\barr F_1 (Q^2)\\ 
F_2 (Q^2) \earr  = \barr 1\\
\kappa \earr - \frac{Q^2}{\pi} \int_{t_0}^\infty \frac{\dd t}{ t (t+Q^2) }
\im \barr  F_1(t) \\ 
F_2 (t) \earr, \eqlab{F12DR}
\eea
with $t_0$ being the lowest particle-production threshold.
The use of the DRs makes the rest of the derivation 
to be very much analogous to Schwinger's method of calculating the Uehling (vacuum polarization) effect \cite{Schwinger1949}. The Breit potential for the
Uehling effect was considered in, e.g., Refs.~\cite{Pachucki:1996zza,Karshenboim:2012wv}. 
  
At leading order (in $1/c$) we obtain the following terms
for the  Breit potential in momentum space:
\begin{subequations}
\eqlab{Vmomentumspace}
\bea
V_{e\mathrm{FF}}(Q) &=&4Z\al  \int_{t_0}^\infty\frac{\dd t}{t}\,\frac{\im G_E(t)}{t+Q^2},\\
V^{l=0}_{m\mathrm{FF}}(Q) &=&\frac{4\pi Z\al}{3mM} \left[F(F+1)-\thalf\right] \bigg\{1+\kappa- 
\frac{Q^2}{\pi} \int_{t_0}^\infty\frac{\dd t}{t}\,\frac{\im G_M(t) }{t+Q^2} \bigg\},
\eea
\end{subequations}
where the magnetic (spin-dependent)
part is only given for the
$S$-states ($l=0$). The imaginary part
(discontinuity along the branch cuts) of the electric and magnetic Sachs FFs is straightforwardly related to the one of Dirac and Pauli FFs:
\begin{subequations}
\bea
 \im G_E(t) &=& \im F_1(t) + \frac{t}{(2M)^2} \im F_2(t), \\
\im G_M(t) &=& \im F_1(t) + \im  F_2(t).
\eea 
\end{subequations}
The potential
in coordinate space is obtained via Fourier transform,
\beq 
V(r) = \frac{4\pi}{(2\pi)^3 r} \int_0^\infty \! \dd Q\,  Q  \,  V(Q)\, \sin Q r,
\eeq
\begin{subequations}
with the following result:
\bea
V_{e\mathrm{FF}}(r)&=&\frac{Z \al}{\pi r} \int_{t_0}^\infty\frac{\dd t}{t} e^{-r\sqrt{t}} \im G_E(t),\eqlab{Yukawa}\\
 V^{l=0}_{m\mathrm{FF}}(r)& = &\frac{4\pi Z\al}{3 mM}\left[F(F+1)-\thalf\right]  (1+\kappa)\, \rho_M(r),\eqlab{HFSb} 
\eea
\end{subequations}
where the magnetization density $\rho_M$ (the Fourier transform of $G_M$) is a Laplace-type of transform of $\im G_M$: 
 \beq
 \rho_M (r) =  \frac{1}{(2\pi)^2\, r} \int^\infty_{t_0}  \! \dd t\, \frac{\im G_M(t)}{1+\kappa}\,  e^{-r\sqrt{t}} .
 \eeq
The latter definition shows explicitly  that 
a spherically symmetric density is a Lorentz
invariant quantity.
An analogous definition, but in terms of $\im G_E$, 
applies to the charge density
$\rho_E(r)$. 


The FF effect can now be worked out using time-independent perturbation theory. 
For example, the energy shift of the $nl$-level due to a spherically  symmetric correction $V_\de (r)$  to the Coulomb potential $V_C(r) = -Z\al/r $ is to 1\textsuperscript{st} order given by:
 \beq
 E^{\langle \de \rangle }_{nl} \equiv \left<nlm|\,  V_\de \, | nlm\right> = \frac{1}{2\pi^2}
 \int_0^\infty\! \dd Q\, Q^2\, w_{nl}(Q)\, 
 V_\de (Q)
 = \int_0^{\infty}\! \dd r\, r^2 R_{nl}^2(r) \,  V_\de (r),
 \eeq
 where the momentum-space expression contains
 the convolution of the momentum-space wave functions,
 \beq
  w_{nl}(Q) = 
  \int \dd \bp \,
\vfi_{nlm}^\ast (\bp+\bQ) \, \vfi_{nlm} (\bp),
\eqlab{wfconvolution}
 \eeq
while the coordinate-space one contains the 
radial wave functions $R_{nl}(r)$. The explicit forms
of the wave functions can be found in,  e.g., \citet{BetheSalpeter}. For completeness we
give here the expressions for $1S$, $2S$, and $2P$
states:
\beq 
\eqlab{WF}
\begin{aligned}
 R_{10}(r) &= \frac{2}{a^{3/2}} \, e^{-r/a},\\
 R_{20}(r) &= \frac{1}{\sqrt{2} \, a^{3/2}} \left(1- \frac{r}{2a} \right) 
 e^{-r/2a},\\
 R_{21}(r) &= \frac{1}{\sqrt{3} \, (2a)^{3/2}} \frac{r}{a}
 e^{-r/2a},
\end{aligned}\qquad
\begin{aligned}
w_{1S}(Q) &= \frac{16}{\big(4+(a Q)^2\big)^2},\\
w_{2S}(Q) &= \frac{\big(1-(aQ)^2\big)\big(1-2(aQ)^2\big)}{\big(1+(aQ)^2\big)^4},\\
w_{2P}(Q) &= \frac{1-(aQ)^2}{\big(1+(aQ)^2\big)^4}.
\end{aligned}
\eeq
For the following discussions it is useful to note the
asymptotic behavior of $w$ for large $Q$:
\beq 
w_{nS}(Q) \stackrel{Q\to\infty}{=}  \frac{16\pi}{aQ^4} \phi_n^2, 
\eeq
where $\phi_n = 1/\sqrt{\pi a^3 n^3}$ is the coordinate-space wave function at the origin, $r=0$. 
\subsubsection{Lamb Shift}
Consider first the correction due to the electric FF 
($G_E$), as
given by  \Eqref{Yukawa}. 
At 1\textsuperscript{st} order, it yields the
following correction  to the $2P-2S$ Lamb shift:
\begin{subequations}
 \bea
 E^{\langle {e\mathrm{FF}}\rangle}_\mathrm{LS} & =& -\frac{Z\al}{2\pi a^3} \int_{t_0}^\infty \!\dd t \, \frac{\im G_E (t)}{(\sqrt{t}+Z\al m_r)^4}, 
\eqlab{rmsLSa}\\
&=& -\frac{\pi Z\al}{3a^3}  \int_0^\infty\! \dd r \, r^4 e^{-r/a } \rho_E (r), \eqlab{rmsrhoE}\\
&=& -\frac{Z\al}{12 a^3} \sum_{k=0}^\infty 
 \frac{(-Z\al m_r)^{k}}{k!} \langle r^{k+2}\rangle_E = -\frac{(Z\al)^4 m_r^3}{12} \Big( \langle r^{2}\rangle_E -Z\al m_r \langle r^3\rangle_E \Big) +\ldots ,\quad\eqlab{rmsLSaEXP}
\eea
\end{subequations}
where in the last steps we have expanded in the moments of the charge distribution:
\beq
\eqlab{rmsdef}
\langle r^N\rangle_E \equiv  4\pi \int_0^\infty \dd r\, r^{N+2}
\rho_E(r)  =
\frac{(N+1)!}{\pi}\int_{t_0}^\infty\! \dd t \, \frac{\im  G_E(t)}{t^{N/2+1 }  }.
\eeq
One should keep in mind though that the expansion in moments is not necessarily convergent. For instance, \Eqref{rmsLSa} tells us that the expansion is applicable 
when the nearest particle-production threshold 
is well above the inverse Bohr radius, i.e.:
 $Z\al m_r \ll \sqrt{t_0}$. 

Incidentally, one of the early proposals 
for solving the proton-radius puzzle \cite{DeRujula:2010dp} does not
work out precisely because the expansion in moments
is not applicable for the choice of $\rho_E$ proposed
therein. The fine-tuning of $\rho_E$  affected mainly 
the region $r>a$, enhancing the Friar radius by almost a factor of 3, and 
thus achieving a huge impact on the Lamb shift, according to 
\Eqref{LambShift}. On the other hand,
according to the exact formula \eref{rmsrhoE}, 
the region above the Bohr radius makes a negligible 
impact on the Lamb shift, which was verified  explicitly 
by us~\cite{Hagelstein:2015aa} for the model of 
Ref.~\cite{DeRujula:2010dp}.

It is also useful  
to have an expression in terms of $G_E$ itself:
  \beq
 E^{\langle {e\mathrm{FF}}\rangle}_\mathrm{LS}=
 -\frac{2Z\al}{\pi}\int_0^\infty \!\dd Q \, w_{2P-2S}(Q)\,G_E(Q^2), \quad \text{with }\,\, w_{2P-2S}(Q)= 2(Z\al m_r)^4 Q^2
\frac{(Z\al m_r)^2-Q^2}{\left[(Z\al m_r)^2+Q^2\right]^4}. \eqlab{LSexact}
  \eeq
Using the expression for the weighting function, $w$, in terms of the wave functions, cf.~\Eqref{wfconvolution},
it is easy to see that for $G_E(Q^2)= \mathrm{const}$
the effect vanishes. That is, 
the charge normalization drops out. 
  Note also that this FF effect is still  of $O(\al^4)$, 
  despite the  $O(\al^5)$ overall prefactor.
  The naive expansion of
 $w$ in $\al$ does not work, as
  the resulting integral is infrared divergent. As 
  seen below more explicitly,  this
  infrared enhancement yields in the end 
  the correct charge radius contribution of $O(\al^4)$.

  To complete the derivation of the standard formulae for the FSE in 
the Lamb shift to $O(\al^5)$, we
take the potential to the 2\textsuperscript{nd} order in perturbation theory. The 2\textsuperscript{nd}-order contribution at $O(\al^5)$ comes from the
continuum states only and amounts to:
\beq
 E_\mathrm{LS}^{\langle {e\mathrm{FF}}\rangle\langle {e\mathrm{FF}}\rangle}
=  
\frac{Z\al}{a^4} \, \frac{2}{\pi}\int_0^\infty \!\frac{\dd Q}{Q^4}\, \Big[ G_E(Q^2) -1 \Big]^2+O(\al^6)
=-\frac{Z\al}{12a^4}\, \, \Big[
\langle r^3\rangle_E - \frac12 \langle r^3\rangle_{E(2)} \Big]+O(\al^6).
\eeq 
Adding this result to the $O(\al^5)$ term from the 1\textsuperscript{st} order, \Eqref{rmsLSaEXP}, we 
can see that the 
2\textsuperscript{nd}-order effect replaces the 3\textsuperscript{rd} charge radius $\langle r^3\rangle_E$ by the 3\textsuperscript{rd} Zemach moment $\langle r^3\rangle_{E(2)} $, resulting in \Eqref{LambShift}.

\subsubsection{Consistency of the Charge Radius Extraction}
\seclab{consistency}
  The consistency problem, recently addressed by \citet{Karshenboim:2014maa},
  is basically that {\em in order to compute the Friar 
  radius \eref{3rdZmoment},
  and hence its contribution to the Lamb shift \eref{LambShift},
  one must know the charge radius $R_E$, which in turn
  needs to be extracted from the Lamb shift.}
  Presently the $\mu$H extraction of $R_E$ uses
  as input the value of $R_\mathrm{F}$ obtained
  from the empirical FF, which has a different $R_E$.
  This obviously is not consistent and leads to
  a systematic uncertainty.
  
To see the origin of this problem, let us examine the
exact (unexpanded in moments) FSE  
to 2\textsuperscript{nd} order in perturbation theory:
\beq 
E^{\mathrm{FSE}}_\mathrm{LS}\equiv  E^{\langle {e\mathrm{FF}}\rangle}_\mathrm{LS}+E_\mathrm{LS}^{\langle {e\mathrm{FF}}\rangle\langle {e\mathrm{FF}}\rangle}
= -\frac{Z\al}{\pi} \int_0^\infty \!\dd Q \, w_{2P-2S}(Q)\,G_E^2 (Q^2), \eqlab{LSexactGE2}
\eeq 
where the weighting function is given in \Eqref{LSexact}.
This form clearly shows that the Lamb shift is
a functional of the FF. Ideally, one needs to find
$G_E$ which fits the $ep$ and atomic data simultaneously.
This, however, has not yet been realized. 

\begin{wrapfigure}{l}{0.35\textwidth}
\centering
\includegraphics[scale=0.3]{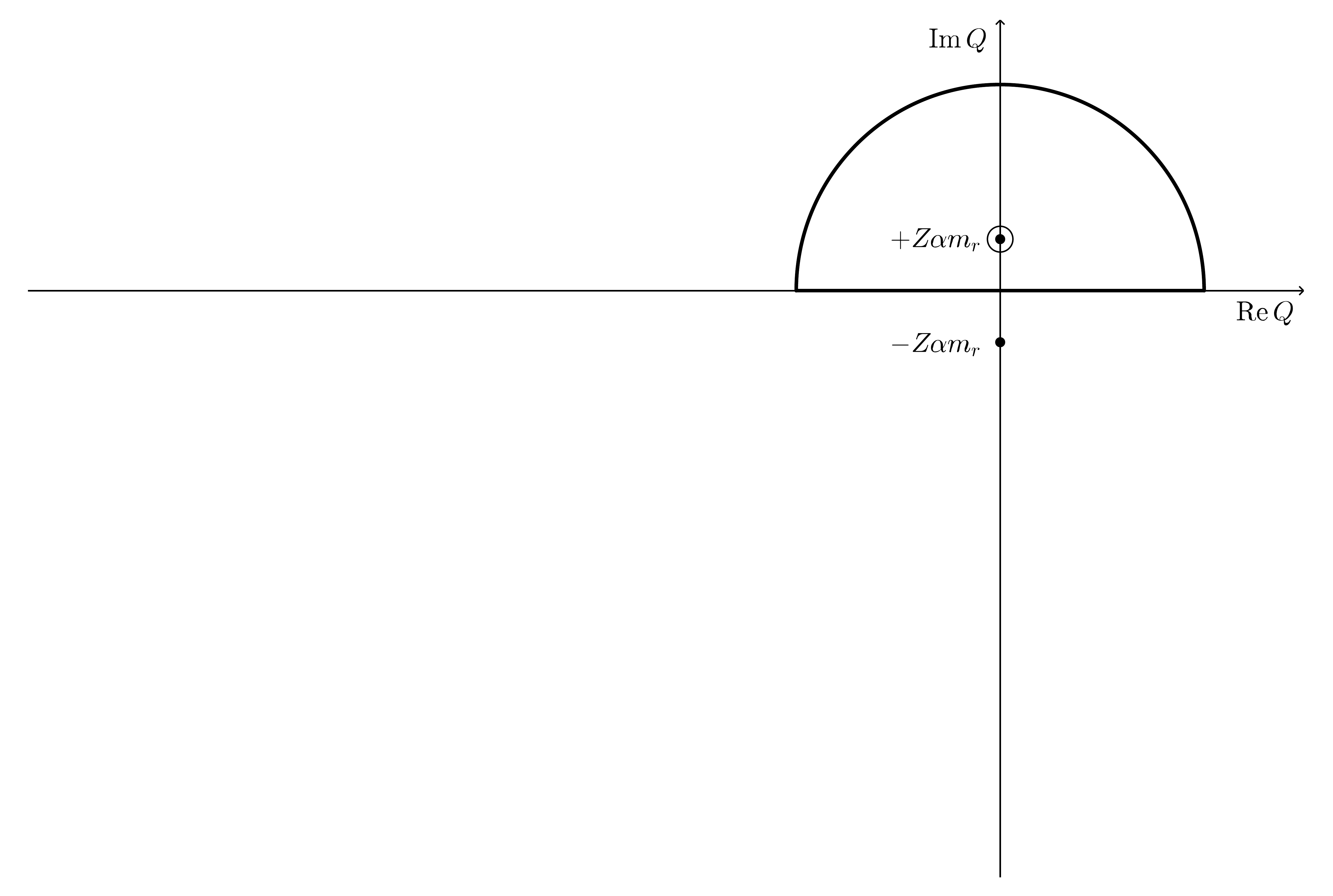}
\caption{Poles of $w_{2l}$, and the contour in the complex $Q$ plane. \figlab{complexQ}}
\end{wrapfigure}
Let us 
  evaluate the integral by the residue method. 
  For this we consider the complex $Q$ plane, \Figref{complexQ}. 
  The poles of $w(Q)$ are at 
  \beq 
  Q=\pm i Z\al m_r.
  \eeq 
  The singularities of $G_E(Q^2)$ also lie on the imaginary axis, since it obeys the DR of the type \eref{F12DR}, see \Eqref{GEMDR}.
  The integrand is even in $Q$ and hence we may
  extend the integration to negative $Q$. We then close the
  contour in the upper half-plane, use the residue theorem, and neglect
  the integral over the big semicircle to arrive at
  \beq 
  \eqlab{res}
    \int_{0}^\infty \!\dd Q \, w(Q)\,G^2_E(Q^2)
  = \pi i \,\mbox{Res}\big[w(Q)\,G^2_E(Q^2)\big]_{Q=i Z\al m_r}
  + \ldots, \eeq 
 where the dots stand for the residue of $G_E$ poles. 
 Those can be computed using the DR or an explicit anzatz
 (e.g., an empirical parametrization) for $G_E$. When using
 the DR of \eref{GEMDR} one simply obtains the Friar radius contribution
 written in terms of $\im G_E$, cf.~\Eqref{imFriar}.
 
 The residue over the pole of $w$ evaluates to:
 \beq 
  \mbox{Res}[w(Q)\,g(Q^2)]_{Q=iZ\al m_r} =
  i (Z\al m_r)^3 \left[ \nicefrac{1}{4}\, g'(Q^2)+ 
 Q^2 g''(Q^2) + \nicefrac13\, Q^4 g'''(Q^2)\right]_{Q^2=-(Z\al m_r)^2}, 
 \eqlab{resres}
 \eeq 
 where we have introduced for a moment $g\equiv G_E^2$, and
 the primes denote the derivatives over
 $Q^2$. Obviously, the first term dominates (lowest in $\al$) 
 and yields the usual $R_E^2$ contribution of 
 $O((Z\al)^4 m_r^3)$, cf.~\Eqref{LambShift}.
 
 It is interesting to observe that \Eqref{resres}
 is only dependent on the derivatives of $G_E$, and thus
 does not involve any of the odd moments, which are integrals
 of $G_E$, see \Eqref{Gmomentsodd}. 
 The contribution of odd moments, and in particular the one of the Friar radius, comes from the singularities of $G_E$. 
 Thus, the consistency problem in question is
 absent if the rms charge radius and
  the poles of $G_E$ are {\it uncorrelated}. 
  We, however, are not aware at the moment 
  of an empirical parametrization in which 
  the charge radius and the poles are not
  correlated. Quite the opposite, the correlation 
  is usually strong. The simplest example is 
  provided by the dipole form,
  $G_E(Q^2)=(1+Q^2/\Lambda^2)^{-2}$. Both the radius and
  the pole positions are given by the mass parameter $\Lambda$: the radius is $12/\Lambda^2$, 
  while the pole is at $Q=\pm i\Lambda$. 
  An empirical parametrization with weak correlation between
  the value of $G_E'(0)$ and the position of its poles 
  would be preferred from this point of view.
  
 General constraints on the FF parametrizations 
 have recently been discussed at length by 
 Sick et al.~\cite{Sick:2011zz,Sick:2012zz,Sick:2014sra,Sick:2014kna}. One
 finds in particular that certain parametrizations have unphysical poles which result in weird charge or magnetization distributions. To take the advantage of studying the $r$-space simultaneously, it has been suggested to parametrize the FF in a basis with analytic Fourier transform, e.g., with a sum of Gaussians. 
It would be interesting to see if these kind of parametrizations
lead to weaker correlation between the rms radius and the FF
poles.


\subsubsection{Hyperfine Splitting}
The HFS, introduced in \Eqref{HFS}, 
receives at first only the magnetic contribution:
\begin{subequations}
\bea 
E^{\langle m\mathrm{FF}\rangle}_{\mathrm{ HFS}}(nS)
 &=& \frac{4Z\al}{3\pi mM} \int_0^\infty\! \dd Q\, Q^2\, 
 w_{nS}(Q)\, G_M(Q^2), \eqlab{HFSPTQ}\\
&=& \frac{8\pi Z\al}{3 mM}\,(1+\kappa) \int_0^\infty \dd r\, r^2 R_{n0}^2(r) \,\rho_M(r),\eqlab{HFSPTr}
\eea  
\end{subequations}
where $w_{nS}$ is given by \Eqref{wfconvolution}.
Setting in this expression $G_M = 1+\kappa$, or equivalently $  \rho_M(r) = \de(r)/4\pi r^2$, yields the Fermi energy, $E_F(nS)$, given in \Eqref{EFermi}. Here, the expansion in $\al$
works straightforwardly (we can expand under the integrals), and we obtain:
\beq 
\frac{E^{\langle m\mathrm{FF}\rangle}_{\mathrm{ HFS}}(nS)}{E_F(nS)}=1 -
\frac{2}{a} \langle r\rangle_M+O(\al^2),
\eeq
where the first moment of $\rho_M(r)$ 
can equivalently be written as: 
\beq
\langle r\rangle_M = -\frac{4}{\pi}
\int_0^\infty\! \frac{\dd Q}{Q^2} \, \left[\frac{G_M(Q^2)}{1+\kappa}-1\right].
\eeq

At the 2\textsuperscript{nd} order in perturbation 
theory, we obtain the interference between the potentials of the
electric and magnetic term:
\beq
\frac{E^{\langle{e\mathrm{FF}}\rangle\langle{m\mathrm{FF}}\rangle}_{\mathrm{ HFS}}(nS)}{E_F(nS)}=\frac{8}{a\pi}\int_0^\infty \frac{\dd Q}{Q^2}\left[G_E(Q^2)-1\right]\frac{G_M(Q^2)}{1+\kappa}+O(\al^2).
\eeq
Adding up the 1\textsuperscript{st}- and 2\textsuperscript{nd}-order contributions,
we obtain the well-known FSE in the
HFS given by \Eqref{HFS}.

Taking $V^{l=0}_{m\mathrm{FF}}$ to 2\textsuperscript{nd} order gives rise to~$E^{\langle{m\mathrm{FF}}\rangle\langle{m\mathrm{FF}}\rangle}$,  which is a higher-order recoil effect.
This, and some other, recoil effects are treated 
more properly within the approach we consider next.

\subsection{Structure Effects through Two-Photon Exchange}
\seclab{poleffect}

\begin{figure}[tbh]
\centering
\begin{minipage}{0.49\textwidth}
\centering
       \includegraphics[width=7cm]{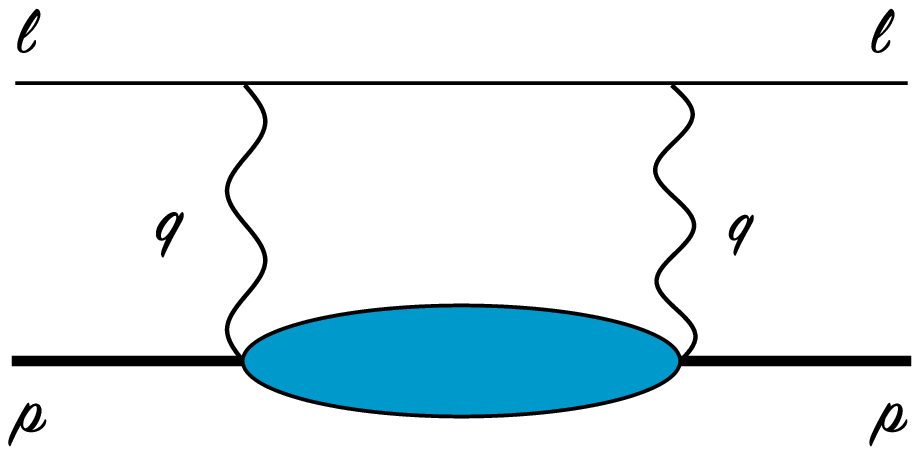}

      (a)
    \end{minipage}\hfill
\begin{minipage}{0.49\textwidth}
\centering
  \includegraphics[width=7cm]{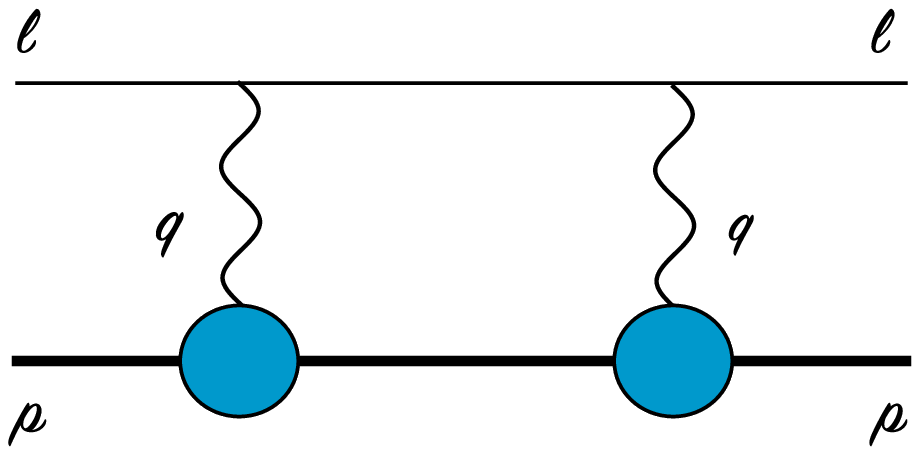}
  
  (b)
\end{minipage}
\caption{(a) TPE diagram in forward kinematics: the horizontal lines correspond to the lepton and the proton (bold), where the `blob' represents all possible excitations. (b) Elastic  contribution to the TPE. The crossed diagrams are not drawn.\label{fig:TPE}}
\end{figure}

Having obtained the standard FSE of \eref{FSEs} using
the Breit potential, we consider here a different approach.
We consider the correction, to the Coulomb potential,
due to the TPE, see \Figref{TPE}~(a).
A great advantage of this approach is that one can access
the inelastic effects of the proton structure \cite{Iddings:1965zz,Drell:1966kk}.
On the other hand, it is less systematic and cannot be used 
without matching to a systematic expansion. We shall only use
it to 1\textsuperscript{st}-order perturbation
theory and match the elastic part, \Figref{TPE}~(b),
with the FSE
derived from the Breit potential.  

Let us note right away that the TPE 
contains an iteration
of the Coulomb potential present in the wave functions.
However, we need not to worry about the double-counting.
The Coulomb interaction by itself is harmless, as it gives no
contribution to the Lamb shift or HFS.

To $O(\al^5)$ it will be sufficient to evaluate the TPE amplitude at zero energy ($p\cdot \ell=mM$)
and momentum transfer ($t=0$). The resulting amplitude yields 
a constant potential in momentum space, which of course translates to
a $\de$-function potential in coordinate space. The energy shift 
is thus proportional to the wave-function at the origin, and hence
only the $S$-levels are affected.

The forward TPE amplitude is a one-loop integral
involving the doubly-virtual Compton scattering (VVCS) 
amplitude. The latter has been discussed in \Secref{SRs}.
According to \Eqref{fVVCS}, its tensor structure decomposes
into a symmetric, spin-independent part parametrized 
by the scalar amplitudes $T_{1,2}(\nu,Q^2)$, and 
an antisymmetric, spin-dependent part parametrized by
$S_{1,2}(\nu,Q^2)$.  The HFS obviously depends on the latter, while the Lamb shift depends on the former. 

More specifically, 
the shift of the $nS$-level is given by:
\beq
\Delta E(nS)= 8\pi \al m \,\phi_n^2\,
\frac{1}{i}\int_{-\infty}^\infty \!\frac{\dd\nu}{2\pi} \int \frac{\dd \bq}{(2\pi)^3} \,  \frac{\left(Q^2-2\nu^2\right)T_1(\nu,Q^2)-(Q^2+\nu^2)\,T_2(\nu,Q^2)}{Q^4(Q^4-4m^2\nu^2)},\eqlab{VVCS_LS}
\eeq
with $\phi_n^2=1/(\pi n^3 a^3)$ the wave function squared at the origin, and $\nu=q_0$, $Q^2 = \bq^2 -q_0^2$. The correction to the HFS is given by:
\beq
\frac{E_{\mathrm{HFS}}(nS)}{E_F(nS)}=\frac{4m}{\mu}\frac{1}{i}\int_{-\infty}^\infty \!\frac{\dd\nu}{2\pi} \int \frac{\dd \bq}{(2\pi)^3}\,\frac{1}{Q^4-4m^2\nu^2}\left\{\frac{\left(2Q^2-\nu^2\right)}{Q^2}S_1(\nu,Q^2)+3\frac{\nu}{M}S_2(\nu,Q^2)\right\}.\eqlab{VVCS_HFS} \qquad
\eeq

To proceed further, one performs a
Wick rotation, i.e.\ changes the integration 
over $q_0$  to $Q_0= -i q_0$. Note that this is only possible at vanishing 
external energy (threshold) of lepton-proton scattering. 
At finite energy one needs to take
care of the poles moving across the imaginary $q_0$ axis
(see, e.g., Ref.~\cite{Pascalutsa:1999pv}). 

After the Wick rotation, the integration four-momentum
is Euclidean and we can evaluate it in hyperspherical
coordinates, $$Q^\mu = Q (\cos\chi, \, \sin\chi \sin\th\cos\vfi,\, \sin\chi \sin\th\sin\vfi, \,\sin\chi\cos\th). $$ 
The integrals over $\th$ and $\vfi$ can be done right
away, yielding a factor of $4\pi$. 
The integral over $\nu = iQ\cos\chi$ can be done after
substituting the DRs for the VVCS amplitudes, \Eqref{genDRs}.  Introducing the ``lepton velocity''
$v_l = \sqrt{1+4m^2/Q^2}$, we obtain
the following expression for the $S$-level shift:
\bea
\Delta E(nS)&=&\frac{16(Z\al)^2m}{M}\,\phi_n^2\,\int_0^\infty \frac{\dd Q}{Q^3}\,\int_0^1\dd x\,\frac{1}{v_l+\sqrt{1+x^2\tau^{-1}}}\nn\\
&\times &\Bigg\{ \frac{f_1(x,Q^2)}{x} 
- \frac{f_2(x,Q^2)}{2\tau} + \, \frac{1}{(1+v_l)(1+\sqrt{1+x^2\tau^{-1}})}\bigg(\frac{2f_1(x,Q^2)}{x}+ \frac{f_2(x,Q^2)}{2\tau}\bigg) \Bigg\},
\eqlab{LSMaster}
\eea
and the HFS:
\bea
\frac{E_{\mathrm{HFS}}(nS)}{E_F(nS)}&=&\frac{16Z\al m M}{\pi (1+\kappa)}\int_0^\infty\frac{\dd Q}{Q^3}\int_0^1\dd x\,\frac{1}{v_l+ \sqrt{1+x^2\tau^{-1}}} \nn\\
&\times &   \left\{ \left[1+\frac{1}{2(v_l+1)(1+ \sqrt{1+x^2\tau^{-1}})}\right]2g_1(x,Q^2)+3g_2(x,Q^2) \right\}.\eqlab{fullHFS}
\eea
These are the master formulae containing all the structure
effects to $O(\alpha^5)$. One ought to be careful though
in matching the contribution of the elastic structure functions~\eref{elstructure} to the standard FSE of \eref{FSEs}. In the non-relativistic 
(heavy-mass) limit we obtain:\footnote{For the expansion we use,
$1/(v_l+v) \simeq \frac{Q}{2(m+M)} \big( 1 - Q^2/8mM \big)$, $[2(v_l+1)(v+1)]^{-1}\simeq Q^2/8mM$, where $v=\sqrt{1+\tau^{-1}}$.}
\begin{subequations}
\bea
\Delta E^{\mathrm{el}}(nS)&=& 
- 16(Z\al)^2m_r\,\phi_n^2\,\int_0^\infty \frac{\dd Q}{Q^4}\,
G_E^2 (Q^2), 
\eqlab{NRelasticLS}
\\
E_{\mathrm{HFS}}^{\mathrm{el}}(nS)&=&\frac{64(Z\al)^2 m_r}{3mM}\,\phi_n^2
 \int_0^\infty\frac{\dd Q}{Q^2}\, G_M(Q^2)  G_E(Q^2) .
\eqlab{NRelasticHFS}
\eea
\end{subequations}
The correct matching is achieved by regularizing the
infrared divergencies with the convoluted wave functions, 
i.e., $w(Q)$ in \Eqref{wfconvolution}. For example, to obtain
the charge radius contribution to the $2P-2S$ Lamb shift
one should replace $\phi_n^2$ in \Eqref{NRelasticLS} with: 
$- a w_{2P-2S}(Q)\,Q^4/16\pi$. For the HFS, the
replacement in \Eqref{NRelasticHFS} is: $\phi_n^2\to a w_{nS}(Q)\,Q^4/16\pi$,
yielding the correct Fermi energy and Zemach radius contributions.
The infrared-safe contributions, such as the recoil
and polarizability corrections, need no regularization.
In what follows we only consider those kind of effects.

\subsection{Empirical Evaluations}
\seclab{empir}
\subsubsection{Lamb Shift}
\label{section:polcontrLS}
The $O(\alpha^5)$ effects of proton structure in the Lamb shift are usually divided into the
effect of (i) the Friar radius, (ii) finite-size recoil, and (iii) polarizabilities. The first two 
are sometimes combined into (i') the `elastic' TPE contribution. The `polarizability'
effect is often split between  (ii') the `inelastic' TPE and (iii') a `subtraction' term, i.e., the contribution
of $\ol T_1(0,Q^2)$.

The elastic and inelastic TPE contributions are well-constrained by the available
empirical information on, respectively, the proton FFs and unpolarized structure functions, whereas the subtraction contribution must be modeled. It certainly helps to know
that 
\beq
\lim_{Q^2\rightarrow0}\overline T_1(0,Q^2)/Q^2=4\pi \beta_{M1},\eqlab{sub1}
\eeq
but otherwise, the $Q^2$ behavior of this amplitude 
leaves room to imagination.
For example, \citet{Pachucki:1999zza} and later \citet{Martynenko:2005rc} use: 
\beq
\ol T_1(0,Q^2)=4 \pi \beta_{M1}\,Q^2/\big(1+Q^2/\Lambda^2\big)^4\, ,\eqlab{sub2}
\eeq
with $\Lambda^2=0.71\,\mathrm{GeV}^{2}$, whereas 
\citet{Carlson:2011zd} and 
 \citet{Birse:2012eb} use more sophisticated forms, inspired
 by chiral loops. The leading-order [$O(p^3)$] $\chi$PT calculation contains a genuine prediction for the subtraction
 function, as well as for the whole 
 polarizability effect, see \Secref{sec66} for more details. 
 
An early study of the electric polarizability effect on the $S$-level shift in electronic and muonic atoms can be found in Ref.~\cite{Bernabeu:1982qy}. That work exploited an unsubtracted DR for the longitudinal amplitude $T_L(\nu,Q^2)=(1+\nu^2/Q^2)\,T_2(\nu,Q^2)-T_1(\nu,Q^2)$, as introduced in Ref.~\cite{Bernabeu:1973uf}. As we have
discussed in the previous section, such a DR is not 
valid for the proton. 

A first standard dispersive calculation of the TPE effect was done by~\citet{Pachucki:1999zza}, see also \cite{Pachucki:1996zza, Veitia:2004zz}. The
most recent updates can be found in Refs.~\cite{Carlson:2011zd,Birse:2012eb}.
Presently, the recommended value is that
of \citet{Birse:2012eb}.
A somewhat different dispersive evaluation has recently been done by 
\citet{Gorchtein:2013yga}.
There,  the high-energy behavior of  the subtraction function is related to the fixed $J=0$ Regge pole \cite{Creutz:1968ds} through a finite-energy sum rule (see Eq.~(29) in Ref.~\cite{Gorchtein:2013yga}). 

Table~\ref{Table:Summary1} summarizes the dispersive evaluations of the TPE effects in the $\mu$H Lamb shift,
while the $\chi$PT predictions can be found in Table \ref{Table:Summary2}. 
The corresponding `polarizability' and `elastic' TPE results are represented in the summary plots, see \Figref{LSSummary} and \Figref{LSSummaryelastic}.
Table~\ref{Table:Summary1} also shows the value
of the magnetic polarizability used in the evaluations,
since this is the main source of discrepancy among them.

Other frameworks, different from DR and $\chi$PT, 
for calculating the TPE effects in the
Lamb shift can be found in \cite{Hill:2011wy}
and \cite{Mohr:2013axa}. The values
obtained in these works are generally in agreement
with the dispersive results. For example, 
\citet{Mohr:2013axa} quote:
\beq
\Delta E^\mathrm{inel}(2S)=-17\,\upmu\mathrm{eV}, \qquad \Delta E^\mathrm{el}(2S)=-20\,\upmu\mathrm{eV}.
\eqlab{elMohr}
\eeq

\begin{table}
\footnotesize
\centering
\caption{Summary of available dispersive calculations for the TPE correction to the $2S$-level in $\mu$H. Energy shifts are given in $\upmu\mathrm{eV}$, $\beta_{M1}$ is given in $ 10^{-4} \,\mathrm{fm}^3$.
\label{Table:Summary1}}
%
\begin{minipage}{\linewidth} 
\footnotesize
\centering
\begin{tabular}{|p{0.1\linewidth}| p{.14\linewidth}p{.11\linewidth}p{.18\linewidth}p{.13\linewidth}p{.17\linewidth}|}
\hline                  
& Pachucki& Martynenko&  Carlson \& & Birse \&&   Gorchtein et al.~\cite{Gorchtein:2013yga}\footnote{Adjusted values;
the original values of Ref.~\cite{Gorchtein:2013yga},  $\Delta E^\mathrm{subt}(2S)=3.3$ and $\Delta E^\mathrm{el}(2S)=-30.1$, are based on a different decomposition into the elastic and polarizability contributions.}\\
& \cite{Pachucki:1999zza}& \cite{Martynenko:2005rc}& Vanderhaeghen \cite{Carlson:2011zd}\footnote{In this work a separation of the amplitude into `pole' and `non-pole', rather than `Born' and `non-Born', was chosen. It is pointed out in Ref.~\cite{Birse:2012eb} that the `pole' decomposition applied in \cite{Carlson:2011zd} is inconsistent with the standard definition of the magnetic polarizability used ibidem.} & McGovern \cite{Birse:2012eb}& \\

\hline
\hline
$\beta_{M1}$&$1.56(57)$\cite{Tonnison:1998mi}&$1.9(5)$ \cite{Eidelman20041}&$3.4(1.2)$\cite{Beane:2002wn,Beane:2004ra}&$3.1(5)$\cite{Griesshammer:2012we}&\\
\hdashline[0.5pt/5pt]
$\Delta E^{\mathrm{subt}}(2S)$        & $1.9$                                                             &   $ 2.3$                                                         & $5.3(1.9)$                                       &$4.2(1.0)$                            &  $ -2.3 (4.6)$                                     \\
$\Delta E^{\mathrm{inel}}(2S)$            &  $-13.9$ \cite{Brasse:1976bf, Abramowicz:1997}                                                         &    $ -16.1$                                                      & $-12.7(5)$\cite{Christy:2011,Capella:1994cr}                               &$-12.7(5)$\footnote{Value taken from Ref.~\cite{Carlson:2011zd}.}   & $-13.0(6)$ \cite{Christy:2011,Capella:1994cr, Gorchtein:2011mz}                                       \\ 
\hdashline
$\Delta E^{\mathrm{pol}}(2S)$              &$-12(2)$                                                           &   $-13.8(2.9)$                                              & $ -7.4  (2.0)$                               &$-8.5(1.1)$                                     & $ -15.3(4.6)$                                     \\
\hline
 $\Delta E^{\mathrm{el}}(2S)$&$-23.2(1.0)$\cite{Simon1980381}&&$\begin{cases}
-27.8\text{\cite{Kelly:2004hm}}\\
\bold{-29.5(1.3)}\text{\cite{Arrington:2007ux}}\\
-30.8\text{\cite{Bernauer:2010wm,Vanderhaeghen:2010nd}}
\end{cases}$&$-24.7(1.6)$\footnote{Result taken from Ref.~\cite{Carlson:2011zd} (FF \cite{Arrington:2007ux}) with reinstated `non-pole' Born piece.}&$-24.5(1.2)$ \cite{Bernauer:2010wm, Kelly:2004hm, Arrington:2007ux}\\
\hline
\hline
 $\Delta E(2S)$&$-35.2(2.2)$&&$-36.9(2.4)$&$-33(2)$&$-39.8(4.8)$\\
\hline
\end{tabular}
\end{minipage}
\end{table}

\subsubsection{Hyperfine Splitting}

The leading-order HFS is given by the Fermi energy of the $nS$-level, cf.\ \Eqref{EFermi}. The full HFS is divided
into the following contributions:\footnote{A review of polarizability corrections to the hydrogen HFS can be found in  Ref.~\cite[Sect.~3]{Carlson:2007aa}. A detailed formalism of the structure-dependent corrections to the $2S$ HFS in both H and $\mu$H is given in Ref.~\cite{Carlson:2011af}, with comments on various conventions.}
\beq
E_{\mathrm{HFS}}(nS)=\left[1+\Delta_\mathrm{QED}+\Delta_\mathrm{weak}+\Delta_\mathrm{structure}\right]E_F(nS).
\eeq
We are interested in the proton-structure correction, which is split into three terms: Zemach radius, recoil, and polarizability contribution,
\beq
\Delta_\mathrm{structure}=\Delta_Z+\Delta_\mathrm{recoil}+\Delta_\mathrm{pol}. \eqlab{FS_HFS}
\eeq
All of these can be deduced from \Eqref{VVCS_HFS}, and are thus given in terms of the spin-dependent VVCS amplitudes $S_1$ and $S_2$, which satisfy the DRs of Eqs.~\eref{S1DR}\footnote{The validity of the unsubtracted DRs is based on Regge theory \cite{Abarbanel:1967zza}, see also Refs.~\cite{Carlson:2006yj,Carlson:2006mw} for a discussion of the no-subtraction assumption.} and \eref{DRS2woBC}.
This means the entire TPE contribution is given by 
the spin structure functions $g_1$ and $g_2$. To separate out the polarizability contribution, one can 
write the DRs for the non-Born (polarizability) part of the amplitudes only:
\begin{subequations}
\bea
\ol S_1(\nu,Q^2)&=&\frac{2\pi Z^2\al}{M} F_2^2(Q^2)+ \frac{16\pi Z^2\al M}{Q^2} \int_{0}^{x_0} 
\!\dd x\, 
\frac{g_1 (x, Q^2)}{1 - x^2 (\nu/\nu_{\mathrm{el}})^2},\eqlab{S1g1}\\
\nu\ol S_2(\nu,Q^2) &=& \frac{64\pi Z^2\al M^4\nu^2}{Q^6} \int_{0}^{x_0} 
\!\dd x\,x^2 
\frac{g_2 (x, Q^2)}{1 - x^2 (\nu/\nu_{\mathrm{el}})^2}.\eqlab{S2g2}
\eea 
\end{subequations}
One should not be perplexed by the Pauli FF term, $F_2^2$,
appearing in $\ol S_1$. 
Its purpose is to cancel the elastic contribution 
of the GDH integral $I_1(Q^2)$, such that 
$\ol S_1$ is indeed proportional to polarizabilities
alone.

Let us now specify the decomposition of the structure-dependent correction into the three terms of  \Eqref{FS_HFS}. The first one is the Zemach contribution \cite{Zemach:1956}:
\beq
\Delta_Z=\frac{8Z\al m_r}{\pi}\int_0^\infty \frac{\dd Q}{Q^2} \left[\frac{G_E(Q^2)G_M(Q^2)}{1+\kappa}-1\right]\equiv-2Z\al m_r R_\mathrm{Z}.
\eeq
The second one is the remaining elastic TPE contribution,
which is a recoil-type of correction to the Zemach term:
\bea
\Delta_\mathrm{recoil}&=&\frac{Z \al}{\pi (1+\kappa)}\int_0^\infty \frac{\dd Q}{Q}\left\{\frac{G_M(Q^2)}{Q^2}\frac{8m M}{v_l+ v }   \left(2F_1(Q^2)+\frac{F_1(Q^2)+3F_2(Q^2)}{(v_l+1)(v+1)}  \right)\right.\nn\\
&&\left.-\,\frac{8m_r\,G_M(Q^2)G_E(Q^2)}{Q}-\frac{m\,F_2^2(Q^2)}{M}\frac{5+4v_l}{(1+v_l)^2}\right\} \\
&\approx & \frac{Z \al m_r}{\pi (1+\kappa)mM}\int_0^\infty \!\dd Q 
\left[ \Big(3+\frac{2m}{M}\Big) F_1(Q^2) 
+ F_2(Q^2) \right] F_2(Q^2), \quad\mbox{with $v=\sqrt{1+\tau^{-1}}$} \nn
\eqlab{recoilHFS}.
\eea
Finally, the polarizability contribution is written as:
\begin{subequations}
\eqlab{POL}
\beq
\Delta_\mathrm{pol}=\frac{Z\al m}{2\pi (1+\kappa) M}\left[\delta_1+\delta_2\right]=\Delta_1+\Delta_2,
\eeq
with the separate contributions due to $g_1$ and $g_2$ 
given by:
\bea
\delta_1&=&2\int_0^\infty\frac{\dd Q}{Q}\left(\frac{5+4v_l}{(v_l+1)^2}\left[4I_1(Q^2)+F_2^2(Q^2)\right]+\frac{8M^2}{Q^2}\int_0^{x_0}\dd x\, g_1(x,Q^2)\right.\nn\\
&\times &   \left.\left\{ \frac{4}{v_l+ \sqrt{1+x^2\tau^{-1}}}\left[1+\frac{1}{2(v_l+1)(1+ \sqrt{1+x^2\tau^{-1}})}\right]-\frac{5+4v_l}{(v_l+1)^2} \right\}\right),\eqlab{Delta1}\\
\delta_2&=&96M^2\int_0^\infty\frac{\dd Q}{Q^3}\int_0^{x_0}\dd x\,g_2(x,Q^2) \left\{\frac{1}{v_l+ \sqrt{1+x^2\tau^{-1}}}-\frac{1}{v_l+1} \right\}.\eqlab{Delta2}
\eea
\end{subequations}

As emphasized before, our decomposition into $\Delta_\mathrm{recoil}$ and $\Delta_\mathrm{pol}$ corresponds with the decomposition into the Born and non-Born part. In this way, the decomposition is consistent with \citet{Bodwin:1987mj, Pachucki:1996zza,Carlson:2011af} and different from \citet{Faustov:2001pn,Martynenko:2004bt}. In the latter works,
the $F_2^2$ term was shared differently between
the elastic and polarizability contributions.
The conversion between the two decompositions can be found in Ref.~\cite{Carlson:2011af}.

\begin{table}
\footnotesize
\centering
\caption{Summary of available dispersive calculations for the TPE correction to the $2S$ HFS of $\mu$H. 
}
\label{Table:Summary3}
\begin{minipage}{\linewidth}  
\renewcommand{\thefootnote}{\alph{footnote}}
\footnotesize
\centering
\begin{tabular}{|p{0.19\linewidth}| p{.07\linewidth}p{.05\linewidth}|p{.05\linewidth}| p{.06\linewidth}p{.06\linewidth}p{.06\linewidth}|p{.09\linewidth}p{.14\linewidth}|}
\hline                
 Reference&  $R_Z$ [$\mathrm{fm}$]&$\Delta_Z$ [$\mathrm{ppm}$]&$\Delta_\mathrm{recoil}$ [$\mathrm{ppm}$]&$\Delta_\mathrm{pol}$ [$\mathrm{ppm}$]&$\Delta_1$ [$\mathrm{ppm}$]&$\Delta_2$ [$\mathrm{ppm}$]&$\Delta_\mathrm{structure}$ [$\mathrm{ppm}$]&$E_{2S\,\mathrm{HFS}}$ [$\mathrm{meV}$]\\
 \hline
Carlson et al.\ \cite{Carlson:2008ke}\footnotemark[1]& $1.080$&$-7703$&$931$&$351(114)$&$370(112)$&$-19(19)$&$-6421(140)$&$22.8146(49)$ \cite{Carlson:2011af}\\
Faustov  et al.~\cite{Faustov:2006ve}&&&&$470(104)$&$518$&$-48$&&\\
Martynenko et al.~\cite{Faustov:2001pn}&$1.022$&$-7180$&&$460(80)$&$514$&$-58$&&$22.8138(78)$\\
Experiment \cite{Antognini:1900ns}&$1.082(37)$\footnotemark[2]&&&&&&&$22.8089(51)$\\
\hline
\end{tabular}
\footnotetext[1]{QED and
structure-independent corrections are taken from~\citet{Martynenko:2004bt}. The Zemach term includes radiative corrections: $\Delta_Z=-2\al m_r R_\mathrm{Z} (1+\delta^\mathrm{rad}_Z)$, with $\delta^\mathrm{rad}_Z$
of Refs.~\cite{Bodwin:1987mj,Karshenboim:1996ew}.}
\footnotetext[2]{Extraction based on the recoil and polarizability 
corrections from Ref.~\cite{Carlson:2008ke} (1st row of the Table).}
\end{minipage}
\end{table}

Let us now consider the numerical results. Early works mainly studied the proton structure corrections to the ground-state HFS in H \cite{Iddings:1965zz, Drell:1966kk, DeRafael:1971mc,Gnaedig:1973qt}. More recent evaluations of the polarizability contribution to the H HFS can be found in Refs.~\cite{Faustov:2000xu, Faustov:2002yp,Nazaryan:2005zc,Carlson:2006yj,Carlson:2006mw}, radiative corrections are calculated in Ref.~\cite{Karshenboim:1996ew}. The most recent calculations of the polarizability contribution
to the HFS in H are:
\begin{subequations}
\bea
\text{\citet{Carlson:2008ke}}: \quad \Delta^{\mathrm{H}}_\mathrm{pol}&=&1.88 \pm 0.64\, \mathrm{ppm} ,\\
\text{\citet{Faustov:2006ve}}: \quad \Delta^{\mathrm{H}}_\mathrm{pol}&=&2.2\pm 0.8\, \mathrm{ppm}.
\eea
\end{subequations}

The available dispersive calculations for the $2S$ TPE correction to the HFS in $\mu$H are listed in Table \ref{Table:Summary3} and, in a more illustrative form, \figref{HFS}. Some of the results are given in terms of  the $2S$ Fermi energy in $\mu$H:
$E_F(2S)=22.8054$ meV.

Most of the calculations show
a relatively small effect from $g_2$, see 
$\Delta_2$ in Table \ref{Table:Summary3}. It seems to
be well within the uncertainty of the $g_1$ contribution $(\Delta_1)$.
However, it is important to note here that the
spin structure function $g_2$ of the proton
has not been measured experimentally in the low-$Q$ region, relevant to the atomic
calculations. 
The above evaluations are either modeling $g_2$, or make use of the Wandzura-Wilczek relation \cite{Wandzura:1977qf} to express it in terms of $g_1$:\footnote{This relation automatically satisfies the BC sum rule, i.e., $\int_0^1 \dd x\, g_2^\mathrm{WW}(x,Q^2)=0$,  as 
easily seen via the Fubini rule.}
\beq
\eqlab{WWrelation}
g_2^\mathrm{WW}(x,Q^2)=-g_1(x,Q^2)+\int_x^1 \frac{\dd x'}{x'} \,g_1(x',Q^2).
\eeq
The latter relation is for asymptotically large $Q^2$. It is
certainly violated for low $Q^2$--- it's only a question of how badly.
The ongoing JLab measurement of proton $g_2$ \cite{E08-027} is extremely
important for answering that question. 

Information on the structure function $g_1$ is available for momentum-transfers larger than $Q_\mathrm{min.}^2\sim 0.05\, \mathrm{GeV}^{2}$ 
\cite{Dharmawardane:2006zd}. Below this threshold, the $Q^2$-integrand of \Eqref{Delta1} is interpolated by exploiting the sum rules. In the case of H, where the electron mass can safely be neglected, the slope of the integrand is fixed by the GDH sum rule \eref{GDH}. In $\mu$H the dependence on the muon mass is not negligible, and the GTT sum rule \eref{FSP}
proves to be useful, cf.~\citet{Carlson:2008ke}. 

\subsection{Chiral EFT Evaluations}
\seclab{sec66}
Below $O(p^4)$, $\chi$PT provides a genuine prediction
for the TPE effects. At $O(p^4)$ there is a
number of low-energy constants (LECs), entering through
the effective lepton-lepton-nucleon-nucleon ($\ell\ell NN$) coupling, whose values are presently unknown. 
Therefore, the predictive power is lost at this order.  
Here
we only consider the "predictive orders", i.e., 
$O(p^3)$ and $O(p^{7/2})$. These will be called the
leading (LO) and next-to-leading (NLO) order, respectively.
\footnote{Technically, the leading order is $O(p^2)$, but it is included
in the Coulomb interaction.}

\begin{figure}[bth] 
 \centering 
  \includegraphics[scale=0.8]{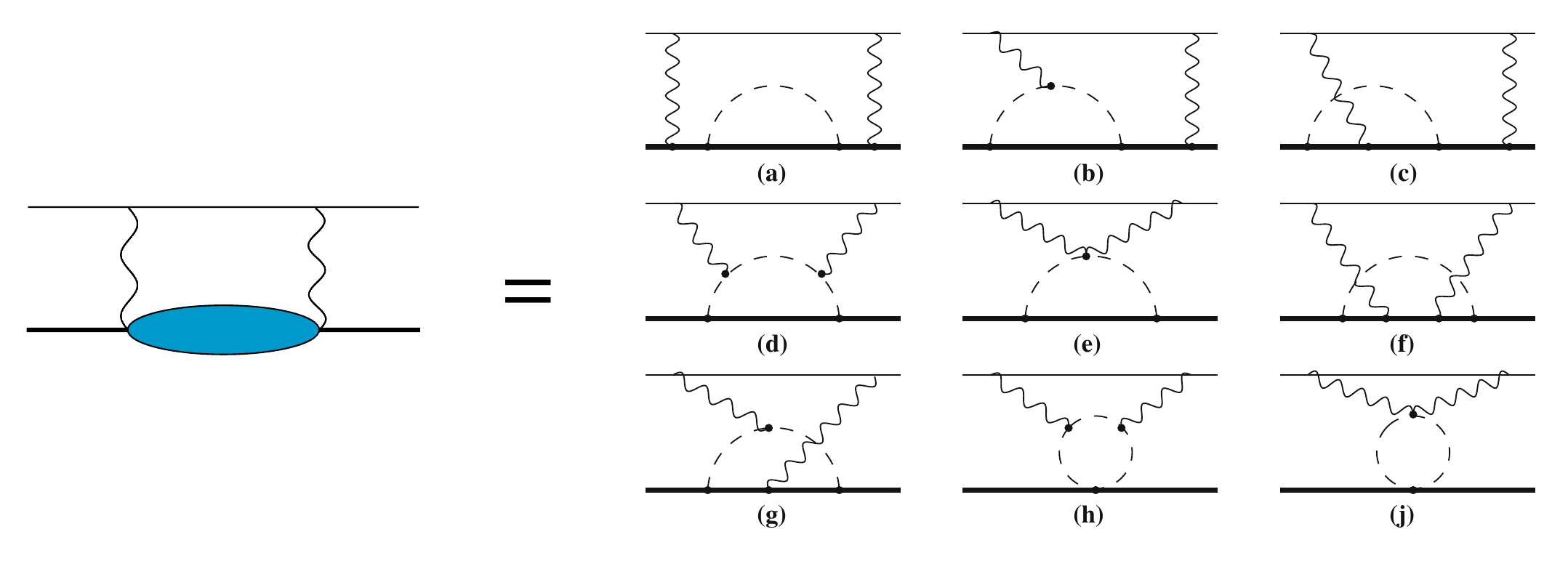}
                     \caption{The TPE diagrams of elastic lepton-nucleon scattering to $O(p^3)$ in $\chi$PT. Diagrams obtained from these by crossing and time-reversal symmetry are not drawn. \label{fig:Pascalutsa}}
\end{figure}
\subsubsection{Lamb Shift}
The leading-order [$O(p^3)$] calculations of the
$\mu$H Lamb shift have been done in both 
heavy-baryon (HB$\chi$PT) and baryon (B$\chi$PT)
frameworks \cite{Nevado:2007dd,Alarcon:2013cba}.
The diagrams arising in these calculations are
shown in \Figref{Pascalutsa}. The LO HB$\chi$PT
result for the polarizability contribution 
to the $2S$-level shift is well described by 
the following simple formula \cite{Alarcon:2013cba}:
\beq 
\De E^{\mathrm{pol}}_{\mathrm{HB\chi PT}} (2S) = \frac{\al^5 m_r^3  g_A^2}{4(4\pi f_\pi)^2} \, \frac{m_\mu}{m_\pi}
\Big( 1- 10G + 6 \ln 2\Big) \simeq -16.1 \,\, \upmu\mbox{eV},
\eeq 
where $G\simeq 0.9160$ is the Catalan constant; other
parameters are defined in \Secref{definitions}.
The LO B$\chi$PT result is somewhat smaller in 
magnitude, see\ Table \ref{Table:Summary2}. 
This is mainly because of the smaller value
of the proton electric polarizability $\al_{E1}$
arising in B$\chi$PT at leading order, cf.\ \Secref{theory2}. The $\pi\De$ loops at
$O(p^{7/2})$ are expected to correct this situation.

\begin{figure}
\centering 
\includegraphics[scale=0.5]{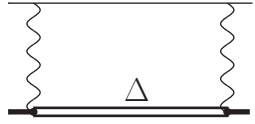}
 \caption{TPE diagram with $\Delta(1232)$, represented
 by the double line.}
 \figlab{TPEdelta}
 \end{figure}

At next-to-leading order [$O(p^{7/2})$], 
the diagrams with the $\De(1232)$-isobar arise, 
of which the one in \Figref{TPEdelta}
dominates the magnetic polarizability of the proton. 
In the Lamb shift, however, the magnetic polarizability
is suppressed, and this particular diagram is suppressed too~\cite{Alarcon:2013cba}. 

The calculations including the $\Delta$ have thus far
been done in HB$\chi$PT only \cite{Peset:2014jxa}.
The resulting NLO polarizability contribution is larger
in magnitude than the LO, see Table \ref{Table:Summary2} and \Figref{LSSummary}.
This is simply because in HB$\chi$PT the $\De$ gives too large of
a correction to the polarizabilities, cf.~\Eqref{abHBChPT}.
On the other hand, the HB$\chi$PT underpredicts the elastic TPE contribution,
see Table \ref{Table:Summary2} and \Figref{LSSummaryelastic},
because the Friar radius comes out to be smaller than the empirical value.  
The total value for the TPE effect in HB$\chi$PT happens to be in agreement with the empirical expectations. 

\begin{table}[tbh]
\caption{Summary of available $\chi$PT calculations for the TPE effect in the $2S$-level shift of $\mu$H  (in $\upmu\mathrm{eV}$).}
\label{Table:Summary2}
\centering
\footnotesize
\begin{minipage}{\linewidth}  
\centering
\begin{tabular}{|p{0.1\linewidth}| p{.17\linewidth}p{.15\linewidth}p{.15\linewidth}|}            
\hline      
& Nevado \& Pineda    & Alarc\'on et al.  &Peset \& Pineda\\
& LO HB$\chi$PT \cite{Nevado:2007dd}& LO B$\chi$PT \cite{Alarcon:2013cba}& NLO HB$\chi$PT \cite{Peset:2014yha}\\
\hline
$\Delta E^\mathrm{pol}(2S)$                                        &    $-18.5(9.3)$                                                      & $-8.2 (^{+1.2}_{-2.5})$ &$-26.2(10.0)$\\
$\Delta E^\mathrm{el}(2S)$ &$-10.1(5.1)$&&$-8.3(4.3)$\\
\hline
\end{tabular}
\end{minipage}
\end{table}

\subsubsection{Hyperfine Splitting}

\begin{figure}[tb]
\centering
\includegraphics[scale=0.5]{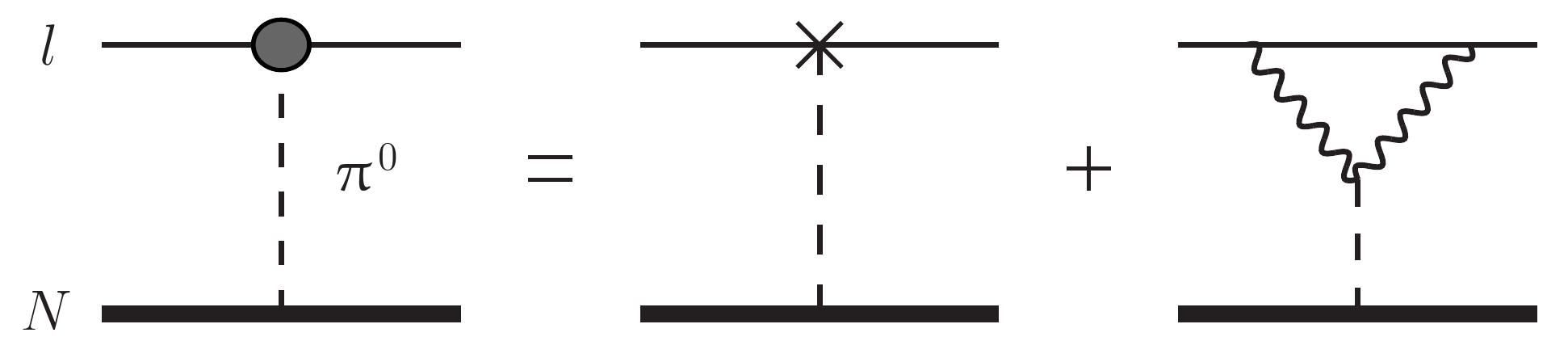}
\caption{Pion-exchange in hydrogen. \label{fig:PionExchange}}
\end{figure}

The LO B$\chi$PT calculation of the HFS should 
in addition to the diagrams in \Figref{Pascalutsa}
include the neutral-pion exchange, \Figref{PionExchange}.
The latter effect, however, turned out to be consistent
with 0, at least for the $\mu$H 2S HFS~\cite{Hagelstein:2015b}:
\beq
E_{\mathrm{HFS}}^{\langle \pi^0\rangle}(2S)=0.02\pm 0.04\, \upmu\mathrm{eV},
\eqlab{PE}
\eeq
where the uncertainty comes from the experimental
error of the $\pi^0 \to e^+ e^-$ decay width.
In retrospect this is not so surprising, since the pion-exchange vanishes in the forward kinematics, and as such
becomes suppressed by an additional $\al$.

A substantially larger pion-exchange effect
has recently been found in Refs.~\cite{Zhou:2015bea,Huong:2015naj}. The calculation of \citet{Zhou:2015bea} suffers from a technical mistake, as pointed
out in \cite{Huong:2015naj}. On the other hand, \citet{Huong:2015naj}  neglect the $Q^2$-dependence
of the pion coupling to leptons which is not
a good
approximation for the reason explained below.\footnote{Note added in proof: The $Q^2$-dependence is taken into account in the revised
version of Ref.~\cite{Huong:2015naj}. Their revised value is in
agreement with \Eqref{PE}.}

The non-relativistic limit of the pion-exchange potential
reads (for the $S$-waves):
\beq 
V_{\pi^0}^{l=0}(Q) = -\frac{F(F+1)-\nicefrac32}{3m M}
\frac{Q^2 }{Q^2+m_{\pi}^2} \,g_{\pi NN} F_{\pi\ell\ell}(Q^2),
\eeq 
where $F=0$ or 1 is the eigenvalue of the total angular momentum, $m_\pi$ is the neutral-pion mass, $g_{\pi NN}$ is the pion-nucleon coupling constant, and
$F_{\pi\ell\ell}$ is the FF describing the pion coupling
to leptons. The latter satisfies the well-known 
(once-subtracted) DR \cite{Drell59}:
\beq
 F_{\pi\ell\ell}(Q^2)   = F_{\pi\ell\ell}(0) - \frac{Q^2}{\pi} \int_0^\infty \!\frac{\dd s}{s}\,  \frac{\im F_{\pi\ell\ell}(s)}{s+Q^2}  , \quad \mbox{with}\,\,
  \im F_{\pi\ell\ell}(s) = -\frac{\al^2 m \arccosh(\sqrt{s}/2m)}{2\pi f_\pi \sqrt{1-4m^2/s}},
  \eqlab{pillFF}
\eeq  
where $f_\pi$ is the pion-decay constant, $m$ is the lepton mass. This decomposition into the subtraction constant and
the effect of the $2\gamma$ loop is illustrated in 
\Figref{PionExchange}. The subtraction constant can be
extracted from the experimental value of 
the $\pi^0\to e^+ e^-$ decay width, which in terms of the
FF is given by:
\beq
\Gamma(\pi^0\to e^+ e^-) = \frac{m_\pi}{8\pi} \sqrt{1-\frac{4m_e^2}{m_\pi^2}} \, \big| F_{\pi e e} (-m_\pi^2) \big|^2.
\eeq

Now, the point is that the FF in \Eqref{pillFF} does not
admit a good Taylor expansion around $Q^2=0$, because of the branch cut starting at 0. Hence,
in contrast to the $\pi NN$ FF, we cannot neglect its $Q^2$-dependence. A straightforward calculation yields
the following result for the HFS effect:
\beq
E_{\mathrm{HFS}}^{\langle\pi^0 \rangle}(nS)= - E_F(nS)
\frac{g_{\pi NN} m_r }{2\pi(1+\kappa) m_{\pi}}\left[F(0)+\frac{\al^2 m}{2\pi^2 f_\pi}\, 
I\big(\mbox{$\frac{m_{\pi}}{2m}$}\big)\right],
\eqlab{pionHFS}
 \eeq
 where we introduce the following integral,
 \beq
 I(\gamma)\equiv 2 \int_0^\infty \frac{\dd \xi}{1+(\xi/\gamma)}\frac{\arccos \xi}{\sqrt{1-\xi^2}}.
 \eeq
For H, $\ga\gg 1$, and one can make use of
the expansion: 
$ I(\gamma ) = 7\pi^2/12 +\ln^2(2\gamma)  + O(1/\gamma)$.
For the more general situation, $\gamma =\sin\th \ge 0$, we have:
\bea
I(\sin\th ) = \tan\th \left[ \mbox{Cl}_2(2\th) - \pi \ln \tan(\th/2) \right], 
\eea
where 
$
\mbox{Cl}_2(\th) =- \int_0^\th  \dd t \, \ln\big(2\sin\nicefrac{t}{2} \big)
=\frac{ i}{2} \left[ \mbox{Li}_2\big(e^{-i\th}\big)-\mbox{Li}_2\big(e^{i\th}\big)\right] $
is the the Clausen integral;
Li$_2(x)$ is the Euler dilogarithm.
The numerical values for the electron and muon, respectively, are:
$ I( m_\pi/2m_e ) \simeq 36.8316$, $ I ( m_\pi/2m_\mu ) \simeq 3.4634 $.

We find that
in H and $\mu$H alike, there is a large cancellation
between the two terms in \Eqref{pionHFS}, or 
equivalently between the two diagrams in \Figref{PionExchange}.
The resulting $\mu$H value is the one
quoted above, in 
\Eqref{PE}.

A preliminary calculation~\cite{Hagelstein:2015b}
shows that in total the LO B$\chi$PT effects
amount to the following polarizability contribution
to the $2S$ HFS of $\mu$H:
\beq
O(p^3):\quad E_{\mathrm{HFS}}^{\mathrm{pol}}(2S)=0.87\pm 0.42 \, \upmu\mathrm{eV}.
\eqlab{LOHFS}
\eeq
This is about an order of magnitude smaller than the
effect obtained in the empirical dispersive calculations,
cf.~Table~\ref{Table:Summary3}. However, it can be expected that 
the $\De$-excitation mechanism of \Figref{TPEdelta}
can play an important role here. It remains to be seen
whether this NLO effect restores the agreement
between the $\chi$PT and dispersive results.

\section{Summary Plots and Conclusions}
\seclab{conclusion}

To summarize and conclude we have compiled the following 
summary plots surveying the 
recent results for nucleon polarizabilities and for their
contribution to the $2S$-levels of muonic hydrogen.

\subsection{Scalar Polarizabilities}
\begin{figure}[htb] 
  \centering 
\begin{minipage}[t]{0.49\textwidth}
    \centering 
       \includegraphics[width=\textwidth]{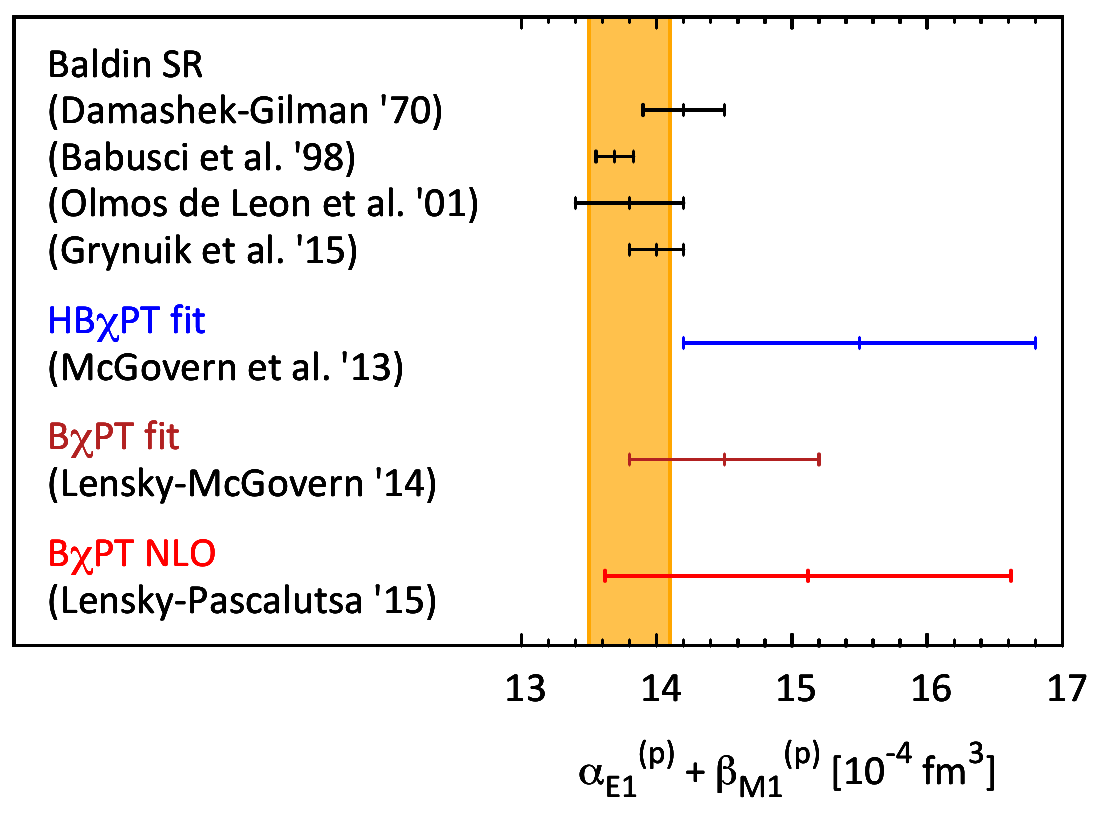}
\end{minipage}
\begin{minipage}[t]{0.49\textwidth}
    \centering 
      \raisebox{0.04cm}{ \includegraphics[width=\textwidth]{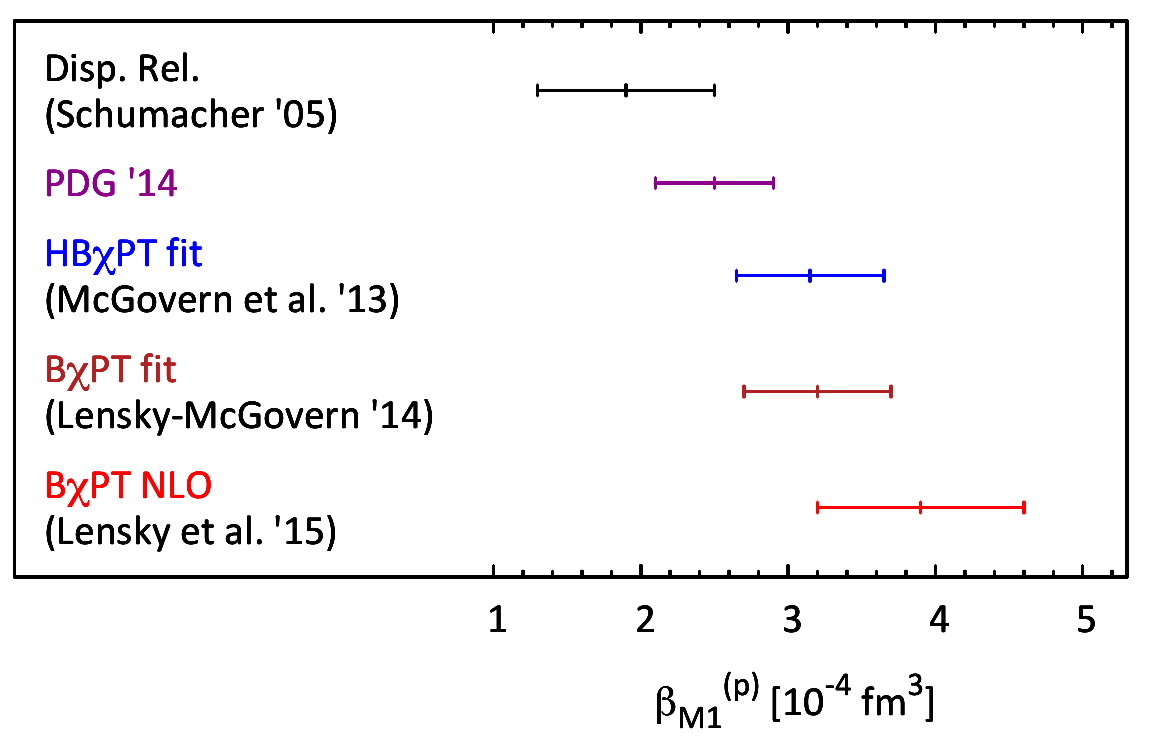}}
       \end{minipage}
 \caption{Left panel: sum of the electric and magnetic dipole polarizabilities of the proton. Right panel: the magnetic dipole polarizability of the proton. The orange band is the weighted average over the Baldin sum rule evaluations listed in Table~\ref{Tab:sumruletest_proton}. The DR prediction for $\beta_{M1}^{(p)}$ can be found in the review of  \citet{Schumacher:2005an}. ``Lensky-Pascalutsa '15'' refers to Ref.~\cite{Lensky:2014dda,Alarcon2015}, whereas ``Lensky et al.\ '15'' refers to Ref.~\cite{Lensky:2015awa}. All other references and declarations are given in the text. \label{fig:alphabeta_p}}
\end{figure}

\begin{figure}[htb] 
  \centering 
\begin{minipage}[t]{0.52\textwidth}
    \centering 
 \includegraphics[width=\textwidth]{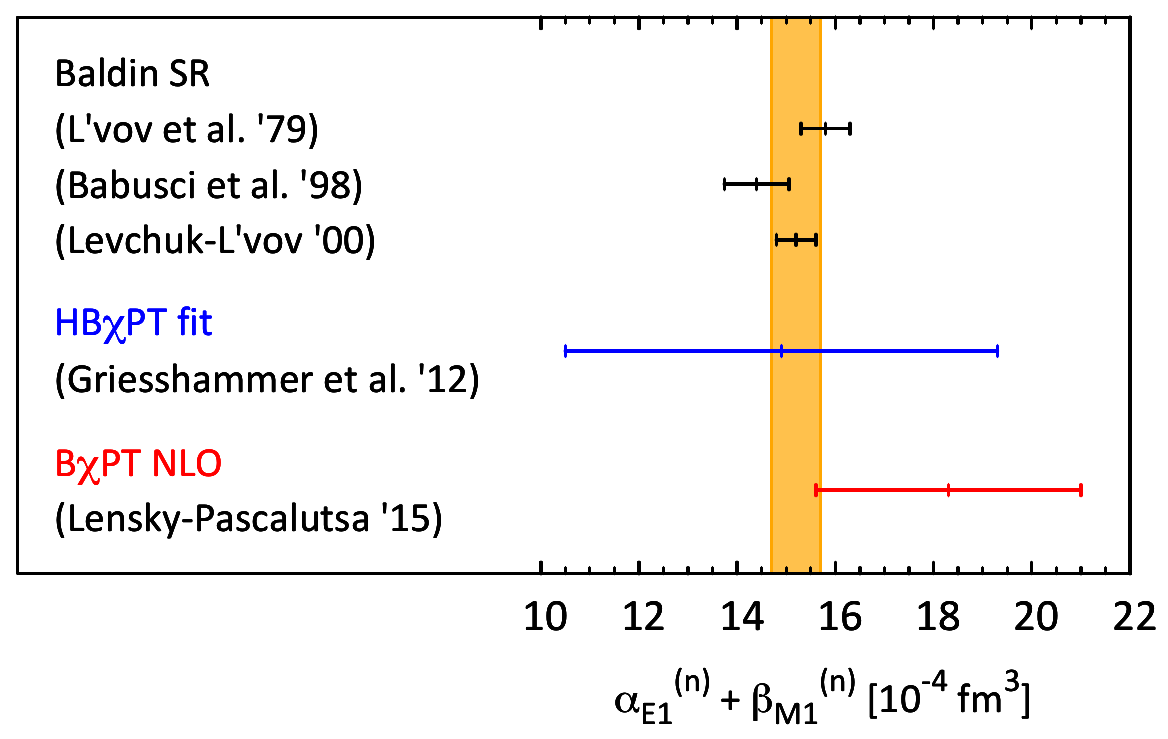}
\end{minipage}
\begin{minipage}[t]{0.46\textwidth}
    \centering 
     \raisebox{0.04cm}{  \includegraphics[width=\textwidth]{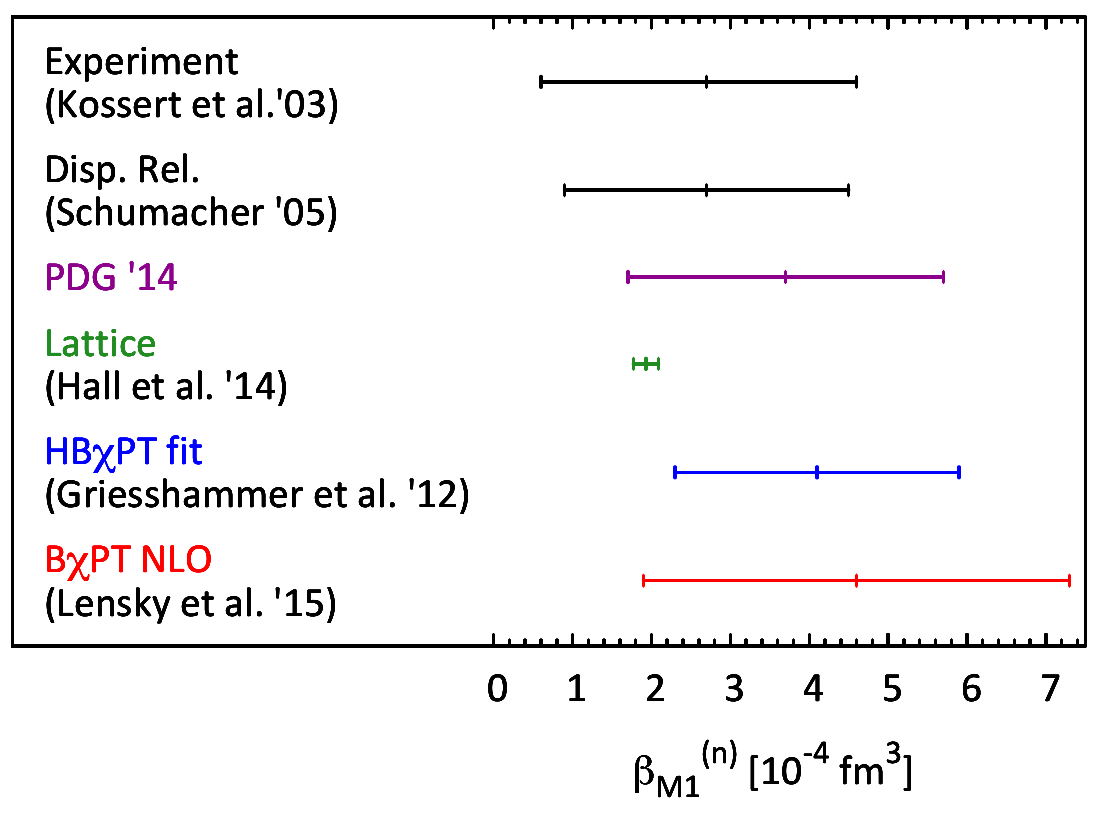}}
       \end{minipage}
 \caption{Left panel: sum of the electric and magnetic dipole polarizabilities of the neutron. Right panel: the magnetic dipole polarizability of the neutron. The orange band is the weighted average over the Baldin sum rule evaluations listed in Table~\ref{Tab:sumruletest_proton}.
The experimental results for $\beta_{M1}^{(n)}$ are from Refs.~\cite{Kossert:2002jc,Kossert:2002ws} and \cite{Schumacher:2005an}. Other references are given in the text.   \label{fig:alphabeta_n}} 
\end{figure}

Figures \ref{fig:alphabeta_p} and \ref{fig:alphabeta_n} present the situation for $\alpha_{E1} + \beta_{M1}$ and $\beta_{M1}$ of the proton and neutron, respectively. 
In the top of the left panels we have the results
of the Baldin sum-rule evaluations considered in Table \ref{Tab:sumruletest_proton}. The orange band indicates the weighted-average of these evaluations. 

It appears that there is a substantial tension in the value of the proton magnetic polarizability,
cf.~the right panel of \Figref{alphabeta_p}.
An emerging objective in this area is to reduce the uncertainty on $\beta_{M1}^{(p)}$ by approximately $50\%$ through a measurement technique that is ideally independent of the Baldin sum rule.  The utilization of photon beams with high intensity and high linear polarization will be a key part of these investigations.  Exploratory measurements are currently underway at HIGS and Mainz. 

In the area of the neutron scalar polarizabilities, the recent Lund publication \cite{Myers14} of elastic CS on the deuteron is an important milestone.   For the first time, relatively high statistics and wide kinematic coverage elastic data are available, and the data are analyzable with state-of-the-art effective-field theory calculations. With the unfortunate discontinuation of the CS program at Lund, the focus will now likely shift to other labs and different nuclear targets.  At Mainz an experiment to measure elastic CS on $^4$He is in preparation. 

\begin{figure}[tb] 
  \centering 
\begin{minipage}[t]{0.49\textwidth}
    \centering 
       \raisebox{0.00cm}{\includegraphics[width=0.9\textwidth]{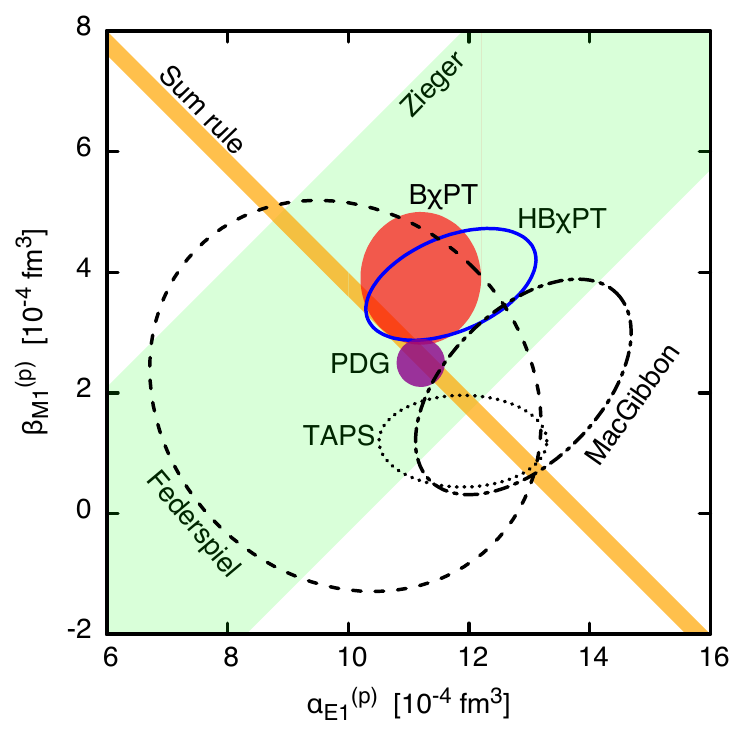}}
\end{minipage}
\begin{minipage}[t]{0.49\textwidth}
    \centering 
       \raisebox{0.1cm}{\includegraphics[width=0.9\textwidth]{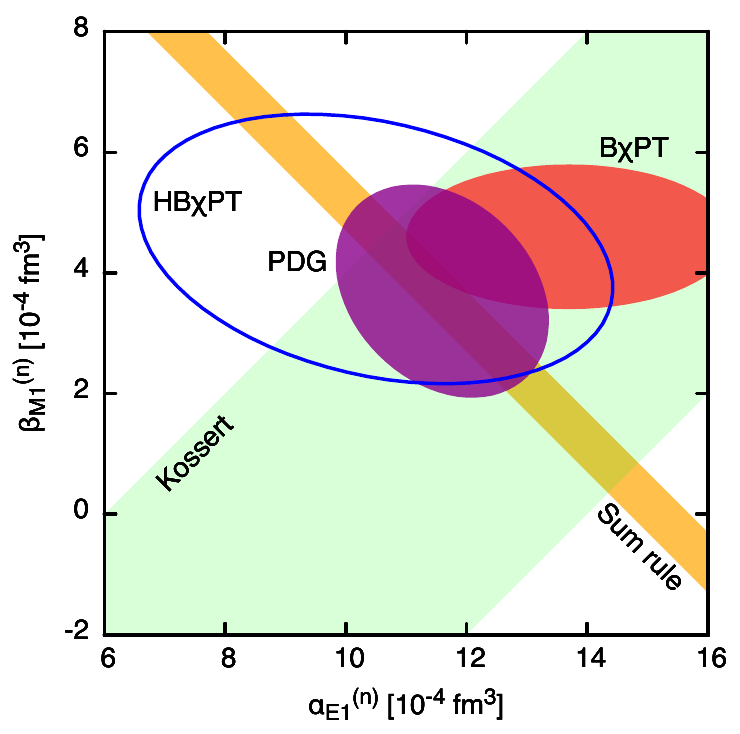}}
       \end{minipage}
 \caption{Plot of $\alpha_{E1}$ versus $\beta_{M1}$ for the proton (left panel) and neutron (right panel), respectively. The orange band is the average over the Baldin sum rule evaluations listed in Table~\ref{Tab:sumruletest_proton}. References and declarations are given in the text, cf.\ also Fig.~\ref{fig:alphabeta_p} and \ref{fig:alphabeta_n}. \label{fig:alphaVSbeta}}
\end{figure}

Another graphical representation of the experimental and theoretical results for the dipole polarizabilities, $\alpha_{E1}$ and $\beta_{M1}$, is shown in \Figref{alphaVSbeta}. The orange band again represents the constraint by the Baldin sum rule. The light green bands show experimental constraints on the difference of dipole polarizabilities, i.e., $\alpha_{E1}-\beta_{M1}$, cf.\ \citet{Kossert:2002jc,Kossert:2002ws} and \citet{Zieger:1992jq}. For the proton, other experimental constraints are shown by black lines: \citet{MacGibbon95, Fed91} and TAPS \cite{Olm01}. The B$\chi$PT constraint is from Ref.~\cite{Lensky:2015awa}. The HB$\chi$PT constraint is from Ref.~\cite{McG13}, in case of the proton, and \cite{Griesshammer:2012we}, in case of the neutron. Obviously the knowledge of the neutron polarizabilities is less precise than for the proton. This is mainly due to the lack of free neutron targets. 

Note that in these plots we have used the most recent PDG values \cite{Agashe:2014kda}:
\begin{subequations}
\bea
\al_{E1}^{(p)}=(11.2\pm 0.4) \times 10^{-4}\,\mathrm{fm}^3, \quad \beta_{M1}^{(p)}=(2.5\mp 0.4) \times 10^{-4}\,\mathrm{fm}^3,\\
\al_{E1}^{(n)}=(11.6\pm 1.5) \times 10^{-4}\,\mathrm{fm}^3, \quad \beta_{M1}^{(n)}=(3.7\mp 2.0) \times 10^{-4}\,\mathrm{fm}^3.
\eea
\end{subequations}
They differ for the proton from the 2012 and earlier editions by inclusion
of the global data fit analysis \cite{McG13}.\footnote{The 2015 PDG online edition has also changed the values for the neutron.}

Concerning VCS, what has emerged from the low-$Q^2$ studies, see \Figref{VCS_GP_alpha}, is interesting and provocative; there may well be a non-dipole-like structure in $\alpha_{E1}(Q^2)$ at $Q^2 \approx 0.33 \,\mathrm{GeV}^{2}$. If correct, this would indicate that the proton has a pronounced structure in its induced polarization at large
transverse distances,  $0.5$ to $1\,\mathrm{fm}$, cf.\ \Figref{induced_polarization}. New data are required to confirm this.  The Mainz A1 collaboration have taken VCS data at $Q^2 \approx  0.1$, $0.2$ and $0.5\, \mathrm{GeV}^{2}$, and this data is currently under analysis. Formulating a connection between VCS
and VVCS polarizabilities at finite $Q^2$ is a future task and could be of interest in this context.

\subsection{Spin Polarizabilities}
\begin{figure}[t] 
  \centering 
       \includegraphics[width=0.7\textwidth]{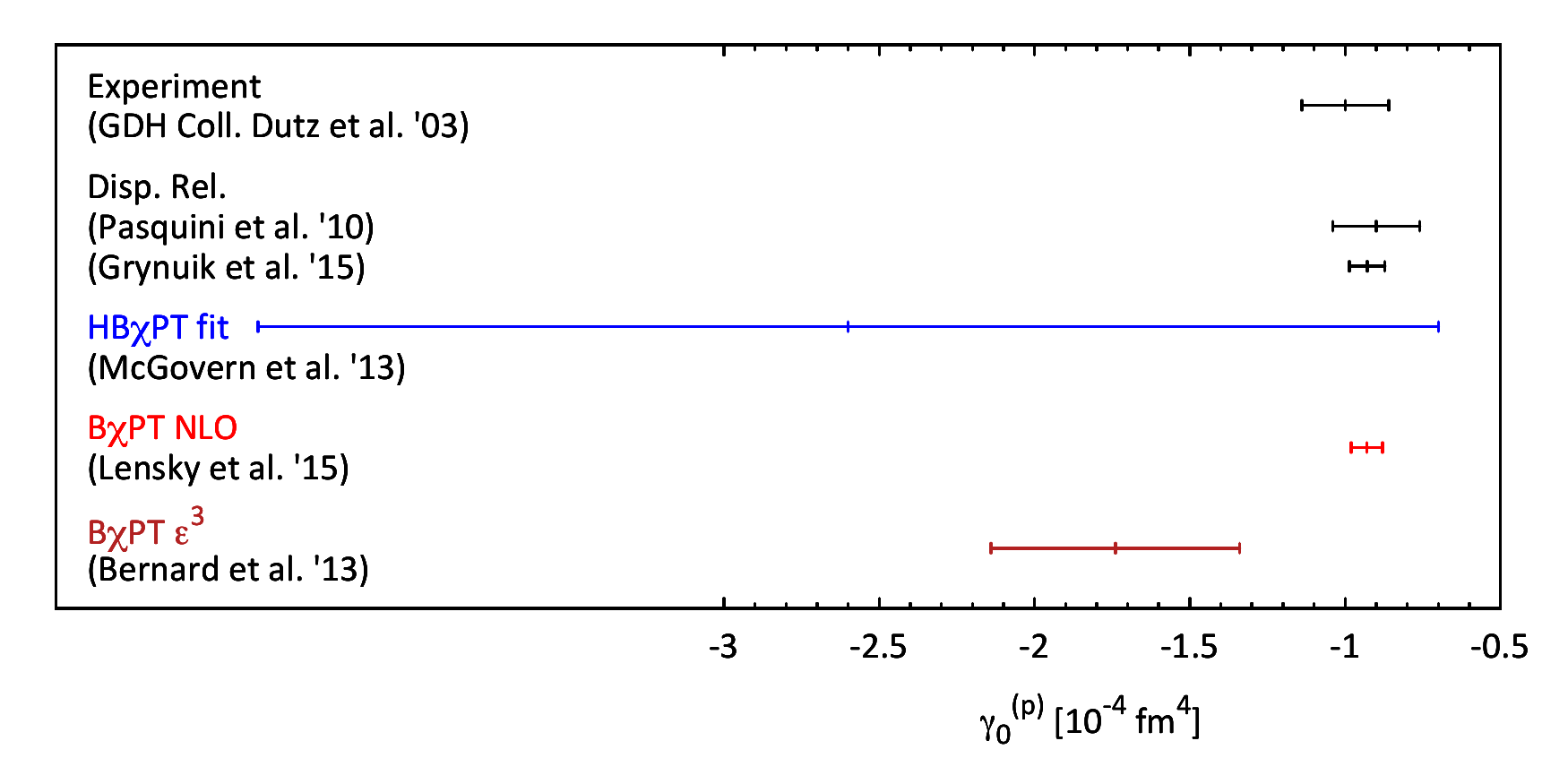}
       \includegraphics[width=0.7\textwidth]{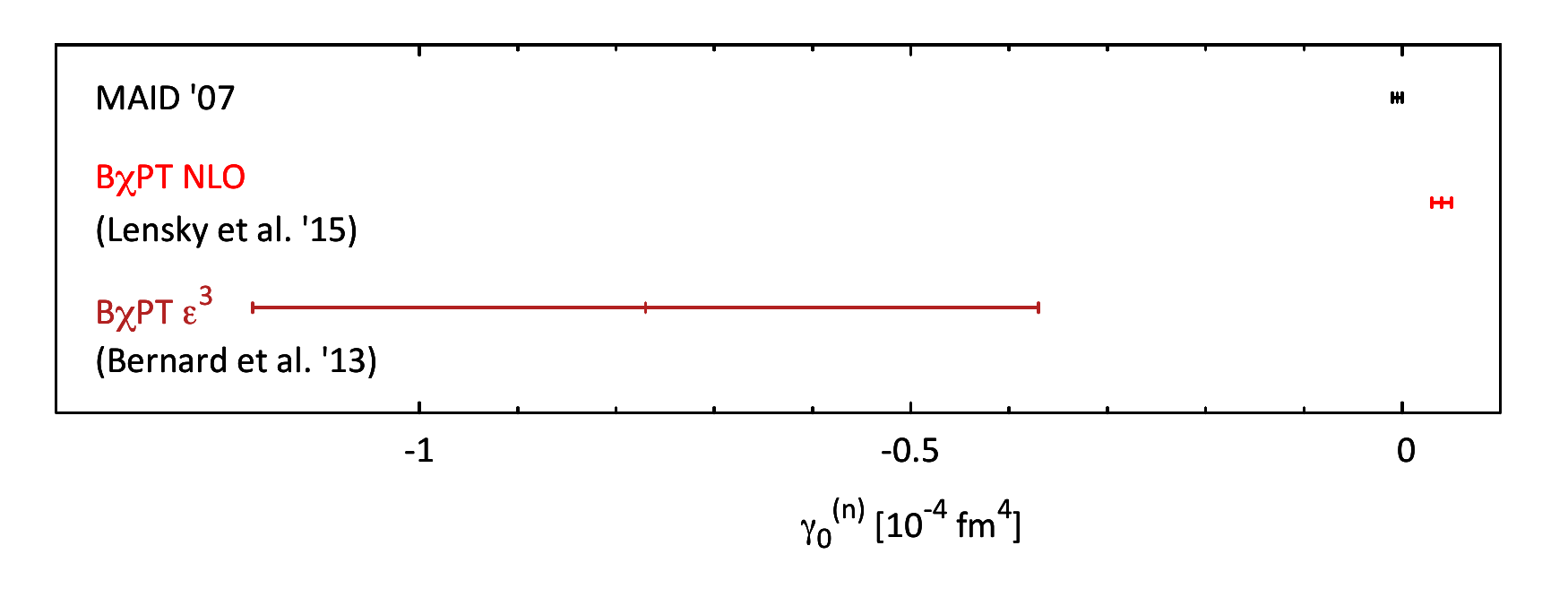}
 \caption{Forward spin polarizability, $\gamma_0$, of the proton (top panel) and neutron (bottom panel). Shown are the  experimental value from the GDH-Collaboration \cite{Dutz:2003mm}, the DR 
 result of Pasquini et al.~\cite{Pasquini:2010zr}, the HB$\chi$PT fit of McGovern et al.~\cite{McG13}, the B$\chi$PT predictions of Lensky et al.~\cite{Lensky:2014dda,Alarcon2015} and \citet{Bernard:2012hb}. All other references and declarations are given in the text or the Tables \ref{Tab:SPs_theory} and \ref{Tab:GDH}.\label{fig:gamma0}}
\end{figure}

\begin{wrapfigure}[18]{r}{0.52\textwidth}
    \centering 
       \includegraphics[width=0.48\textwidth]{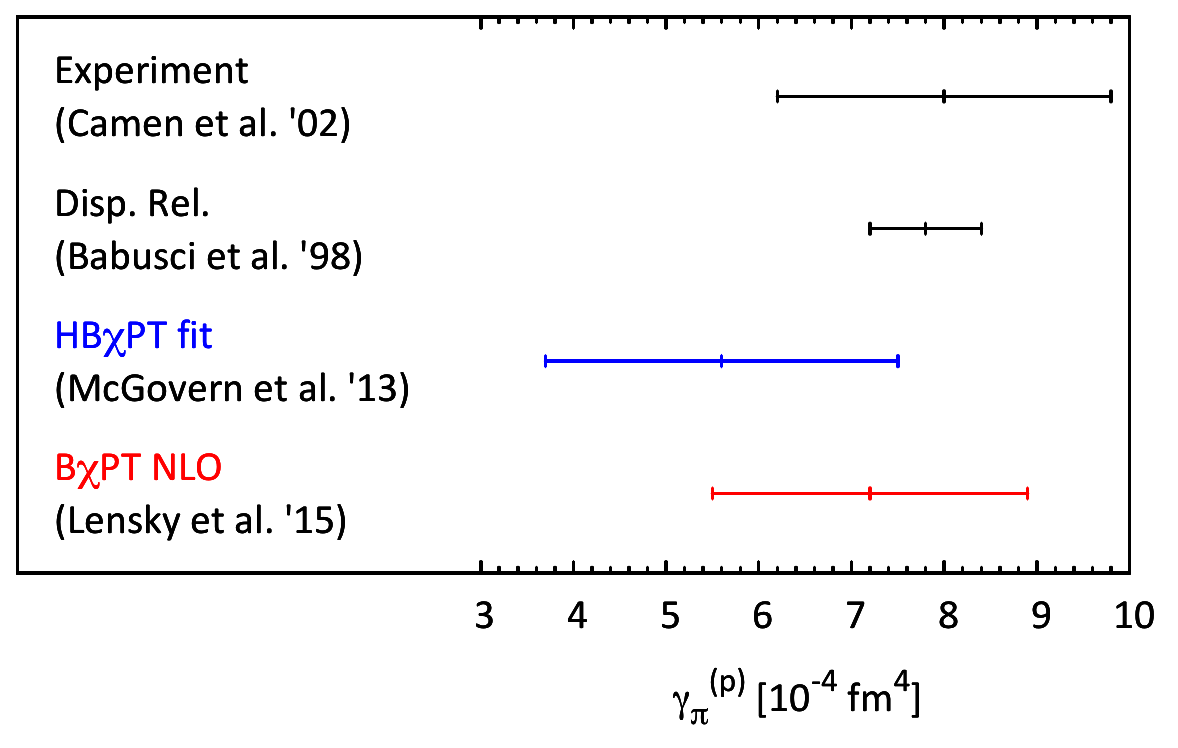}
                     \caption{Backward spin polarizability, $\gamma_\pi$, of the proton. We show  the experimental value from Camen \cite{Cam02}, cf.\ Eq.~(\ref{Eq:gamma_pi_value}), the fixed-$t$ DR result of \citet{Bab98}, the HB$\chi$PT fit of McGovern et al.~\cite{McG13} and the B$\chi$PT prediction of \citet{Lensky:2015awa}. \label{fig:gammaPi}}
\end{wrapfigure}

The theoretical and experimental results for
the proton spin polarizabilities have been
presented in Table~\ref{Tab:SPs_theory}.
Figure~\ref{fig:gamma0} summarizes the situation for the forward spin polarizability of the proton and neutron.
The sum rule evaluations therein are from Table \ref{Tab:GDH}.
Results for the backward spin polarizability of the proton are shown in \figref{gammaPi}.


A milestone in this area has been the recent Mainz publication of double polarized CS data for the $\Si_{2x}$ asymmetry, and the global analysis of CS asymmetry data, leading to the first measurement of all four spin polarizabilities, $\gamma_{E1E1}$, $\gamma_{M1M1}$, $\gamma_{E1M2}$, and $\gamma_{M1E2}$, cf.\ Table \ref{Tab:SPs_theory} \cite{Martel15}. At Mainz new data have been taken on the linear polarization asymmetry $\Si_3$, and the double polarization asymmetry with longitudinally polarized target $\Si_{2z}$.

An attainable goal in this area is to reduce the uncertainties in spin polarizabilities, currently at $\approx \pm 1\times10^{-4}\,\mathrm{fm}^4$, by approximately $50\%$. Given that CS count rates for a $2\,\mathrm{cm}$ long frozen spin butanol target are very low compared to a $10\,\mathrm{cm}$ liquid hydrogen target, long running times on polarized targets may not be the best approach to drive down errors. Another strategy is to combine a global analysis of the asymmetry data, $\Si_3$, $\Si_{2x}$ and $\Si_{2z}$, with measurements of backward angle CS cross sections in the $\Delta(1232)$ region. Because the $\Si_{2z}$ asymmetry for backward angle CS approaches $1$ (see \figref{sigma2z_sensitivity}), unpolarized CS preferentially selects one initial target polarization, and there is reasonable sensitivity to the spin polarizabilities in the unpolarized cross sections.  

 \begin{figure}[t] 
  \centering 
\begin{minipage}[t]{0.49\textwidth}
    \centering        \includegraphics[width=\textwidth]{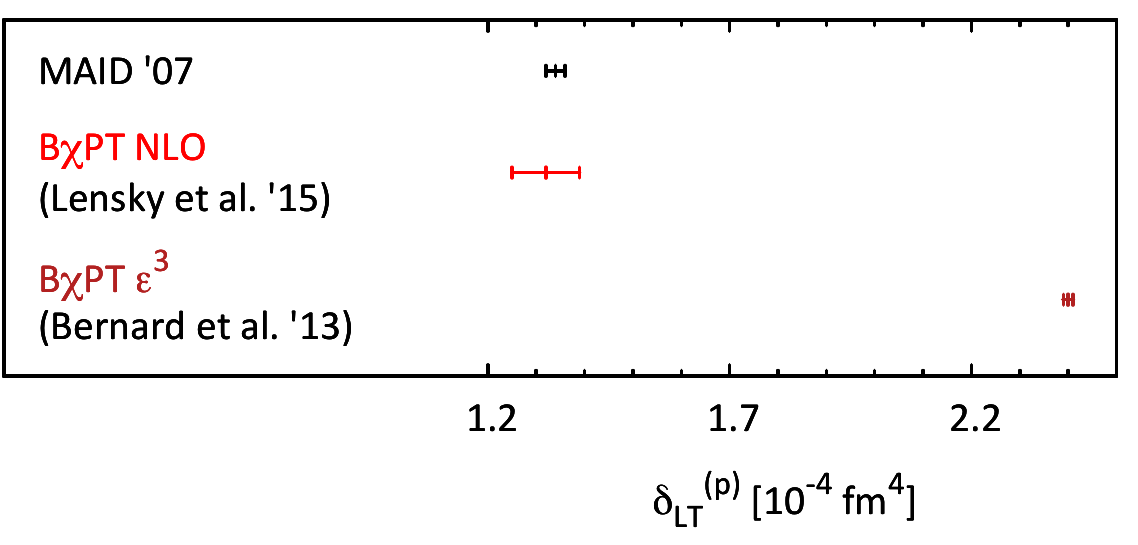}
\end{minipage}
\begin{minipage}[t]{0.49\textwidth}
    \centering 
       \raisebox{0.0cm}{ \includegraphics[width=\textwidth]{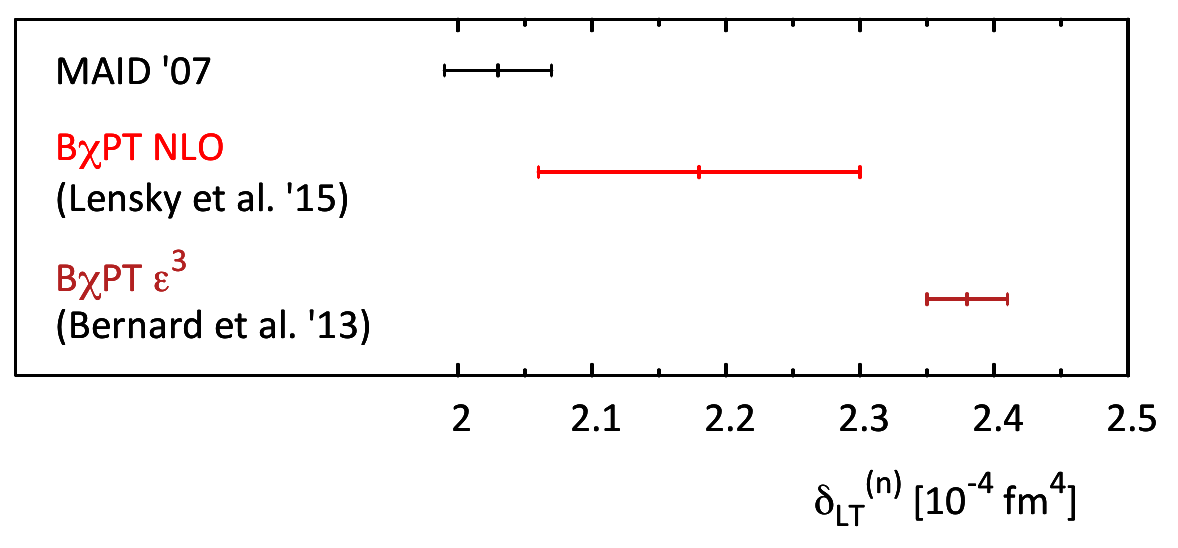}}
       \end{minipage}
 \caption{Longitudinal-transverse polarizability, $\delta_{LT}$, for the proton (left panel) and neutron (right panel), respectively. We show the B$\chi$PT predictions of Lensky et al.~\cite{Lensky:2014dda,Alarcon2015} and \citet{Bernard:2012hb}, and a result from MAID \cite{MAID}. \label{fig:deltaLT}}
\end{figure}

\begin{figure}[h!] 
    \centering        \includegraphics[width=0.4\textwidth]{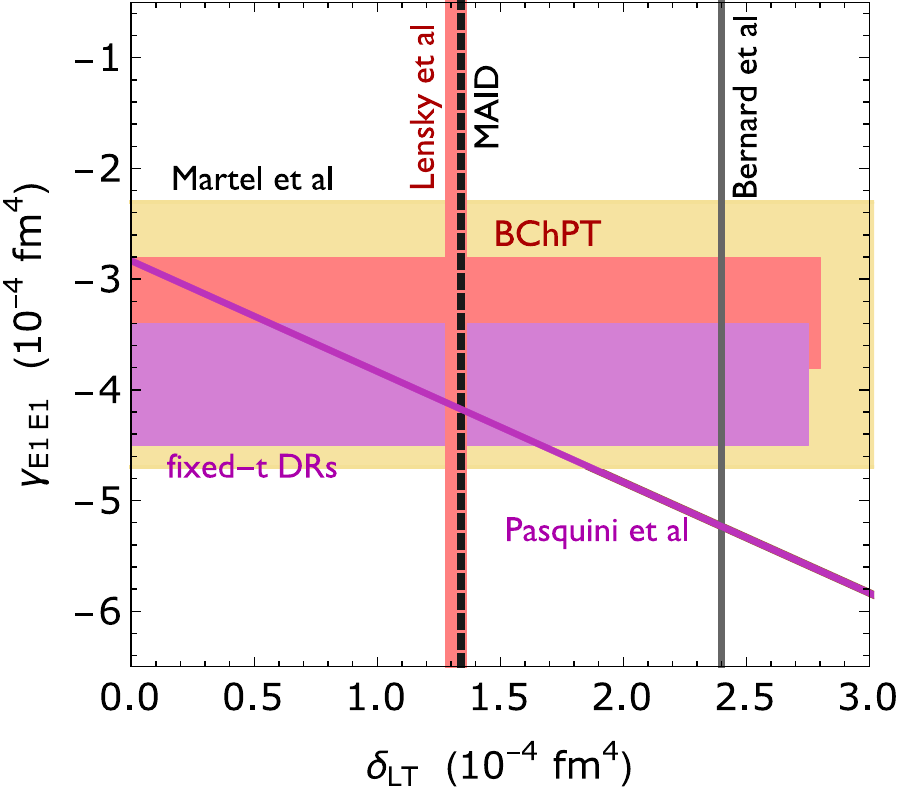}
                     \caption{Spin polarizabilities,
                     $\ga_{E1E1}$ versus $\delta_{LT}$,  for the proton.  Results for 
 $\ga_{E1E1}$ (horizontal bands) are from: the experiment of \citet{Martel15} (beige),
 the B$\chi$PT calculation of \citet{Lensky:2015awa} (red),
 and the fixed-$t$ DR calculation 
 of \cite{Holstein:1999uu,Bab98} (purple). Results for 
  $\delta_{LT}$ (vertical bands) are from:
  MAID \cite{MAID} (dashed line), 
  \citet{Lensky:2014dda} (red), and \citet{Bernard:2012hb} (gray).
  The line across is based on the relation of
  (\ref{S2sr3}) using the values of GPs from
  the DR calculation of \citet{Pasquini:2001yy}.
  \label{fig:acrossSR}}
\end{figure}

The longitudinal-transverse polarizability, $\delta_{LT}$, and the forward spin polarizability, $\ga_0$,
deserve a special attention. Their values at
the real-photon point are shown in 
\Figref{deltaLT} and \ref{fig:gamma0} for the proton and neutron, respectively. On the theory side, the baryon $\chi$PT yields genuine
predictions for the spin polarizabilities, cf.\ \citet{Bernard:2012hb}
and Lensky et al.~\cite{Lensky:2014dda,Alarcon2015}. On the empirical side, we have for instance the results from the latest
version of the MAID partial-wave analysis (MAID'07), which 
is based on the empirical knowledge
of the single-pion photoproduction cross section
$\si_{LT}$. Especially for $\delta_{LT}^{(p)}$, one
B$\chi$PT result is in significant contradiction with MAID,
while the other one is in agreement. The two B$\chi$PT calculations are done in
different counting schemes for the $\De$-isobar
contributions, cf.~\Secref{theory2}. As result,
the $\eps^3$ calculation includes in addition 
the graphs with 
several $\De$-propagators, and in particular
the one where the photons couple minimally to the
$\De$ inside the chiral loop. The latter graph is
allegedly making up all the difference
\cite{KrebsPrivateCom}. This would mean
the $\pi\De$ channel is extremely important
for this quantity and that is why the MAID
estimate would be inadequate. 
New JLab data \cite{E08-027} for $\delta_{LT}^{(p)}$ down to virtualities of $0.02\,\mathrm{GeV}^2$ are currently at a final stage of analysis and will shed a further light on this ``$\delta_{LT}$ puzzle''. Complementary, as a check one
could simply study the effect of this graph
for the sum of the scalar polarizabilities,
$\alpha_{E1}+\beta_{M1}$. There the empirical number
is known very well 
from the Baldin sum rule. 
At the moment, however, this discrepancy is an open problem
of B$\chi$PT and is
yet another reincarnation of the ``$\delta_{LT}$ puzzle''.

Another view of this problem is presented in
\Figref{acrossSR}. The vertical lines
clearly show the discrepancy in $\de_{LT}$.
The red bands are from the NNLO calculations of
\citet{Lensky:2014dda,Lensky:2015awa} using the $\de$-expansion.
They are consistent with the DR approach.
It would be interesting to see the $\eps$-expansion 
result of \citet{Bernard:2012hb} for $\gamma_{E1E1}$ too, 
because it seems it would contradict with the empirical results in either the GP's slope [cf.~Eq.~(\ref{S2sr3})] or $\gamma_{E1E1}$ itself.

\subsection{Status of the Proton-Radius Puzzle}
\seclab{puzzle}

Currently, the value of
the proton rms charge radius extracted
from H spectroscopy disagrees with the $\mu$H value by nearly five standard deviations:\\[-1cm]
\begin{wrapfigure}[15]{r}{3.3in}
\vspace{13mm}
\centering
\includegraphics[scale=0.82]{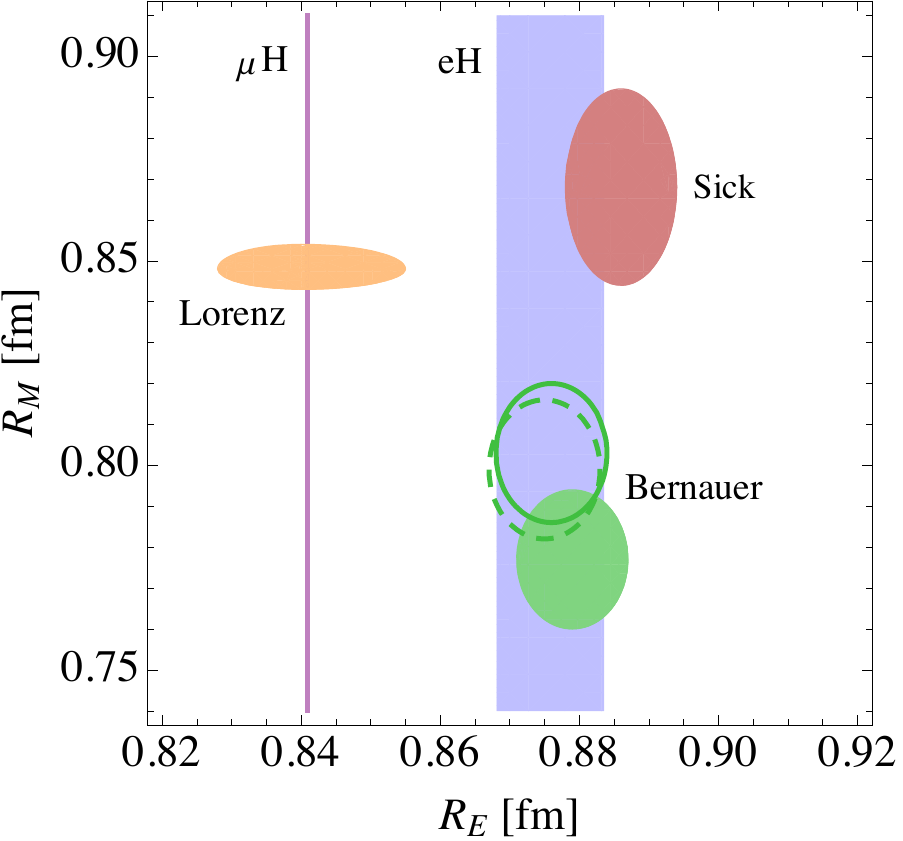}
\caption{Determination of the proton's electric and magnetic radii. The shown values are given in the text. The green lines display the Bernauer fit with TPE corrections: TPE,a (solid), TPE,b (dashed). The different uncertainties given in Ref.~\cite{Bernauer:2014} are added in quadrature.\label{fig:RERM}}
\end{wrapfigure} 
\begin{itemize}[$\square$]
\item H \cite{Mohr:2012aa}: $R_E=0.8758(77) \,\mathrm{fm}$;\footnote{Based
on H and D spectroscopy. From H alone (neglecting the isotope-shift measurements) $R_E=0.8796(56)\, \mathrm{fm}$.} 
\item $\mu$H \cite{Antognini:2012ofa,Antognini:1900ns}: $R_E=0.84087(39)\, \mathrm{fm}$.
\end{itemize}

On the other hand, the elastic electron-proton ($ep$)
scattering, which is the classic way of accessing the charge radius, yields conflicting results on the charge and magnetic rms radii: 
\begin{itemize}[$\square$]
\item Sick \cite{Sick:2012zz}: $R_E=0.886(8)\,\mathrm{fm}$, 
$R_M=0.868(24)\,\mathrm{fm}$;
\item Lorenz et  al.~\cite{Lorenz:2014yda}:
$R_E=0.840 \left[0.828\ldots 0.855\right]\,\mathrm{fm}$, $R_M=0.848 \left[0.843\ldots 0.854\right]\,\mathrm{fm}$;
\item Bernauer et  al.~\cite{Bernauer:2014} (world data):
\bea
R_E&=&0.879(5)_\mathrm{stat}(4)_\mathrm{syst}(2)_\mathrm{model}(4)_\mathrm{group} \,\mathrm{fm},\nn\\
R_E^\mathrm{TPE,a}&=&0.876(5)_\mathrm{stat}(4)_\mathrm{syst}(2)_\mathrm{model}(5)_\mathrm{group} \,\mathrm{fm},\nn\\
R_E^\mathrm{TPE,b}&=&0.875(5)_\mathrm{stat}(4)_\mathrm{syst}(2)_\mathrm{model}(5)_\mathrm{group} \,\mathrm{fm},\nn\\
R_M&=&0.777(13)_\mathrm{stat}(9)_\mathrm{syst}(5)_\mathrm{model}(2)_\mathrm{group} \,\mathrm{fm},\nn\\
R_M^\mathrm{TPE,a}&=&0.803(13)_\mathrm{stat}(9)_\mathrm{syst}(5)_\mathrm{model}(3)_\mathrm{group} \,\mathrm{fm},\nn\\
R_M^\mathrm{TPE,b}&=&0.799(13)_\mathrm{stat}(9)_\mathrm{syst}(5)_\mathrm{model}(3)_\mathrm{group} \,\mathrm{fm},\nn
\eea
where the superscript refers to the set of applied TPE corrections: TPE,a \cite{PhysRevC.75.038202}, TPE,b \cite{Arrington2011782,PhysRevC.72.034612}.
\end{itemize}

The current situation is illustrated in \Figref{RERM}.
The CODATA 2010 recommended value, which combines the H and some of the $ep$ scattering results, is \cite{Mohr:2012aa}:
\beq 
R_E(\mathrm{H}+ep) = 0.8775(51) \,\mathrm{fm} ,\label{CODATA}
\eeq
which is in $7 \si$ disagreement with the $\mu$H result. This value does
not include the interpretation of the $ep$ scattering data based on dispersive approaches \cite{Lorenz:2014yda,Dubnicka:2003,Adamuscin:2005aq}. 

Further details can be found in dedicated reviews~\cite{Carlson:2015jba,Karshenboim:2015}.
A nice overview of the current and future experimental activities called to resolve the puzzle has
recently been given  by \citet{Antognini:2015moa}.

\subsection{Proton Structure in Muonic Hydrogen}
\seclab{FSEsummary}
 \begin{figure}[htb] 
    \centering 
       \includegraphics[width=0.75\linewidth]{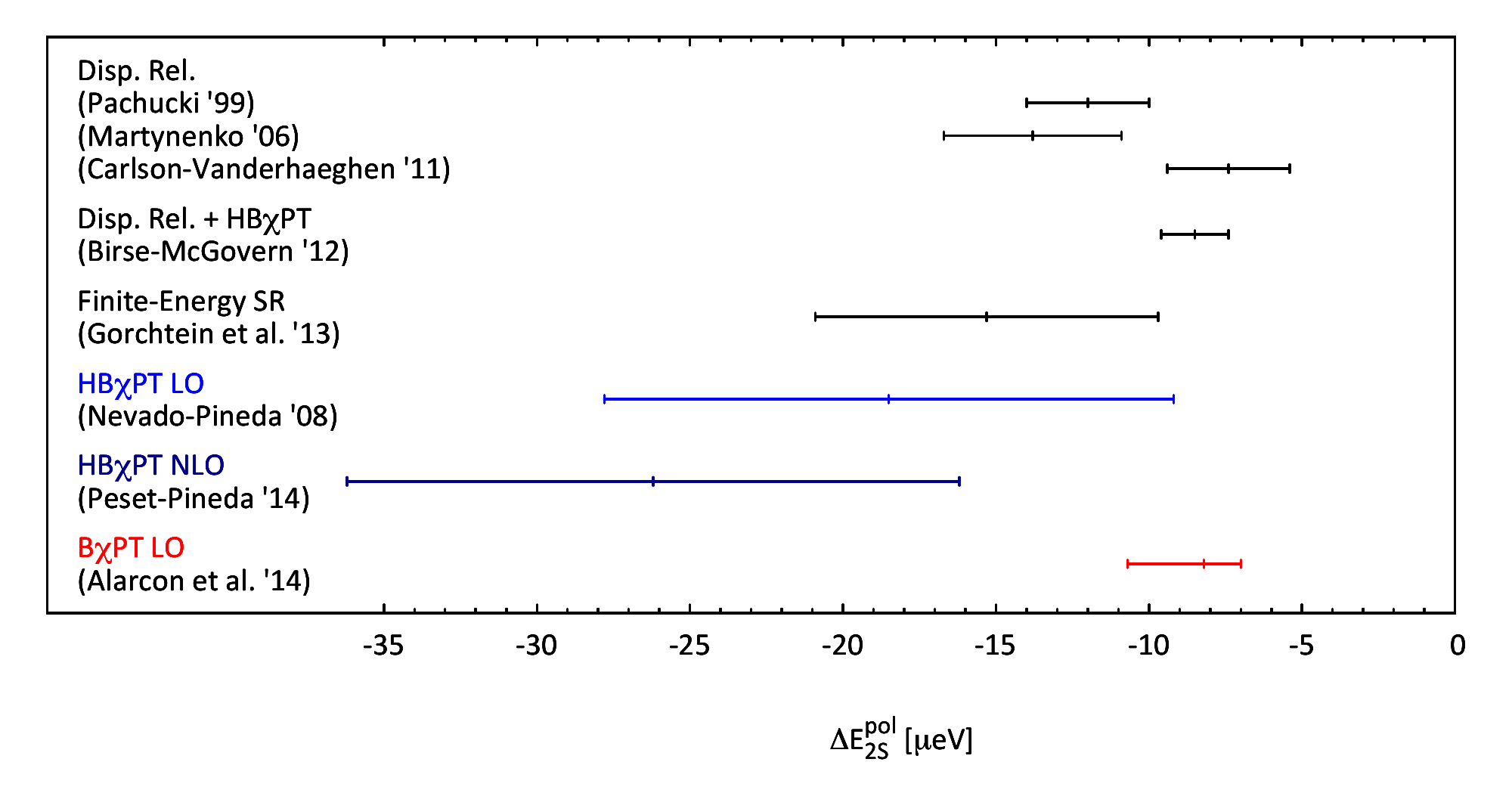}
                     \caption{Proton polarizability effect in the $2S$-level shift of $\mu$H. \label{fig:LSSummary}}
\end{figure}

\begin{figure}[h!]
\begin{minipage}[t]{0.49\textwidth}
   \centering 
      \includegraphics[scale=0.7]{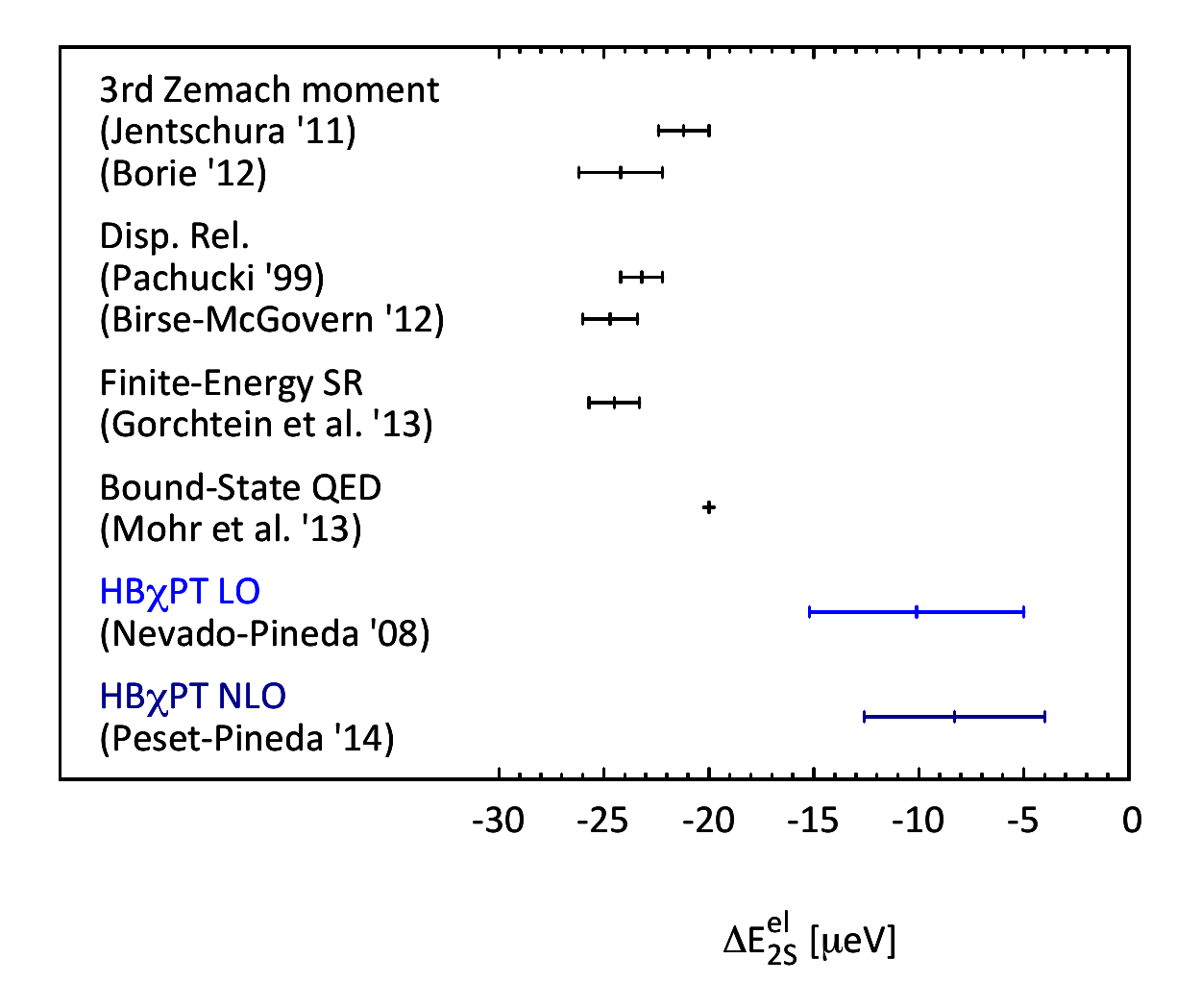}
                   \caption{`Elastic' TPE effect in the $2S$-level shift of $\mu$H. \label{fig:LSSummaryelastic}}
                \end{minipage} 
                           \hfill
                \begin{minipage}[t]{0.49\textwidth}
                \centering
                       \includegraphics[scale=0.7]{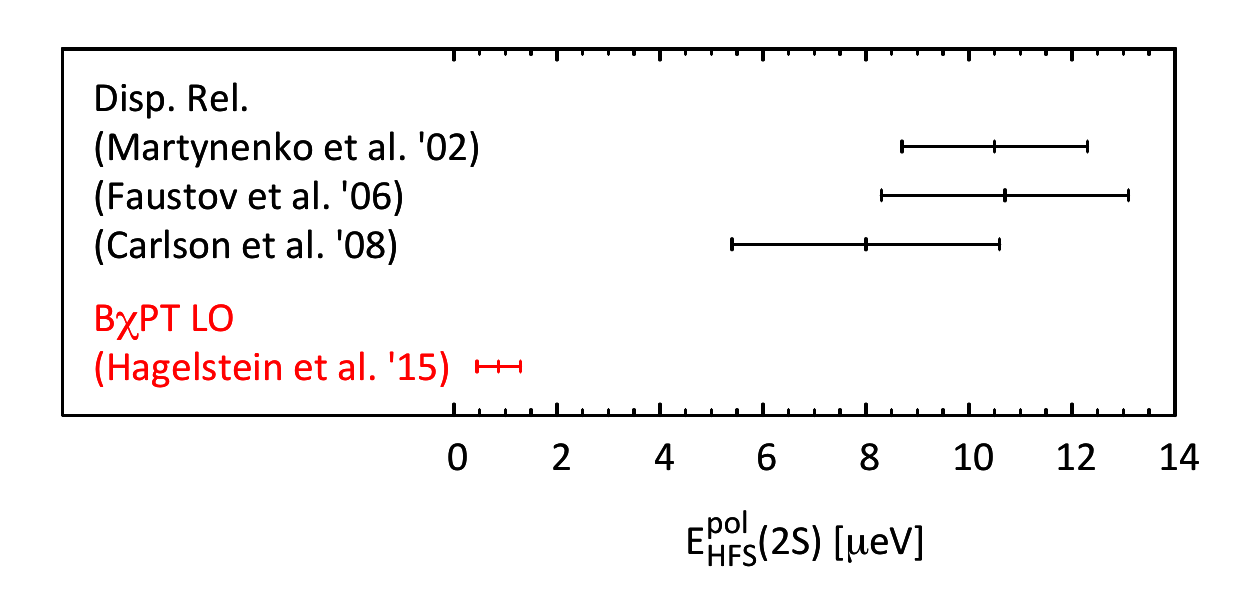}
                      \caption{Proton polarizability effect in the $2S$ HFS of $\mu$H. \label{fig:HFS}}
                       \end{minipage} 
\end{figure}

Figures \ref{fig:LSSummary} and \ref{fig:LSSummaryelastic} display the various results of the  dispersive and $\chi$PT calculations for the `polarizability' and `elastic' contributions of the TPE correction to the Lamb shift in $\mu$H. The corresponding values for the dispersive calculations are listed in Table \ref{Table:Summary1}, whereas the B$\chi$PT and HB$\chi$PT predictions are
summarized in Table~\ref{Table:Summary2}. In \figref{LSSummaryelastic}, we also show the contribution of the Friar radius (3\textsuperscript{rd} Zemach moment) from \citet{Jentschura:2010ej} and \citet{Borie:2012zz}. We also quote the result from the bound-state QED approach of \citet{Mohr:2013axa}, cf.\ \Eqref{elMohr}.

Figure \ref{fig:LSSummary} shows an overall agreement among the dispersive and B$\chi$PT calculations of the proton polarizability correction.  The dispersive results 
involve the modeling of the `subtraction' contribution which 
rely on the empirical value of proton  $\beta_{M1}$, cf.\ Table \ref{Table:Summary1}, and Eqs.~\eref{sub1}, \eref{sub2}. Given this model dependence, the agreement
with the leading-order B$\chi$PT prediction is quite remarkable. 

Figure \ref{fig:LSSummaryelastic} shows the situation for the `elastic' contribution. Since this contribution is completely determined by empirical FFs, the dispersive calculations agree very well. The bound-state QED approach of Ref.~\cite{Mohr:2013axa} gives a compatible result. The HB$\chi$PT results are not in good agreement with 
the empirical evaluations because the FFs are not well reproduced in these calculations.  

Concerning the HFS, the polarizability corrections  
to the $2S$ HFS are summarized in \figref{HFS}. 
The leading-order $\chi$PT prediction is quoted from \Eqref{LOHFS}.
It is rather small compared with the dispersive calculations, which are
taken from the $\Delta_{\mathrm{pol}}$ column of Table \ref{Table:Summary3} 
and converted to $\upmu\mathrm{eV}$ (multiplying the number in the column with 
$E_F(2S)\times 10^{-6}\simeq 0.0228054\, \upmu\mathrm{eV}$). 
Account of the $\De(1232)$-resonance excitation in $\chi$PT is
expected to improve the agreement. The different dispersive
calculations are in agreement with each other and serve as input for the
recent extraction of the proton Zemach radius \cite{Antognini:1900ns}.

Precise calculations of proton structure effects 
will be required to enable a direct measurement
of the $\mu$H ground-state HFS (see, e.g., Sect.\ 6 of Ref.~\cite{Antognini:2015aa}). 
The corresponding transition is much narrower than the observed $2S - 2P$ transitions, and hence is much harder to find. Quantitative theory guidance will be very important here. 
Once found, the HFS transition
will greatly amplify the precision of our understanding of the 
proton structure in general, and of proton polarizabilities in particular. 

Until then, in the words of the title of this paper,  
the {\em nucleon polarizabilities} are taken {\em from
Compton scattering} and serve as input {\em to hydrogen atom}.
We look forward to the times when the reverse is possible.

\section*{Acknowledgements}
\addcontentsline{toc}{section}{Acknowledgements}
We thank A.\ Alexandru, A.\ Antognini, C.\ Collicott, J.~M.~M.\ Hall, H.\ Fonvieille, P.~P.\ Martel, and  V.\ Sokhoyan for providing figures and useful insights into their work. We thank Jeremy Green, Misha Gorchtein,  Vadim Lensky, Anatoly L'vov, Barbara Pasquini, Randolf Pohl and Marc Vanderhaeghen for reading 
the manuscript and their valuable remarks. 
 R.~M. thanks and acknowledges the Mainz A2 Compton collaboration for their critical support. V.~P.\ gratefully acknowledges the inspiring discussions with 
Aldo Antognini, Carl Carlson, Misha Gorchtein, Savely Karshenboim, 
Marc Vanderhaeghen, and Thomas Walcher, which found their vague reflection in the pages of this manuscript.

This work was supported by the Deutsche Forschungsgemeinschaft (DFG) through the Collaborative Research Center SFB 1044 [The Low-Energy Frontier of the Standard Model], and the Graduate School DFG/GRK 1581
[Symmetry Breaking in Fundamental Interactions], and by the U.S. Department of Energy under grant DE-FG02-88ER40415.\\



 \appendix
\begin{small}

\section{Born Contribution in RCS and VVCS amplitudes}
\seclab{appBorn}

For RCS, the Born term is given by the tree-level graphs with the photon
coupling to the nucleon charge $\zZ e$ and the anomalous magnetic moment $\varkappa$.
The invariant amplitudes of the overcomplete tensor decomposition of \Eqref{CovarAmp}
are given by
\cite{Pascalutsa:2003aa}:
\beq 
\label{eq:crossing_symmetric_amp}
\scA_i^{\mathrm{Born}} (s, t)= \left\{ \begin{array}{ll}
\scA_i^s (\nu, t) + \scA_i^s (-\nu', t), & \mbox{for $i=1, 2, 8$} \\
\scA_i^s (\nu, t) - \scA_i^s (-\nu', t), & \mbox{for $i=3, \dots, 7$},
\end{array}
\right.
\eeq
with $\scA_i^s$ being the contribution of the $s$-channel graph:
\beq
\begin{aligned}
\label{eq:cA_i_Born}
\scA^s_1(\nu, t) &= -\frac{1}{2M} \left[\zZ^2+\frac{t}{4 M\nu}(\zZ+\varkappa )^2 +\frac{1}{2} \varkappa ^2
   \left(\frac{\nu}{M}+\frac{t}{4M^2}\right)\right], \\
\scA_2^s(\nu, t) &=\frac{\varkappa}{2 M^2 \nu}  \left[\zZ+\frac{1}{2} \varkappa  \left(1-\frac{\nu}{2M}-\frac{t}{8M^2}\right)\right], \\
\scA_3^s(\nu, t) &= \scA_1^s(\nu, t),\\
\scA_4^s(\nu, t) &= -\frac{1}{4 M^2\nu} \left[(\zZ+\varkappa )^2+\frac{\nu}{2M}\varkappa ^2 \right],  \\
\scA_5^s(\nu, t) &= \frac{(\zZ+\varkappa )^2}{4 M^2 \nu},\\
\scA_6^s(\nu, t) &= -\frac{\zZ (\zZ+\varkappa )}{4 M^2 \nu},\\
\scA_7^s(\nu, t) &= \frac{\varkappa ^2}{16 M^4 \nu},\\
\scA_8^s (\nu, t) &= -\scA_4^s(\nu, t),
\end{aligned}
\eeq
Adding it up, we obtain:
\beq
\begin{aligned}
\label{eq:BornA}
& \scA_1^{\mathrm{Born}} = -\frac{\zZ^2}{M} -
\frac{  (\zZ+\varkappa)^2 \xi_0^2 }{M( \xi^2-\xi_0^2) },\quad 
\scA_2^{\mathrm{Born}} = -\frac{\varkappa^2}{4M^3} + 
\frac{  2M(2\zZ\varkappa+\varkappa)\xi_0 -\varkappa^2\xi_0^2   }{ 4M^3(\xi^2-\xi_0^2 )}
,\\ 
&   \scA_3^{\mathrm{Born}} =-2M\xi\left[\frac{\varkappa^2}{4M^3} + \frac{  (\zZ+\varkappa)^2 \xi_0 }{2M^2( \xi^2-\xi_0^2) } \right] = -2M\xi \scA_8^{\mathrm{Born}},\\  
& \scA_4^{\mathrm{Born}} = -\frac{  (\zZ+\varkappa)^2 \xi }{2M^2( \xi^2-\xi_0^2) }
=-\scA_5^{\mathrm{Born}} ,  \quad \scA_6^{\mathrm{Born}} = -\frac{\zZ (\zZ+\varkappa )\xi}{2 M^2 ( \xi^2-\xi_0^2)}, 
 \quad  \scA_7^{\mathrm{Born}} = \frac{\varkappa ^2\xi}{8 M^4 ( \xi^2-\xi_0^2) }, 
\end{aligned}
\eeq
where $\xi_0 = -q\cdot q'/2M= t/4M$ and $\xi^2-\xi_0^2=\nu\nu'$ for real photons. 

To obtain the Born contribution to the
forward VVCS amplitudes, \Eqref{T12Born}, one may use 
\begin{subequations}
\bea 
&& T_1 =  e^2 \scA_1, \quad  T_2 =  \frac{e^2 Q^2}{\nu^2} \big( \scA_1
+ Q^2 \scA_2 \big), \\
&& S_1 =  \frac{e^2 M}{\nu} \big[ \scA_3
+ Q^2 \big(\scA_5+\scA_6\big) \big] ,\quad 
S_2=e^2 M^2  \big(\scA_5+\scA_6\big),
\eea 
\end{subequations}
with $\xi=\nu$ and $\xi_0=-q^2/2M=Q^2/2M$, and replace 
\beq
\zZ \to F_1(Q^2), \quad \varkappa \to F_2(Q^2).
\eeq 

Often, the $\pi^0$-exchange contribution is considered to be a part of the Born
contribution. The only non-vanishing amplitude for the $\pi^0$-exchange graph is:
\beq 
\scA_8^{(\pi^0)} = -\frac{(2\zZ - 1) g_A}{(2 \pi f_{\pi})^2} \frac{M}{m_{\pi^0}^2-t}\,.
\eeq

\section{Derivation of a Dispersion Relation}
\seclab{disprel}

Consider $f(\nu)$, an analytic function in the entire complex $\nu$ plane except
for the branch cut on the real axis, starting at $\nu_0$ and extending to infinity, 
as shown in \Figref{Cuts} (a). In the case
when there are left- and right-hand branch cuts, located symmetrically around
$\nu=0$, corresponding to  \Figref{Cuts} (b),  we can assume that 
\beq
\eqlab{twocut}
f(\nu) = f_s (\nu) \pm f_s(-\nu),
\eeq 
where $f_s$ has
only the right-hand cut. Hence, for our purpose it is sufficient to only consider the case 
of \Figref{Cuts} (a). 

\begin{figure}[tbh]
\centering
\begin{minipage}{0.49\textwidth}
\centering
       \includegraphics[width=0.7\textwidth]{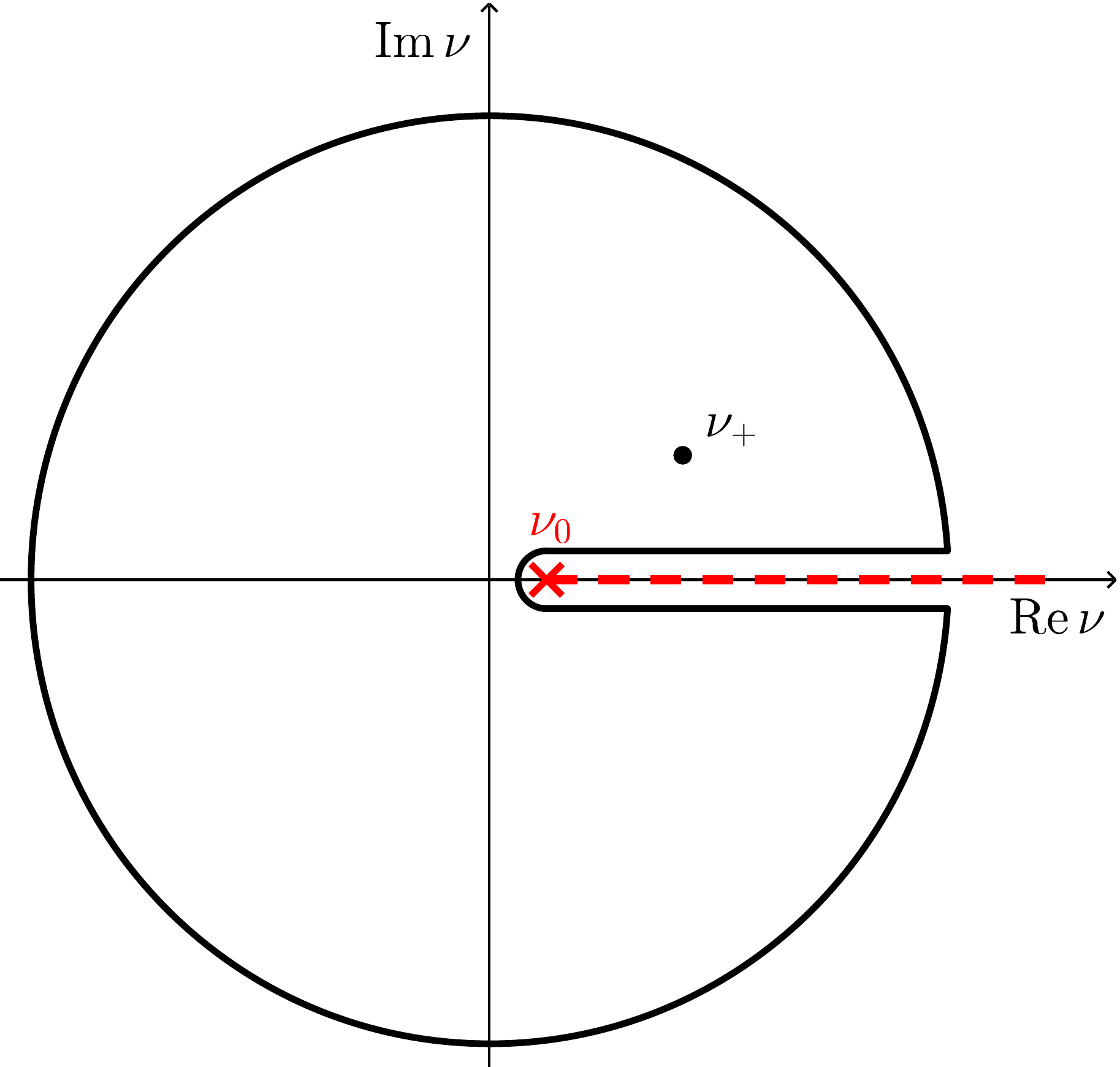}
       
   \centerline{{\small(a)}}
\end{minipage}
\hfill
\begin{minipage}{0.49\textwidth}
\centering
\includegraphics[width=0.7\textwidth]{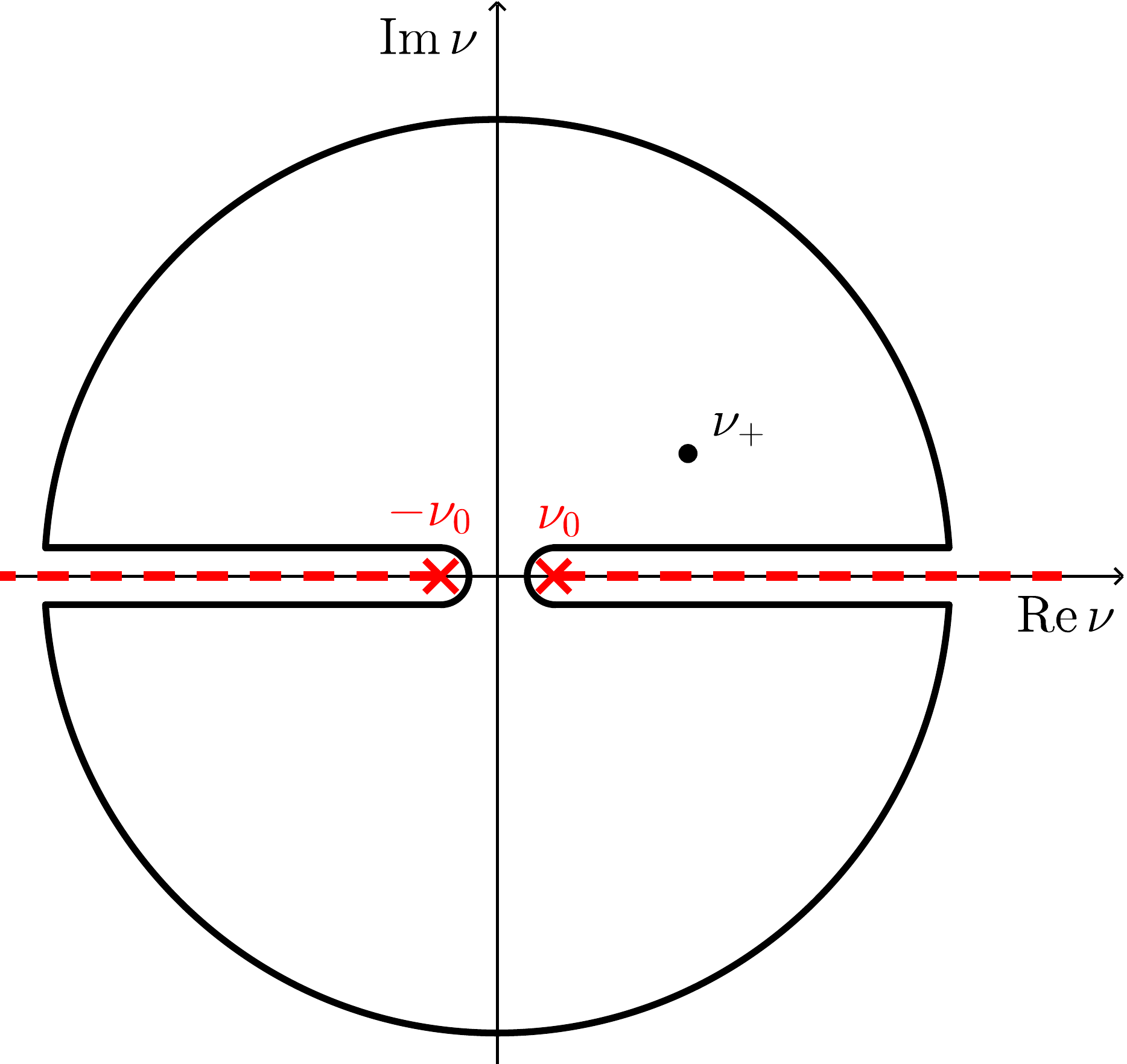}

\centerline{{\small(b)}}
\end{minipage}
\caption{Contours in the complex $\nu$ plane for (a) a single cut along the positive real axis, and (b) branch cuts along the positive and negative real axes.\label{fig:Cuts}}
\end{figure}

The starting point is Cauchy's formula for analytic functions:  
\beq
\eqlab{Ctheorem}
f(\nu_+)  =  \frac{1}{2\pi i }\oint_C \dd \xi \, \frac{f(\xi)}{\xi - \nu_+ },\\
\eeq
where the contour $C$ avoids the branch cut on the real axis, as seen in \Figref{Cuts},
and $\nu_+$ is in the region of analyticity (i.e., inside the contour). We choose
$\nu_+ = \nu + i \de$, where $\nu$ and $\de$ are real. We will take the limit
$\de\to 0$ in the end. 

Next, we assume that $f$ drops to zero for all large $|\nu_+ |$, and it does so sufficiently fast, such that 
the integral over the big semicircle can be neglected.\footnote{When this is not the case,
we could replace $f(\nu_+)$ with $f(\nu_+)/\nu_+^n$, provided
we do not introduce another pole by doing that. So, for small $\nu_+$, $f$ should go as
$\nu_+^n$. If $f$ does not have any obvious zeros, one 
can make a subtraction. The subtracted function, by definition, has a zero at the subtraction point.}
We are then left with only the integrals around the cut:
\beq
f(\nu_+)  =  \lim_{\eps\to 0^+} \frac{1}{2\pi i } \int_{\nu_0}^\infty \dd\xi \, \left[
 \frac{f(\xi+i\eps)}{\xi - \nu_++i\eps}- \frac{f(\xi-i\eps)}{\xi - \nu_+-i\eps}\right] ,
\eeq
where $\eps$ is the gap between the contour and the real axis. The integral over the
small semicircle around $\nu_0$ vanishes in the limit $\eps\to 0$, since $f$ has no
poles at $\nu_0$. 

Employing the Schwarz
reflection principle for analytic functions:
\beq
f^\ast (s) = f(s^\ast),
\eeq
and substituting $\nu_+ = \nu + i \de$,  we have
\beq
f(\nu_+) = \lim_{\eps \to 0^+} \frac{1}{2\pi i } \int_{\nu_0}^\infty \dd\xi \, \left[
 \frac{f(\xi+i\eps)}{\xi - \nu-i(\de-\eps)}- \frac{f^\ast(\xi+i\eps)}{\xi - \nu -i(\de+\eps)}\right] .
 \eeq
With $\eps<\de$, the limit $\eps \to 0$ can now be taken explicitly, since $\de$ takes over the signs of the
imaginary parts, hence
 \bea
f(\nu_+) &=& \frac{1}{2\pi i } \int_{\nu_0}^\infty \dd\xi \, 
 \frac{f(\xi)-f^\ast(\xi)}{\xi - \nu-i\de} =  \frac{1}{\pi } \int_{\nu_0}^\infty \dd\xi  \, 
  \frac{\im f(\xi)}{\xi - \nu-i\de}, \nn\\
 &=& \frac{1}{\pi } \int_{\nu_0}^\infty \dd\xi \, \left[
\frac{ \xi-\nu }{(\xi - \nu)^2+\de^2}+
 \frac{i\delta }{(\xi - \nu)^2+\de^2} \right] \im f(\xi). 
\eea
Taking the limit to the real axis, $\de\to 0$, we obtain the sought DR:
\beq
f(\nu) = \lim_{\de \to 0^+} f(\nu_+) = 
 \frac{1}{\pi } \int_{\nu_0}^\infty \dd\xi \, 
\frac{\im f(\xi) }{\xi -\nu -i0^+} = \frac{1}{\pi } \, \fint_{\nu_0}^\infty \dd\xi \, 
\frac{\im f(\xi) }{\xi - \nu} + i \im f(\nu).
\eeq

Coming back to the case when $f$ is given as in \Eqref{twocut}, with $f_s$ satisfying the above DR,
we then obviously have
\beq
f(\nu) = \frac{1}{\pi } \int_{\nu_0}^\infty \dd\xi \, 
\left[\frac{1}{\xi -\nu -i0^+} \pm \frac{1}{\xi +\nu -i0^+}\right] \im f_s(\xi) 
= \frac{2}{\pi } \int_{\nu_0}^\infty \dd\xi \, 
\left\{ {\xi \atop \nu}\right\} \frac{\im f(\xi)}{\xi^2 -\nu^2 -i0^+} ,
\eeq
where in the last step we have used $\im f(\xi)=\im f_s(\xi)$, which is 
true if $\nu_0\geq 0$. The situation with negative $\nu_0$ is in principle treatable but is beyond the present scope. 

\section{Collection/Index of Formulae }
\seclab{appindex}

\begin{itemize}[$\square$]
\item Kinematical invariants and relations for real Compton scattering (RCS)
\begin{equation}
\eta=\frac{M^4-su}{M^2}, \; \nu = \frac{s-M^2}{2M}, \; \nu' = \frac{M^2-u}{2M}.
\tag{\ref{eq:invdefs1}} 
\end{equation}
\begin{equation}
 \w = \frac{s-M^2}{2\sqrt{s}}=\frac{M\nu}{\sqrt{s}} , \;
\w_B = \frac{s-u}{2\sqrt{4M^2 -  t}},\; \eps_B = \frac12 \sqrt{4M^2 -  t}.
\eqlab{kin1}
 \end{equation}
 \begin{equation}
\xi=\frac{s-u}{4M}= \frac{\nu+\nu'}{2} = \frac{\w_B \eps_B}{M}, \; 4\xi^2-
\frac{t^2}{4M^2}= \eta -  t = 4\nu\nu' .
\eqlab{kin2}
\end{equation}
 \beq
 \eqlab{kin3}
\begin{aligned}
& t = 2M(\nu'-\nu) = -2\nu\nu'(1-\cos\vartheta) 
= - 2\w^2(1-\cos\th) = - 2 \w_B^2(1-\cos\th_B), \\
& \dd t = 2M \dd \nu' = (1/\pi)\nu^{\prime\, 2} \dd\varOmega_L
= (1/\pi)\w^{2} \dd\varOmega_{cm},\\
& \cos^2 (\vartheta/2)  =\eta/(4\nu\nu'), \;
\sin^2 (\vartheta/2) = -t/(4\nu\nu') , \; \sin \vartheta = \sqrt{-t\eta}/(2\nu\nu'), \\
& \cos^2 (\th/2)  = \eta/(2\nu)^2, \;
\sin^2 (\th/2) = -t/(2\w)^2 , \; \sin \th = \sqrt{-t\eta s }/(2M\nu^2),\\
& \cos^2 (\th_B/2)  = \eta/(2\xi)^2, \;
\sin^2 (\th_B/2) = -t/(2\w_B)^2 , \; \sin \th_B = \sqrt{-t\eta  }/(2\xi\w_B),\\
& \tan (\vartheta/2)  = ( \nicefrac{-t}{\eta})^{\nh} = \nicefrac{M}{\sqrt{s}} 
\,\tan (\th/2) = \nicefrac{M}{\eps_B} 
\tan (\th_B/2).
\end{aligned}
\eeq
\item Relations for the forward doubly-virtual Compton scattering (VVCS)
\beq 
\nu  = \w_B = \xi = \half\eta^{1/2}= \frac{s- u}{4M} = \frac{s-M^2+Q^2}{2M} = \frac{Q^2}{2Mx} = \frac{2M\tau}{x}, \quad \nu_{\mathrm{el}} = \frac{Q^2}{2M} = 2M\tau
\eeq
\beq
\int_{\frac{Q^2}{2M} }^\infty \frac{\dd\nu}{\nu^n} f(\nu,Q^2)  = \left(\frac{2M}{Q^2}\right)^{n-1} \! \int_0^1 \dd x\, 
x^{n-2} f(x,Q^2) \tag{\ref{eq:xnuConv}} 
\eeq
\item Elastic structure functions
\begin{align*} 
f_1^{\mathrm{el}}(x,Q^2) & = \frac{1}{2}\,  G^2_M(Q^2)\,  \delta( 1 - x ),\tag{\ref{eq:f1elastic}} \\
f_2^{\mathrm{el}}(x,Q^2) & = \frac{1}{1 + \tau} \, \big[G^2_E(Q^2) + \tau G^2_M(Q^2) \big]\, \delta( 1 - x), \tag{\ref{eq:f2elastic}}\\
g_1^{\mathrm{el}}(x,Q^2) & =  \frac{1}{2}\,  F_1(Q^2) \,G_M(Q^2) \, \delta(1 - x), \tag{\ref{eq:g1elastic}}\\
g_2^{\mathrm{el}}(x,Q^2) & = -  \frac{\tau}{2}\, F_2(Q^2)\, G_M(Q^2)\,  \delta(1 - x)\tag{\ref{eq:g2elastic}}.
\end{align*}
\item Nucleon-pole and Born contributions to the forward VVCS amplitudes
\begin{align}
T_1^{\mathrm{pole}}(\nu, Q^2) &=\frac{4\pi \alpha}{M}\frac{\nu_{\mathrm{el}}^2\,  G_M^2(Q^2)}{\nu_{\mathrm{el}}^2-\nu^2
-i 0^+ }= T_1^{\mathrm{Born}}(\nu, Q^2) + \frac{4\pi\al}{M}F_1^2(Q^2), 
\tag{\ref{eq:T1pole}}\\
T_2^{\mathrm{pole}}(\nu, Q^2)&=\frac{8\pi \alpha\,  \nu_{\mathrm{el}} }{\nu_{\mathrm{el}}^2-\nu^2
-i 0^+ }\frac{G^2_E(Q^2) + \tau G^2_M(Q^2)}{1 + \tau} = T_2^{\mathrm{Born}}(\nu, Q^2) ,
\tag{\ref{eq:T2pole}}
\\
S_1^{\mathrm{pole}}(\nu, Q^2) &= \frac{4 \pi \alpha \, \nu_{\mathrm{el}}}{\nu_{\mathrm{el}}^2-\nu^2
-i 0^+ } F_1(Q^2)\, G_M(Q^2)= S_1^{\mathrm{Born}}(\nu, Q^2) + \frac{2\pi\al}{M}F_2^2(Q^2),\tag{\ref{eq:S1pole}}\\
S_2^{\mathrm{pole}}(\nu, Q^2)&=-\frac{2 \pi \alpha \nu}{\nu_{\mathrm{el}}^2-\nu^2
-i 0^+ } F_2(Q^2)\, G_M(Q^2) = 
S_2^{\mathrm{Born}}(\nu, Q^2) \,. \eqlab{S2subpole}
\end{align}
\item Sum rules
\begin{align*}
&\text{Baldin:}\quad
\alpha_{E1} + \beta_{M1} = \frac{1}{2\pi^2} \int_{\nu_0}^\infty \frac{\dd\nu}{\nu^2}\si(\nu)
\tag{\ref{eq:BaldinSR}} \\
&\text{$4^\mathrm{th}$-order:}\quad
\alpha_{E1\nu} + \beta_{M1\nu} + \nicefrac{1}{12}\,(\alpha_{E2} + \beta_{M2}) = \frac{1}{2\pi^2} \int_{\nu_0}^\infty \frac{\dd\nu}{\nu^4}\si(\nu)
\tag{\ref{4thSR}} \\
&\text{Gerasimov--Drell--Hearn (GDH):}\quad
-\frac{ \al }{M^2}\varkappa^2=\frac{1}{2\pi^2}\int _{\nu_0}^\infty \frac{\dd\nu}{\nu }\left[\si_{1/2}(\nu)-\si_{3/2}(\nu)\right]
\tag{\ref{eq:GDH}}\\
&\text{Gell-Mann--Goldberger--Thirring (GTT):}\quad
\gamma_0=\frac{1}{4\pi^2} \int_{\nu_0}^\infty \frac{\dd\nu}{\nu^3}\left[\si_{1/2}(\nu)-\si_{3/2}(\nu)\right]
\tag{\ref{eq:FSP}}\\
&\text{$5^\mathrm{th}$-order:}\quad
\bar\gamma_0=\frac{1}{4\pi^2} \int_{\nu_0}^\infty \frac{\dd\nu}{\nu^5}\left[\si_{1/2}(\nu)-\si_{3/2}(\nu)\right]
\tag{\ref{eq:HOFSP}}\\
&\text{Burkhardt--Cottingham (BC):}\quad
0 = \int_{0}^{1}\dd x\,  g_{2}(x,\,Q^2) =\frac{\tau}{2}\left[4\,I_2(Q^2)-F_2(Q^2)G_M(Q^2)\right] \tag{\ref{eq:BCsumrule}}\\
&\text{spin-GP sum:}\quad \frac{\dd}{\dd \bq^2 } \big[
 P^{ (M1, M1)1}(\bq^2) +  P^{ (L1, L1)1}(\bq^2 )  \big]_{\bq^2 =0}
 = \frac{\gamma_{E1M2}}{3\al M} - \frac{2}{3M^3} 
 \frac{\dd}{\dd Q^2} \big[ \quarter F_2^2(Q^2) + I_1(Q^2) \big]_{Q^2=0}
 \tag{\ref{S1sr3}}\\
&\text{spin-GP difference:}\quad \frac{\dd}{\dd \bq^2 } \big[
 P^{ (M1, M1)1}(\bq^2) -  P^{ (L1, L1)1}(\bq^2 )  \big]_{\bq^2 =0}
 = \frac{\gamma_{E1E1} + \delta_{LT}}{3\al M} \tag{\ref{S2sr3}}
\end{align*}
\item Dispersion relations for the Sachs FFs [for Dirac and Pauli FFs, see \Eqref{F12DR}]
\begin{align}
\barr G_E (Q^2)\\ 
G_M (Q^2) \earr  = \barr 1\\
1+\kappa \earr - \frac{Q^2}{\pi} \int_{t_0}^\infty \frac{\dd t}{ t (t+Q^2) }
\im \barr  G_E(t) \\ 
G_M (t) \earr. 
\eqlab{GEMDR}
\end{align}
\item Moments of the (spherically-symmetric) charge distribution, for any $N$:
\beq
\langle r^N\rangle_E  \equiv4\pi \int_0^\infty \dd r\, r^{N+2}
\rho_E(r)=
\frac{\Gamma(N+2)}{\pi}\int_{t_0}^\infty\! \dd t \, \frac{\im  G_E (t)}{t^{N/2+1 }  },
\tag{\ref{eq:rmsdef}}
\eeq
with the normalization $\langle r^0\rangle_E=1$. Equivalently, for integer $N$:
\bea
\eqlab{Gmoments}
\langle r^{2N}\rangle_E &=&(-1)^{N} \,\frac{(2N+1)!}{N!}\, G^{(N)}_E(0),\\
\eqlab{Gmomentsodd}
\langle r^{2N-1}\rangle_E &=&(-1)^{N}\,(2N)!\,\frac{2}{\pi}\int _0^\infty \frac{\dd Q}{Q^{2N}} \Big[ G_E(Q^2) -\sum_{k=0}^{N-1} \frac{Q^{2k}}{k!} G_E^{(k)} (0) \Big], \nn\\
&=& (-1)^{N}\,(2N)!\,\frac{2}{\pi}\int _0^\infty \frac{\dd Q}{Q^{2N}} \Big[ G_E(Q^2) -\sum_{k=0}^{N-1} \frac{(-Q^2)^k}{(2k+1)!} \langle r^{2k}\rangle_E\Big].
\eea
The moments of the magnetization distribution,  $\langle r^N\rangle_M$, are defined similarly, replacing $G_E$ with $G_M/(1+\kappa)$.
\item Friar radius (or, the 3\textsuperscript{rd} Zemach moment of the charge  distribution)
\begin{align}
R_{\mathrm{F} } =\sqrt[3]{\langle r^3\rangle_{E(2)} }, \quad \langle r^3\rangle_{E(2)} &\equiv  \frac{48}{\pi} \int_0^\infty \!\frac{\dd Q}{Q^4}\,
\Big[ G_E^2(Q^2) -1 +\third R^2_E\, Q^2\Big], \tag{\ref{eq:3rdZmoment}}\\
&=
2\langle r^3\rangle_{E} +\frac{24}{\pi^2}\int_{t_0}^\infty\! \dd t \int_{t_0}^\infty\! \dd t' \, \frac{\im  G_E(t)\, \im  G_E(t')}{(t't)^{3/2} (\sqrt{t'}+\sqrt{t})  }. \eqlab{imFriar}
\end{align}
\item Zemach radius
\begin{align}
R_{\mathrm{Z}}&\equiv  -\frac{4}{\pi}\int_0^\infty \frac{\dd Q}{Q^2}\left[\frac{G_E(Q^2)G_M(Q^2)}{1+\kappa}-1\right], \tag{\ref{eq:RZ}}\\
&= \langle r\rangle_{E} + \langle r\rangle_{M} - \frac{2}{\pi^2}\int_{t_0}^\infty\! \frac{\dd t}{t} \,\frac{\im  G_M(t')}{1+\kappa} \int_{t_0}^\infty\! \frac{\dd t'}{t'} \, \frac{\im  G_E(t')}{\sqrt{t'}+\sqrt{t}}.
\end{align}
\item Finite-size effects
\begin{enumerate}[a)]
\item $2P-2S$ Lamb shift:
  \begin{align*}
E_\mathrm{LS}
&= -\frac{Z\al}{\pi} \int_0^\infty \!\dd Q \, w_{2P-2S}(Q)\,G_E^2 (Q^2), \tag{\ref{eq:LSexactGE2}}\\
&=-\frac{Z\al}{12a^3} \left[ R_E^2 - (2a)^{-1} R^3_{\mathrm{F} }\right] + O(\al^6).\tag{\ref{eq:LambShift}}
  \end{align*}
\item $nS$ hyperfine splitting (HFS): 
\begin{align*}
E_{\mathrm{HFS}}(nS)
 &= \frac{4Z\al}{3\pi mM} \int_0^\infty\! \dd Q\, Q^2\, 
 w_{nS}(Q)\, G_E(Q^2)G_M(Q^2)\tag{\ref{eq:NRelasticHFS}},\\
&=E_F(nS)\left[1-2a^{-1} R_{\mathrm{Z}}\right]+ O(\al^6), \tag{\ref{eq:HFS}}
\end{align*}
with the convolution of momentum-space wave functions:
\beq
\tag{\ref{eq:WF}}
w_{1S}(Q) = \frac{16}{\big(4+(a Q)^2\big)^2},\; w_{2S}(Q) = \frac{\big(1-(aQ)^2\big)\big(1-2(aQ)^2\big)}{\big(1+(aQ)^2\big)^4},\; w_{2P-2S}(Q) =
\frac{2(aQ)^2 \big(1-(aQ)^2\big)}{\big(1+(aQ)^2\big)^4},
\eeq
and the Fermi energy:
\beq
E_F(nS)=\frac{8Z \al}{3a^3}\frac{1+\kappa}{mM} \frac{1}{n^3}.\tag{\ref{eq:EFermi}}
\eeq
\end{enumerate}
\end{itemize}

\end{small}



\small 
\bibliographystyle{model1a-num-names}

\bibliography{lowQ}







\end{document}